\documentclass[reprint,twocolumn,superscriptaddress,amsmath,amssymb,aps,prx,floatfix]{revtex4-2}

\usepackage{float}
\usepackage{amsmath}
\usepackage{amssymb}
\usepackage{amsthm}

\usepackage{amsfonts} \usepackage{nicematrix}
\usepackage{longtable}

\usepackage{tabularx}
\newcolumntype{C}{>{\centering\arraybackslash}X}
\newcolumntype{R}{>{\raggedleft\arraybackslash}X}
\newcolumntype{L}{>{\raggedright\arraybackslash}X}

\usepackage{physics}
\usepackage{bbold}
\usepackage{chemformula}
\usepackage{graphicx}
\usepackage{xcolor} 
\usepackage{hyperref}
\hypersetup{
    colorlinks = true,
    linkcolor  = blue,
    citecolor  = red,
    urlcolor   = purple,
    }

\usepackage{cleveref} \crefname{appendix}{Appendix}{Appendices}
\crefname{equation}{Eq.}{Eqs.}
\crefname{figure}{Fig.}{Figs.}
\crefname{table}{Table}{Tables}
\crefname{section}{Section}{Sections}
\crefname{paragraph}{paragraph}{paragraphs}
\crefname{enumi}{case}{cases}
\creflabelformat{appendix}{[#2#1#3]}

\usepackage{silence}
\newcommand{\nc}{\newcommand}

\nc{\webirvsp}{\href{https://github.com/zjwang11/irvsp}{\texttt{IRVSP} }}
\nc{\webirtb}{\href{https://github.com/zjwang11/irvsp}{\texttt{IR2TB} }}
\nc{\webBRdecomp}{\href{http://tm.iphy.ac.cn/UnconvMat.html}{\texttt{BRdecomp} }}
\nc{\webposabr}{\href{https://github.com/zjwang11/UnconvMat/blob/master/src_pos2aBR.tar.gz}{\texttt{POS2ABR} }}

\nc{\red}[1]{{\color{red}{#1}}}
\nc{\green}[1]{{\color{teal}{#1}}}
\nc{\blue}[1]{{\color{blue}{#1}}}

\nc{\YJ}[1]{\textcolor{blue}{YJ: #1}}
\nc{\djz}[1]{\red{\;\text{JD:}\;#1}}

\nc{\eq}{=&\;}
\nc{\eqv}{\equiv&\;}

\nc{\cre}[2]{\hat{#1}^{\dagger}_{#2}}
\nc{\des}[2]{\hat{#1}_{#2}}

\def\qq{\mathbf{q}}
\def\kk{\mathbf{k}}

\def\rr{\mathbf{r}}

\begin{document}

\title{Theory of Superconductivity in \ch{LaRu3Si2} 
\\
and Predictions of New Kagome Flat Band Superconductors}

\author{Junze Deng}
\thanks{These authors contributed equally to this work.}
\affiliation{Department of Applied Physics, Aalto University School of Science, FI-00076 Aalto, Finland}
\affiliation{Beijing National Laboratory for Condensed Matter Physics, and Institute of Physics, Chinese Academy of Sciences, Beijing 100190, China}
\affiliation{University of Chinese Academy of Sciences, Beijing 100049, China}

\author{Yi Jiang}
\thanks{These authors contributed equally to this work.}
\affiliation{Donostia International Physics Center (DIPC), Paseo Manuel de Lardiz\'{a}bal. 20018, San Sebasti\'{a}n, Spain}

\author{Tiago F. T. Cerqueira}
\thanks{These authors contributed equally to this work.}
\affiliation{CFisUC, Department of Physics, University of Coimbra, Rua Larga, 3004-516 Coimbra, Portugal}

\author{Haoyu Hu}
\affiliation{Donostia International Physics Center (DIPC), Paseo Manuel de Lardiz\'{a}bal. 20018, San Sebasti\'{a}n, Spain}

\author{Eeli O. Lamponen}
\affiliation{Department of Applied Physics, Aalto University School of Science, FI-00076 Aalto, Finland}

\author{Dumitru C\u{a}lug\u{a}ru}
\affiliation{Department of Physics, Princeton University, Princeton, NJ 08544, USA}

\author{Hanqi Pi}
\affiliation{Donostia International Physics Center (DIPC), Paseo Manuel de Lardiz\'{a}bal. 20018, San Sebasti\'{a}n, Spain}

\author{Zhijun Wang}
\affiliation{Beijing National Laboratory for Condensed Matter Physics, and Institute of Physics, Chinese Academy of Sciences, Beijing 100190, China}
\affiliation{University of Chinese Academy of Sciences, Beijing 100049, China}

\author{Maia G. Vergniory}
\affiliation{Donostia International Physics Center (DIPC), Paseo Manuel de Lardiz\'{a}bal. 20018, San Sebasti\'{a}n, Spain}
\affiliation{D\'{e}partement de Physique et Institut Quantique, Universit\'{e} de Sherbrooke, Sherbrooke, J1K 2R1 Qu\'{e}bec, Canada}

\author{Emilia Morosan}
\affiliation{Department of Physics and Astronomy, Rice University, Houston, TX 77005, USA}
\affiliation{Department of Physics and Astronomy, Rice Center for Quantum Materials (RCQM), Rice University, Houston, Texas 77005, USA}
\affiliation{Smalley-Curl Institute, Rice University, Houston, TX 77005, USA}

\author{Titus Neupert}
\affiliation{Department of Physics, University of Zurich, Winterthurerstrasse 190, 8057 Zurich, Switzerland}

\author{S. Blanco-Canosa}
\affiliation{Donostia International Physics Center (DIPC), Paseo Manuel de Lardiz\'{a}bal. 20018, San Sebasti\'{a}n, Spain}
\affiliation{IKERBASQUE, Basque Foundation for Science, 48013 Bilbao, Spain}

\author{Claudia Felser}
\affiliation{Max Planck Institute for Chemical Physics of Solids, 01187 Dresden, Germany}

\author{Kristjan Haule}
\affiliation{Center for Materials Theory, Department of Physics and Astronomy, Rutgers University, Piscataway, NJ 08854, USA}

\author{Miguel A. L. Marques}
\affiliation{Research Center Future Energy Materials and Systems of the University Alliance Ruhr and Interdisciplinary Centre for Advanced Materials Simulation, Ruhr University Bochum, Universit\"{a}tsstra{\ss}e 150, D-44801 Bochum, Germany}

\author{P\"{a}ivi T\"{o}rm\"{a}}
\email{paivi.torma@aalto.fi}
\affiliation{Department of Applied Physics, Aalto University School of Science, FI-00076 Aalto, Finland} 

\author{B. Andrei Bernevig}
\email{bernevig@princeton.edu}
\affiliation{Donostia International Physics Center (DIPC), Paseo Manuel de Lardiz\'{a}bal. 20018, San Sebasti\'{a}n, Spain}
\affiliation{Department of Physics, Princeton University, Princeton, NJ 08544, USA}
\affiliation{IKERBASQUE, Basque Foundation for Science, 48013 Bilbao, Spain}

\begin{abstract}
Kagome materials exhibit a large range of intriguing physical properties due to the emergence of exotic electronic phases. In this study, we present a comprehensive investigation of the flat-band kagome superconductor \ch{LaRu3Si2}, which has recently been reported to host charge density wave (CDW) order above room temperature ($T_{\mathrm{CDW}} \simeq 400 \,\mathrm{K}$). The stable crystal structure above the CDW transition is identified via soft phonon condensation and confirmed to be harmonically stable through \textit{ab initio} calculations, consistent with recent X-ray diffraction refinements.
The electron-phonon coupling (EPC) in \ch{LaRu3Si2} is found to be mode-selective, primarily driven by strong interactions between Ru-$B_{3u}$ phonons (local $x$-direction, pointing toward the hexagon center) and Ru-$A_g$ electrons (local $d_{x^2-y^2}$ orbital) within the kagome lattice. Using a spring-ball model, we identify this mode-selective EPC as a universal feature of kagome materials. 
Employing the newly developed Gaussian approximation of the hopping parameters, 
we derive an analytical expression for the EPC and demonstrate that superconductivity in \ch{LaRu3Si2} is mostly driven by the coupling between the kagome $B_{3u}$ phonons and the $A_g$ electrons. 
The impact of doping is also investigated, revealing that light hole doping (approximately one hole per unit cell) significantly enhances the superconducting critical temperature $T_c$ by 50\%, whereas heavy doping induces structural instability and ferromagnetism.
Furthermore, high-throughput screening identifies 3063 stable 1:3:2 kagome materials, of which 428 are predicted to exhibit superconductivity with $T_c > 1\,\mathrm{K}$, and the highest $T_c$ reaching 15 K. These findings establish \ch{LaRu3Si2} and related materials as promising platforms for exploring the interplay among kagome flat bands, EPC, and superconductivity. Additionally, they may offer valuable insights into potential limitations on the $T_c$ of flat-band superconductivity in real materials.
\end{abstract}
\maketitle

\textit{Introduction}.~Kagome materials, characterized by their lattice geometry and unique electronic structure, provide a natural platform for exploring emergent quantum phenomena. 
Their distinctive lattice geometry gives rise to flat bands (FBs), Dirac cones, and van Hove singularities (vHSs) at certain electron fillings, enabling both nontrivial band topology~\cite{Ma2020Spin-Orbit-InducedA, Kang2020DiracA, Ortiz2019New-kagomeA, Ortiz2020CsV3Sb5A, Ortiz2021SuperconductivityA} and interaction-driven phenomena such as charge density waves (CDWs)~\cite{Ortiz2019New-kagomeA, Ortiz2020CsV3Sb5A, Ortiz2021SuperconductivityA, Ortiz2021FermiA, Liang2021Three-DimensionalA, Jiang2021UnconventionalA, Luo2022ElectronicA, Kang2022TwofoldA, Guguchia2023TunableA, Scammell2023ChiralA, Han2023Orbital-Hybridization-DrivenA, Subires2023Order-disorderA, Hu2022CoexistenceA, Zhao2021CascadeA}, superconductivity (SC)~\cite{Ortiz2019New-kagomeA, Ortiz2020CsV3Sb5A, Ortiz2021SuperconductivityA, Yin2021SuperconductivityA, Mu2021S-WaveA, Mielke2022Time-reversalA, Zhong2023NodelessA, Duan2021NodelessA},
and quantum spin-liquid states~\cite{Shores2005A-StructurallyA, Balents2010SpinA, Mendels2010QuantumA, Han2012FractionalizedA, Depenbrock2012NatureA, Liao2017GaplessA, Khuntia2020GaplessA}. 

Among recently reported kagome lattice materials, the 1:3:5, 1:1, and 1:6:6 families have garnered significant attention.
The vanadium-based kagome metal \ch{\textit{A}V3Sb5} ($A = \text{K, Rb, Cs}$), a prominent representative of the 1:3:5 family, is one of the most extensively studied materials~\cite{Ortiz2019New-kagomeA, Ortiz2020CsV3Sb5A, Ortiz2021SuperconductivityA, Ortiz2021FermiA, Liang2021Three-DimensionalA, Jiang2021UnconventionalA, Luo2022ElectronicA, Kang2022TwofoldA, Scammell2023ChiralA, Han2023Orbital-Hybridization-DrivenA, Subires2023Order-disorderA, Yin2021SuperconductivityA, Mu2021S-WaveA, Mielke2022Time-reversalA, Zhong2023NodelessA, Denner2021AnalysisA, Wu2021NatureA, Deng2023Two-elementaryA, Gutierrez-Amigo2024PhononA, Li2024IntertwinedA, Li2023UnidirectionalA, Guo2022SwitchableA, Jiang2022KagomeA, Hu2022CoexistenceA, Zhao2021CascadeA, Hu2022CoexistenceA, Zhao2021CascadeA, Christensen2021TheoryA, Romer2022SuperconductivityA, Ritz2023SuperconductivityA, Guguchia2023TunableA, Duan2021NodelessA, Chen2024CascadeA}. \ch{\textit{A}V3Sb5} has been reported to exhibit $\mathbb{Z}_2$ nontrivial band topology~\cite{Ortiz2019New-kagomeA, Ortiz2020CsV3Sb5A, Ortiz2021SuperconductivityA}, a CDW transition in the temperature range of $T_{\mathrm{CDW}} \simeq 80 \sim 104 \,\mathrm{K}$~\cite{Ortiz2021FermiA, Liang2021Three-DimensionalA, Jiang2021UnconventionalA, Luo2022ElectronicA, Kang2022TwofoldA, Guguchia2023TunableA, Scammell2023ChiralA, Han2023Orbital-Hybridization-DrivenA, Subires2023Order-disorderA, Hu2022CoexistenceA, Zhao2021CascadeA}, and SC transition temperature $T_c \simeq 0.9 \sim 2.5 \,\mathrm{K}$~\cite{Yin2021SuperconductivityA, Mu2021S-WaveA, Mielke2022Time-reversalA, Zhong2023NodelessA, Duan2021NodelessA}. Despite significant experimental progress, the mechanisms underlying these phenomena remain under active debate, with numerous theoretical models proposed~\cite{Denner2021AnalysisA, Wu2021NatureA, Deng2023Two-elementaryA, Gutierrez-Amigo2024PhononA, Li2024IntertwinedA, Christensen2021TheoryA, Romer2022SuperconductivityA, Ritz2023SuperconductivityA}. 
Element substitution within this family has been successfully implemented in compounds such as the titanium-based \ch{CsTi3Bi5}~\cite{Wang2023FlatA, Yang2023ObservationA, Zhou2023PhysicalA, Li2023ElectronicA, Yang2024SuperconductivityA} and the chromium-based \ch{CsCr3Sb5}~\cite{Liu2024SuperconductivityB, Liu2024ChargeA, Li2024CorrelatedA, Peng2024FlatA, Guo2024UbiquitousA}. Notably, \ch{\textit{A}V3Sb5} and \ch{CsTi3Bi5} exhibit SC~\cite{Yin2021SuperconductivityA, Mu2021S-WaveA, Mielke2022Time-reversalA, Zhong2023NodelessA, Yang2023ObservationA, Zhou2023PhysicalA, Li2023ElectronicA, Yang2024SuperconductivityA}, but their Fermi energy ($E_F$) is positioned near the vHS rather than the FB~\cite{Kang2022TwofoldA, Guguchia2023TunableA, Wang2023FlatA, Yang2023ObservationA}. In contrast, \ch{CsCr3Sb5} exhibits FBs near $E_F$~\cite{Guo2024UbiquitousA}, yet SC with a maximum critical temperature ($T^{\mathrm{max}}_c \simeq 6.4\,\mathrm{K}$) is observed only under applied pressures of $3.65$ to $8.0\,\mathrm{GPa}$~\cite{Liu2024SuperconductivityB}. However, the electronic structure of \ch{CsCr3Sb5} remains largely unexplored.

Unlike the 1:3:5 family which includes members reported to exhibit SC, the 1:1 and 1:6:6 kagome families are primarily reported to exhibit CDW and magnetic transitions~\cite{Yin2020Quantum-limitA, Teng2022DiscoveryA, Yin2022DiscoveryA, Arachchige2022ChargeA, Cao2023CompetingA, Hu2025FlatA, Korshunov2023SofteningA, Tan2023AbundantA, Pokharel2023FrustratedA, Teng2023MagnetismA, Korshunov2024PressureA, Subires2024FrustratedA, Bonetti2024CompetingA, Wen2024UnconventionalA, Ortiz2024StabilityA, Pokharel2021ElectronicA}. 
A representative example is \ch{ScV6Sn6}, which has been extensively studied for its CDW behavior~\cite{Arachchige2022ChargeA, Cao2023CompetingA, Hu2025FlatA, Korshunov2023SofteningA, Tan2023AbundantA, Pokharel2023FrustratedA}.
Similarly, members of the 1:1 family, such as \ch{FeGe}, showcase various pressure-sensitive CDW orders~\cite{Teng2022DiscoveryA, Subires2024FrustratedA, Korshunov2024PressureA, Jiang2023KagomeA, Bonetti2024CompetingA, Wen2024UnconventionalA}. 
The electronic band structures of the 1:1 and 1:6:6 families can be understood using a LEGO-like building block approach, as demonstrated in \ch{FeGe}~\cite{Jiang2023KagomeA}, where kagome $d$ orbitals have been decoupled into smaller groups that coupled with other orbitals.  
Materials in the 1:1 and 1:6:6 families can accommodate various chemical substitutions~\cite{Feng2024CatalogueA}, with the $E_F$ often positioned near either the vHS or the FB. However, when the FBs are located close to $E_F$, the system almost always exhibits magnetization, leading to the polarization of the FBs and shifting them away from $E_F$~\cite{Teng2022DiscoveryA, Teng2023MagnetismA, Feng2024CatalogueA}.

Besides the 1:3:5 and 1:6:6 families,
another kagome metal that shares a similar LEGO-like~\cite{Jiang2023KagomeA} structural framework with the 1:1 family is the 1:3:2 family.
In this family, multiple compounds have been reported to exhibit SC~\cite{Barz1980TernaryA, Ku1980SuperconductingA, Chevalier1983SuperconductingA, Rauchschwalbe1984SuperconductivityA, Godart1987CoexistenceA, Escorne1994Type-IIA, Li2011AnomalousA, Mielke2021NodelessA, Gong2022SuperconductivityA, Gui2022LaIr3Ga2A, Chaudhary2023RoleA, Liu2024SuperconductivityA, Ma2024Dome-ShapedA, Ushioda2024Two-gapA, Li2012DistinctA, Li2016ChemicalA, Chakrabortty2023EffectA, Kishimoto2002MagneticA}, e.g., \ch{CeRu3Si2} with $T_c \simeq 1 \,\mathrm{K}$~\cite{Rauchschwalbe1984SuperconductivityA}, \ch{YRu3Si2} with $T_c\simeq 3\,\mathrm{K}$~\cite{Gong2022SuperconductivityA}, \ch{ThRu3Si2} with $T_c \simeq 3.8\,\mathrm{K}$~\cite{Liu2024SuperconductivityA} and \ch{LaIr3Ga2} with $T_c \simeq 5.2\,\mathrm{K}$~\cite{Gui2022LaIr3Ga2A}.
The kagome metal \ch{LaRu3Si2}, also part of the 1:3:2 family, has been identified as hosting both SC and CDW orders~\cite{Barz1980TernaryA, Ku1980SuperconductingA, Chevalier1983SuperconductingA, Rauchschwalbe1984SuperconductivityA, Godart1987CoexistenceA, Escorne1994Type-IIA, Kishimoto2002MagneticA, Li2011AnomalousA, Mielke2021NodelessA, Gong2022SuperconductivityA, Gui2022LaIr3Ga2A, Chaudhary2023RoleA, Liu2024SuperconductivityA, Ma2024Dome-ShapedA, Mielke2024ChargeA, Plokhikh2024DiscoveryA, Mielke2024MicroscopicA, Ushioda2024Two-gapA, Li2012DistinctA, Li2016ChemicalA, Chakrabortty2023EffectA}. Notably, \ch{LaRu3Si2} (and its siblings in the 1:3:2 family) stands out as the only known kagome metal that superconducts at ambient pressure with an FB near $E_F$ and exhibits no magnetic order. 
Among kagome materials, it also exhibits the highest reported $T_c$ ($\simeq 7.8\,\mathrm{K}$)~\cite{Barz1980TernaryA, Ku1980SuperconductingA, Chevalier1983SuperconductingA, Rauchschwalbe1984SuperconductivityA, Godart1987CoexistenceA, Escorne1994Type-IIA, Kishimoto2002MagneticA, Li2011AnomalousA, Mielke2021NodelessA, Gong2022SuperconductivityA, Gui2022LaIr3Ga2A, Chaudhary2023RoleA, Liu2024SuperconductivityA, Ma2024Dome-ShapedA, Ushioda2024Two-gapA, Li2012DistinctA, Li2016ChemicalA, Chakrabortty2023EffectA}.
Its CDW transition temperature is remarkably high, occurring at approximately $400 \,\mathrm{K}$, with a wave vector of $(1/4, 0, 1/2)$ relative to the high-temperature Brillouin zone (BZ). At lower temperatures at around $80 \sim 170 \,\mathrm{K}$, an additional CDW order emerges with a wave vector of $(1/6, 0, 1/2)$~\cite{Mielke2024ChargeA, Plokhikh2024DiscoveryA}.
First-principles calculations reveal that \ch{LaRu3Si2} hosts quasi FBs near $E_F$~\cite{Vergniory2019A-completeA}, raising the intriguing question of whether its relatively high superconducting $T_c$ to other kagome is associated with the presence of this FB. Exact FBs with nontrivial quantum geometry could significantly enhance $T_c$~\cite{Peotta2015SuperfluidityA, Herzog-Arbeitman2022SuperfluidA, Herzog-Arbeitman2022Many-BodyA}. 
However, most FBs at $E_F$ tend to magnetize and shift away from $E_F$, thereby suppressing SC.
Hence, in realistic circumstances, flat bands that are fractionally filled will generally not appear but magnetize.
Experimentally, \ch{LaRu3Si2} is reported to host a nodeless $s$-wave SC gap~\cite{Mielke2021NodelessA} and a two-gap structure~\cite{Ushioda2024Two-gapA}. The mechanisms underlying both the SC and the CDW phases in \ch{LaRu3Si2} remain unresolved, necessitating a comprehensive explanation.

In this study, we conduct a thorough theoretical investigation of \ch{LaRu3Si2}, utilizing both \textit{ab-initio} and analytical methods to explore its electronic structure, structural stability, and electron-phonon interactions under both undoped and doped conditions. Our results reveal that \ch{LaRu3Si2} exhibits strong mode-selective electron-phonon interaction, predominantly driven by coupling between Ru $B_{3u}$ (local $x$, pointing toward the hexagon center) phonons and Ru $A_g$ (local $d_{x^2-y^2}$) orbitals in the kagome lattice. This coupling behavior reflects a universal property of kagome systems, underscoring its potential implications for identifying novel superconducting materials within the 1:3:2 family. While doping into the FB induces instabilities and magnetic orders, we demonstrate that weak doping (less than one hole per unit cell, when the system remains paramagnetically stable) enhances the EPC strength and increases the superconducting $T_c$ by approximately $50\%$.
Additionally, we perform a high-throughput screening within the 1:3:2 family, identifying 3063 stable compounds sharing the same crystalline structure of \ch{LaRu3Si2}. Among these kagome materials, 428 are predicted to exhibit SC with $T_c > 1\,\mathrm{K}$ with the highest estimated $T_c \simeq 15\,\mathrm{K}$.

\begin{figure*}[!t]
    \centering
    \includegraphics[width=0.9\linewidth]{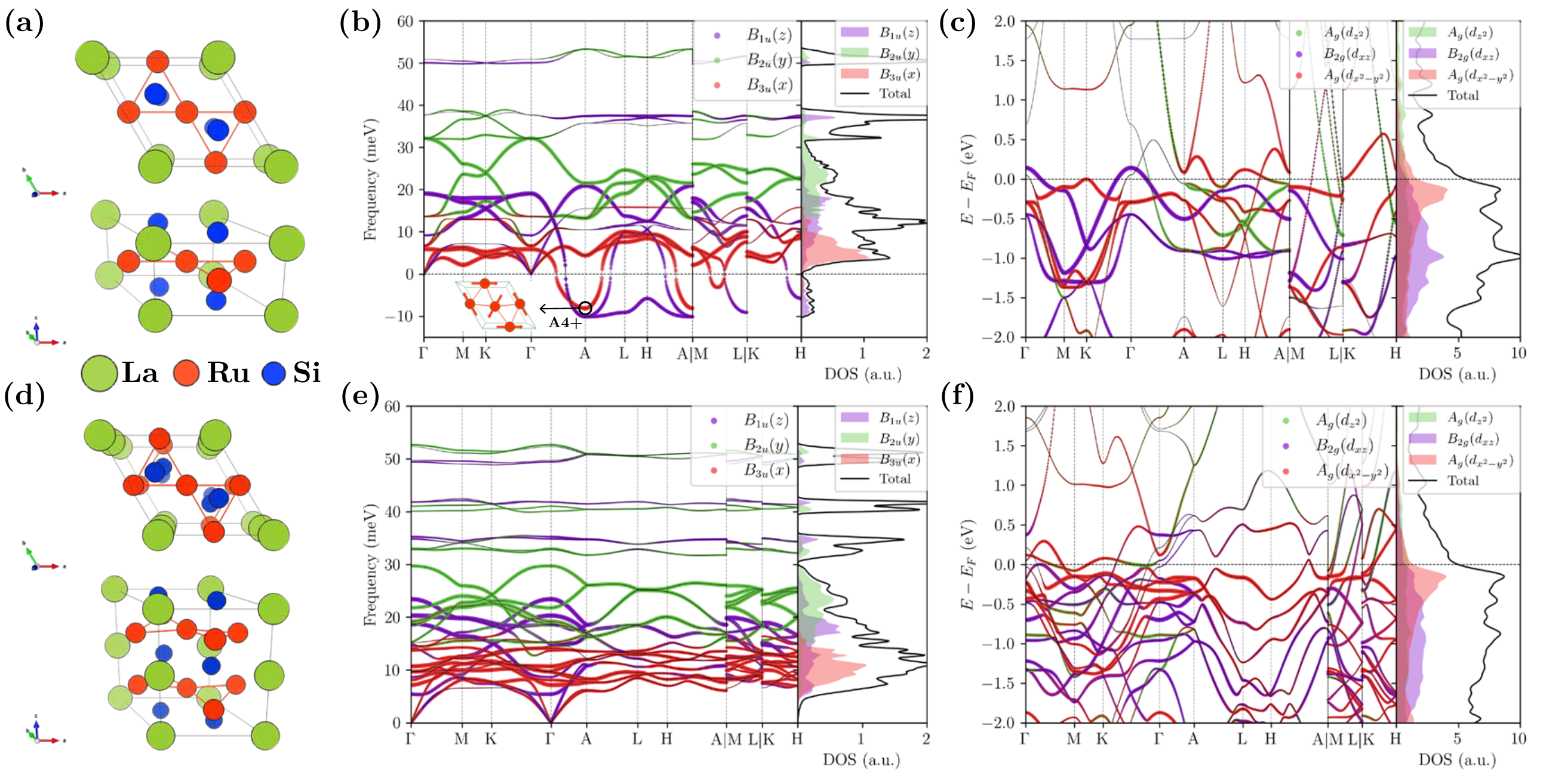}
    \caption{Structural, phonon, and electronic properties of \ch{LaRu3Si2} in different phases. 
    (a) Crystal structure, (b) phonon dispersion, and (c) band structure of \ch{LaRu3Si2} in the $P6/mmm$ phase; (d), (e), and (f) show the same but for the $Cccm$ phase. Due to the $1\times 1\times 2$ reconstruction of the $Cccm$ phase, the original bands at $k_z=0$ and $k_z=\pi/c$ planes in the $P6/mmm$ phase are folded to the $k_z=0$ plane in the $Cccm$ phase.}
    \label{fig:ph}
\end{figure*}

\textit{Crystal structure}.~The compound \ch{LaRu3Si2} undergoes multiple structural phase transitions as a function of temperature, as evidenced by X-ray diffraction (XRD) analysis~\cite{Plokhikh2024DiscoveryA, Mielke2024ChargeA}. At temperatures exceeding $620\,\mathrm{K}$, \ch{LaRu3Si2} adopts its most symmetric crystal structure in space group (SG) 191 ($P6/mmm$). As shown in \cref{fig:ph}{(a)}, the SG 191 structure is composed of a kagome lattice of Ru atoms at $z = c/2$ ($c$ being the lattice constant in the $z$-direction), a honeycomb lattice of Si atoms at $z = 0$, and a triangular lattice of La atoms also at $z = 0$, occupying the Wyckoff positions (WPs) $3g$, $2c$, and $1a$, respectively. However, the first-principles harmonic phonon spectrum shown in \cref{fig:ph}{(b)} indicates that this phase is dynamically unstable at low temperatures, with imaginary phonon frequencies on the $q_z = \pi$ plane. This instability persists at high temperatures (see \cref{fig:Ph191HighT} in \cref{app:ResultsDFTStructure191}).

We then identify the irreducible representations (irreps) of the imaginary phonon modes at the high-symmetry point $A (0,0,1/2)$ in SG 191, and the possible distortion patterns with the corresponding subgroups for the resulting low-temperature phase (\cref{tab:subgroup}). 
The one-dimensional (1D) irrep $A_4^+$ (following the convention on \textit{Bilbao Crystallographic Server} (BCS)~\cite{Aroyo2006BilbaoA, Aroyo2006BilbaoB}) is associated with the $B_{3u}$ phonon, which is given by the local $x$ phonon that pointing toward the hexagon center ($1a$ position). It plays a dominant role in driving in-plane torsional distortions of the kagome lattice, as illustrated in the inset of \cref{fig:ph}{(b)}. The highest-symmetry (and the only) subgroup associated with this mode is SG 193 ($P6_3/mcm$)~\cite{Aroyo2006BilbaoA, Aroyo2006BilbaoB}. However, powder XRD measurements have reported a lower-symmetry phase, SG 176 ($P6_3/m$)~\cite{Barz1980TernaryA}, which exhibits similar in-plane distortions of the kagome lattice as SG 193. The difference between the two phases is that the bond length of the corner-sharing triangles of the kagome lattice is identical in SG 193 phase while different in SG 176 phase, i.e., kagome layers in SG 193 have the in-plane $C_{2x}$ symmetry. Interestingly, SG 176 is a subgroup of SG 193, which can be obtained starting from SG 193 by breaking the in-plane non-symmorphic operation $\tilde{C}_{2x}\equiv\Bqty{C_{2x}|0,0,1/2}$.
First-principles structural relaxations starting from SG 176 converge to a final structure with SG 193 symmetry, consistent with the total energy comparison in \cref{tab:total_en}. Thus SG 193 is energetically favored over SG 176. However, phonon spectra reveal that neither SG 176 nor SG 193 is dynamically stable (\cref{fig:Ph191and193and176}{(b) and (c)} in \cref{app:ResultsDFTStructure191}). Consequently, these two structures will not be the focus of this study. Detailed structural descriptions and phonon spectra for both phases are provided in \cref{app:ResultsDFTStructure193}. 

Besides the 1D soft mode  $A_4^+$, there is a two-dimensional (2D) soft mode with $A_5^+$ irrep in SG 191. Condensation of this irrep leads to potential subgroups SG 66 ($Cccm$), SG 65 ($Cmmm$), or SG 10 ($P2/m$) by breaking the $C_{3z}$ symmetry. Upon full structural relaxation in \textit{ab initio} calculations, the distorted structure stabilizes in the SG 66 symmetry. This is consistent with the experimentally refined structure observed at intermediate temperatures ($400 < T < 620 \,\mathrm{K}$), as recently reported in Refs.~\cite{Plokhikh2024DiscoveryA, Mielke2024ChargeA}.
Additionally, the phonon spectrum of this $Cccm$ structure exhibits harmonic stability at low temperatures [\cref{fig:ph}{(e)}] and has the lowest total energy among all reported $1\times1\times2$ structures (\cref{tab:total_en}).

In the SG 66 phase, the 1D soft mode $A_4^+$ in SG 191 also becomes hardened. Upon cooling (i.e., decreasing the electronic temperature in \textit{ab initio} calculation, see details in \cref{app:ResultsDFTStructure66}), a 1D mode at $\Gamma$, identified as the $\Gamma_3^+$ irrep, softens (but is still positive). The atomic displacement associated with this 1D mode in the $Cccm$ phase is identical to that of the $A_4^+$ mode in the SG 191 phase (\cref{fig:Ph66HighT} in \cref{app:ResultsDFTStructure66}).
It is also noteworthy that, similar to the softened modes responsible for the CDW phases in 1T-transition metal dichalcogenide (1T-TMD) materials -- an observation that suggests the presence of strong electron-phonon coupling (EPC)~\cite{Weber2011ExtendedA, Weber2013OpticalA} -- the 1D $A_4^+$ mode in the SG 191 phonon spectrum hardens with increasing electronic temperature (see \cref{fig:Ph191HighT} in \cref{app:ResultsDFTStructure191}). This behavior indicates that the instability of this softened mode may be driven by significant EPC. A more detailed investigation into the role of EPC in \ch{LaRu3Si2} will be presented in part \textit{Electron-phonon coupling}.

\begin{table}[b!]
    \centering
    \caption{Possible subgroups of SG 191 $P6/mmm$ driven by imaginary phonon modes at high symmetry point $A$ in \ch{LaRu3Si2}.}
    \begin{tabularx}{\linewidth}{C|C|C}
        \hline\hline
        $\omega (\text{meV})$ & Irrep ($\rho$) & Subgroup  \\
        \hline
        $-8.384$ & $A_4^+$ (1D) &    SG 193 $P6_3/mcm$ \\ 
        $-10.342$ & $A_5^+$ (2D) &   SG 66 $Cccm$ \\
        \hline\hline
    \end{tabularx}
    \label{tab:subgroup}
\end{table}

\begin{table}[b!]
    \centering
    \caption{Total energy of \ch{LaRu3Si2} in different structures, 
with $\Delta E = E_{\mathrm{SG}} - E_{\mathrm{SG 191}} \,(\mathrm{meV}/\mathrm{atom})$.}
    \begin{tabularx}{\linewidth}{c|c|c|c|c}
        \hline\hline
SG & $P6/mmm$ & $P6_3/m$ & $P6_3/mcm$ & $Cccm$ \\
        \hline
        $\Delta E (\text{meV}/\text{atom})$ & 0 & $+13.690$ & $-2.539$ & $-13.738$ \\
        \hline\hline
    \end{tabularx}
    \label{tab:total_en}
\end{table}

At even lower temperatures (below 400 K), two CDW orders have been reported in \ch{LaRu3Si2} based on XRD data: a $4 \times 4 \times 2$ phase, which emerges above room temperature (around $400 \,\mathrm{K}$), and a $6 \times 6 \times 2$ phase, detected at $80 \,\mathrm{K}$ ($170 \,\mathrm{K}$ in Fe-doped samples)~\cite{Plokhikh2024DiscoveryA, Mielke2024MicroscopicA, Mielke2024ChargeA}. 
The precise symmetries of both CDW phases remain under investigation and require further refinement. Structural analysis indicates that atomic displacements from the $1\times1\times2$ ($Cccm$) phase to the $4 \times 4 \times 2$ phase are relatively minor compared to those from the SG 191 ($P6/mmm$) phase to the $1 \times 1 \times 2$ ($Cccm$) phase (\cref{fig:CrystStructAll}{(i) and (j)} in \cref{app:ResultsDFTStructure}). 
Since the symmetry and refined crystal structure of these CDW phases at lower temperatures remain ambiguous and require further investigation, and as our findings indicate they are unlikely to be directly related to superconductivity, we do not explore these phases in detail in this work.

\textit{Electronic structure}.~Since the SG 191 phase hosts the highest-symmetry crystalline structure, we first examine its electronic structure. As previously described, the three sublattices are located at WPs $1a$ (triangular lattice), $2c$ (honeycomb lattice), and $3g$ (kagome lattice), respectively. The corresponding site symmetry groups are $D_{6h} (6/mmm)$ for $1a$, $D_{3h} (-6m2)$ for $2c$, and $D_{2h} (mmm)$ for $3g$.
The bands near the $E_F$ are mainly contributed by the Ru-$4d$ orbitals and Si-$3p$ orbitals. The site symmetry group $D_{2h}$ at WP $3g$ splits the Ru-$4d$ orbitals into four distinct irreps, classified according to the eigenvalues of the three two-fold rotational symmetries, as summarized in \cref{tab:pos2abr} in \cref{app:ResultsDFTElectronFB}.
The band dispersion, presented in \cref{fig:ph}{(c)}, highlights the contributions of different orbitals (irreps), along with the corresponding density of states (DOS) obtained from first-principles calculations. The total DOS exhibits a peak approximately $150 \,\mathrm{meV}$ below $E_F$, primarily due to the quasi-FB formed by the $A_g$ (local Ru-$d_{x^2-y^2}$ specifically) orbitals (see the orbital-resolved band structure and DOS of Ru $d_{x^2-y^2}$ in \cref{fig:ph}{(c)} and \cref{fig:Bands191} in \cref{app:ResultsDFTElectronFB}). 
The Ru-$d_{x^2-y^2}$ orbital forms an extensive Fermi surface (\cref{fig:FS191} in \cref{app:ResultsDFTElectronFB}), which, as we will demonstrate in the following, plays a crucial role in SC. 

To investigate the electron correlation effects, dynamical mean field theory (DMFT) calculations were performed using a combination of \texttt{WIEN2k} and the \texttt{Rutgers eDMFT} code~\cite{Blaha2020WIEN2k:A, Haule2010DynamicalA, Haule2015FreeA, Haule2016ForcesA, Haule2007QuantumA}, with interaction parameters $U = 6\,\mathrm{eV}$ and $J_{\mathrm{H}} = 0.9 \,\mathrm{eV}$. The spectral function obtained at $T = 100\,\mathrm{K}$ closely resembles the \textit{ab initio} band structure, as does the calculated Fermi surface, as shown by \cref{fig:SOCandDMFT191}{(b) and (c)} in \cref{app:ResultsDFTElectronDMFT}. 
These results indicate that \ch{LaRu3Si2} is a highly itinerant system.
Furthermore, the largest mass renormalization factor was found to be $Z \simeq 0.66$, which is relatively small. This suggests that SC in \ch{LaRu3Si2} is primarily mediated by EPC rather than electronic correlations. Consequently, correlation effects are neglected in the following calculations.

As the temperature decreases, \ch{LaRu3Si2} undergoes a structural phase transition, driven by the condensation of the 2D softened phonon mode $A_5^+$, as discussed in the previous paragraph. This transition results in a doubling of the unit cell along the $z$-axis and a reduction in SG from $P6/mmm$ to $Cccm$ [\cref{fig:ph}{(d)}].
In the resulting $1 \times 1 \times 2$ ($Cccm$) phase, the ideal kagome lattice is distorted along the $z$ direction, breaking the $C_{3z}$ rotational symmetry present in the SG 191 phase. Consequently, the WP $3g$ in $P6/mmm$ splits into $4b$ and $8k$ in $Cccm$, with their site symmetry groups reduced to $222$ ($D_2$) and $m$, respectively.
Although the symmetry breaking induces hybridization between different irreps of the original $D_{2h}$ group, the fundamental electronic features remain largely unchanged. Notably, the quasi-FB formed by the $A_g$ (local Ru-$d_{x^2-y^2}$) orbitals, as well as the associated DOS peak, persist in the $Cccm$ phase. Such a feature is demonstrated in the band structure and orbital-resolved DOS shown in \cref{fig:ph}{(f)} and further confirmed by the unfolded band structures presented in \cref{fig:UnfoldedAll} in \cref{app:ResultsDFTElectronFB}.
Since the $C_{3z}$ symmetry-breaking distortion can be treated as a perturbation to the ideal kagome lattice, we use the same local axis definitions from the $P6/mmm$ phase in the $Cccm$ phase for consistency.

\textit{Electron-phonon coupling}.~An exact FB, whose flatness arises from the lattice geometry~\cite{Calugaru2022GeneralA}, has been shown to exhibit superfluid weight proportional to its quantum geometry under attractive interactions~\cite{Peotta2015SuperfluidityA}. In dispersive bands, the topology and quantum geometry of the Fermi surface can also contribute to EPC~\cite{Yu2024Non-trivialA}, further influencing superconducting properties.
The study of three-dimensional (3D) partially filled FB SC remains crucial, as partially filled FBs are prone to magnetization due to repulsive Coulomb interactions~\cite{Herzog-Arbeitman2022Many-BodyA, Feng2024CatalogueA}, which can suppress SC.
Given that \ch{LaRu3Si2} holds the highest $T_c \sim 7.8 \,\mathrm{K}$ among known kagome materials~\cite{Barz1980TernaryA, Ku1980SuperconductingA, Chevalier1983SuperconductingA, Rauchschwalbe1984SuperconductivityA, Godart1987CoexistenceA, Escorne1994Type-IIA, Li2011AnomalousA, Mielke2021NodelessA, Gong2022SuperconductivityA, Gui2022LaIr3Ga2A, Chaudhary2023RoleA, Liu2024SuperconductivityA, Ma2024Dome-ShapedA, Ushioda2024Two-gapA}, and features a quasi-FB approximately $150 \,\mathrm{meV}$ below $E_F$, we explore whether this FB is responsible for this $T_c$.

\begin{figure}[!t]
    \centering
    \includegraphics[width=0.9\linewidth]{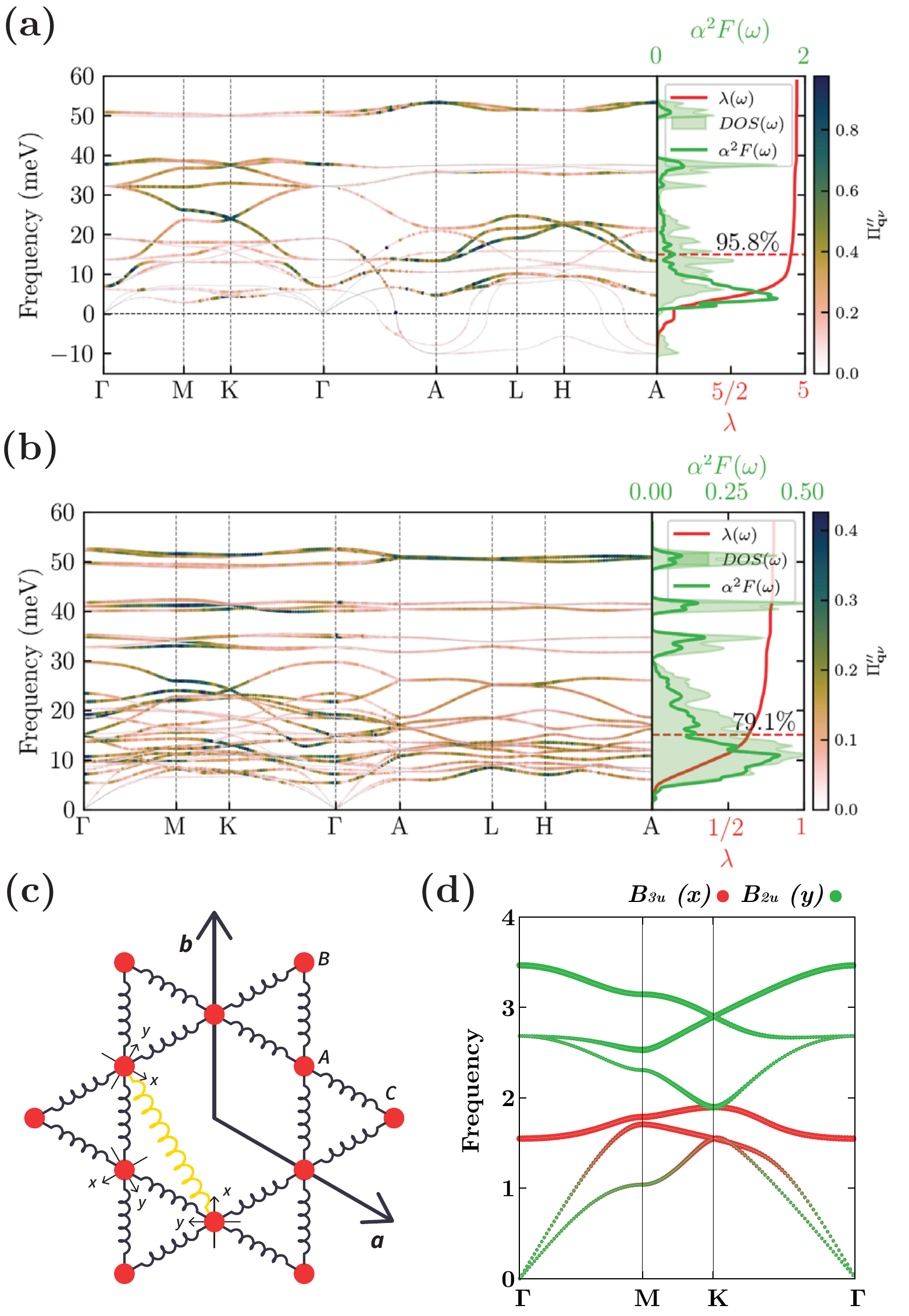}
    \caption{Phonon linewidth $\Pi''_{\vb{q}\nu}$, Eliashberg function $\alpha^2F(\omega)$, and electron-phonon coupling strength $\lambda(\omega)$ of \ch{LaRu3Si2} in $P6/mmm$ phase (a) and $Cccm$ phase (b). (c) The illustration figure of the kagome spring-ball model. (d) The mode-resolved phonon spectrum of the kagome spring-ball model. The size of the colored circles represents the mode weight.}
    \label{fig:epc}
\end{figure}

To address this question, we first examine the EPC in \ch{LaRu3Si2} using \textit{ab initio} methods. EPC calculations were performed for both the $P6/mmm$ and $Cccm$ phases using the \texttt{EPW} code~\cite{Giustino2007Electron-phononA, Noffsinger2010EPW:A, Ponce2016EPW:A, Lee2023Electron-phononA}.
It should be noted that, due to the presence of imaginary phonon modes in the SG 191 ($P6/mmm$) phase, the EPC strength and $T_c$ values derived from this phase are not physically meaningful. However, we will later derive these values for the stable SG 66 phase. Nevertheless, analyzing the EPC properties in \ch{LaRu3Si2} within the high-symmetry SG 191 ($P6/mmm$) phase remains insightful, as the $C_{3z}$-breaking distortion in the low-temperature phase introduces only a perturbative effect on the electronic structure.
As shown in \cref{fig:epc}{(a) and (b)}, despite notable differences in structural stability between these two phases, key EPC-related quantities -- including the phonon linewidth $\Pi_{\vb{q},\nu}''$, the Eliashberg function $\alpha^2F(\omega)$, and the EPC strength $\lambda(\omega)$ -- remain consistent. In both cases, the EPC strength $\lambda$ is predominantly ($95.8\%$ in $P6/mmm$ and $79.1\%$ in $Cccm$) contributed by low-frequency phonons (below $15 \,\mathrm{meV}$). Further phonon-mode-resolved EPC analysis reveals that $\lambda$ is mainly contributed by the $B_{3u}$ (local $x$) phonon of the kagome lattice (\cref{app:ResultsDFTElectronPhononLaRu3Si2}). However, the phonon linewidth $\Pi''_{\vb{q},\nu}$, shown in \cref{fig:epc}{(a) and (b)}, indicates that the linewidth remains relatively uniform across both $B_{3u}$ (local $x$) and $B_{2u}$ (local $y$) phonons.
The EPC between the Ru $A_g$ electrons and the $B_{3u}/B_{2u}$ phonon modes in the kagome lattice was further examined by projecting the \textit{ab-initio} EPC tensor $g_{mn,\nu}(\vb{k},\vb{q})$ from \texttt{EPW}~\cite{Giustino2007Electron-phononA, Noffsinger2010EPW:A, Ponce2016EPW:A, Lee2023Electron-phononA} to the Wannier basis, i.e., obtaining the real-space EPC tensor $g_{mn,\nu}(\vb{R}_e,\vb{R}_p)$. The results indicate that the coupling strengths of Ru $A_g (d_{x^2-y^2})$ electrons to both $B_{3u}$ and $B_{2u}$ phonons are nearly identical, being $0.0201 \,\mathrm{Ry}/\mathrm{Bohr}$ for $B_{3u}$ and $0.0204 \,\mathrm{Ry}/\mathrm{Bohr}$ for $B_{2u}$. 

To better understand this behavior, we parameterize the hopping terms using Slater-Koster (SK) hopping integrals~\cite{Slater1954SimplifiedA} within the Gaussian approximation (GA) (\cref{app:ModelsGaussianEPC})~\cite{Yu2024Non-trivialA}. 
Specifically, the hopping term between $A_g (d_{x^2-y^2})$ orbitals on different sublattices $m, n$ (see \cref{app:ModelsGaussianEPCHoppingSK} for details) is expressed as
\begin{equation}\label{eq:hop_sk}
    t_{mn}(\vb{r}) = f_{mn}(\vb{r}) V^d(\vb{r}) - \frac{1}{2} V_{\pi}(\vb{r}).
\end{equation}
Here $\vb{r}$ is the hopping vector connecting two sublattices $m,n$, and $f_{mn}(\vb{r})$ is a function of the direction cosines of $\vb{r}$ (i.e., $r_\mu/\abs{\vb{r}}, \mu\in\Bqty{x,y,z}$). The terms $V_{\pi}(\vb{r})$ and $V^d(\vb{r})=\bqty{V_{\delta}(\vb{r})-4V_{\pi}(\vb{r})+3V_{\sigma}(\vb{r})}/8$ correspond to SK hopping integrals for $d$ orbitals, as defined in Ref.~\cite{Slater1954SimplifiedA} (see \cref{app:ModelsGaussianEPCHoppingSK} for details).
The GA assumes these hopping integrals decay as a Gaussian function as $V_i(\vb{r}) = V_{i,0}\mathrm{exp}(-\gamma_i \abs{\vb{r}}^2/2)$, where $i\in\Bqty{\sigma,\pi,\delta}$, $V_{i,0}$ represents the initial overlap integral strength, $\gamma_i$ is a decay parameter specific to each bond type.
With \cref{eq:hop_sk} established, we proceed to analyze the EPC.
First, due to mirror symmetry $\mathcal{M}_z$, out-of-plane atomic motions do not couple to electrons within the perfect kagome layer in the $P6/mmm$ phase. Consequently, the EPC $g_{mn,\nu}(\vb{k},\vb{q})$ for \ch{LaRu3Si2} is non-zero only for in-plane atomic motions, i.e., $\nu = x, y$.
For the nearest-neighbor intra-sublattice hopping terms between $A_g$ orbitals, the EPC can be decomposed into two components: (i) the radial contribution, which corresponds to the partial derivatives of the hopping integrals $\pdv*{V_i(\vb{r})}{r_{\nu}}$, and (ii) the angular contribution, associated with the partial derivatives of the function $f_{mn}(\vb{r})$ (see \cref{app:ModelsGaussianEPCinSK} for details), 
\begin{equation}\label{eq:epc_sk}
\begin{aligned}
    g_{mn;\nu}(\vb{r}) \eq \pdv{t_{mn}(\vb{r})}{r_{\nu}} \\ 
    \eq \pdv{f_{mn}(\vb{r})}{r_{\nu}} V^d(\vb{r}) + f_{mn}(\vb{r}) \pdv{V^d(\vb{r})}{r_{\nu}} - \frac{1}{2} \pdv{V_{\pi}(\vb{r})}{r_{\nu}}.
\end{aligned}
\end{equation}
For the coupling between $A_g$ electrons and the $B_{3u}$ and $B_{2u}$ atomic motions, the radial contribution (the second and third terms in \cref{eq:epc_sk}) is proportional to the projection of the hopping vector along each axis. Specifically, the coupling strength to the $B_{2u}$ mode is $\sqrt{3}$ times greater than that to the $B_{3u}$ mode, where $\sqrt{3}=\tan(\frac{\pi}{3})$ is determined by the geometry of the kagome lattice. In contrast, for the angular component (the first term in \cref{eq:epc_sk}), the relationship is inverted, with the coupling to the $B_{3u}$ mode being $\sqrt{3}$ times weaker than that to the $B_{2u}$ mode.
This relationship between the two modes complicates a direct analysis of the amplitude corresponding to each mode. In practice, for \ch{LaRu3Si2}, the nearest-neighbor SK parameters and the decaying factors are determined from first-principles calculations (\cref{app:ModelsGaussianEPCinSK}).  
By applying these \textit{ab-initio} parameters to the expressions in \cref{eq:epc_sk}, the calculated coupling amplitudes of $A_g$ electrons to the $B_{3u}$ and $B_{2u}$ atomic motions are $0.0192\,{\mathrm{Ry}}/{\mathrm{Bohr}}$ and $0.0334\,{\mathrm{Ry}}/{\mathrm{Bohr}}$, respectively. These values closely approximate the \textit{ab initio} results of $0.0201\,{\mathrm{Ry}}/{\mathrm{Bohr}}$ and $0.0204\,{\mathrm{Ry}}/{\mathrm{Bohr}}$, indicating the validity of this rough approximation (see \cref{app:ModelsGaussianEPCinSK} for details).

The bulk EPC constant~\cite{McMillan1968TransitionA}, $\lambda = 2 \int \dd{\omega} \alpha^2F(\omega)/\omega$, is derived from the Eliashberg function $\alpha^2F(\omega)$~\cite{Eliashberg1960InteractionsA}. It can alternatively be expressed as $\lambda = 2 \frac{D(\mu)}{N}\frac{\hbar \ev*{g^2}}{\hbar^2\ev*{\omega^2}}$, where $D(\mu)$ represents the DOS at the chemical potential $\mu$, $N$ is the number of lattice sites, and $\ev*{\omega^2}$ is the McMillan mean-squared phonon frequency~\cite{McMillan1968TransitionA}.
In this formulation of $\lambda$, mode information is embedded within both $\ev*{g^2}$ and $\ev*{\omega^2}$. Since the mode dependence of the average phonon linewidth $\ev*{g^2}$ has been analyzed above using the Gaussian approximation (GA), we now shift our focus to the averaged phonon frequency $\ev*{\omega^2}$.

To quantitatively analyze the kagome phonon spectrum, we constructed a spring-ball model consisting of springs connecting nearest-neighbor atoms (black curves in \cref{fig:epc}{(c)}). As shown in \cref{fig:epc}{(d)}, the mode-resolved phonon spectrum reveals that the $B_{3u}$ phonon (red dots) consistently exhibits a lower frequency than the $B_{2u}$ phonon (green dots) throughout the BZ. The accidental degeneracy at the $K$ point between the two EBRs, $B_{3u}@3g$ and $B_{2u}@3g$, can be lifted by considering long-range springs, such as next-nearest-neighbor springs (yellow curve in \cref{fig:epc}{(c)}; see details in \cref{fig:SpringBall} in \cref{app:ModelsSpringBall}).
The energy difference between the $B_{2u}@3g$ and $B_{3u}@3g$ EBRs originates from intrinsic variations in the potential energy associated with these modes. Vertical displacements corresponding to the $B_{2u}$ phonon experience stronger restoring forces, leading to higher potential energy, whereas horizontal displacements associated with the $B_{3u}$ phonon involve weaker restoring forces, resulting in lower energy. Detailed derivations are provided in \cref{app:ModelsSpringBall}.
This characteristic of the phonon spectrum, derived from the spring-ball model, represents a universal property of kagome materials, as further validated by high-throughput calculations presented in \cref{app:Prediction}. Our model thus provides a simple yet quantitative framework for understanding the kagome phonon spectrum.

\begin{figure}[!t]
    \centering
    \includegraphics[width=\linewidth]{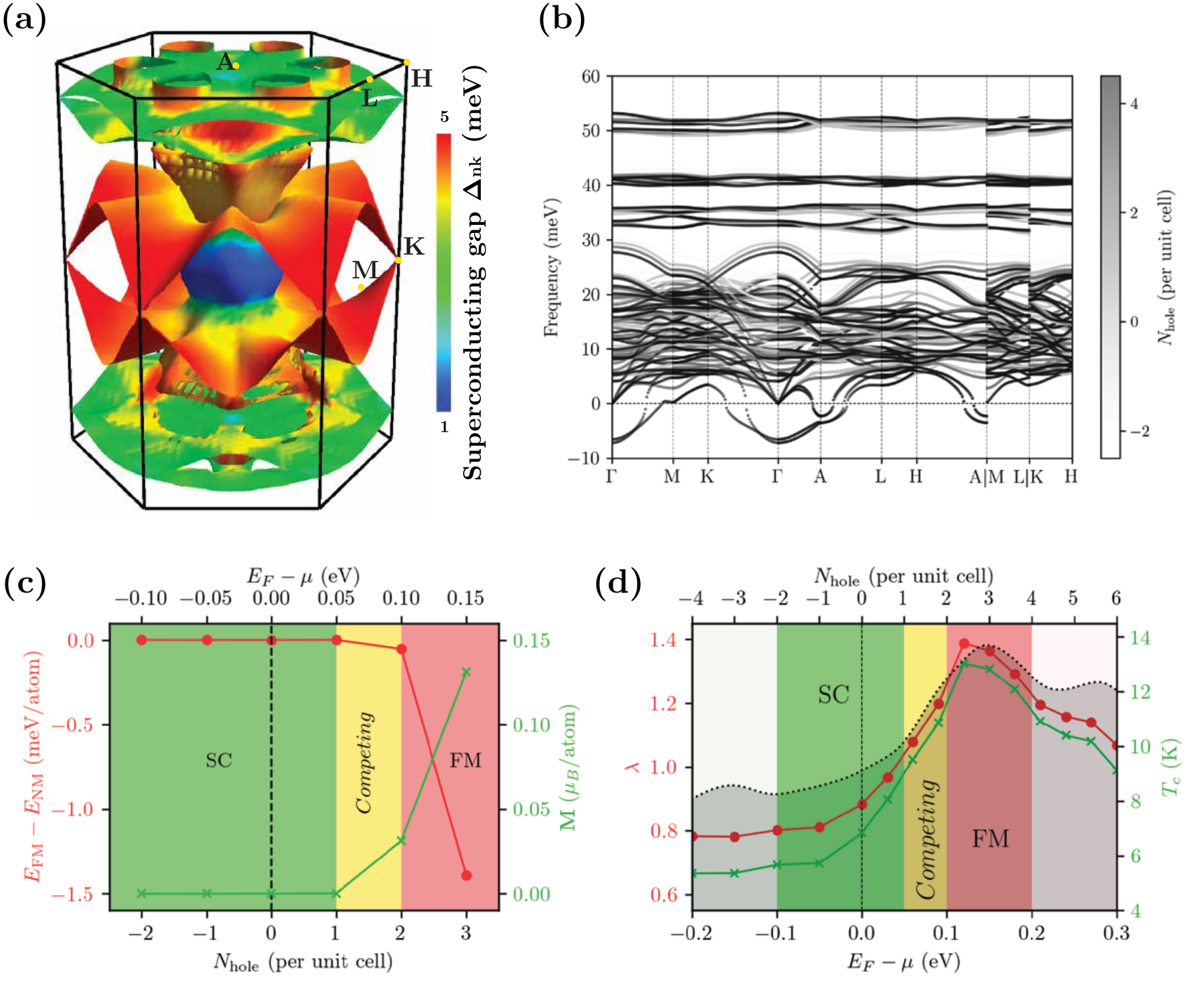}
    \caption{Superconducting properties of \ch{LaRu3Si2} under doping. 
    (a) Superconducting gap function $\Delta_{\vb{k}n}$ on the Fermi surface of \ch{LaRu3Si2} in $P6/mmm$ phase. (b) Phonon spectra, (c) total energy difference ($\mathrm{meV}/\mathrm{atom}$) between paramagnetic phase and ferromagnetic phase and magnetic momentum ($\mu_{\mathrm{B}}/\mathrm{atom}$) as a function of hole doping level ($N_{\text{hole}}/\text{unit cell}$) of \ch{LaRu3Si2} in the stable $Cccm$ phase. Instability occurs when $N_{\mathrm{hole}}/\mathrm{unit\,cell} > 2$. (d) The EPC constant $\lambda$ and superconducting $T_c$ as a function of chemical potential $\mu \, \text{(eV)}$ of \ch{LaRu3Si2} under rigid band approximation, the gray background is the total DOS of \ch{LaRu3Si2}.}
    \label{fig:sc}
\end{figure}

\textit{Superconductivity and doping stability}.~We investigate the SC induced by EPC in \ch{LaRu3Si2}.
As shown in \cref{fig:sc}{(a)}, the SC gap function $\Delta_{\vb{k} n}$ on the Fermi surface exhibits non-uniform behavior, with a prominent superconducting gap of approximately $5 \,\mathrm{meV}$ appearing on the Fermi surface. This superconducting gap is primarily contributed by the Ru $A_g (d_{x^2-y^2})$ orbitals, as demonstrated by the orbital-weighted Fermi surfaces in \cref{fig:FS191} in \cref{app:ResultsDFTElectronFB}. 
Notably, this gap value is about three times larger than the gaps on other Fermi surfaces, consistent with the recently reported two-gap structure~\cite{Ushioda2024Two-gapA}. 
This difference is likely due to the high DOS and the large Fermi surface from the quasi-FB formed by the Ru $A_g (d_{x^2-y^2})$ orbitals. Additionally, since the superconducting gap is smaller than the bandwidth of the nearly FB, the conventional contribution to the superfluid weight dominates over the quantum geometric component (see \cref{app:SFW}).

Since the DOS peak from this quasi-FB is located around $150 \,\mathrm{meV}$ below the $E_F$, we explored whether doping towards the FB could enhance EPC and increase the superconducting $T_c$. To investigate the effects of doping on structural stability and SC, additional calculations on the $1\times 1\times 2 ~(Cccm)$ phase were performed. As shown in \cref{fig:sc}{(b)}, the system remains stable under weak hole doping (less than 2 holes per unit cell), although some phonon modes exhibit slight softening, suggesting that excessive doping toward the FB could destabilize the structure.
When the $E_F$ aligns with the FB, the system tends to develop magnetic orders~\cite{Herzog-Arbeitman2022Many-BodyA, Feng2024CatalogueA}, which we also observe in our calculations. The total energy difference between the ferromagnetic and paramagnetic states, along with the evolution of the magnetic moment (per atom) as a function of the doping level, is illustrated in \cref{fig:sc}{(c)}. These results suggest that doping towards the FB could induce ferromagnetism ($0.15 \, \mu_{\mathrm{B}}/\mathrm{atom}$).
On the other hand, in \cref{fig:sc}{(d)}, we show the variation of the superconducting $T_c$ as a function of doping level under the rigid band approximation. Hole doping is observed to enhance $T_c$, likely due to an increased total DOS, suggesting its potential to strengthen EPC and SC in \ch{LaRu3Si2}. In contrast, electron doping leads to a decrease in $T_c$, consistent with experimental results~\cite{Li2012DistinctA, Li2016ChemicalA, Chakrabortty2023EffectA}.
Therefore, under weak hole doping -- such as removing fewer than one electron per unit cell from the $Cccm$ phase -- the system remains both stable and paramagnetic, with the superconducting $T_c$ expected to be maximally enhanced by approximately 50\%.

\textit{Prediction of new kagome superconductors}.~To assess the potential for doping through chemical substitution, a high-throughput structural search was performed for compounds sharing the same crystal structure and stoichiometry as \ch{LaRu3Si2} within the 1:3:2 family. 
The high-throughput search results indicate that, within an energy range of up to $100 \,\mathrm{meV}/\mathrm{atom}$ above the convex hull, 3063 compounds possess the same crystalline structure as the SG 191 ($P6/mmm$) phase of \ch{LaRu3Si2}. Among these compounds, 428 exhibit $T_c > 1 \,\mathrm{K}$, with 58 of them predicted to show SC at $T_c > 5\,\mathrm{K}$.
Comprehensive lists of thermodynamically (distance to convex hull $E_{\mathrm{hull}}<100\,\mathrm{meV/atom}$) and dynamically stable materials with $T_c > 1 \,\mathrm{K}$ are provided in \cref{tab:StablePredictionTypeI,tab:StablePredictionTypeII,tab:PredictionTypeI,tab:PredictionTypeII} in \cref{app:Prediction}. Detailed information on the electronic structures, phonon dispersions, and Eliashberg functions of compounds with $T_c > 1 \,\mathrm{K}$ and $E_{\mathrm{hull}} < 30,\mathrm{meV/atom}$ can be found in \cref{fig:type-i-1,fig:type-ii-1} in \cref{app:Prediction}.

The predicted kagome superconductors can be categorized into two distinct types (see \cref{app:Prediction}).
The first type consists of materials similar to \ch{LaRu3Si2}, sharing comparable electronic band structures but with different filling levels and exhibiting EPC dominated by low-frequency phonons (\cref{fig:type-i-1} in \cref{app:PredictionTypeI}). These materials feature a kagome lattice formed by transition metal elements. However, most compounds in this category represent electron-doped analogs of \ch{LaRu3Si2} and display lower $T_c$. Among the 65 promising compounds in this category with $E_{\mathrm{hull}} < 30$ meV/atom (\cref{tab:StablePredictionTypeI}), notable candidates include \ch{SrIr3Ni2}, with an estimated $T_c \sim 7 \,\mathrm{K}$ and \ch{CaRh3Ga2}, with an estimated $T_c\sim 6$~K. In contrast, hole-doped analogs, such as \ch{\textit{R}Ru3B2} ($R=\mathrm{Tm,Er,Dy}$), tend to magnetize and are unstable in the paramagnetic phase in experiment~\cite{Ku1980SuperconductingA}, consistent with the expectation that doping toward the FB can induce ferromagnetism and structural instability.

In addition to the first type, a new class of kagome superconductors is predicted (\cref{tab:StablePredictionTypeII,tab:PredictionTypeII} in \cref{app:PredictionTypeII}), featuring a kagome lattice composed of post-transition metal elements, with kagome bands primarily contributed by the $p$ electrons. Interestingly, the EPC in this class also exhibits a low-frequency-phonon-dominated feature (see \cref{fig:type-ii-1} in \cref{app:Prediction}). Among the 20 promising compounds in this category with $E_{\mathrm{hull}} < 30$ meV/atom (\cref{tab:StablePredictionTypeII}), notable candidates include \ch{\textit{X}Be3Ni2} ($X = \mathrm{Zr, Sc}$), with an estimated $T_c \sim 2 \,\mathrm{K}$.

\textit{Discussion}.~In summary, we have systematically investigated the kagome superconductor \ch{LaRu3Si2}, distinguished as the only known material with a FB near $E_F$ and exhibiting SC at the ambient pressure with the highest reported transition temperature ($T_c \simeq 7.8 \,\mathrm{K}$) among kagome materials. Our study focuses on its structural stability, EPC, and superconducting properties.
Contrary to earlier reports suggesting a structure in SG 176 ($P6_3/m$)~\cite{Barz1980TernaryA}, our \textit{ab initio} calculations and analysis demonstrate that the stable structure of \ch{LaRu3Si2} above the CDW transition temperature is stabilized by the condensation of the 2D $A_5^+$ soft phonon mode. This finding aligns with the recently refined experimental structure obtained from XRD studies~\cite{Plokhikh2024DiscoveryA}.
The EPC in \ch{LaRu3Si2} is strongly mode-selective, primarily driven by the pronounced coupling between the $B_{3u}$ (local $x$ direction, pointing toward the hexagon center) phonon and the $A_g$ (local $d_{x^2-y^2}$) electrons within the kagome lattice. This behavior originates from a universal characteristic of kagome systems: the $B_{3u}$ phonon consistently exhibits a significantly lower frequency than the $B_{2u}$ (local $y$ direction, pointing toward the triangle center) phonon. Our spring-ball model and high-throughput structure search calculations suggest that this mode-selective EPC is not unique to \ch{LaRu3Si2} but is likely a general feature of kagome materials.
Doping calculations for the $Cccm$ phase indicate that heavy hole doping induces both ferromagnetism and structural instability. However, under light hole doping (approximately one hole per unit cell), the system remains stable and paramagnetic, resulting in a substantial enhancement of both the EPC constant $\lambda$ and the superconducting transition temperature $T_c$ by approximately 50\%.
Furthermore, high-throughput screening identifies 3063 stable compounds within the 1:3:2 family that share the same crystalline structure as \ch{LaRu3Si2}, with 428 of them predicted to exhibit superconductivity with $T_c > 1 \,\mathrm{K}$. Additionally, a new class of kagome superconductors is predicted, characterized by kagome bands primarily contributed by $p$ electrons. Promising candidates with relatively high $T_c$ include \ch{SrIr3Ni2} ($T_c \sim 7 \,\mathrm{K}$) , \ch{CaRh3Ga2} ($T_c\sim 6$ K), and \ch{\textit{X}Be3Ni2} ($X = \mathrm{Zr, Sc}$, with $T_c \sim 2 \,\mathrm{K}$).

\textit{Acknowledgement}.~B.A.B, K.H., M.A.L.M., and P.T. were supported by a grant from the Simons Foundation (SFI-MPS-NFS-00006741-01, B.A.B.; SFI-MPS-NFS-00006741-06, K.H.; SFI-MPS-NFS-00006741-13, M.A.L.M.; SFI-MPS-NFS-00006741-12, P.T.) in the Simons Collaboration on New Frontiers in Superconductivity.
B.A.B., C.F., K.H., M.A.L.M, and P.T. belong to the SuperC collaboration.
J.D., E.O.L., and P.T. were supported by Jane and Aatos Erkko Foundation, Keele Foundation, and Magnus Ehrnrooth Foundation as part of the SuperC collaboration. J.D. acknowledges the computational resources provided by the Aalto Science-IT project. 
J.D. and Z.W. were supported by the National Natural Science Foundation of China (Grants No. 11974395, No. 12188101), the Strategic Priority Research Program of Chinese Academy of Sciences (Grant No. XDB33000000), National Key R\&D Program of Chain (Grant No. 2022YFA1403800), and the Center for Materials Genome.
Y.J. and H.H. were supported by the European Research Council (ERC) under the European Union's Horizon 2020 research and innovation program (Grant Agreement No. 101020833) as well as by IKUR Strategy. 
T.F.T.C. acknowledges the financial support from FCT through project CEECINST/00152/2018/CP1570/CT0006, and computing resources provided by the project 2023.14294.CPCA.A3, platform Deucalion.
D.C\u{a}l. was supported by the ERC under the European Union's Horizon 2020 research and innovation program (Grant Agreement No. 101020833) and by the Simons Investigator Grant No. 404513. 
S.B-C. acknowledges financial support from the MINECO of Spain through the project PID2021-122609NB-C21.
B.A.B was supported by the Gordon and Betty Moore Foundation through Grant No. GBMF8685 towards the Princeton theory program, the Gordon and Betty Moore Foundation’s EPiQS Initiative (Grant No. GBMF11070), Office of Naval Research (ONR Grant No. N00014-20-1-2303), Global Collaborative Network Grant at Princeton University, BSF Israel US foundation No. 2018226, NSF-MERSEC (Grant No. MERSEC DMR 2011750), and the Schmidt Foundation at the Princeton University.


%

\clearpage
\onecolumngrid
\appendix
\pagestyle{plain}
\setcounter{page}{1}
\numberwithin{equation}{section}
\numberwithin{figure}{section}
\numberwithin{table}{section}
\tableofcontents
\clearpage

\section{\label{app:DetailsDFT}First-principles calculations details}

First-principles calculations were performed based on density functional theory (DFT), using the projector augmented wave (PAW) method~\cite{Blochl1994ProjectorA, Kresse1999FromA}, as implemented in the Vienna \emph{ab initio} Simulation Package (\texttt{VASP})~\cite{Kresse1996EfficiencyA, Kresse1996EfficientA} and the \textsc{Quantum ESPRESSO} package~\cite{Giannozzi2009QUANTUMA, Giannozzi2017AdvancedA}. The generalized gradient approximation (GGA) was used, incorporating the Perdew-Burke-Ernzerhof (PBE) exchange-correlation functional~\cite{Perdew1996GeneralizedA}. 
The internal atomic positions within the structures were relaxed until the forces on each atom were reduced to less than $0.1\,\mathrm{meV}/\mathring{\mathrm{A}}$. A kinetic energy cutoff of 500 eV or 100 Ry was set for the plane-wave basis set. The Brillouin zone (BZ) was sampled using the Monkhorst-Pack method~\cite{Monkhorst1976SpecialA} with a $\Gamma$-centered $\vb{k}$-point grid of $9 \times 9 \times 12$ and $9\times9\times6$ for \ch{LaRu3Si2} in the SG 191 and SG 66 ($1 \times 1 \times 2$) phase respectively. 

Phonon spectra were calculated via \texttt{VASP} using the frozen phonon method in conjunction with the \texttt{PHONOPY} package~\cite{Togo2023ImplementationA, Togo2023First-principlesA} and via the \textsc{Quantum ESPRESSO} package in the framework of density functional perturbation theory (DFPT).
Irreducible representations (irreps) were determined using \texttt{IRVSP}~\cite{Gao2021IrvspA, Zhang2023LargeA}. Wannier-based calculations were done with \textsc{WannierTools}~\cite{Wu2018WannierToolsA}. The band representation (BR) analysis was conducted on the website \webBRdecomp. Furthermore, a Wannier-based tight-binding Hamiltonian was constructed using \textsc{Wannier90}~\cite{Mostofi2008Wannier90A, Pizzi2020Wannier90A}, with La-$d$, Ru-$s,d$ and Si-$p$ orbitals.

Electron-phonon coupling (EPC) properties were computed using the \texttt{EPW} package~\cite{Giustino2007Electron-phononA, Noffsinger2010EPW:A, Ponce2016EPW:A, Lee2023Electron-phononA}, included within \textsc{Quantum ESPRESSO}~\cite{Giannozzi2009QUANTUMA, Giannozzi2017AdvancedA}. Electron-phonon matrix elements were first computed on coarse $\vb{k}$-point and $\vb{q}$-point meshes ($9\times9\times12$ and $9\times9\times6$ for the SG 191 and SG 66 phases, respectively) and then interpolated onto finer grids using maximally localized Wannier functions.

To study electronic correlation effects, first-principles calculations combining DFT and embedded dynamical mean-field theory (eDMFT) were performed using \texttt{WIEN2k} and the \texttt{Rutgers eDMFT} code~\cite{Blaha2020WIEN2k:A, Haule2010DynamicalA}. The local-density approximation (LDA)~\cite{Perdew1992AccurateA} was employed, as it provides optimal results for lattice properties when combined with eDMFT~\cite{Haule2015FreeA, Haule2016ForcesA}. Coulomb repulsion and Hund’s coupling parameters were set to $U = 6.0\,\mathrm{eV}$ and $J_{\mathrm{H}} = 0.9\,\mathrm{eV}$, respectively. The auxiliary impurity problem in eDMFT was solved using a continuous-time quantum Monte Carlo impurity solver~\cite{Haule2007QuantumA}.

\section{\label{app:ResultsDFT}First-principles results}

\subsection{\label{app:ResultsDFTIntroLEGO} Kagome materials from LEGO building blocks}

Kagome materials in space group (SG) 191 crystallize in various structures, including the $1:1$, $1:6:6$, $1:3:5$, and $1:3:2$ families, with representative materials such as FeGe, \ch{MgFe6Ge6}, \ch{CsV3Sb5}, and \ch{LaRu3Si2}, respectively. Despite their distinct properties, the crystal and even electronic band structures of these materials can be understood through the $1:1$ (or $3:3$) family, which serves as a foundational ``LEGO'' building block~\cite{Jiang2023KagomeA}. 

\begin{figure}[H]
    \centering
\includegraphics[width=0.7\linewidth]{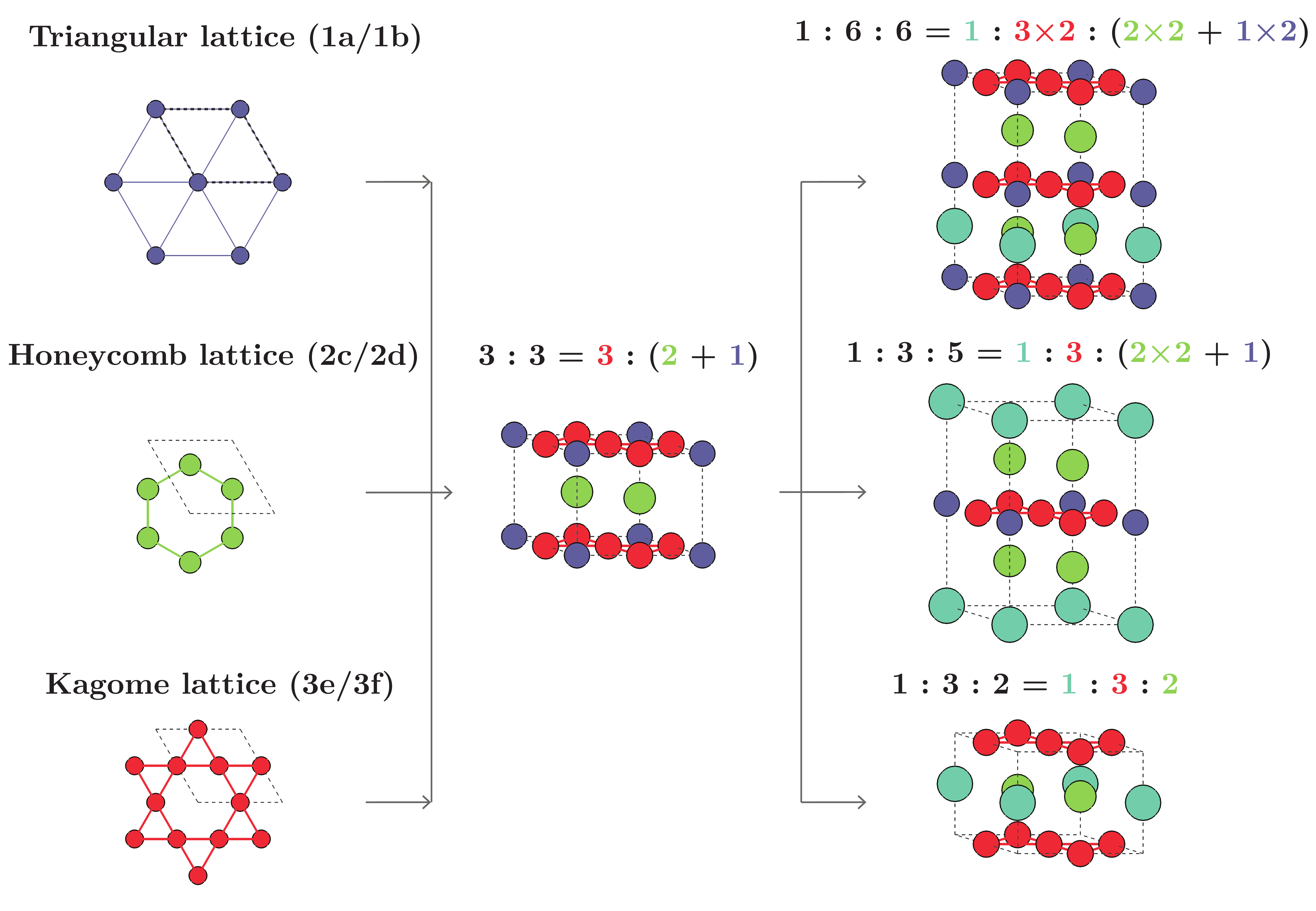}
    \caption{Crystal structures of various kagome material families and their connection to the $1:1$ structure. The $1:1$ (or $3:3$) structure acts as a fundamental LEGO building block for different kagome materials.  }
    \label{fig:IllustrationLEGO}
\end{figure}

We begin by examining the crystal structure in different kagome families, as illustrated in \cref{fig:IllustrationLEGO}. 
\begin{itemize}
    \item In the $1:1$ family, the unit cell contains an equal number of honeycomb, triangular, and kagome layers, with the triangular and kagome lattices lying on the same plane (sharing the same $z$-coordinate).
    \item The $1:6:6$ structure can be obtained by doubling the unit cell of the $1:1$ structure along the $z$-direction and adding an additional atom (triangular lattice) to one of the honeycomb layers. Thus we have the relation $(1:6:6)=2\times(0:3:3)+(1:0:0)$.

    \item In the $1:3:5$ family, the unit cell contains one kagome, two honeycomb, and two triangular layers. The kagome layers are sandwiched between two honeycomb layers within the same unit cell. One of the triangular lattices lies in the same plane as the kagome lattice, while the other triangular lattice, composed of alkali metals, is positioned between two honeycomb layers from adjacent unit cells. Starting from the $3:3$ structure, one can first double the unit cell along the $z$-direction to obtain the $6:6$ structure. Then, one layer formed by the kagome and triangular lattices is replaced by a distinct triangular lattice composed of alkali metals. Thus we have the relation $(1:3:5)=2\times(0:3:3) - (0:3:1)+(1:0:0)$. 
    
    \item The $1:3:2$ family contains the same number of atoms as the $1:1$ family but features triangular and honeycomb lattices formed by different elements. Additionally, unlike the $1:1$ structure, where the triangular lattice lies in the same plane as the kagome lattice, the $1:3:2$ structure places the triangular and honeycomb lattices on the same plane. Thus we have the relation $(1:3:2)=(0:3:3)-(0:0:1)+(1:0:0)$. 
\end{itemize}

In addition to the structural similarity, the electronic band structures of different kagome families also originate from similar LEGO building blocks. In these kagome metals, the kagome lattice is consistently formed by transition metal elements, producing kagome bands near the Fermi level primarily contributed by the $d$-electrons. 
As demonstrated in Ref.~\cite{Jiang2023KagomeA}, the complex \emph{spaghetti}-like electronic structure of the $1:1$ family, exemplified by FeGe, can be understood by partitioning the kagome $d$-orbitals into distinct groups based on the symmetry and chemical analysis. Each group of $d$ orbitals couples to specific $p$ or $s$ orbitals from neighboring lattices, forming separate LEGO building blocks. These blocks effectively describe the kagome band structures and have been successfully extended to the $1:6:6$ and $1:3:5$ families~\cite{Jiang2023KagomeA, Deng2023Two-elementaryA}. In the following \cref{app:ResultsDFTElectronFB}, we show that the $1:3:2$ family, represented by \ch{LaRu3Si2}, also shares similar LEGO building blocks of $d$ orbitals in its band structure.

\subsection{\label{app:ResultsDFTStructure}Crystal structure and phonon properties of \ch{LaRu3Si2}}

\ch{LaRu3Si2} is a prominent member of the kagome $1:3:2$ family. It has been identified as a paramagnetic type-II superconductor with a superconducting transition temperature ($T_c$) of 7.8 K~\cite{Ku1980SuperconductingA, Kishimoto2002MagneticA, Li2011AnomalousA, Mielke2021NodelessA, Mielke2024MicroscopicA, Ushioda2024Two-gapA}. Notably, \ch{LaRu3Si2} stands out among kagome metals due to its superconductivity at ambient pressure, exhibiting the highest reported $T_c$ in this category. It also features a flat band close to the Fermi level without magnetization.

\begin{figure}\centering
    \includegraphics[width=0.7\linewidth]{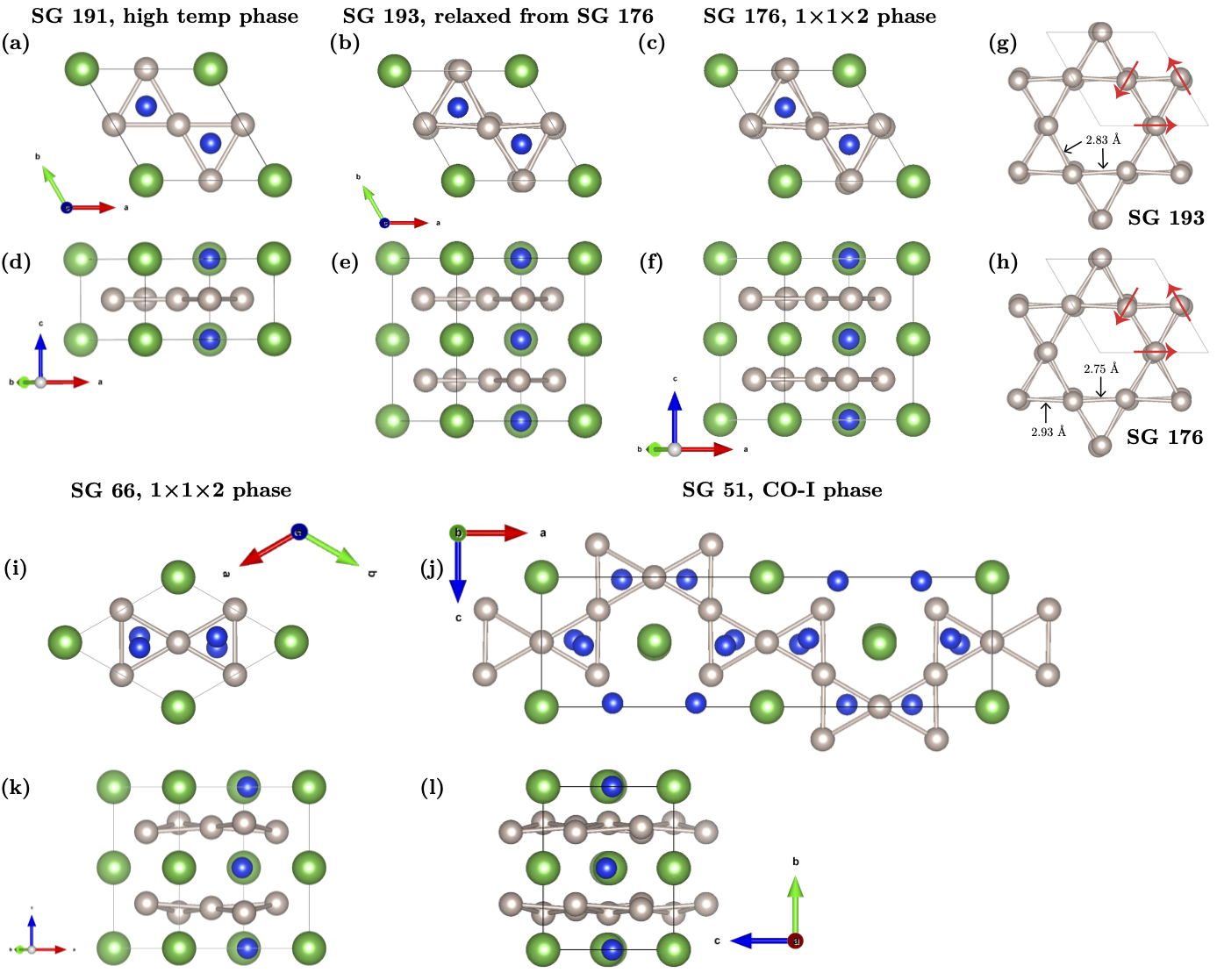}
    \caption{Structure of \ch{LaRu3Si2} in  SG 191 [(a), (d)], 193 [(b), (e)], 176 [(c), (f)], 66 [(i), (k)], and 51 [(j), (l)], where the first row is the top view and the second row is the side view. 
    SG 193, 176, and 66 are possible $1\times1\times2$ structures of the high-temperature structure in SG 191, while SG 51 is a $4\times4\times2$ CDW structure. The SG 193 and 176 structures are nearly identical, with the largest displacements being the inplane movements of Ru atoms. The SG 66 and 51 phases, however, have Ru atoms slightly moved in-plane but largely moved out-of-plane, while Si atoms displace oppositely. 
    Structures are obtained from Ref.~\cite{Plokhikh2024DiscoveryA}. 
    }
    \label{fig:CrystStructAll}
\end{figure}

\begin{figure}[htbp]
    \centering
    \includegraphics[width=0.8\linewidth]{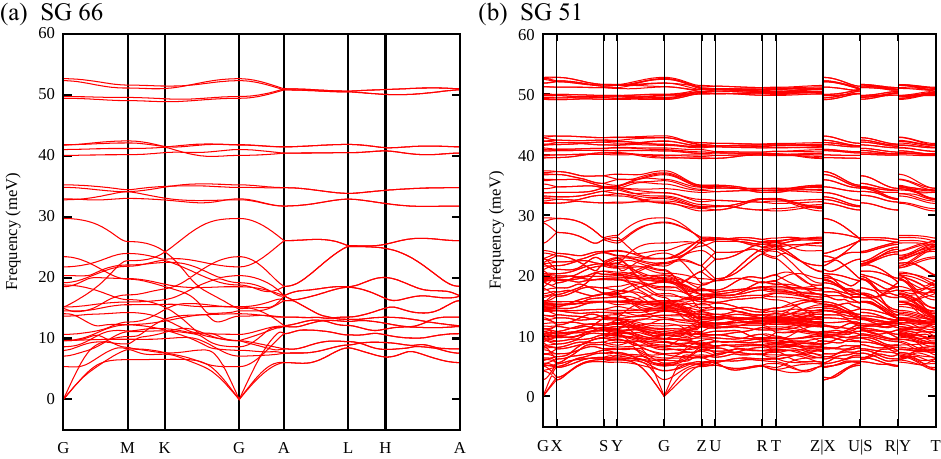}
    \caption{Phonon spectrum in (a) the $1\times1\times2$ CDW phase SG 66 and (b) the $4\times4\times2$ CDW phase SG 51. All three soft phonon modes in SG 191 are eliminated. The phonon spectrum in SG 51 closely resembles that of SG 66 when the Brillouin zone is folded.}
    \label{fig:Ph66and51}
\end{figure}

Muon spin rotation ($\mu$SR) experiments have provided compelling evidence for nodeless $s$-wave superconductivity in \ch{LaRu3Si2}~\cite{Mielke2021NodelessA, Mielke2024MicroscopicA, Ushioda2024Two-gapA}. Additionally, multiple distinct crystal structures of \ch{LaRu3Si2} have been identified, with temperature serving as a critical factor in driving structural transitions. Below is a summary of the key structural phases, as detailed in Ref.~\cite{Plokhikh2024DiscoveryA}:

\begin{itemize}
\item Above 620 K: \ch{LaRu3Si2} crystallizes in space group 191 ($P6/mmm$) as shown in \cref{fig:CrystStructAll}(a)(d), representing the most symmetric structure. In this phase, the unit cell comprises a kagome layer (Wyckoff position $3g$) formed by Ru atoms, a triangular lattice (Wyckoff position $1a$) formed by La atoms, and a honeycomb lattice (Wyckoff position $2c$) formed by Si atoms. We call this phase the SG 191 phase through this manuscript.

\item Between 400 and 620 K: The compound was first reported to adopt space group 176 ($P6_3/m$)~\cite{Ku1980SuperconductingA} (\cref{fig:CrystStructAll}(c)(f)), where the unit cell doubles along the $z$-axis. In this phase, in-plane displacements of Ru atoms in the kagome layer break the $C_{6z}$ symmetry, although a $C_{6z}$-screw symmetry $\tilde{C}_{6z}\equiv\Bqty{C_{6z}|0,0,1/2}$ is retained. DFT calculations show that the SG 176 structure relaxes to SG 193 ($P6_3/mcm$) (\cref{fig:CrystStructAll}(b)(e)). SG 176 is a subgroup of SG 193. Starting from SG 193, SG 176 can be obtained by breaking the in-plane non-symmorphic symmetry $\tilde{C}_{2x}\equiv\Bqty{C_{2x}|0,0,1/2}$. In contrast to report~\cite{Ku1980SuperconductingA}, the structure could not be fully refined in SG 176, but refinement was successfully achieved in SG 66 ($Cccm$) as reported in Ref.~\cite{Plokhikh2024DiscoveryA} (\cref{fig:CrystStructAll}(i)(k)). 
Since the phonon spectra (\cref{fig:Ph191and193and176}{(b) and (c)}) for both SG 193 and 176 phases still manifest soft modes throughout the $q_z=0$ plane, while SG 66 is harmonically stable (\cref{fig:Ph66and51}{(a)}), we will use the SG 66 structure as the structure for this mid-temperature phase.
We refer to this phase with SG 66 structure as the $1 \times 1 \times 2$ ($Cccm$) phase in this work.

\item Around 400 K: The first charge-ordered phase, CO-I, emerges in the super space group $Cmmm(0b0)s00$, with modulation vectors $\vb{q} = (1/4, 0, 0)$, $(0, 1/4, 0)$, and $(1/4, -1/4, 0)$ relative to the $1\times1\times2$ BZ~\cite{Plokhikh2024DiscoveryA, Mielke2024ChargeA}. The crystal structure of this phase is illustrated in \cref{fig:CrystStructAll}(j)(l).

\item A second charge-ordered phase, CO-II, emerges alongside the CO-I phase and is most pronounced in Fe-doped samples ($x = 0.01$). This phase is characterized by modulation vectors $\vb{q} = (1/6, 0, 0)$, $(0, 1/6, 0)$, and $(1/6, -1/6, 0)$ relative to the $1\times1\times2$ BZ~\cite{Plokhikh2024DiscoveryA}. In undoped \ch{LaRu3Si2}, CO-II is observed at a lower transition temperature ($T_c \sim 80 ,\mathrm{K}$)~\cite{Plokhikh2024DiscoveryA, Mielke2024ChargeA}.
\end{itemize}

\begin{figure}[htbp]
    \centering
    \includegraphics[width=0.45\linewidth]{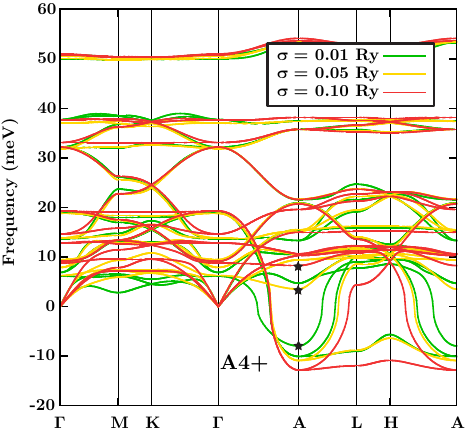}
    \caption{Phonon spectrum of the SG 191 phase at various electronic temperatures. The electronic temperature is controlled by adjusting the smearing parameter in the Fermi-Dirac distribution function during DFT calculations. The one-dimensional mode, marked by a black star, hardens with increasing electronic temperature.}
    \label{fig:Ph191HighT}
\end{figure}

\subsubsection{\label{app:ResultsDFTStructure191} SG 191 ($P6/mmm$) phase}

\begin{figure}[htbp]
    \centering
    \includegraphics[width=\linewidth]{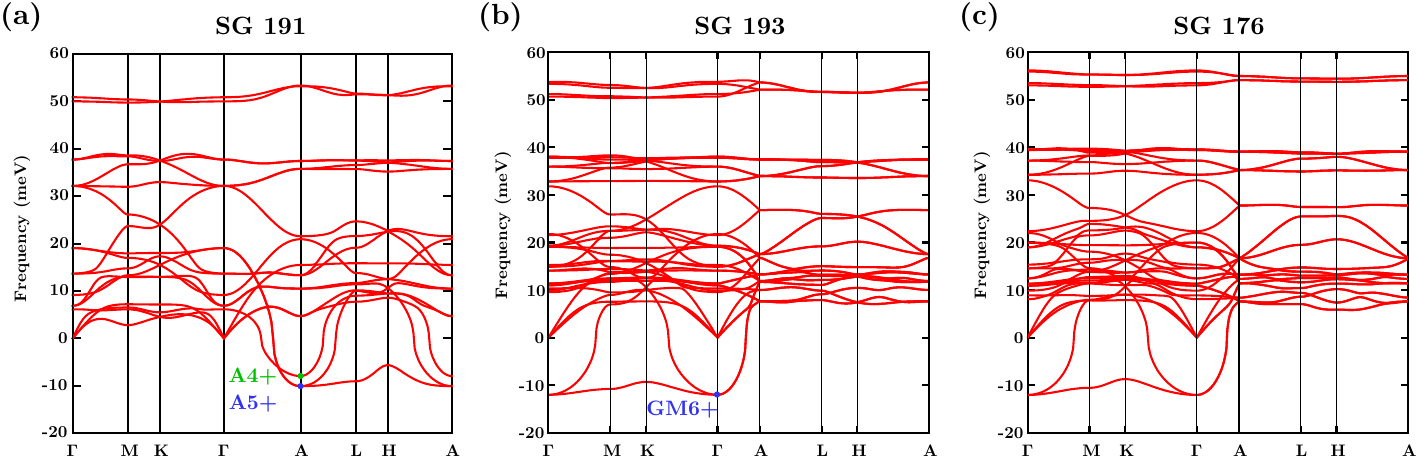}
    \caption{Phonon spectrum of \ch{LaRu3Si2} in (a) high-temperature SG 191 phase and the $1\times1\times2$ phase in SG (b) 193 and (c) 176. In SG 191, there are three soft phonons on the $k_z=\pi$ plane, with a 1D irrep $A_4^+$ and a 2D irrep $A_5^+$ at $A$.
    In the SG 193 and 176 phase, the 1D soft phonon is eliminated but the 2D one persists, even with higher electron temperature (up to 1 eV or 0.3 Ry).}
    \label{fig:Ph191and193and176}
\end{figure}

\begin{figure}[htbp]
    \centering
    \includegraphics[width=\linewidth]{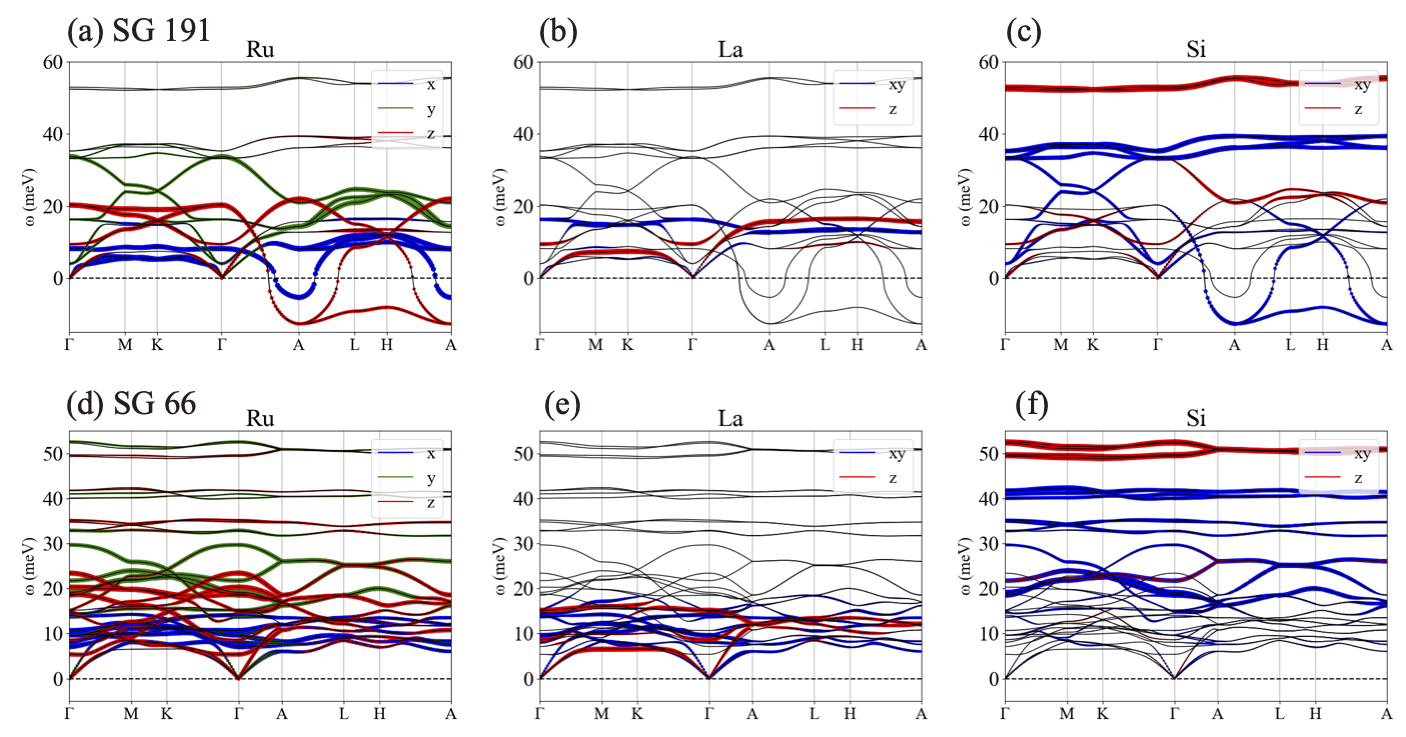}
    \caption{Local-orbital resolved phonon spectrum in SG 191 from (a) Ru, (b) La, and (c) Si. The 2D soft mode at $A$ is mainly from the Ru-$z$ phonon, while the 1D soft mode at $A$ is from the Ru-$x$ phonon. The 1D soft phonon can lead to the SG 193/176 structure, while the 2D soft mode can lead to the SG 66 phase. (d)-(f) are the same but for the SG 66 structure, where the soft modes are eliminated.}
    \label{fig:PhOrbWeight191and66}
\end{figure}

In the SG 191 phase, the structure comprises a kagome lattice of Ru atoms at $z = c/2$ plane ($c$ being the lattice constant in the $z$-direction), a honeycomb lattice of Si atoms at $z = 0$ plane, and a triangular lattice of La atoms also at $z = 0$ plane, occupying Wyckoff positions $3g$, $2c$, and $1a$, respectively [\cref{fig:CrystStructAll}{(a)}]. To explore the stability of this phase at lower temperatures, we performed density functional perturbation theory (DFPT) calculations within the harmonic approximation. The phonon spectrum, presented in \cref{fig:Ph191HighT} and \cref{fig:Ph191and193and176}{(a)}, exhibits significant dynamical instability. Imaginary phonon frequencies spanning the $q_z = \pi$ plane indicate that the high-symmetry phase is unstable. The irreducible representations (irreps) of these unstable phonon modes at the $A$ point were identified, as labeled in \cref{fig:Ph191and193and176}{(a)} (following the convention in Bilbao Crystallographic Server~\cite{Aroyo2006BilbaoA, Aroyo2006BilbaoB}). Among the three unstable modes at $A$, the one-dimensional $A_4^+$ mode is primarily associated with the in-plane motion of Ru atoms within the kagome lattice, specifically directed toward the centers of the hexagons (aligned with the local-$x$ direction of the kagome site). This characteristic is further corroborated by the phonon orbital weights depicted in \cref{fig:PhOrbWeight191and66}(a). 
The $A_4^+$ mode exhibits a strong dependence on electronic temperature. As shown in \cref{fig:Ph191HighT}, increasing the smearing parameter in the Fermi-Dirac distribution within the \textit{ab initio} calculations results in the hardening of this mode. In contrast, the two-dimensional $A_5^+$ mode remains largely unaffected by changes in electronic temperature. The selective hardening of the $A_4^+$ mode while the $A_5^+$ mode persists suggests that the $A_4^+$ phonon softening is likely driven by electron-phonon coupling (EPC). This behavior closely resembles the phonon softening observed in 1T-phase transition metal dichalcogenide (TMD) materials~\cite{Weber2011ExtendedA, Weber2013OpticalA}. 

\subsubsection{\label{app:ResultsDFTStructure193} $1 \times 1 \times 2$ phase in SG 193}

Experimental studies~\cite{Barz1980TernaryA, Ku1980SuperconductingA, Mielke2024ChargeA, Plokhikh2024DiscoveryA} indicate that the room-temperature (above 400 K) structure of \ch{LaRu3Si2} undergoes a $1 \times 1 \times 2$ reconstruction relative to the SG 191 unit cell. To investigate the underlying phonon instabilities, we generated crystal structures based on the one-dimensional (1D) and two-dimensional (2D) softened phonon modes at the $A (0,0,1/2)$ point, corresponding to a $1 \times 1 \times 2$ supercell.
As depicted in \cref{fig:CrystStructAll}{(b) and (e)}, the structure derived from the 1D $A_4^+$ mode belongs to SG 193. However, experimental refinements report the structure in SG 176 ($P6_3/m$, shown in \cref{fig:CrystStructAll}{(c) and (f)})~\cite{Ku1980SuperconductingA}. Notably, SG 176 is a subgroup of SG 193, and it can be obtained from SG 193 by breaking the in-plane symmetry $\tilde{C}_{2x} \equiv \Bqty{C_{2x}|0,0,1/2}$.
Both SG 193 and SG 176 structures involve in-plane displacements of Ru atoms toward the hexagon centers, as illustrated in \cref{fig:CrystStructAll}{(g) and (h)}. The key distinction lies in the nearest-neighbor (NN) bond lengths of Ru atoms: SG 176 exhibits inequivalent NN bonds (2.75 \AA and 2.93 \AA, \cref{fig:CrystStructAll}{(h)}), whereas SG 193 maintains uniform NN bond lengths (2.83 \AA, \cref{fig:CrystStructAll}{(g)}). This structural difference implies that the kagome layers in SG 193 retain in-plane $C_{2x}$ symmetry, while those in SG 176 do not.
Despite these variations, phonon spectra in \cref{fig:Ph191and193and176}{(b) and (c)} reveal that the 2D softened mode from the SG 191 phase remains present in both SG 193 and SG 176, folding onto the $q_z = 0$ plane. This suggests that the phonon instability persists in these two lower-symmetry structures.

\subsubsection{\label{app:ResultsDFTStructure66}$1 \times 1 \times 2$ phase in SG 66}

\begin{figure}[H]
    \centering
    \includegraphics[width=0.7\linewidth]{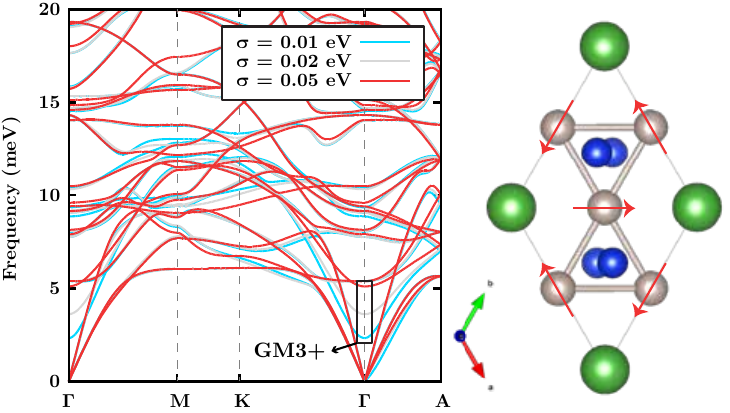}
    \caption{Phonon spectrum of the SG 66 phase at different electronic temperatures, controlled by varying the smearing parameter ($\sigma$) in the Fermi-Dirac distribution. The one-dimensional mode $\Gamma_3^+$ softens as the electronic temperature decreases. The right panel illustrates the atomic displacements corresponding to the $\Gamma_3^+$ eigenvector.}
    \label{fig:Ph66HighT}
\end{figure}

The alternative $1 \times 1 \times 2$ superstructure in SG 66 arises from the condensation of the two-dimensional softened $A_5^+$ phonon mode. 
According to the group-subgroup relationship on \textit{Bilbao Crystallographic Server}~\cite{Aroyo2006BilbaoA, Aroyo2006BilbaoB}, condensing the $A_5^+$ irrep in SG 191 could lead to structure in SG 66, SG 65, or SG 10. 
After full structural relaxation in DFT, the condensed phonon mode leads to a structure belonging to SG 66 ($Cccm$). In addition to the previously reported SG 176 structure from powder X-ray diffraction (XRD)~\cite{Ku1980SuperconductingA}, recent experimental studies have identified an alternative $1 \times 1 \times 2$ structural phase in SG 66~\cite{Plokhikh2024DiscoveryA, Mielke2024ChargeA}, which is consistent with our theoretical analysis. This phase is primarily characterized by out-of-plane ($z$-direction) displacements of Ru atoms, as illustrated in \cref{fig:CrystStructAll}{(k)}. Notably, the SG 66 structure is harmonically stable at low temperatures, as confirmed by the phonon spectrum shown in \cref{fig:Ph66and51}{(a)}. 
Upon further cooling (modeled by decreasing the smearing parameter in the Fermi-Dirac distribution function in DFT from $0.05 \,\mathrm{eV}$ to $0.01 \,\mathrm{eV}$), a 1D phonon mode at $\Gamma$ softens but does not become imaginary, as indicated by the black rectangle in \cref{fig:Ph66HighT}. This mode corresponds to the same atomic displacements associated with the $A_4^+$ mode in the SG 191 phonon spectrum -- namely, the in-plane movement of Ru atoms within the kagome lattice toward the centers of the hexagons (\cref{fig:Ph66HighT}). The softening reflects the strong electron-phonon coupling of this mode. 

\begin{figure}[H]
    \centering
    \includegraphics[width=0.6\textwidth]{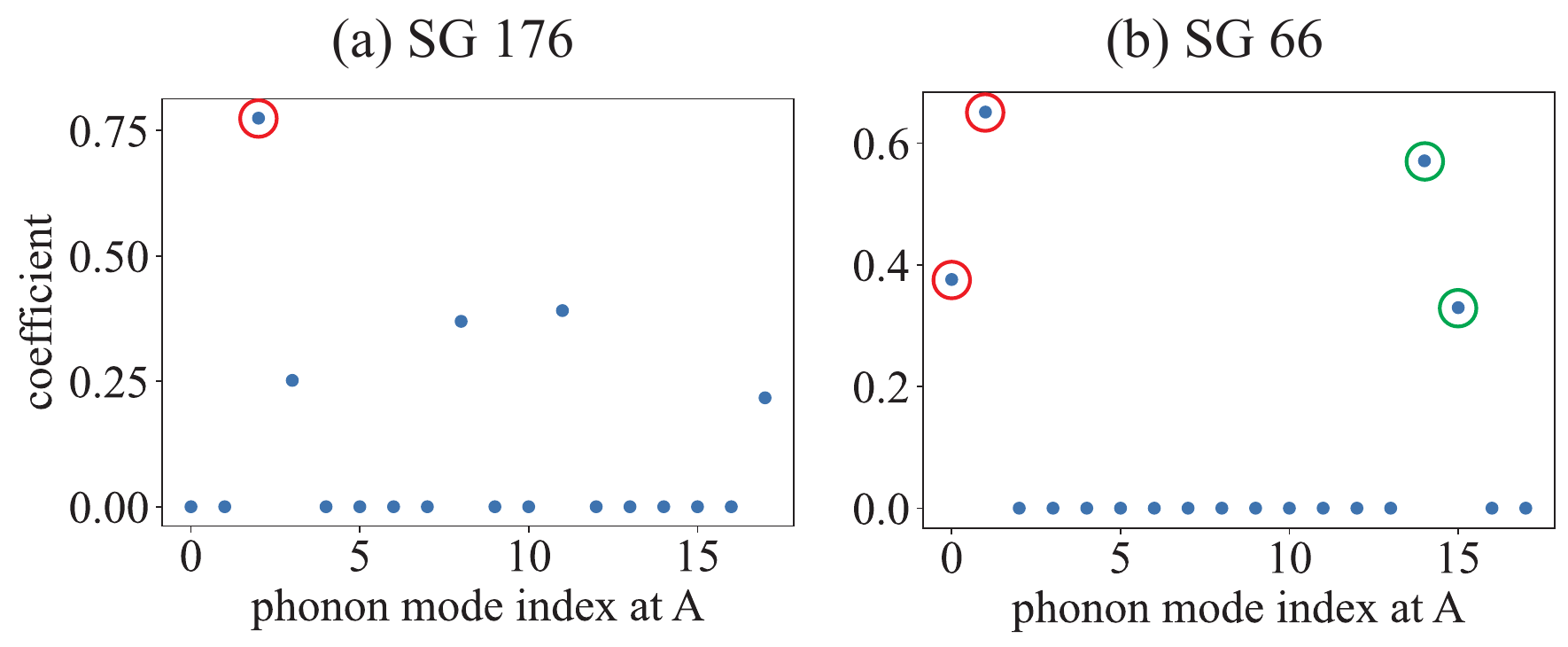}
    \caption{(a) Decomposition of the CDW structure in SG 176 using the phonon eigenmodes at $A$ from SG 191. The largest coefficient comes from the 1D soft phonon, i.e., the phonon mode with index 3 (marked with red circle). (b) Same as (a) but for CDW structure in SG 66. The 2D soft phonon has the largest coefficient (phonon mode index 0, 1, marked by red), along with another 2D optical phonon (index 14 and 15, marked by green). In the plot, the phonon mode indices 0-17 represent the 18 phonon modes at A in SG 191, as shown in \cref{fig:PhOrbWeight191and66}{(a)-(c)}. 
    \label{fig:CondensedPhonon} 
    }
\end{figure}

To further verify that the $1 \times 1 \times 2$ structures originate from the condensation of the soft phonons in the high-temperature SG 191 phase, we compare the experimental $1 \times 1 \times 2$ structure in SG 176 and SG 66 with the eigenmodes of the soft phonon at the $A$ point in SG 191. The steps are as follows:
\begin{itemize}
\item Starting from the experimental structure in SG 191, we make a $1 \times 1 \times 2$ supercell and use $\mathbf{u}_{i\mu}^{191}$ to denote the atomic positions, where $i,\mu$ are the atom and coordinate indices. Similarly, $\mathbf{u}_{i\mu}^{176/66}$ denotes the atomic positions in SG 176 and 66. Then the displacement vector $\Delta\mathbf{u}_{i\mu}^{176/66}=\mathbf{u}_{i\mu}^{176/66}-\mathbf{u}_{i\mu}^{191}$. 

\item We then compute the phonon eigenmodes, $\{\mathbf{v}^n_{i\mu}\}$ ($n$ being the phonon mode index), at $A$ in SG 191, which form a complete basis for displacements in the $1 \times 1 \times 2$ unit cell.

\item Expand $\Delta\mathbf{u}_{i\mu}^{176/66}$ as a linear combination of the eigenmodes: $\Delta\mathbf{u}_{i\mu}^{176/66} = \sum_n c_n \mathbf{v}^n_{i\mu}$, where $c_n$ is the expansion coefficient.

\item Compute $c_n$ to quantify the contribution of each eigenmode. 
\end{itemize}
The computed absolute values of the $c_n$ coefficients are shown in \cref{fig:CondensedPhonon}. 
For the SG 176 phase, the largest coefficient comes from the 1D soft mode, which is mainly formed by the local $x$ movements of Ru. We also compute the overlap between the displacement vector $\Delta\mathbf{u}^{176}$ and this 1D soft mode $\mathbf{v}^{n}$, i.e., $\sqrt{\Delta\mathbf{u}^{176}\cdot \mathbf{v}^{n}}$, which gives a large overlap of $93.1\%$. For the SG 66 phase, the largest coefficients come from the 2D soft phonon [i.e., the phonon modes with index 0 and 1 in \cref{fig:CondensedPhonon}{(b)}] together with a 2D optical phonon [i.e., the phonon modes with index 14 and 15 in \cref{fig:CondensedPhonon}{(b)}, which are the optical modes at about 40 meV in \cref{fig:PhOrbWeight191and66}{(c)}]. The 2D soft phonon is mainly given by the $z$-directional phonon of Ru and the $xy$ phonon of Si, while the 2D optical phonon is given by the $xy$ phonon of Si at about 40 meV. The overlap between the CDW displacement vector and the 2D soft mode (combined using decomposition coefficients) is $97.4\%$. Thus we conclude that the $1\times1\times2$ structures in SG 176 and 66 can be understood from the condensation of the 1D and 2D soft phonons in the SG 191 phase, respectively. The SG 193 phase is very close to the SG 176 phase, thus we omit it for simplicity.

\subsubsection{\label{app:ResultsDFTStructureCDW} Charge density wave phases of $4 \times 4 \times 2$ and $6 \times 6 \times 2$ reconstruction}

At lower temperatures, two charge density wave (CDW) phases have been identified in \ch{LaRu3Si2} through X-ray diffraction (XRD) measurements: the $4 \times 4 \times 2$ phase, which emerges below 400 K, and the $6 \times 6 \times 2$ phase, which appears below 80 K (or below 170 K in Fe-doped samples)~\cite{Plokhikh2024DiscoveryA, Mielke2024ChargeA}.
While the $4 \times 4 \times 2$ phase has been reported to exhibit symmetry within the superspace group $Cmmm(0b0)s00$~\cite{Plokhikh2024DiscoveryA, Mielke2024ChargeA} (note that the notation of superspace groups is different from SGs), the precise symmetries of both CDW phases remain ambiguous and require further refinement. Structural analysis reveals that atomic displacements associated with the transition from the $Cccm$ phase to the $4 \times 4 \times 2$ phase are relatively minor (0.05 \AA) compared to the more significant displacements observed in the transition from the high-symmetry $P6/mmm$ phase to the $Cccm$ phase (0.2 \AA). Moreover, as illustrated in \cref{fig:Ph66and51}{(b)}, the phonon spectrum of the $4 \times 4 \times 2$ phase is harmonically stable. Given the negligible impact of these distortions, it is reasonable to overlook the effects of CDW-induced lattice distortion on SC or other less detailed properties in subsequent discussions. Moreover, due to computational constraints, the electron-phonon coupling and superconductivity properties have primarily been analyzed for the $P6/mmm$ phase and the $Cccm$ $1 \times 1 \times 2$ phase.

\subsection{\label{app:ResultsDFTElectron} Electron properties}

In this section, we explore the electronic structure of \ch{LaRu3Si2} across its various CDW phases. The key characteristic of the system is the presence of a quasi-flat band below the Fermi level ($E_F$), which leads to a pronounced peak in the density of states (DOS). 

\subsubsection{\label{app:ResultsDFTElectronFB} Quasi-flat band and Fermi surfaces}
\begin{figure}[H]
    \centering
    \includegraphics[width=0.5\linewidth]{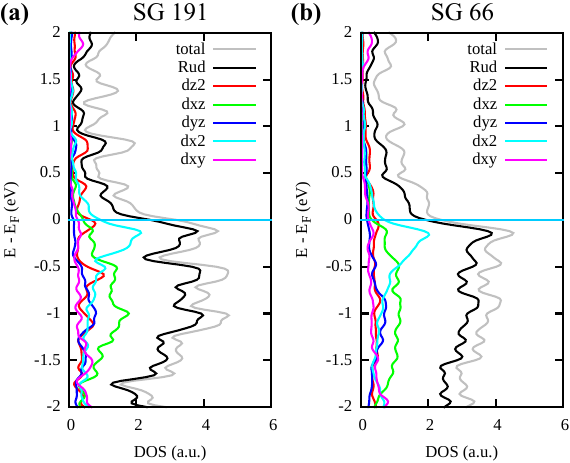}
    \caption{Total density of states (DOS) and local orbital resolved DOS of (a) SG 191 and (b) SG 66. The peak below the $E_F$ at about $\sim150\,\mathrm{meV}$ is mainly contributed by local $d_{x^2-y^2}$ in both phases. In SG 66, the DOS at the $E_F$ is reduced by about 25\% due to the structurll reconstruction. For better comparison with SG 191, the DOS in SG 66 is divided by 2 because of the lattice doubling along the $z$ direction. The DOS is computed without SOC and should be times 2 if the spin degree of freedom is considered. 
    }
    \label{fig:DOS191and66}
\end{figure}

\begin{figure}[H]
    \centering
    \includegraphics[width=0.8\linewidth]{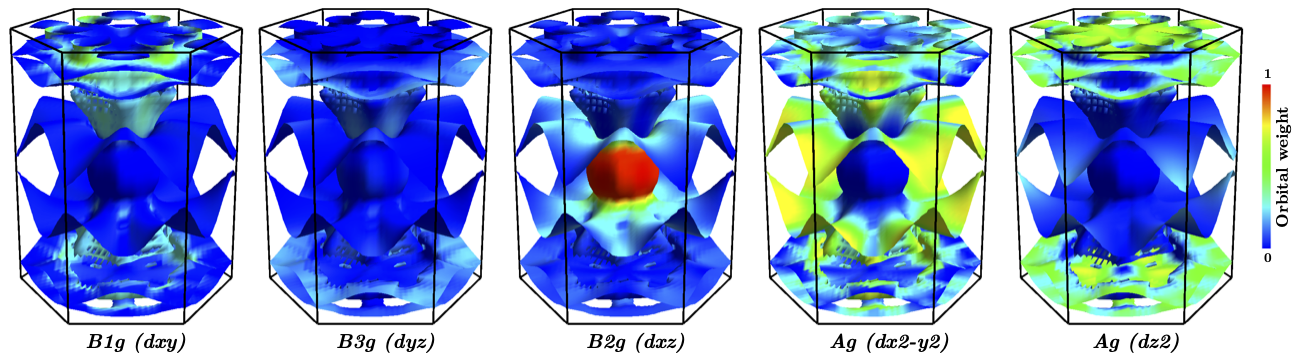}
\caption{Orbital-resolved Fermi surfaces of the high-temperature pristine phase of \ch{LaRu3Si2} in SG 191, with the color denoting the orbital weight of Ru $d$.}
    \label{fig:FS191}
\end{figure}

We first tabulate the atomic Wyckoff positions and the induced elementary band representations (EBRs)~\cite{Bradlyn2017TopologicalA, Cano2018BuildingA, Elcoro2021MagneticA} in \cref{tab:pos2abr} in the high-temperature SG 191 phase of \ch{LaRu3Si2}.
The electronic structure is dominated by a quasi-flat band below the Fermi level $E_F$, originating primarily from the $d_{x^2-y^2}$ orbitals of the kagome Ru atoms, as seen in the orbital-resolved band structures in \cref{fig:Bands191}. This quasi-flat band has a maximum at the $K$ point on the $k_z=0$ plane and significantly contributes to the DOS at approximately $-150 \,\mathrm{meV}$, as shown in \cref{fig:DOS191and66}{(a)}. Additionally, a partial flat band from the Ru $d_{z^2}$ orbitals is present on the $k_z=\pi$ plane, contributing to a smaller DOS peak at approximately $50 \,\mathrm{meV}$ below the $E_F$. The Fermi surfaces, as shown in \cref{fig:FS191}, are primarily contributed by the $d_{x^2-y^2}$, $d_{z^2}$, and $d_{xz}$ orbitals of the Ru atoms.
The $d_{xz}$ orbitals contribute to a more dispersive band with a hole pocket near $\Gamma$, which is less relevant to the SC in \ch{LaRu3Si2} (for example, \cref{app:ResultsDFTElectronPhononLaRu3Si2} shows the SC gap function which is small on the $d_{xz}$ FS). 
We remark that the $d$ orbitals in \ch{LaRu3Si2} have similar distributions in the band structure as the 1:1 family FeGe, which could be understood from the LEGO building blocks in FeGe~\cite{Jiang2023KagomeA}.

\begin{figure}[H]
    \centering
\includegraphics[width=0.3\linewidth]{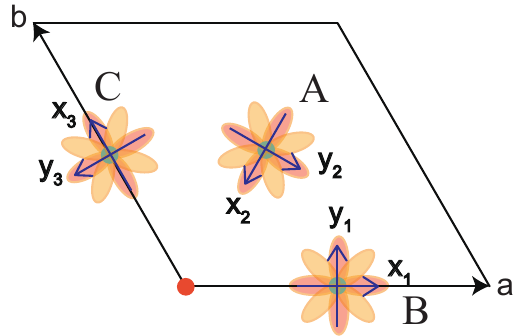}
    \caption{Illustration of the Ru $d$ orbitals defined on local coordinate systems on the kagome lattice. The local axes are shown by blue arrows. }
    \label{fig:KagomeLocalOrbs}
\end{figure}

\begin{figure}[H]
    \centering
\includegraphics[width=\linewidth]{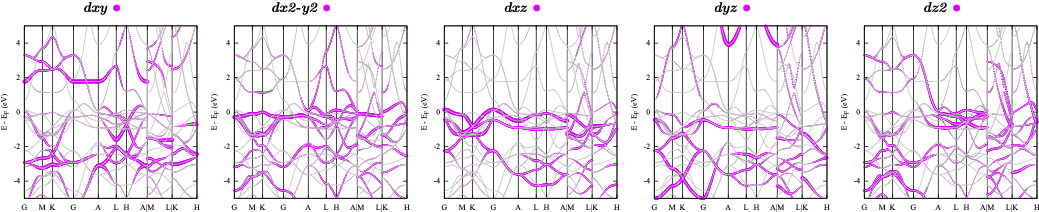}
    \caption{Orbital weights of five $d$ orbitals of kagome Ru in the band structure of SG 191. The orbitals are defined in the local coordinates in \cref{fig:KagomeLocalOrbs}. A quasi-flat band close to the $E_F$ on the $k_z=0$ plane is mainly contributed by the $d_{x^2-y^2}$ orbital. }
    \label{fig:Bands191}
\end{figure}

\begin{figure}[H]
    \centering
    \includegraphics[width=\linewidth]{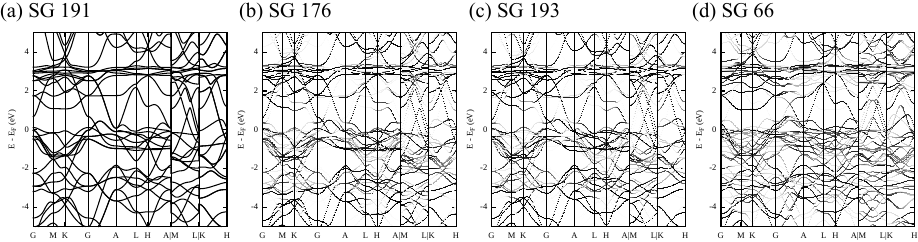}
    \caption{The electronic structure of \ch{LaRu3Si2} along high symmetry $\vb{k}$ path in (a) SG 191, and unfolded band structure from (b) SG 176, (c) SG 193, and (c) SG 66 to the same BZ in SG 191. The unfolded bands in three $1\times1\times2$ phases (b)-(d) are similar to the SG 191 bands in (a), with weak reconstructions at $E_F$ that slightly reduce the DOS at $E_F$.}
    \label{fig:UnfoldedAll}
\end{figure}

\begin{figure}\centering
\includegraphics[width=0.9\linewidth]{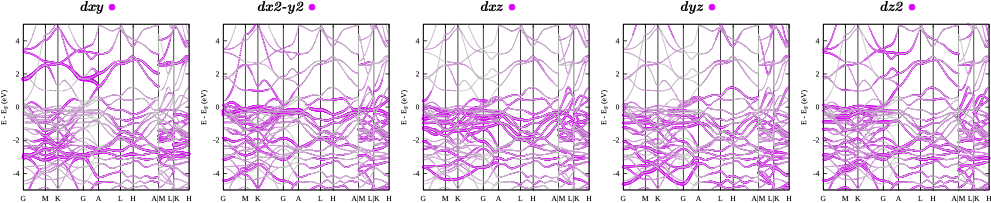}
    \caption{Orbital weights of five $d$ orbitals of kagome Ru in the band structure of $1\times1\times2$ phase in SG 66 $Cccm$. Note that although the $C_3$ symmetry is broken in SG 66, we use the same local coordinates defined in \cref{fig:KagomeLocalOrbs}. Compared with the orbital weights in SG 191 (\cref{fig:Bands191}), the weights in SG 66 remain mostly unchanged after a twofold folding along the $k_z$-direction, due to the doubled unit cell. 
    }
    \label{fig:Bands66}
\end{figure}

The unfolded band structures for the $1\times1\times2$ phases in SG 176, SG 193, and SG 66 are shown in \cref{fig:UnfoldedAll}{(b)-(d)}, demonstrating a close resemblance to those in the high-symmetry SG 191 phase.
In \cref{fig:Bands66}, the DFT-calculated band structures and orbital contributions for the $1\times1\times2$ SG 66 phase are presented, revealing only minor modifications compared to SG 191, apart from a twofold band folding along the $k_z$ direction.
Furthermore, a comparison of the orbital-resolved DOS in SG 191 and SG 66 is provided in \cref{fig:DOS191and66}. In SG 66, the DOS at $E_F$ is slightly reduced by about 25\% due to the structural reconstruction, decreasing from 2.855 states per unit cell per eV in SG 191 to 2.206 states per unit cell per eV in SG 66 (after normalization for direct comparison).
Despite these structural differences, the primary DOS peak at approximately $-150$ meV persists in both SG 191 and SG 66. This peak is primarily attributed to the Ru $d_{x^2-y^2}$ orbitals, defined according to the local coordinate system illustrated in \cref{fig:KagomeLocalOrbs}.

\begin{table}[t!]
    \centering
    \caption{The atomic valence electrons and their induced band representations of \ch{LaRu3Si2} in the pristine phase SG 191.}
    \label{tab:pos2abr}
    \begin{tabularx}{0.6\linewidth}{cccCrLc}
        \hline\hline  
        Atom & WKP ($q$) & Symm. & Conf. & & Irreps ($\rho$) & ABRs ($\rho@q$) \\ 
        \hline
        La & $1a$ & $D_{6h}$ & $s^2d^1$ & $A_{1g}$:&$s$ & \\
        & & & & $A_{2u}$:&$p_z$ & \\
        & & & & $E_{1u}$:&$p_x,p_y$ & \\
        & & & & $A_{1g}$:&$d_{z^2}$ & $A_{1g}@1a$ \\
        & & & & $E_{2g}$:&$d_{x^2-y^2},d_{xy}$ & $E_{2g}@1a$ \\
        & & & & $E_{1g}$:&$d_{xz},d_{yz}$ & $E_{1g}@1a$ \\
        \hline
        Ru & $3g$ & $D_{2h}$ & $s^1d^7$ & $A_g$:&$d_{z^2}$ & $A_g@3g$ \\
        & & & & $A_g$:&$d_{x^2-y^2}$ & $A_g@3g$ \\
        & & & & $B_{1g}$:&$d_{xy}$ & $B_{1g}@3g$ \\
        & & & & $B_{2g}$:&$d_{xz}$ & $B_{2g}@3g$ \\ 
        & & & & $B_{3g}$:&$d_{yz}$ & $B_{3g}@3g$ \\
        \hline
        Si & $2c$ & $D_{3h}$ & $s^2p^2$ & $A_1'$:&$s$ & $A_1'@2c$ \\
        & & & & $A_2''$:&$p_z$ & $A_2''@2c$ \\
        & & & & $E'$:&$p_x,p_y$ & $E'@2c$ \\
        \hline\hline  
    \end{tabularx}
\end{table}

\begin{table}[t!]
    \centering
    \caption{Irreps of site symmetry group $mmm (D_{2h})$ for WKP $3g$ in SG 191 ($P6/mmm$).}
    \label{tab:SiteSymmetryGroupD2h}
    \begin{tabular}{c|c|c|c|c|c|c|c|c}
        \hline\hline
        Irrep ($\rho$) & $C_{2z}$ & $C_{2y}$ & $C_{2x}$ & $\mathcal{P}$ & $\mathcal{M}_{z}$ & $\mathcal{M}_{y}$ & $\mathcal{M}_{x}$ & orbitals \\
        \hline
        $A_{g}$ & $1$ & $1$ & $1$ & $1$ & $1$ & $1$ & $1$ & $s, d_{x^2-y^2}, d_{z^2}$ \\
        $B_{1g}$ & $1$ & $-1$ & $-1$ & $1$ & $1$ & $-1$ & $-1$ & $d_{xy}$ \\
        $B_{2g}$ & $-1$ & $1$ & $-1$ & $1$ & $-1$ & $1$ & $-1$ & $d_{xz}$ \\
        $B_{3g}$ & $-1$ & $-1$ & $1$ & $1$ & $-1$ & $-1$ & $1$ & $d_{yz}$ \\
        $A_{u}$ & $1$ & $1$ & $1$ & $-1$ & $-1$ & $-1$ & $-1$ & \\
        $B_{1u}$ & $1$ & $-1$ & $-1$ & $-1$ & $-1$ & $1$ & $1$ & $p_{z}$ \\
        $B_{2u}$ & $-1$ & $1$ & $-1$ & $-1$ & $1$ & $-1$ & $1$ & $p_{y}$ \\
        $B_{3u}$ & $-1$ & $-1$ & $1$ & $-1$ & $1$ & $1$ & $-1$ & $p_x$ \\
        \hline\hline
    \end{tabular}
\end{table}

\subsubsection{\label{app:ResultsDFTElectronDMFT} Correlation effect}

\begin{figure}[H]
    \centering
    \includegraphics[width=0.8\linewidth]{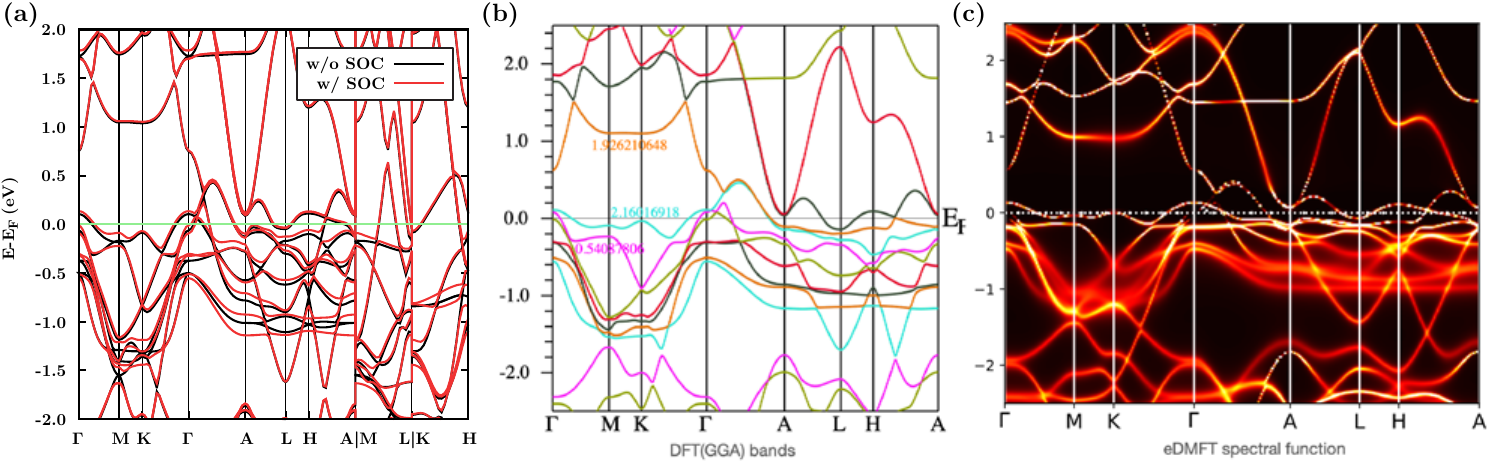}
    \caption{(a) The DFT band structure with and without SOC, indicating SOC has negligible effects on the overall band structure and the Fermi surfaces. (b) The DFT bands with SOC and (c) eDMFT spectral function at $T=100 \,\mathrm{K}$. The eDMFT bands are very close to DFT bands, with the largest mass renormalization $Z$ being 0.66, showing a weak correlation effect in \ch{LaRu3Si2}.  
    }
    \label{fig:SOCandDMFT191}
\end{figure}

From the comparison of band structures with and without spin-orbit coupling (SOC) in \cref{fig:SOCandDMFT191}{(a)}, one can observe that SOC has a minor impact on the overall band structure. Consequently, all subsequent calculations are performed without considering SOC effects. However, band crossings below $E_F$ at about $-0.2$ eV without SOC are observed on the $k_z=0$ plane between the $d_{x^2-y^2}$ and $d_{xz}$ bands, protected by $\mathcal{M}_z$ symmetry. When SOC is included, this crossing opens a gap, leading to a gapped quasi-flat band on the $k_z = 0$ plane.
To further explore the role of correlation effects in \ch{LaRu3Si2}, dynamical mean-field theory (DMFT) calculations were conducted using a combination of density functional theory (DFT) and embedded DMFT (eDMFT), as implemented in \textsc{WIEN2k} and the \textsc{Rutgers eDMFT} code~\cite{Blaha2020WIEN2k:A, Haule2010DynamicalA, Haule2015FreeA, Haule2016ForcesA, Haule2007QuantumA}. The spectral function obtained at $T = 100\,\mathrm{K}$ from the eDMFT calculations aligns closely with the DFT band structure [\cref{fig:SOCandDMFT191}{(b) and (c)}], as does the Fermi surface, indicating weak electron correlation effects and supporting the conclusion that this material is highly itinerant. 
The estimated mass renormalization factor, $Z \approx 0.66$, remains relatively small, further supporting the conclusion that superconductivity in \ch{LaRu3Si2} is predominantly mediated by electron-phonon coupling rather than strong electronic correlations. Given these results, correlation effects are not considered in the subsequent analysis of superconductivity in this study.

\subsection{\label{app:ResultsDFTElectronPhonon} Electron-phonon coupling and superconductivity}

In this section, we explore the electron-phonon coupling and superconducting properties of \ch{LaRu3Si2}. 
We begin by providing a brief overview of the electron-phonon coupling mechanism and its relationship with key superconducting quantities, including the critical temperature $T_c$ and the electron-phonon coupling constant $\lambda$. 
Following this, we calculate the EPC strength $\lambda$ and the corresponding $T_c$ for both the high-temperature SG 191 phase and the intermediate-temperature $1 \times 1 \times 2$ phase in SG 66 ($Cccm$). These calculations are performed for both undoped and doped conditions to comprehensively assess the impact of doping on the superconducting properties of \ch{LaRu3Si2}. Notably, while the SG 191 phase exhibits soft phonon modes, we still analyze its superconducting properties by neglecting these modes, allowing for a direct comparison with the results in SG 66. 

\subsubsection{\label{app:IntroEPCinDFT} EPC from \textit{ab-initio} calculation}
First, we review the EPC from first principles.
The first-order electron-phonon matrix element of a generic electron-phonon-coupled system can be computed using the density functional perturbation theory (DFPT) as:
\begin{equation}\begin{aligned}\label{eq:app-gmnnu}
    \bqty{\tilde{g}(\vb{k},\vb{k}')}_{n'n,\nu} \eq \frac{1}{\sqrt{2\omega_{\vb{k}'-\vb{k},\nu}}} \bqty{g(\vb{k},\vb{k}')}_{n'n,\nu}  
    = \frac{1}{\sqrt{2\omega_{\vb{k}'-\vb{k},\nu}}} \mel{\psi_{\vb{k}',n'}}{\partial_{\vb{k}'-\vb{k},\nu} \hat{V}}{\psi_{\vb{k},n}}
\end{aligned}\end{equation}
which quantifies a scattering process between the Kohn-Sham eigenstates $\ket{\psi_{\vb{k}' n'}}$ and $\ket{\psi_{\vb{k} n}}$ computed by density functional theory (DFT), where $n$ is the band index. The matrix elements can be conveniently expressed as $\bqty{\tilde{g}(\vb{k}, \vb{q})}_{n'n,\nu}$ instead of $\bqty{\tilde{g}(\vb{k}, \vb{k}+\vb{q})}_{n'n,\nu}$ using the phonon wavevector $\vb{q} = \vb{k}' - \vb{k}$:
\begin{equation}\begin{aligned}\label{eq:epcDefinitionAbinitio}
    \bqty{\tilde{g}(\vb{k},\vb{q})}_{n'n,\nu} \eq \frac{1}{\sqrt{2\omega_{\vb{q},\nu}}} \mel{\psi_{\vb{k}+\vb{q},n'}}{\partial_{\vb{q},\nu} \hat{V}}{\psi_{\vb{k},n}}.
\end{aligned}\end{equation}
In \cref{eq:epcDefinitionAbinitio}, 
the operator $\partial_{\vb{q},\nu} \hat{V}$ is the derivative of the self-consistent Born-Oppenheimer potential (obtained from DFT, with both electronic and ionic contributions) with respect to a collective ionic displacement corresponding to a phonon with branch index $\nu$ and momentum $\vb{q}$.
The nucleus masses are \emph{already} included in the phonon eigenmodes.

The electron ($\Sigma$) and phonon ($\Pi$) self-energies at the temperature $T$ for metals and doped semiconductors are given by~\cite{Giustino2017Electron-phononA, Noffsinger2010EPW:A}
\begin{equation}\begin{aligned}\label{eq:epcElectronSelfen}
    \Sigma_{\vb{k},n}(\omega, T) \eq \sum_{n',\nu} \int_{\text{BZ}} \frac{\dd{\vb{q}}}{\Omega_{\text{BZ}}} \abs{\bqty{\tilde{g}(\vb{k}, \vb{q})}_{n'n,\nu}}^2 \bqty{
    \frac{n_{\vb{q},\nu}(T) + f_{\vb{k}+\vb{q},n'}(T)}{\omega - \pqty{\epsilon_{\vb{k}+\vb{q},n'} - E_F} + \omega_{\vb{q},\nu} + i \eta} 
    + \frac{n_{\vb{q},\nu}(T) + 1 - f_{\vb{k}+\vb{q},n'}(T)}{\omega - \pqty{\epsilon_{\vb{k}+\vb{q},n'} - E_F} - \omega_{\vb{q},\nu} + i \eta},
    }
\end{aligned}\end{equation}
\begin{equation}\begin{aligned}\label{eq:epcPhononSelfen}
    \Pi_{\vb{q},\nu}(\omega_{\qq,\nu}, T) \eq -2 \sum_{n', n} \int_{\text{BZ}} \frac{\dd{\vb{k}}}{\Omega_{\text{BZ}}} \abs{\bqty{\tilde{g}(\vb{k}, \vb{q})}_{n'n,\nu}}^2
    \frac{f_{\vb{k},n}(T) - f_{\vb{k}+\vb{q},n'}(T)}{\epsilon_{\vb{k}+\vb{q},n'} - \epsilon_{\vb{k},n} - \omega_{\vb{q},\nu} - i \eta}.
\end{aligned}\end{equation}
In \cref{eq:epcPhononSelfen}, the factor $2$ accounts for the spin degeneracy, $E_F$ is the Fermi energy, $n_{\vb{q},\nu}(T) = \bqty{\exp( \omega_{\vb{q},\nu}/T ) - 1}^{-1}$ is the Bose-Einstein distribution, $f_{\vb{k},n}(T)$ is the electronic (Fermi-Dirac) occupation at wavevector $\vb{k}$ and band $n$, and $\eta$ is a small positive real parameter which ensures the correct analytical structure of the self-energies and avoids numerical instabilities. The physical interpretation of $\eta$ is related to the finite lifetime of the electronic states. The integrals extend over the Brillouin Zone (BZ) of volume $\Omega_{\text{BZ}}$.

The electron and phonon linewidths can be obtained from the imaginary part of the electron ($\Sigma''=\Im \Sigma$) and phonon ($\Pi''=\Im \Pi$) self-energies~\cite{Giustino2007Electron-phononA}:
\begin{equation}\begin{aligned}
    \Sigma_{\vb{k},n}''(\omega, T) \eq \pi \sum_{n',\nu} \int_{\text{BZ}} \frac{\dd{\vb{q}}}{\Omega_{\text{BZ}}} \abs{\bqty{\tilde{g}(\vb{k}, \vb{q})}_{n'n, \nu}}^2 
    \left\{
    \bqty{n_{\vb{q},\nu}(T) + f_{\vb{k}+\vb{q},n'}(T)} \delta(\omega - (\epsilon_{\vb{k}+\vb{q},n'} - E_F) + \omega_{\vb{q}, \nu}) \right. \\
    & + \left.
    \bqty{n_{\vb{q},\nu}(T) + 1 - f_{\vb{k}+\vb{q},n'}(T)} \delta(\omega - (\epsilon_{\vb{k}+\vb{q},n'} - E_F) - \omega_{\vb{q}, \nu}) \right\}
\end{aligned}\end{equation}
\begin{equation}\begin{aligned}
    \gamma_{\vb{q}, \nu} = \Pi_{\vb{q},\nu}''(\omega_{\vb{q},\nu}, T) \eq 2 \pi \sum_{n', n} \int_{\text{BZ}} \frac{\dd{\vb{k}}}{\Omega_{\text{BZ}}} \abs{\bqty{\tilde{g}(\vb{k}, \vb{q})}_{n'n, \nu}}^2 
    \bqty{f_{\vb{k},n}(T) - f_{\vb{k}+\vb{q}, n'}(T)} \delta\pqty{\epsilon_{\vb{k}+\vb{q},n'} - \epsilon_{\vb{k},n} - \omega_{\vb{q},\nu}}.
\end{aligned}\end{equation}
Under the double-delta approximation (see derivation at the end of this section), the phonon linewidth $\gamma_{\vb{q},\nu}$ is expressed as 
\begin{equation}\begin{aligned}
    \gamma_{\vb{q},\nu} \eq 2\pi \omega_{\vb{q},\nu} \sum_{n',n} \int_{\mathrm{BZ}} \frac{\dd{\vb{k}}}{\Omega_{\mathrm{BZ}}} \abs*{\bqty{\tilde{g}(\vb{k},\vb{q})}_{n'n,\nu}}^2 \delta(\epsilon_{\vb{k},n} - E_F) \delta(\epsilon_{\vb{k}+\vb{q},n'} - E_F)
\end{aligned}\end{equation}
Finally, the EPC strength associated with a specific phonon mode $\nu$ and wavevector $\vb{q}$ is:
\begin{equation}\begin{aligned}\label{eq:lambda_qn}
    \lambda_{\vb{q},\nu} \eq \frac{1}{D(E_F) \omega_{\vb{q},\nu}} \sum_{n' n} \int_{\text{BZ}} \frac{\dd{\vb{k}}}{\Omega_{\text{BZ}}} \abs{\bqty{\tilde{g}(\vb{k}, \vb{q})}_{n'n, \nu}}^2
    \delta\pqty{\epsilon_{\vb{k},n} - E_F} \delta\pqty{\epsilon_{\vb{k}+\vb{q},n'} - E_F} \\
    \eq \frac{\gamma_{\vb{q},\nu}}{2 \pi D(E_F) \omega_{\vb{q},\nu}^2},
\end{aligned}\end{equation}
where $D(E_F)$ is the density of states per spin at the Fermi level. 
From \cref{eq:epcDefinitionAbinitio}, $\bqty{\tilde{g}(\vb{k}, \vb{q})}_{n'n, \nu}$ is proportional to $\frac{1}{\sqrt{\omega_{\qq,\nu}}}$, and the integration of two delta-functions gives $D(E_F)^2$, thus
\begin{equation}
    \lambda_{\qq,\nu}\propto \frac{D(E_F)}{\omega_{\qq,\nu}^2}.
    \label{app:eq:lambda_wqv_relation}
\end{equation}
Therefore, a low phonon energy with a finite EPC matrix element $\tilde{g}$ will result in a considerably high EPC $\lambda_{\vb{q},\nu}$. 
The total EPC constant $\lambda$ is the BZ average over the phonon-resolved $\lambda_{\vb{q},\nu}$:
\begin{equation}\begin{aligned}
    \lambda \eq \sum_{\nu} \int_{\mathrm{BZ}} \frac{\dd{\vb{q}}}{\Omega_{\mathrm{BZ}}} \lambda_{\vb{q},\nu}
\end{aligned}\end{equation}
The Eliashberg spectral function $\alpha^2F(\omega)$ is defined as 
\begin{equation}\begin{aligned}
    \alpha^2F(\omega) &= \frac{1}{2}\sum_{\nu} \int_{\mathrm{BZ}} \frac{\dd{\vb{q}}}{\Omega_{\mathrm{BZ}}} \omega_{\vb{q}\nu} \lambda_{\vb{q}\nu} \delta(\omega - \omega_{\vb{q}\nu})\\
    &= \frac{1}{2}\sum_{\nu} \int_{\mathrm{BZ}} \frac{\dd{\vb{q}}}{\Omega_{\mathrm{BZ}}} \frac{\gamma_{\qq,\nu}}{2\pi D(E) \omega_{\qq,\nu}} \delta(\omega - \omega_{\vb{q}\nu}).
\end{aligned}\end{equation}
This function describes the distribution of the electron-phonon coupling over phonon frequencies.
An equivalent definition of total EPC constant $\lambda$ is based on $\alpha^2F(\omega)$:
\begin{equation}\begin{aligned}
    \lambda \eq 2 \int_0^\infty \dd{\omega} \frac{\alpha^2F(\omega)}{\omega}
\end{aligned}\end{equation}

\paragraph{The double-delta approximation of phonon linewidth. }
The double-delta approximation of phonon linewidth makes the assumption that the phonon frequency $\omega_{\qq\nu}$ is small, so one only needs to consider the electron states close to the Fermi level for the SC properties. We start from
\begin{equation}\begin{aligned}
\gamma_{\vb{q}, \nu} = -2\Im \sum_{m, n} \int_{\text{BZ}} \frac{\dd{\vb{k}}}{\Omega_{\text{BZ}}} \abs{\bqty{\tilde{g}(\vb{k}, \vb{q})}_{mn,\nu}}^2
    \frac{f_{\vb{k},n}(T) - f_{\vb{k}+\vb{q},m}(T)}{\epsilon_{\vb{k}+\vb{q},m} - \epsilon_{\vb{k},n} - \omega_{\vb{q},\nu} - i \eta}.
\end{aligned}\end{equation}
The Fermi-Dirac functions can be approximated as 
\begin{equation}\begin{aligned}
f_{\kk,n}(T) - f_{\kk+\qq,m}(T) = -(\epsilon_{\vb{k}+\vb{q},m} - \epsilon_{\vb{k},n}) f'_{\kk,n}(T) 
= (\epsilon_{\vb{k}+\vb{q},m} - \epsilon_{\vb{k},n}) \delta(\epsilon_{\kk,n}-E_F).
\end{aligned}\end{equation}
Then
\begin{equation}\begin{aligned}
\gamma_{\vb{q}, \nu} = -2\Im \sum_{m, n} \int_{\text{BZ}} \frac{\dd{\vb{k}}}{\Omega_{\text{BZ}}} \abs{\bqty{\tilde{g}(\vb{k}, \vb{q})}_{mn,\nu}}^2 
\left(\frac{\epsilon_{\vb{k}+\vb{q},m} - \epsilon_{\vb{k},n}-\omega_{\vb{q},\nu}}{\epsilon_{\vb{k}+\vb{q},m} - \epsilon_{\vb{k},n} - \omega_{\vb{q},\nu} - i \eta} + \frac{\omega_{\vb{q},\nu}}{\epsilon_{\vb{k}+\vb{q},m} - \epsilon_{\vb{k},n} - \omega_{\vb{q},\nu} - i \eta}
\right) \delta(\epsilon_{\kk,n}-E_F)
\end{aligned}\end{equation}
The imaginary part of the first term vanishes when $\eta\rightarrow 0$. Thus  
\begin{equation}\begin{aligned}
\gamma_{\vb{q}, \nu} &\approx 2\pi \sum_{m, n} \int_{\text{BZ}} \frac{\dd{\vb{k}}}{\Omega_{\text{BZ}}} \abs{\bqty{\tilde{g}(\vb{k}, \vb{q})}_{mn,\nu}}^2 
\omega_{\vb{q},\nu} \delta(\epsilon_{\vb{k}+\vb{q},m} - \epsilon_{\vb{k},n} - \omega_{\vb{q},\nu}-E_F)  \delta(\epsilon_{\kk,n}-E_F) \\
&\approx 2\pi \omega_{\vb{q},\nu}
\int_{\text{BZ}} \frac{\dd{\vb{k}}}{\Omega_{\text{BZ}}} \abs{\bqty{\tilde{g}(\vb{k}, \vb{q})}_{mn,\nu}}^2 \delta(\epsilon_{\vb{k}+\vb{q},m} - E_F) \delta(\epsilon_{\kk,n}-E_F)
\end{aligned}\end{equation}
The second approximation follows when $\omega_{\qq,\nu}$ is negligible. Thus we arrive at the expression of phonon linewidth in the double-delta approximation.

\subsubsection{\label{app:SC_properties_formalism} Superconducting properties from EPC}

\textit{Ab-initio} calculations of phonon-mediated superconducting properties are based on the Bardeen–Cooper–Schrieffer (BCS) theory.
The critical temperature $T_c$ at which the phase transition occurs can be estimated with semi-empirical methods like the McMillan formula~\cite{McMillan1968TransitionA}, later refined by Allen and Dynes~\cite{Dynes1972McMillantextquotesinglesA, Allen1975TransitionA} to account for strong electron-phonon coupling:
\begin{equation}\begin{aligned}\label{eq:McMillian-Tc}
    T_c \eq \frac{\omega_{\log}}{1.2} \exp[-\frac{ 1.04(1+\lambda)}{\lambda - \mu^* (1 + 0.62\lambda)}]
\end{aligned}\end{equation}
Here, $\omega_{\log}$ represents the logarithmic average of the phonon frequency, which is defined as:
\begin{equation}\begin{aligned}
    \omega_{\log} \eq \exp[\frac{2}{\lambda} \int \frac{\dd{\omega}}{\omega} \alpha^2 F(\omega) \log\omega],
\end{aligned}\end{equation}
In \cref{eq:McMillian-Tc}, the parameter $\mu^*$, known as the Coulomb pseudopotential, accounts for Coulomb screening effects and typically ranges between $0.1$ and $0.16$, and $\lambda$ is the EPC constant obtained by momentum and mode integration of the EPC strength of \cref{eq:lambda_qn}.
In this context, both the EPC constant $\lambda$ and the Eliashberg spectral function $\alpha^2 F(\omega)$ are dimensionless. The coupling strength $\lambda$ generally takes values below $3$, and $\omega_{\log}$ shares the same unit as the phonon frequency $\omega$, whether expressed in energy or temperature, and has typical values smaller than 250 meV (or about 3000 K).

\subsubsection{\label{app:ResultsDFTElectronPhononLaRu3Si2} Electron-phonon induced superconductivity in \ch{LaRu3Si2}}

In this section, we present the results of electron-phonon-induced superconductivity in \ch{LaRu3Si2}, both in its undoped state and with doping. We begin by analyzing the undoped results.

\begin{figure}[htbp]
    \centering
    \includegraphics[width=0.8\textwidth]{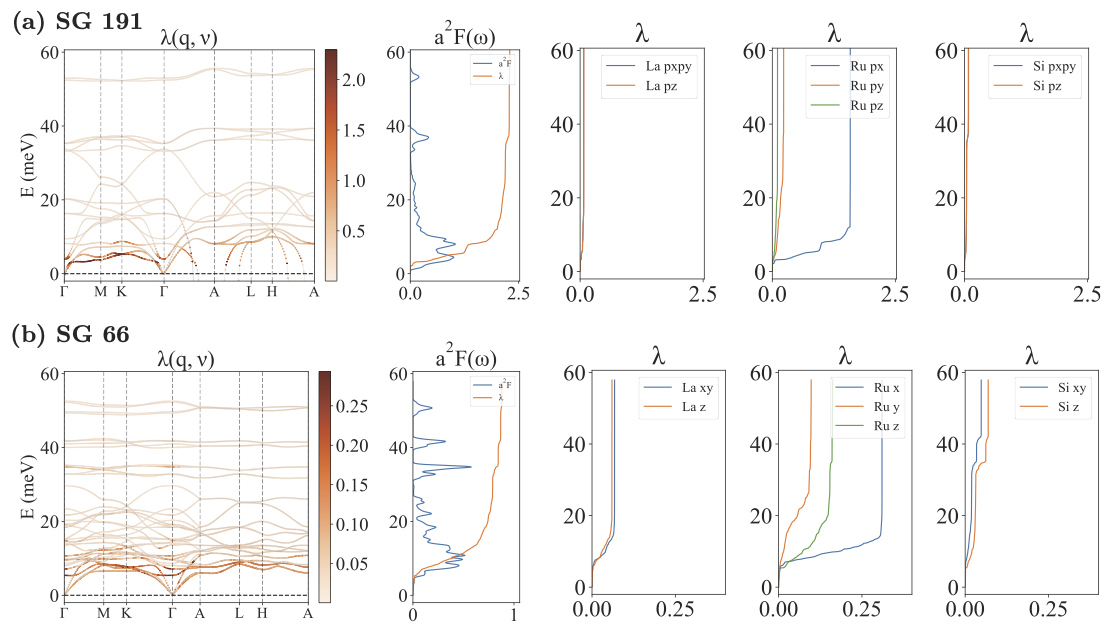}
    \caption{(a) The electron-phonon coupling (EPC) strength $\lambda_{\vb{q},\nu}$ for the phonon along high-symmetry lines, isotropic Eliashberg spectral function $\alpha^2F(\omega)$, the total and orbital-resolved integrated EPC $\lambda$ for SG 191. (b) Same as (a) but for SG 66.}
    \label{fig:EPCUndoped} 
\end{figure}

\paragraph{\label{app:ResultsDFTElectronPhononLaRu3Si2Undoped} \textbf{Undoped case.}}

As shown in \cref{fig:EPCUndoped}, we examine the EPC strength $\lambda_{\vb{q},\nu}$ along high-symmetry lines, the isotropic Eliashberg spectral function $\alpha^2F(\omega)$, and the total and orbital-resolved integrated EPC $\lambda(\omega)$ for both the SG 191 [\cref{fig:EPCUndoped}{(a)}] and SG 66 [\cref{fig:EPCUndoped}{(b)}] phases.

In the SG 191 phase, the EPC strength $\lambda$ is primarily dominated by low-frequency phonons, with the Ru-$x$ phonon (defined in the kagome local coordinate system, as shown in \cref{fig:KagomeLocalOrbs}) being the primary contributor. The three soft phonon modes located on the $q_z = \pi$ plane are excluded from the EPC calculation due to their instability. In the SG 66 phase, where the soft phonons are eliminated, the strong contribution from the Ru-$x$ phonon persists. Although contributions from the Ru-$y$ and Ru-$z$ phonons increase due to hybridization after symmetry breaking, the Ru-$x$ phonon remains the dominant contributor in the orbital-resolved $\lambda(\omega)$.

The presence of soft modes in the SG 191 phase, as seen in \cref{fig:Ph191and193and176}{(a)} and \cref{fig:EPCUndoped}{(a)}, renders the EPC ill-defined. The computed EPC constant $\lambda$ in SG 191 is anomalously large ($\lambda \sim 3$), approximately five times higher than in the SG 66 phase. However, this overestimation is primarily due to the presence of the 1D imaginary phonon mode ($A^+_4$). Given the relation $\lambda \sim 1/\ev*{\omega^2}$, a nearly imaginary mode with $\omega \to 0$ results in an excessively large $\lambda$. As the electronic temperature increases (e.g., $0.15 \,\mathrm{Ry}$ or $0.3 \,\mathrm{Ry}$), the previously soft 1D $A_4^+$ mode in SG 191 hardens (\cref{fig:Ph191HighT}), reducing $\lambda$ to 1.74 and 0.75, respectively. This suggests that the artificially high $\lambda$ in SG 191 is a direct consequence of the soft modes and will decrease to values comparable to those in SG 66 once these modes are fully stabilized.

\cref{tab:TcUndoped} summarizes the integrated EPC strength $\lambda$, logarithmic average phonon frequency $\omega_{\log}$, and the superconducting transition temperature $T_c$, calculated using the Allen-Dynes modified McMillan equation [\cref{eq:McMillian-Tc}]~\cite{McMillan1968TransitionA, Dynes1972McMillantextquotesinglesA, Allen1975TransitionA}. In the dynamically stable SG 66 phase, we obtain a theoretical $T_c$ of $6.8 \,\mathrm{K}$, in excellent agreement with the experimentally reported $T_c \sim 7.8 \,\mathrm{K}$~\cite{Barz1980TernaryA, Ku1980SuperconductingA, Chevalier1983SuperconductingA, Rauchschwalbe1984SuperconductivityA, Godart1987CoexistenceA, Escorne1994Type-IIA, Kishimoto2002MagneticA, Li2011AnomalousA, Mielke2021NodelessA, Chaudhary2023RoleA, Ma2024Dome-ShapedA, Ushioda2024Two-gapA, Li2012DistinctA, Li2016ChemicalA, Chakrabortty2023EffectA}.

\begin{figure}
    \centering
    \includegraphics[width=0.3\linewidth]{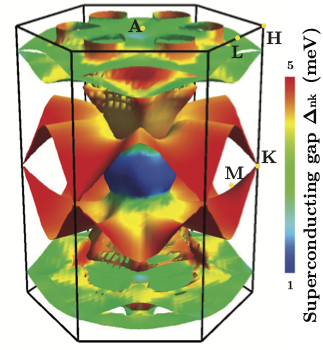}
    \caption{The SC energy gaps of \ch{LaRu3Si2} (expressed in meV) on the Fermi surface for $T = 1$ K.}
    \label{fig:Gap191}
\end{figure}

The superconducting gap function, calculated using the anisotropic Eliashberg theory for the SG 191 phase, is presented in \cref{fig:Gap191}. Due to computational limitations, the gap function from the anisotropic Eliashberg theory for the SG 66 phase is not presented here. The SC gap predominantly arises from the Fermi surface associated with kagome $d_{x^2-y^2}$ electrons [\cref{fig:FS191}], with the gap magnitude being approximately three times larger than on other Fermi surfaces. This observation is consistent with the recently reported two-gap superconducting state in \ch{LaRu3Si2}~\cite{Ushioda2024Two-gapA}.
The large SC gap function suggests that the quasi-FB by kagome $d_{x^2-y^2}$ electrons in \ch{LaRu3Si2} plays a crucial role in SC, likely due to the high DOS it contributes.

\paragraph{\label{app:ResultsDFTElectronPhononLaRu3Si2Doped} \textbf{Doped case.}}

\begin{figure}
    \centering
    \includegraphics[width=0.5\linewidth]{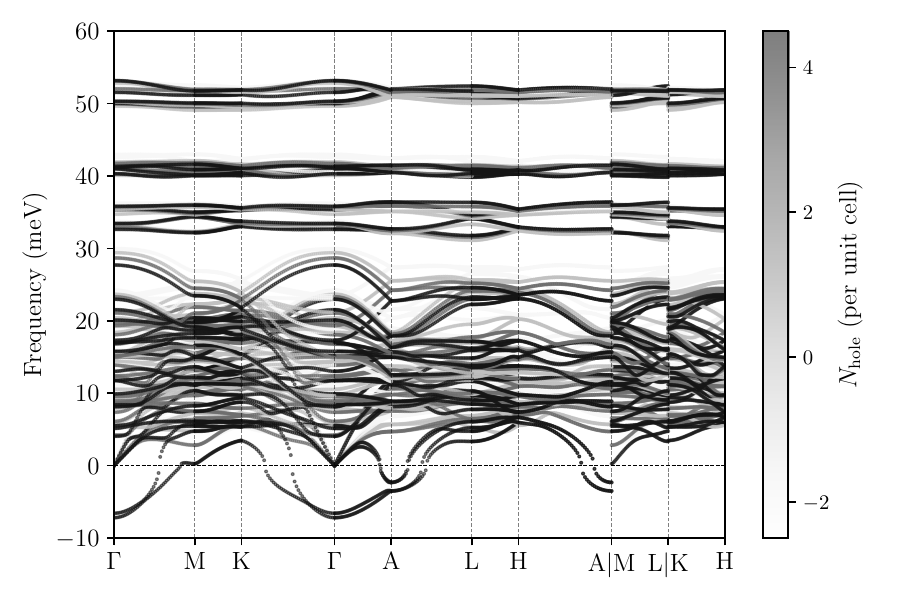}
    \caption{Phonon spectrum of \ch{LaRu3Sh2} in SG 66 phase at different doping level.}
    \label{fig:PhononStability}
\end{figure}

Given that the FB is located just below the $E_F$ at approximately $150 \,\mathrm{meV}$ [\cref{fig:DOS191and66}{(a) and (b)}], it is natural to explore the effects of doping on superconductivity. To investigate this, \textit{ab initio} calculations were performed with a varied number of valence electrons (together with a homogeneous background charge to maintain charge neutrality) to simulate doping.
First, we examine the structural stability of \ch{LaRu3Si2} under hole doping. As shown in \cref{fig:PhononStability}, hole doping generally induces structural instability starting from two holes per unit cell, as indicated by the progressive softening of phonon modes with increasing doping levels. Moreover, the energy difference between the paramagnetic (PM) and ferromagnetic (FM) states as a function of doping level reveals that hole doping beyond two holes per unit cell induces ferromagnetism. This observation aligns with prior studies suggesting that a half-filled FB is prone to magnetization and polarization, as well as with the magnetic instabilities observed in \ch{FeGe} and other kagome materials featuring partially filled flat bands at the Fermi level ($E_F$)~\cite{Feng2024CatalogueA} 
However, at doping levels below one hole per unit cell, although some phonon modes soften, no imaginary modes appear, and the system remains paramagnetic. This suggests that \ch{LaRu3Si2} retains its structural and magnetic stability under light hole doping.

\begin{figure}[htbp]
    \centering
    \includegraphics[width=0.5\linewidth]{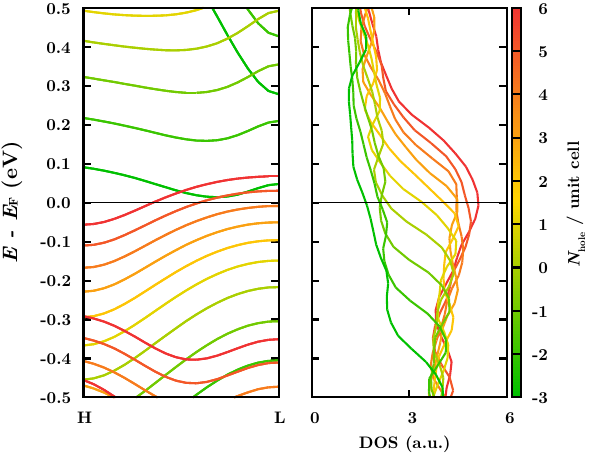}
    \caption{The total electronic band structures (left) and density of states (DOS) (right) of the $1\times1\times2$ ($Cccm$) phase are displayed for various doping levels, with different doping levels indicated by distinct colors in the color bar. The band structures and DOS peaks exhibit nearly rigid shifts as a function of doping level. For simplicity, only the bands along the high-symmetry path $H-L$ are shown in the left panel.}\label{fig:Dope66}
\end{figure}

\begin{figure}[htbp]
    \centering
    \includegraphics[width=0.9\linewidth]{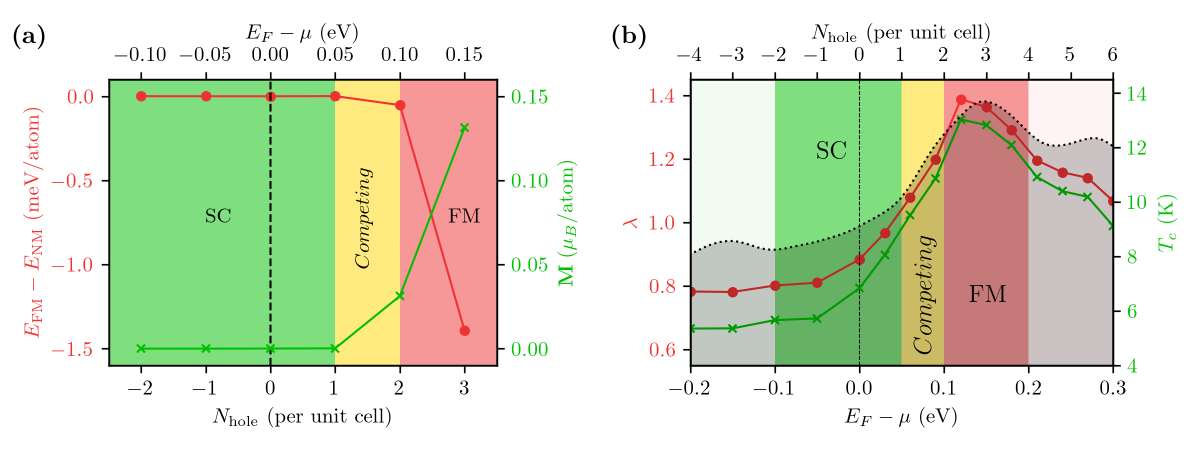}
    \caption{(a) Total energy difference between ferromagnetic state and paramagnetic (nonmagnetic) state of \ch{LaRu3Si2} in $1\times1\times2(Cccm)$ phase, $E_{\mathrm{FM}}-E_{\mathrm{NM}} (\mathrm{meV}/\mathrm{atom})$, and magnetic moment $(\mu_{\mathrm{B}}/\mathrm{atom})$ as a function of doping level $(N_{\mathrm{hole}}/\mathrm{unit\,cell})$, the doping level is also effectively transformed to shifting of the chemical potential $\mu$ according to \cref{fig:Dope66}.}
    \label{fig:DopeStability}
\end{figure}

The doping effects on SC in the SG 66 phase of \ch{LaRu3Si2} are presented in \cref{tab:TcDoped66Ne}, where the filling level is varied by introducing the background charge. As shown in \cref{fig:Dope66}, which depicts the band dispersion and DOS at different filling levels (for simplicity, only the dispersion along the high-symmetry $H-L$ line is shown), changes in the background charge effectively result in a rigid shift of the total band structure and DOS. Specifically, doping one electron or hole shifts $E_F$ by roughly $+50 \,\mathrm{meV}$ or $-50 \,\mathrm{meV}$, respectively.
The full doping curve across an extended range is calculated using the rigid band approximation, as shown in \cref{fig:DopeStability}{(b)}. Here, the shift in the chemical potential $\mu$ is mapped onto the hole doping level, with $50,\mathrm{meV} \simeq 1,\mathrm{hole/unit\,cell}$.
Note that under the rigid band approximation, where both electronic and phononic bands are fixed, the $\lambda$ and $T_c$ calculated here are likely to be underestimated,
This underestimation arises because, as shown in \cref{fig:PhononStability}, the phononic bands tend to soften with increasing levels of hole doping, which would, in reality, enhance $\lambda$ and, consequently, $T_c$.

\begin{figure}[htbp]
    \centering
    \includegraphics[width=0.8\linewidth]{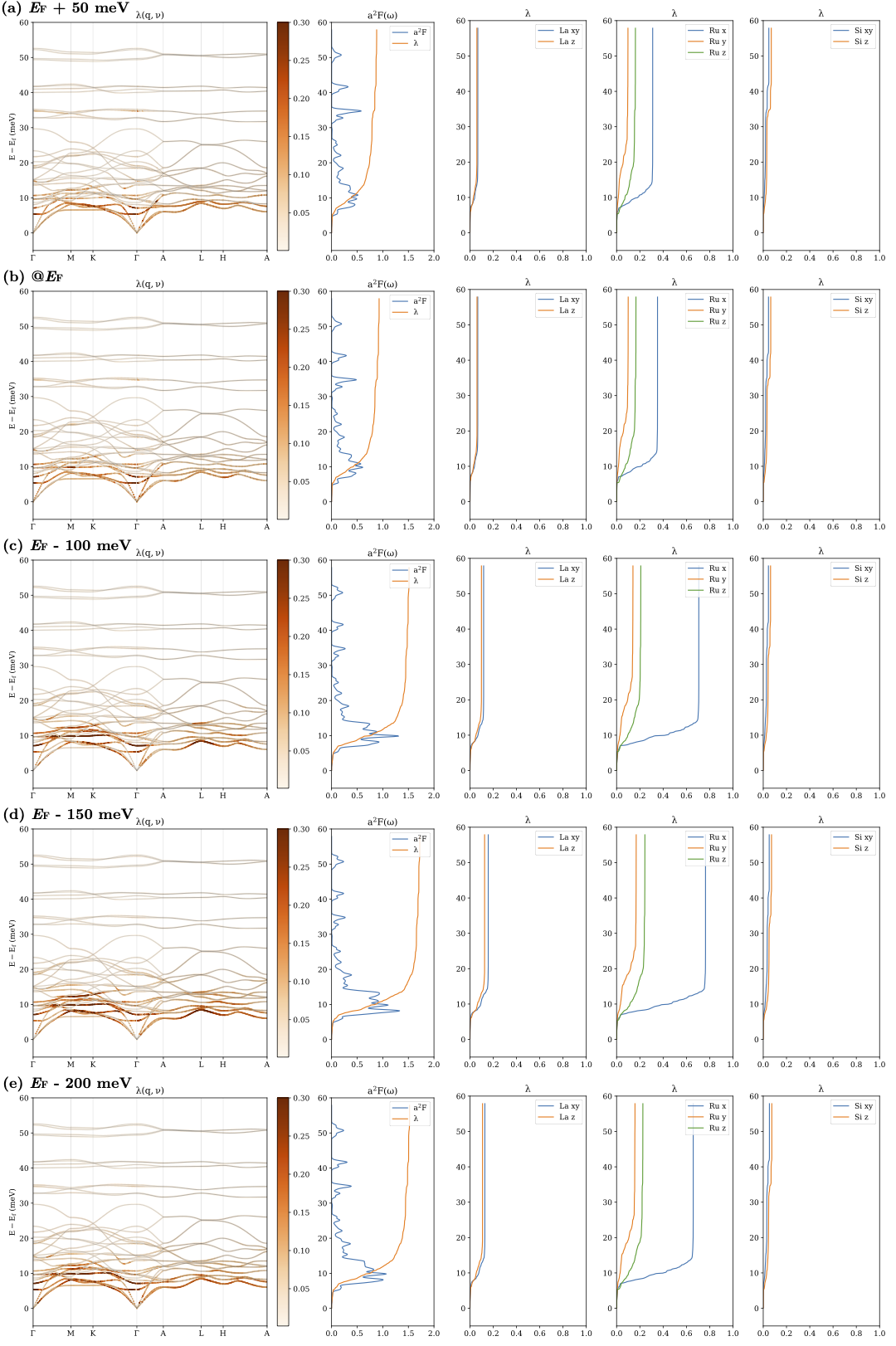}
    \caption{The electron-phonon coupling properties of \ch{LaRu3Si2} in SG 66 phase as a function of doping, where the Fermi level is shifted rigidly by (a) $+50\,\mathrm{meV}$, (b) $0\,\mathrm{meV}$, (c) $-100\,\mathrm{meV}$, (d) $-150\,\mathrm{meV}$, and (e) $-200\,\mathrm{meV}$. Compared with the undoped results shown in \cref{fig:EPCUndoped}{(b)}, the $\lambda$ increases when the Fermi level is shifted closer to the peak in DOS, while decreases when shifted away. The largest $\lambda$ is observed at $E_F-150\,\mathrm{meV}$. 
    }
    \label{fig:EPCDoped}
\end{figure}

A maximal $T_c=12.82$ K is observed at $150\,\mathrm{meV}$ below the $E_F$. This filling is close to the peak of DOS given by the quasi-FB below the $E_F$ (\cref{fig:DOS191and66}{(b)}).  Note that such a large hole doping will result in a FM phase in DFT, as shown in \cref{fig:DopeStability}(b). 
In \cref{fig:EPCDoped}, we show the $\lambda_{\vb{q},\nu}$ along high-symmetry lines and orbital-resolved $\lambda(\omega)$ for different dopings. We observe that the $\lambda$ increases when the Fermi level is shifted closer to the peak in DOS, while it decreases when shifted further from the peak, consistent with the experiment observation~\cite{Li2012DistinctA, Li2016ChemicalA, Chakrabortty2023EffectA}. The phonon orbital contributions remain almost the same. 

Note that while the density of states $D(E_F)$ appears in the denominator of \cref{eq:lambda_qn}, the integration of two delta functions results in an expression proportional to $D(E_F)^2$. Consequently, $\lambda_{\qq,\nu}$ remains proportional to $D(E_F)$, as shown in \cref{app:eq:lambda_wqv_relation}.

\begin{table}[htbp]
    \centering
    \caption{The superconducting relavent quantities of undoped \ch{LaRu3Si2} in different phases. 
    We note that there are soft modes in SG 191, which makes the EPC not well-defined. With higher electronic temperature (such as $0.15\,\mathrm{Ry}$ or $0.3\,\mathrm{Ry}$), the 1D soft mode $A_4^+$ will be eliminated in SG 191 (\cref{fig:Ph191HighT}), and $\lambda$ reduces to $1.74$ or $0.75$. Thus we conclude that the large $\lambda$ in SG 191 is a result of the soft modes, and will be reduced to a similar value as SG 66 once all soft modes are removed. The DOS at $E_F$, $D(E_F)$, is divided by 2 in SG 66 for comparison with the SG 191 phase.}
    \begin{tabular}{c|c|c|c|c}
        \hline\hline
        Phase & $D(E_F)/(\mathrm{states}/\mathrm{spin}/\mathrm{unit\,cell})$ & $\lambda$ & $\omega_{\log}/\mathrm{K}$ & $T_c/\mathrm{K}$ \\
        \hline
        SG 191 ($P6/mmm$) & $2.855$ & $2.47$ & $63.14$ & $10.51$ \\
        \hline
        SG 66 ($Cccm$) & $2.206$ & $0.83$ & $129.75$ & $6.83$ \\
        \hline\hline
    \end{tabular}
    \label{tab:TcUndoped}
\end{table}

\begin{table}[htbp]
    \centering
    \caption{The superconducting property of doped \ch{LaRu3Si2} in $1\times1\times2$ ($Cccm$) phase with background charge. Note that doping -2 holes in the system will result in a FM phase, as shown in \cref{fig:DopeStability}(b).}
    \label{tab:TcDoped66Ne}
    \begin{tabular}{c|c|c|c|c}
        \hline\hline
        $N_{e} - N_{e}^{\mathrm{undoped}}$ & $E_F/\mathrm{eV}$ & $D(E_F)/(\mathrm{states}/\mathrm{spin}/\mathrm{unit\,cell})$ & $\lambda$ & $T_c/\mathrm{K}$ \\
        \hline
        $+1$ & $16.3501$ & $1.991$ & $0.759$ & $6.05$ \\
        \hline
        $0$  & $16.1912$ & $2.206$ & $0.83$ & $6.83$ \\
        \hline
        $-1$ & $16.1262$ & $2.398$ & $1.265$ & $11.03$ \\
        \hline
        $-2$ & $16.0228$ & $3.278$ & $2.605$ & $13.92$ \\
        \hline\hline
    \end{tabular}
\end{table}

\clearpage
\section{\label{app:Models} Analytical results}

In this section, we present an analytical framework to interpret and derive the electron-phonon coupling (EPC) results obtained from the \textit{ab-initio} calculations. 

By projecting the electron-phonon coupling onto the local $x$- and $y$-phonon mode basis, we find that the electron-phonon coupling should be comparable for both modes.
However, as discussed earlier in \cref{app:eq:lambda_wqv_relation}, the mode-resolved electron-phonon coupling $\lambda$ is inversely proportional to the square of the average frequency $\ev*{\omega^2}$ of the respective phonon mode. This implies that small differences in the phonon mode frequencies can lead to considerably large differences in their contributions to the total electron-phonon coupling.

To elucidate this behavior, we employ the spring-ball model, a simple yet effective framework for describing phonon dynamics in kagome lattices. This model reveals that the local $x$-phonon mode consistently exhibits a lower frequency than the local $y$-phonon mode across kagome systems. Given that our \textit{ab-initio} results indicate similar electron-phonon coupling strengths for both phonon modes, the lower frequency of the $x$-phonon leads to a stronger contribution to $\lambda$.

The results of this analytical model suggest that the observed mode-selective EPC, where the local $x$-phonon dominates due to its lower frequency, is not unique to \ch{LaRu3Si2}. Instead, this phenomenon is expected to be a universal feature in a wide class of kagome materials.
In the next section, we extend this analysis by examining high-throughput computational results, which confirm that this mode-selective EPC is prevalent across various kagome-based compounds.

\subsection{\label{app:ModelsGaussianEPC}Electron-phonon coupling of $d_{x^2-y^2}$ orbitals}

\begin{figure}[H]
    \centering    \includegraphics[width=0.6\linewidth]{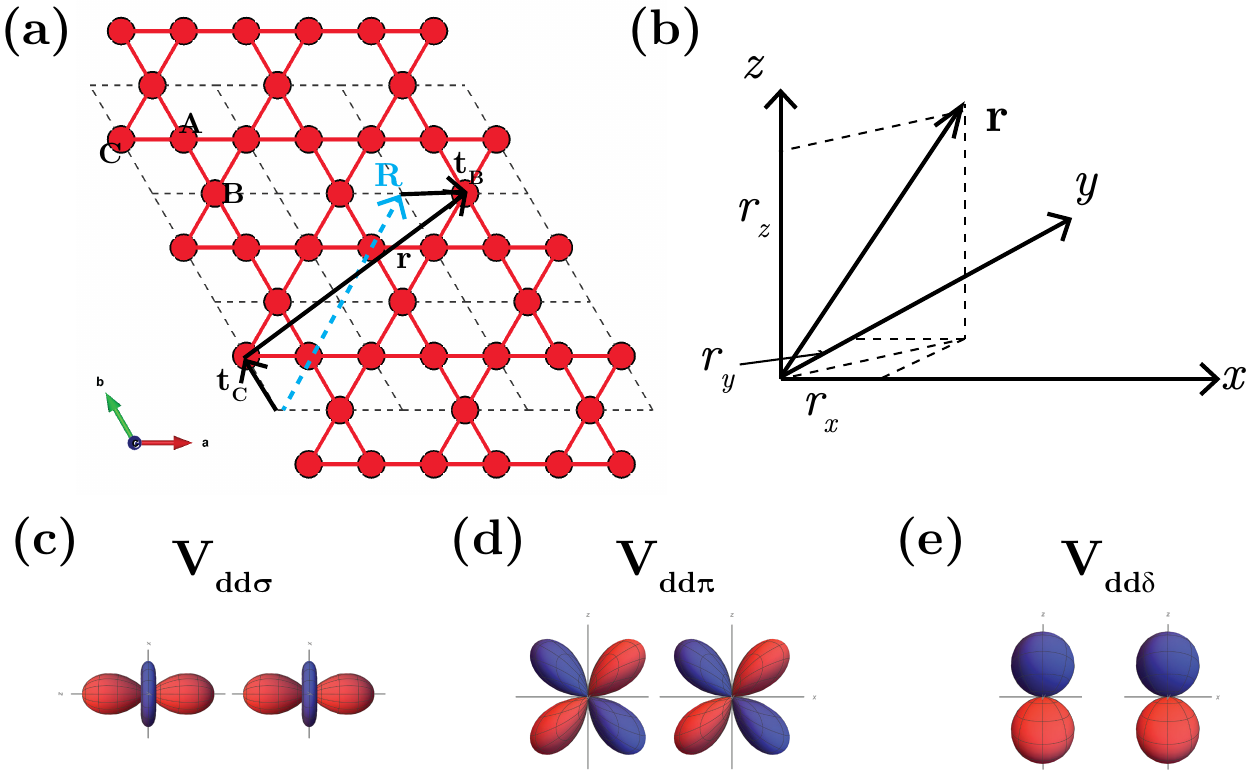}
    \caption{(a) Definition of the hopping vector $\vb{r}=\vb{R}+\vb{t}_\beta-\vb{t}_\alpha$ in \cref{tab:HoppingSK} connecting orbital $\alpha$ to $\beta$, where $\vb{R}$ is a lattice vector and $\vb{t}_{\alpha/\beta}$ are the orbital positions. (b) Illustration of direction cosines in \cref{app:eq:direction_cosine}. (c)-(e) Illustration of three types of $d$ orbital hopping integrals in the Slater-Koster formalism, with (c) being $V_{dd\sigma}$, (d) $V_{dd\pi}$, and (e) $V_{dd\delta}$. 
    }
    \label{fig:def-SK-hopping}
\end{figure}

To further explore why the electron-phonon coupling in \ch{LaRu3Si2} is predominantly contributed by the $B_{3u}$ (local $x$) direction phonon of the kagome lattice, we apply the newly developed Gaussian approximation (GA) in Ref.~\cite{Yu2024Non-trivialA} to the electron hopping terms within the two-center integral framework and the Slater-Koster (SK) formalism~\cite{Slater1954SimplifiedA}.
The GA simplifies the electron hopping terms by assuming that the hopping integrals decay exponentially with the interatomic distance. 
Within the GA~\cite{Yu2024Non-trivialA}, the hopping integrals are assumed to have the following form (without considering the angular part of the electron orbitals):
\begin{equation}\begin{aligned}
    t_{ij}(\vb{r}) \eq t_{0} \exp(-\gamma_{ij}\frac{\abs{\vb{r}}^2}{2})
\end{aligned}\end{equation}
where $t_0$ is the amplitude of the hopping integral, $\vb{r} = \vb{R}_{ij} + \vb{u}$ is the electron position, $\vb{R}_{ij}$ is the distance between atoms $i$ and $j$ at equilibrium position, $\vb{u}$ is the small displacement, and $\gamma_{ij}$ is a parameter describing the spatial decay of $t_{ij}$. 
In the context of the SK formalism, the hopping matrix elements between orbitals are described as a function of the direction cosines of the bond (hopping vector) between two neighboring atoms (\cref{fig:def-SK-hopping}{(b)}). The electron hopping terms are typically parameterized in terms of overlap integrals that account for different types of bonding interactions, i.e., the $\sigma$-, $\pi$-, and $\delta$-bonding (for example the parameterization of $d$ orbitals are explicitly listed in \cref{tab:HoppingSK}).
To incorporate the GA into the SK formalism, we assume that the SK parameters $V_{i}(\vb{r}),\, i\in\Bqty{\sigma,\pi,\delta,...}$ also decay as Gaussian functions:
\begin{equation}\begin{aligned}\label{eq:d_sk_ga}
    V_i(\vb{r}) \eq V_{i,0} \exp(-\gamma_{i} \frac{\abs{\vb{r}}^2}{2}).
\end{aligned}\end{equation}
In \cref{eq:d_sk_ga}, $V_{i,0}$ represents the initial overlap integral strength, $\gamma_i$ is a decay parameter specific to each bond type.
This approximation enables a systematic analysis of how orbital character influences electron-phonon coupling (EPC).

\subsubsection{\label{app:ModelsGaussianEPCHoppingSK} Electron hopping terms of $d_{x^2-y^2}$ orbitals in Slater-Koster formalism}

\begin{table}[!b]
    \centering
    \caption{Hopping elements in terms of SK parameters~\cite{Slater1954SimplifiedA}. $(l,m,n)$ are the direction cosines defined as $l=r_x/\abs{\vb{r}},m=r_y/\abs{\vb{r}},n=r_z/\abs{\vb{r}}$ (\cref{app:eq:direction_cosine}), where $\vb{r}=\vb{R}+\vb{t}_{\beta}-\vb{t}_{\alpha}$ is the hopping vector connecting orbital $\alpha$ in the home unit cell to orbital $\beta$ in unit cell $\vb{R}$. $V_{dd\sigma}, V_{dd\pi}, V_{dd\delta}$ are the bonding integrals used to parameterize the hopping, as illustrated in \cref{fig:def-SK-hopping}(c)-(e). The $d$ orbitals in this table are defined in the global Cartesian coordinates.
    }
    \label{tab:HoppingSK}
    \begin{tabular}{c|c|c}
        \hline\hline
        $\alpha$      & $\beta$       & $t_{\alpha\beta}(\vb{R}+\vb{t}_{\beta}-\vb{t}_{\alpha})$ \\
        \hline
        $d_{xy}$      & $d_{xy}$      & $3l^2m^2 V_{dd\sigma} + (l^2+m^2-4l^2 m^2) V_{dd\pi} + (n^2+l^2m^2) V_{dd\delta}$ \\
        $d_{xy}$      & $d_{yz}$      & $3lm^2n V_{dd\sigma} + ln(1-4m^2) V_{dd\pi} + ln(m^2-1) V_{dd\delta}$ \\
        $d_{xy}$      & $d_{xz}$      & $3l^2mn V_{dd\sigma} + mn(1-4l^2) V_{dd\pi} + mn(l^2-1) V_{dd\delta}$ \\
        $d_{xy}$      & $d_{x^2-y^2}$ & $3/2 lm(l^2-m^2) V_{dd\sigma} + 2lm(m^2-l^2) V_{dd\pi} + lm(l^2-m^2)/2 V_{dd\delta}$ \\
        $d_{xy}$      & $d_{z^2}$     & $\sqrt{3}\{lm[n^2-(l^2+m^2)/2] V_{dd\sigma} - 2lmn^2 V_{dd\pi} + lm(1+n^2)/2 V_{dd\delta}\}$ \\
        \hline 
        $d_{yz}$      & $d_{x^2-y^2}$ & $3/2 mn(l^2-m^2) V_{dd\sigma} - mn[1+2(l^2-m^2)] V_{dd\pi} + mn[1+(l^2-m^2)/2] V_{dd\delta}$ \\
        $d_{yz}$      & $d_{z^2}$     & $\sqrt{3}\{mn[n^2-(l^2+m^2)/2] V_{dd\sigma} + mn(l^2+m^2-n^2) V_{dd\pi} - mn(l^2+m^2)/2 V_{dd\delta}\}$ \\
        \hline 
        $d_{xz}$      & $d_{x^2-y^2}$ & $3/2 nl(l^2-m^2) V_{dd\sigma} + nl[1-2(l^2-m^2)] V_{dd\pi} - nl[1-(l^2-m^2)/2] V_{dd\delta}$ \\
        $d_{xz}$      & $d_{z^2}$     & $\sqrt{3}\{ln[n^2-(l^2+m^2)/2] V_{dd\sigma} + ln(l^2+m^2-n^2) V_{dd\pi} - ln(l^2+m^2)/2 V_{dd\delta}\}$ \\
        \hline 
        $d_{x^2-y^2}$ & $d_{x^2-y^2}$ & $3/4(l^2-m^2)^2 V_{dd\sigma} + [l^2+m^2-(l^2-m^2)^2] V_{dd\pi} + [n^2+(l^2-m^2)^2/4] V_{dd\delta}$ \\
        $d_{x^2-y^2}$ & $d_{z^2}$     & $\sqrt{3}\{(l^2-m^2)[n^2-(l^2+m^2)/2]/2 V_{dd\sigma} + n^2(m^2-l^2) V_{dd\pi} + (1+n^2)(l^2-m^2)/4 V_{dd\delta}\}$ \\
        \hline 
        $d_{z^2}$     & $d_{z^2}$     & $[n^2-(l^2+m^2)/2]^2 V_{dd\sigma} + 3n^2(l^2+m^2) V_{dd\pi} + 3/4(l^2+m^2)^2 V_{dd\delta}$ \\
        \hline\hline
    \end{tabular}
\end{table}

\begin{figure}[H]
    \centering
    \includegraphics[width=0.8\linewidth]{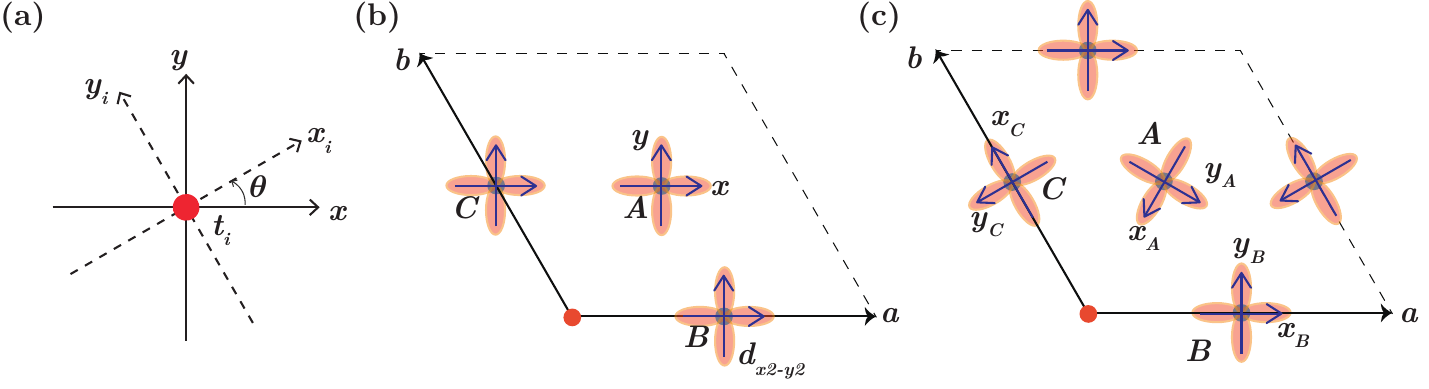}
    \caption{(a) Schematic illustration of a generic rotation of coordinate axes at site $\vb{t}_i$. Here, $x$ and $y$ represent the global Cartesian axes, while $x_i$ and $y_i$ denote the local axes. The rotation angle $\theta$ is defined anticlockwise.
    (b) and (c) Illustrations of coordinate systems on the kagome lattice:
    (b) $d_{x^2-y^2}$ orbitals defined in the global coordinate system.
    (c) $d_{x^2-y^2}$ orbitals defined in the local coordinate systems on the kagome lattice, with the local axes indicated by blue arrows. 
    }
    \label{fig:KagomeLocalAxes}
\end{figure}

The electron hopping matrix of $d$ orbitals under the basis chosen as $\Bqty{d_{xy}, d_{yz}, d_{xz}, d_{x^2-y^2}, d_{z^2}}$ in SK parameters with hopping vector $\vb{r}$, connecting site $i$ and $j$, is 
\begin{equation}\begin{aligned}\label{eq:d_cart_sk}
    h_{ij}(\vb{r}) \eq 
    \begin{pmatrix}
        t_{i1j1}(\vb{r}) & \cdots & t_{i1j5}(\vb{r}) \\ 
         & \ddots & \vdots \\
        * & & t_{i5j5}(\vb{r})
    \end{pmatrix}
\end{aligned}\end{equation}
where we use $\alpha,\beta\in\Bqty{1,\cdots,5}$ to denote $\Bqty{d_{xy},\cdots,d_{z^2}}$ and the explicit formula of each hopping element are listed in \cref{tab:HoppingSK}.
Note that \cref{eq:d_cart_sk} is written in the basis where the $d$ orbitals are aligned to the Cartesian axes. 
The direct lattice vectors are defined as 
\begin{equation}\begin{aligned}\label{eq:direct_lat}
    \pqty{\vb{a}, \vb{b}, \vb{c}} \eq \mqty( a & -\frac{a}{2} & 0 \\ 0 & \frac{\sqrt{3} a}{2} & 0 \\ 0 & 0 & c),
\end{aligned}\end{equation}
and the sublattices are
\begin{equation}\begin{aligned}\label{eq:sub_lat}
    \vb{t}_{A} \eq \pqty{1/2,1/2,1/2}, \quad
    &\vb{t}_{B} \eq \pqty{1/2,0,1/2}, \quad
    &\vb{t}_{C} \eq \pqty{0,1/2,1/2},
\end{aligned}\end{equation}
where the numbers are direct coordinates given under the basis in \cref{eq:direct_lat}. In \cref{fig:KagomeLocalAxes}{(b) and (c)}, we show the direct lattice vectors and three kagome sublattices, together with the local coordinates defined on each sublattice. 

Then we derive the rotation matrix that transforms the global Cartesian coordinate system into the local coordinate system. 
For orbitals situated at an arbitrary site $\vb{t}_i$ and aligned with the global Cartesian coordinate system (denoted by $(x, y, z)$, referred to as global orbitals), we label them as $\alpha$. In this context, we ignore the sublattice indices and use the subscript $\alpha$ solely to distinguish different global orbitals. The creation operator $f^\dag_{i,\alpha}$ is defined as creating a global orbital $\alpha$ at site $\vb{t}_i$.
On the other hand, the creation operators for local orbitals are denoted as $c^\dag_{i,\alpha'}$, where $\alpha'$ represents the local orbitals defined with respect to the local axes (denoted as $(x_i, y_i, z_i)$) specified by the site symmetry group at site $\vb{t}_i$.
The rotation matrix on site $\vb{t}_i$ that transforms the global orbitals $\alpha$ to the local orbitals $\alpha'$ is denoted as $\mathcal{R}^{el}_l$, where the superscript $el$ indicates the electronic system. This matrix defines the relationship between the global and local coordinate systems. Consequently, the local orbitals $c^\dag_{i,\alpha'}$ can be expressed in terms of the global orbitals as:
\begin{equation}
    c^\dag_{i,\alpha'} = \sum_{\alpha} f^\dag_{i,\alpha} \bqty{\mathcal{R}_{i}^{el}}_{\alpha\alpha'},
\end{equation}
where $\bqty{\mathcal{R}_{i}^{el}}_{\alpha\alpha'}$ represents the elements of the rotation matrix $\mathcal{R}^{el}_i$. 
Specifically, if we consider an anticlockwise rotation around the $z$-axis by an angle $\theta$ (as illustrated in \cref{fig:KagomeLocalAxes}{(a)}), the rotation matrices for different types of orbitals are given as follows:

\begin{itemize}
\item For $s$ orbitals:
Since $s$ orbitals are spherically symmetric, the rotation matrix $\mathcal{R}$ (where we omit the site index and superscript $el$ for simplicity) is simply the identity matrix:
\begin{equation}
    \mathcal{R}^{s}(\theta) = \mqty(1).
\end{equation}
\item For $p$ orbitals (ordered as $\Bqty{p_x, p_y, p_z}$):  
The rotation matrix is given by:
\begin{equation}
    \mathcal{R}^{p}(\theta) = 
    \begin{pmatrix}
        \cos \theta & -\sin \theta & 0 \\
        \sin \theta & \cos \theta & 0 \\
        0 & 0 & 1
    \end{pmatrix}.
\end{equation}
This matrix rotates the $p_x$ and $p_y$ orbitals within the $xy$ plane while leaving the $p_z$ orbital unaffected, as expected for a rotation around the $z$-axis.

\item For $d$ orbitals (ordered as $\Bqty{d_{xy}, d_{yz}, d_{xz}, d_{x^2-y^2}, d_{z^2}}$):  
The rotation matrix takes a more complex form due to the different symmetries of $d$ orbitals. It is given by:
\begin{equation}
    \mathcal{R}^{d}(\theta) = 
    \begin{pmatrix}
        \cos 2\theta & 0 & 0 & \sin 2\theta & 0 \\
        0 & \cos \theta & \sin \theta & 0 & 0 \\
        0 & -\sin \theta &  \cos \theta & 0 & 0 \\
        -\sin 2\theta & 0 & 0 &  \cos 2\theta & 0 \\
        0 & 0 & 0 & 0 & 1
    \end{pmatrix}.
\end{equation}
\end{itemize}

This local coordinate system is defined based on the site symmetry group of the kagome lattice. Specifically, at Wyckoff position (WKP) $3g$, the symmetry group is $mmm (D_{2h})$. Define the transformation matrix from the global to local coordinates on three kagome sublattices: 
\begin{equation}\begin{aligned}\label{eq:rotation_matrix_d}
    \mathcal{R}^{el}_{A} = \mathcal{R}^{d}(-\frac{2\pi}{3}) \eq 
    \begin{pmatrix}
        -\frac{1}{2} & 0 & 0 & \frac{\sqrt{3}}{2} & 0 \\
        0 & -\frac{1}{2} & -\frac{\sqrt{3}}{2} & 0 & 0 \\
        0 & \frac{\sqrt{3}}{2} & -\frac{1}{2} & 0 & 0 \\
        -\frac{\sqrt{3}}{2} & 0 & 0 & -\frac{1}{2} & 0 \\
        0 & 0 & 0 & 0 & 1
    \end{pmatrix}, \quad
    &\mathcal{R}^{el}_{B} = \mathcal{R}^{d}(0) \eq \mathbb{1}_{5\times 5}, \quad
    &\mathcal{R}^{el}_{C} = \mathcal{R}^{d}(\frac{2\pi}{3}) \eq \bqty{\mathcal{R}^{el}_{A}}^{-1}.
\end{aligned}\end{equation}
Here, $\mathcal{R}^{el}_{i}, i\in\Bqty{A,B,C}$ are defined as unitary transformations of the creation operators. Specifically, they satisfy the relation $\cre{c}{i,\alpha'} = \sum_{\alpha}\cre{f}{i,\alpha} \bqty{\mathcal{R}^{el}_{i}}_{\alpha\alpha'}$, where $\cre{f}{i,\alpha} (\cre{c}{i,\alpha'})$ are the $\alpha (\alpha')$-th orbital at sublattice $i$ defined in the Cartesian (local) coordinates respectively.
Similarly, the rotation matrices $\mathcal{R}^{ph}_{i},i\in\Bqty{A,B,C}$ for the phonon atom movement $\vb{u} \equiv \pqty{u_{i,x}, u_{i,y}, u_{i,z}}$ (which are equivalent to the rotation matrices for $p$ orbitals $\mathcal{R}^{p}$) are 
\begin{equation}\begin{aligned}\label{eq:rotation_matrix_p}
    \mathcal{R}^{ph}_{A} = \mathcal{R}^{p}(-\frac{2\pi}{3}) \eq 
    \begin{pmatrix}
        -\frac{1}{2} & \frac{\sqrt{3}}{2} & 0 \\
        -\frac{\sqrt{3}}{2} & -\frac{1}{2} & 0 \\
        0 & 0 & 1
    \end{pmatrix}, \quad
    &\mathcal{R}^{ph}_{B} = \mathcal{R}^{p}(0) \eq \mathbb{1}_{3\times 3}, \quad
    &\mathcal{R}^{ph}_{C} = \mathcal{R}^{p}(\frac{2\pi}{3}) \eq \bqty{\mathcal{R}^{ph}_{A}}^{-1}.
\end{aligned}\end{equation}
Accordingly, the hopping matrices transform as
\begin{equation}\begin{aligned}\label{eq:hop_rotation}
    h_{ij}^{c}(\vb{r}) \eq \mathcal{R}_{i}^{el, \dag} h_{ij}^{f}(\vb{r}) \mathcal{R}_{j}^{el}.
\end{aligned}\end{equation}
where the superscript $f/c$ denotes the hopping matrix between global/local orbitals.
For example, for the hopping from sublattice $A$ to sublattice $B$, we have:
\begin{equation}\begin{aligned}
    h_{AB}^{c}(\vb{r}) \eq \mathcal{R}_{A}^{e, \dag} h_{AB}^{f}(\vb{r}) \mathcal{R}_{B}^{e}
\end{aligned}\end{equation}

Due to the $z$-directional mirror ($\mathcal{M}_{z}$) symmetry of a perfect kagome lattice, the five $d$ orbitals can be divided into two distinct groups: the $\mathcal{M}_{z}$-even set, which includes the $d_{x^2-y^2}$, $d_{xy}$, and $d_{z^2}$ orbitals (indexed by $1$, $4$, and $5$), and the $\mathcal{M}_{z}$-odd set, which consists of the $d_{yz}$ and $d_{xz}$ orbitals (indexed by $2$ and $3$).

As a result, the full $5 \times 5$ hopping matrix $h^f_{ij}(\vb{r})$ (explicitly given in \cref{eq:d_cart_sk}) can be decoupled into two separate blocks corresponding to these $\mathcal{M}_{z}$-even and -odd orbitals. Furthermore, since our \textit{ab-initio} calculations indicate that the orbitals of interest are primarily the $d_{x^2-y^2}$ orbitals on the kagome lattice, we can further simplify the expression in \cref{eq:d_cart_sk} to the $3 \times 3$ block associated with the $\mathcal{M}_{z}$ even subset.
In this $\mathcal{M}_{z}$-simplified expression, the full $3 \times 3$ hopping matrix $h_{ij}^{f,\mathcal{M}_z-\mathrm{even}}(\vb{r})$ between site $i$ and $j$, ordered by the basis $\Bqty{d_{xy}(1), d_{x^{2}-y^{2}}(4), d_{z^2}(5)}$, can be rewritten as:
\begin{equation}\begin{aligned}\label{eq:hop_mirror_even}
    h_{ij}^{f,\mathcal{M}_z-\mathrm{even}}(\vb{r}) = 
    \begin{pmatrix}
        t^f_{i1j1}(\vb{r}) & t^f_{i1j4}(\vb{r}) & t^f_{i1j5}(\vb{r}) \\
                       & t^f_{i4j4}(\vb{r}) & t^f_{i4j5}(\vb{r}) \\
        *              &                & t^f_{i5j5}(\vb{r})
    \end{pmatrix},
\end{aligned}\end{equation}
where $t^f_{i\alpha j\beta}(\vb{r}), \alpha,\beta \in\Bqty{1,4,5}$ represents the hopping amplitude between orbital $\alpha$ on site $\vb{t}_i$ and orbital $\beta$ on site $\vb{t}_j$ and the indices $1$, $4$, and $5$ correspond to the $d_{xy}$, $d_{x^{2}-y^{2}}$, and $d_{z^2}$ orbitals defined in global Cartesian coordinate, respectively.
Specifically, the hopping matrix elements for the $\mathcal{M}_z$-simplified $3 \times 3$ hopping matrix can be expressed by taking $n = r_z/|\rr|= 0$ in the expressions from \cref{tab:HoppingSK} as:
\begin{equation}\begin{aligned}\label{eq:tij_cart}
    t^f_{i1j1}(\vb{r}) \eq 3 l^{2} m^{2} V_{\sigma}(\vb{r}) + (l^{2}+m^{2}-4l^{2}m^{2}) V_{\pi}(\vb{r}) + l^{2} m^{2} V_{\delta}(\vb{r}) \\
    t^f_{i1j4}(\vb{r}) \eq \frac{3}{2} lm(l^{2} - m^{2}) V_{\sigma}(\vb{r}) + 2 lm (m^{2} - l^{2}) V_{\pi} + \frac{lm (l^{2} - m^{2})}{2} V_{\delta}(\vb{r}) \\
    t^f_{i1j5}(\vb{r}) \eq \sqrt{3} \Bqty{ \frac{lm (l^{2} + m^{2})}{2} V_{\sigma}(\vb{r}) + \frac{lm}{2} V_{\delta}(\vb{r}) } \\
    t^f_{i4j4}(\vb{r}) \eq \frac{3}{4} (l^{2} - m^{2})^{2} V_{\sigma}(\vb{r}) + \bqty{ l^{2} + m^{2} - (l^{2} - m^{2})^{2} } V_{\pi}(\vb{r}) + \frac{(l^{2} - m^{2})^{2}}{4} V_{\delta} \\
    t^f_{i4j5}(\vb{r}) \eq \sqrt{3} \Bqty{-\frac{(l^2-m^2)(l^2+m^2)}{4} V_{\sigma}(\vb{r}) + \frac{l^2-m^2}{4} V_{\delta}(\vb{r}) } \\
    t^f_{i5j5}(\vb{r}) \eq \frac{1}{4} (l^{2} + m^{2})^{2} V_{\sigma}(\vb{r}) + \frac{3}{4} (l^{2} + m^{2})^{2} V_{\delta}(\vb{r})
\end{aligned}\end{equation}

In systems such as \ch{LaRu3Si2}, where the flat band near $E_F$ is primarily given by Ru $A_g$ (local $d_{x^2-y^2}$) orbitals, the EPC analysis can be focused on the hopping terms among these specific orbitals. 
For $A_g$ ($d_{x^2-y^2}$) orbitals (labeled by $4$), the nearest-neighboring (NN) hopping elements can be obtained from the rotation (unitary transformation) of hopping elements $t^{f}(\vb{r})$ defined in Cartesian coordinate (\cref{eq:tij_cart}). 
Explicitly, starting from \cref{eq:hop_rotation}, and applying \cref{eq:rotation_matrix_d} (only the $\mathcal{M}_z$-even block, denoted by $\mathcal{R}_{i,+}$) to \cref{eq:hop_mirror_even}, we obtain the following relations
\begin{equation}\begin{aligned}
    h^c_{AB}(\vb{r}) \eq \mathcal{R}_{A,+}^{\dagger} h^f_{AB}(\vb{r}) \mathcal{R}_{B,+} \\
    h^c_{AC}(\vb{r}) \eq \mathcal{R}_{A,+}^{\dagger} h^f_{AC}(\vb{r}) \mathcal{R}_{C,+} \\
    h^c_{BC}(\vb{r}) \eq \mathcal{R}_{B,+}^{\dagger} h^f_{BC}(\vb{r}) \mathcal{R}_{C,+}.
\end{aligned}\end{equation}
where the $\mathcal{R}_{i,+}$ are
\begin{equation}\begin{aligned}
    \mathcal{R}_{A,+} \eq
    \begin{pmatrix}
        -\frac{1}{2} & \frac{\sqrt{3}}{2} & 0 \\
        -\frac{\sqrt{3}}{2} & -\frac{1}{2} & 0 \\
        0 & 0 & 1
    \end{pmatrix} \quad
    & \mathcal{R}_{B,+} \eq \mathbb{1}_{3\times 3} \quad
    & \mathcal{R}_{C,+} \eq 
    \begin{pmatrix}
        -\frac{1}{2} & -\frac{\sqrt{3}}{2} & 0 \\
        \frac{\sqrt{3}}{2} & -\frac{1}{2} & 0 \\
        0 & 0 & 1
    \end{pmatrix}
\end{aligned}\end{equation}
Explicitly, the hoppings term between $d_{x^2-y^2}$ orbitals on different sublattices are
\begin{equation}\begin{aligned}\label{eq:tij_local}
    t^{c}_{A4,B4}(\vb{r}) = \bqty{h^c_{AB}(\vb{r})}_{44} \eq  \frac{1}{2} \pqty{ \sqrt{3} t^{f}_{A1B4}(\vb{r}) - t^{f}_{A4A4}(\vb{r}) } \\
    t^{c}_{A4,C4}(\vb{r}) = \bqty{h^c_{AC}(\vb{r})}_{44} \eq \frac{1}{4} \pqty{-3 t^f_{A1C1}(\vb{r}) + t^f_{A4C4}(\vb{r}) } \\
    t^{c}_{B4,C4}(\vb{r}) = \bqty{h^c_{BC}(\vb{r})}_{44} \eq \frac{1}{2} \pqty{- \sqrt{3} t^{f}_{A1B4}(\vb{r}) - t^{f}_{A4A4}(\vb{r}) }.
\end{aligned}\end{equation}
Note that $t^{f}_{i\alpha j\beta}(\vb{r})$ is identical for any pair of sites $i$ and $j$, as the orbitals are defined in the global Cartesian coordinate system. Therefore, by substituting \cref{eq:tij_cart} into \cref{eq:tij_local}, we obtain the explicit forms of the NN hopping terms between $d_{x^2-y^2}$ orbitals defined in the local coordinate system on a kagome lattice within the SK formalism:
\begin{equation}\begin{aligned}\label{eq:d_sk_nn_44}
    t^{c}_{A4,B4}(\vb{r}) \eq -V^{d}(\vb{r}) (l-m)(l+m)(l^2-2\sqrt{3} lm -m^2) - \frac{1}{2} (l^2+m^2) V_{\pi}(\vb{r}), \\
    t^{c}_{A4,C4}(\vb{r}) \eq \frac{V^{d}(\vb{r})}{2} (l^4 - 14 l^2m^2 + m^4) - \frac{1}{2} (l^2+m^2) V_{\pi}(\vb{r}), \\
    t^{c}_{B4,C4}(\vb{r}) \eq -V^{d}(\vb{r}) (l-m)(l+m)(l^2 + 2\sqrt{3} lm -m^2) - \frac{1}{2} (l^2+m^2) V_{\pi}(\vb{r}),
\end{aligned}\end{equation}
where $V^{d}(\vb{r}) \equiv \bqty{V_{\delta}(\vb{r}) - 4V_{\pi}(\vb{r}) + 3V_{\sigma}(\vb{r})}/8$, and $\Bqty{V_{\delta}(\vb{r}), V_{\pi}(\vb{r}), V_{\sigma}(\vb{r})}$ are bonding integrals (SK parameters) of $d$ orbitals (i.e., $V_{dd\sigma/\pi/\delta}$ used in \cref{tab:HoppingSK}), which we assume to be Gaussian (\cref{eq:d_sk_ga}).
Here, the variables $(l,m,n)$ correspond to the direction cosines, as explained in the caption of \cref{tab:HoppingSK}, of the hopping vector $\vb{r}$:
\begin{equation}\begin{aligned}l = \frac{r_x}{\abs{\vb{r}}}, \quad m = \frac{r_y}{\abs{\vb{r}}}, \quad n = \frac{r_z}{\abs{\vb{r}}}
    \label{app:eq:direction_cosine}
\end{aligned}\end{equation}
For a perfect kagome lattice, where the local coordinates are used, these direction cosines can be expressed as $l = \cos \theta = x/\sqrt{x^2+y^2}$ and $m = \sin \theta = y/\sqrt{x^2+y^2}$, with $n=0$ if only intra-layer hopping terms, in this case only the NN intra-layer terms, are considered. Consequently, \cref{eq:d_sk_nn_44} can be simplified as
\begin{equation}\begin{aligned}
    t^{c}_{A4,B4}(\vb{r}) \eq -V^{d}(\vb{r}) (l-m)(l+m)(l^2-2\sqrt{3} lm -m^2) - \frac{1}{2} V_{\pi}(\vb{r}) = f_{A4,B4}(\vb{r}) V^d (\vb{r}) - \frac{1}{2} V_{\pi}(\vb{r}), \\
    t^{c}_{A4,C4}(\vb{r}) \eq \frac{V^{d}(\vb{r})}{2} (l^4 - 14 l^2m^2 + m^4) - \frac{1}{2} V_{\pi}(\vb{r}) = f_{A4,C4}(\vb{r}) V^d (\vb{r}) - \frac{1}{2} V_{\pi}(\vb{r}), \\
    t^{c}_{B4,C4}(\vb{r}) \eq -V^{d}(\vb{r}) (l-m)(l+m)(l^2 + 2\sqrt{3} lm -m^2) - \frac{1}{2} V_{\pi}(\vb{r}) = f_{B4,C4}(\vb{r}) V^d (\vb{r}) - \frac{1}{2} V_{\pi}(\vb{r}),
\end{aligned}\end{equation}

\subsubsection{\label{app:ModelsGaussianEPCinSK}Electron-phonon coupling of $d_{x^2-y^2}$ orbitals in Slater-Koster formalism}

To explore the electron-phonon coupling (EPC) from this SK formalism in GA, it is important to recognize that the EPC strength associated with a given hopping term for a specific electron in relation to a phonon mode can be expressed as~\cite{Yu2024Non-trivialA}:
\begin{equation}\begin{aligned}
\bqty{g(\vb{R}_{ij})}_{i\alpha j\beta;\nu} \eq \eval{\partial_{r_{\nu}} t_{i\alpha j\beta}(\vb{r})}_{\vb{r}=\vb{R}_{ij}}.
\end{aligned}\end{equation}
Here, $g_{\nu}$ represents the EPC constant for the phonon mode $\nu$, and $t(\vb{r})$ is the hopping integral that depends on the electron position $\vb{r}$, as introduced in Ref.~\cite{Yu2024Non-trivialA}. This derivative indicates how sensitive the hopping parameter is to changes in atomic positions due to phonon displacements, capturing the variation of the electronic hopping strength with respect to the lattice distortion induced by the phonon mode $\nu$.
After taking derivatives of \cref{eq:d_sk_nn_44}, we have the following equation
\begin{equation}\begin{aligned}\label{eq:pdv}
    \pdv{t_{mn}(\vb{r})}{r_\nu} \eq \pdv{f_{mn}(\vb{r})}{r_{\nu}} V^d(\vb{r}) + f_{mn}(\vb{r}) \pdv{V^d(\vb{r})}{r_{\nu}} - \frac{1}{2} \pdv{V_{\pi}(\vb{r})}{r_{\nu}}
\end{aligned}\end{equation}
Here, $m,n$ denote the sublattices $A,B,C$ and orbital $d_{x^2-y^2}$.
\cref{eq:pdv} can be divided into two components: the radial part, which consists of the partial derivatives of the Gaussian decaying bonding integrals $V_i(\vb{r})$, and the angular part, which consists of the partial derivatives of the function $f_{mn}(\vb{r})$.
We start with the radial parts in \cref{eq:pdv}. For the partial derivatives of SK hopping integrals $V_i(\vb{r}),\, i\in\Bqty{\sigma,\pi,\delta}$, individually, we have
\begin{equation}\begin{aligned}\label{eq:pdv_vi}
    \pdv{V_i(\vb{r})}{r_\nu} \eq \pdv{V_{i,0} \exp(-\gamma_i \frac{\abs{\vb{r}}^2}{2})}{r_\nu} = -r_{\nu} \gamma_i V_{i,0} \exp(-\gamma_i \frac{r^2}{2} )
\end{aligned}\end{equation}
which leads to
\begin{equation}\begin{aligned}\label{eq:pdv_radial}
    \pdv{V^d(\vb{r})}{r_\nu} \eq \frac{1}{8} \pdv{V_\delta(\vb{r})-4V_\pi(\vb{r})+3V_\sigma(\vb{r})}{r_\nu} \\
    \eq \frac{1}{8} \pdv{V_{\delta,0} \exp(-\gamma_\delta \frac{\abs{\vb{r}}^2}{2})}{r_\nu} - \frac{4}{8}\pdv{V_{\pi,0} \exp(-\gamma_\pi \frac{\abs{\vb{r}}^2}{2})}{r_\nu} +\frac{3}{8} \pdv{V_{\sigma,0} \exp(-\gamma_\sigma \frac{\abs{\vb{r}}^2}{2})}{r_\nu} \\
    \eq -r_{\nu} \bqty{ \gamma_\delta V_{\delta}(\vb{r}) - 4 \gamma_\pi V_\pi(\vb{r}) + 3 \gamma_\sigma V_\sigma(\vb{r}) }/8
\end{aligned}\end{equation}
From \cref{eq:pdv_vi,eq:pdv_radial}, one can determine that the radial component of the electron-phonon coupling (EPC) for the $B_{3u}$ (local $x$) and $B_{2u}$ (local $y$) phonons depends on the projection of the hopping vector $\vb{r}$ onto the corresponding local axes. This dependency leads to the following relation:
\begin{equation}\begin{aligned}
    \frac{\pdv{V_i(\vb{r})}{x}}{\pdv{V_i(\vb{r})}{y}} \eq \frac{x}{y}, \quad i\in\Bqty{\sigma,\pi,\delta}
\end{aligned}\end{equation}
and for the NN hopping vectors on the kagome lattice $\vb{r}_{A4,B4} = (a/4,-\sqrt{3}a/4,0)$, $\vb{r}_{B4,C4} = (a/4,\sqrt{3}a/4,0)$, and $\vb{r}_{A4,C4} = (a/2,0,0)$, we have $y/x=\tan(\pi/3)=\sqrt{3}$ in the local coordinates, meaning
\begin{equation}\begin{aligned}
    \sqrt{3} \pdv{V_i(\vb{r})}{x} \eq \pdv{V_i(\vb{r})}{y}, \quad i\in\Bqty{\sigma,\pi,\delta}
\end{aligned}\end{equation}
for every Gaussian decaying hopping integral $V_i(\vb{r})$ and the same for the linear combination \cref{eq:pdv_radial} of them.

Second, for the angular parts in \cref{eq:pdv},
\begin{equation}\begin{aligned}
    \pdv{f_{A4,B4}(\vb{r})}{x} \eq -y \frac{ 2(\sqrt{3} x^4 + 4 x^3 y - 6 \sqrt{3} x^2 y^2 - 4 x y^3 + \sqrt{3} y^4) }{r^6} \\
    \pdv{f_{A4,B4}(\vb{r})}{y} \eq  x \frac{ 2(\sqrt{3} x^4 + 4 x^3 y - 6 \sqrt{3} x^2 y^2 - 4 x y^3 + \sqrt{3} y^4) }{r^6} \\
    \pdv{f_{A4,C4}(\vb{r})}{x} \eq  y \frac{16 xy (x-y)(x+y) }{r^6} \\
    \pdv{f_{A4,C4}(\vb{r})}{y} \eq -x \frac{16 xy (x-y)(x+y) }{r^6} \\
    \pdv{f_{B4,C4}(\vb{r})}{x} \eq  y \frac{ 2(\sqrt{3} x^4 - 4 x^3 y - 6 \sqrt{3} x^2 y^2 + 4 x y^3 + \sqrt{3} y^4) }{r^6} \\
    \pdv{f_{B4,C4}(\vb{r})}{y} \eq -x \frac{ 2(\sqrt{3} x^4 - 4 x^3 y - 6 \sqrt{3} x^2 y^2 + 4 x y^3 + \sqrt{3} y^4) }{r^6}
\end{aligned}\end{equation}
From the above equations, 
\begin{equation}\begin{aligned}
    \frac{\pdv{f_{mn}(\vb{r})}{x}}{\pdv{f_{mn}(\vb{r})}{y}} \eq -\frac{y}{x} = -\sqrt{3}
\end{aligned}\end{equation}

Thus, we arrive at a relation where the amplitude of the angular component in the partial derivatives of the hopping terms with respect to $x$ is $\sqrt{3}$  times larger than that for $y$, while the opposite holds for the radial component. 
To further check the validity of our approximation, the amplitudes of the EPC are calculated numerically.

\begin{figure}
    \centering
    \includegraphics[width=0.4\linewidth]{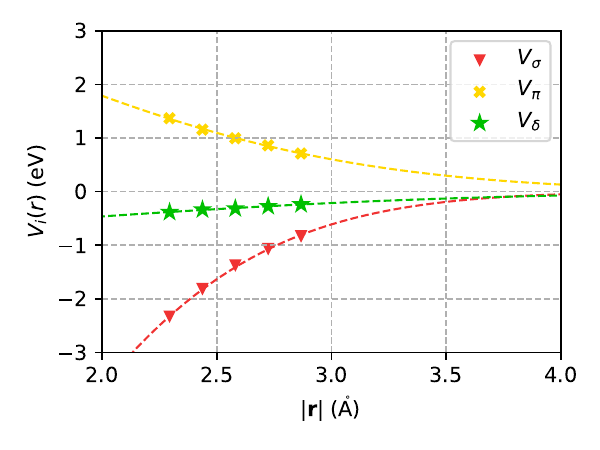}
    \caption{Fitted Gaussian functions for SK hopping integrals $V_i(\vb{r})\;(i=\sigma,\pi,\delta)$. The parameters, defined in \cref{eq:d_sk_ga}, for each $i$ are $\gamma_{\sigma} = 0.7143 \;\mathring{\mathrm{A}}^{-2}$, $V_{\sigma}=-15.2033 \;\mathrm{eV}$; $\gamma_{\pi} = 0.4351\;\mathring{\mathrm{A}}^{-2}$, $V_{\pi}=4.2644\;\mathrm{eV}$; $\gamma_{\delta} = 0.3096\;\mathring{\mathrm{A}}^{-2}$, $V_{\delta}=-0.8584\;\mathrm{eV}$.}
    \label{fig:sk-fit}
\end{figure}

By fitting DFT data (\cref{fig:sk-fit}), the corresponding decaying parameters for the SK hopping integrals $V_i(\vb{r})$ are obtained as, $\gamma_{\sigma} = 0.7143 \;\mathring{\mathrm{A}}^{-2}$, $V_{\sigma}=-15.2033 \;\mathrm{eV}$; $\gamma_{\pi} = 0.4351\;\mathring{\mathrm{A}}^{-2}$, $V_{\pi}=4.2644\;\mathrm{eV}$; $\gamma_{\delta} = 0.3096\;\mathring{\mathrm{A}}^{-2}$, $V_{\delta}=-0.8584\;\mathrm{eV}$.
The corresponding EPC of $A_g (d_{x^2-y^2})$ orbitals coupling to $B_{3u} (x)$ and $B_{2u} (y)$ atom movements are calculated to be $0.0192\;{\mathrm{Ry}}/{\mathrm{Bohr}}$ and $0.0334\;{\mathrm{Ry}}/{\mathrm{Bohr}}$, respectively, which are close to the \textit{ab-initio} results, being $0.0201\;{\mathrm{Ry}}/{\mathrm{Bohr}}$ and $0.0204\;{\mathrm{Ry}}/{\mathrm{Bohr}}$.

\begin{figure}
    \centering
    \includegraphics[width=0.8\linewidth]{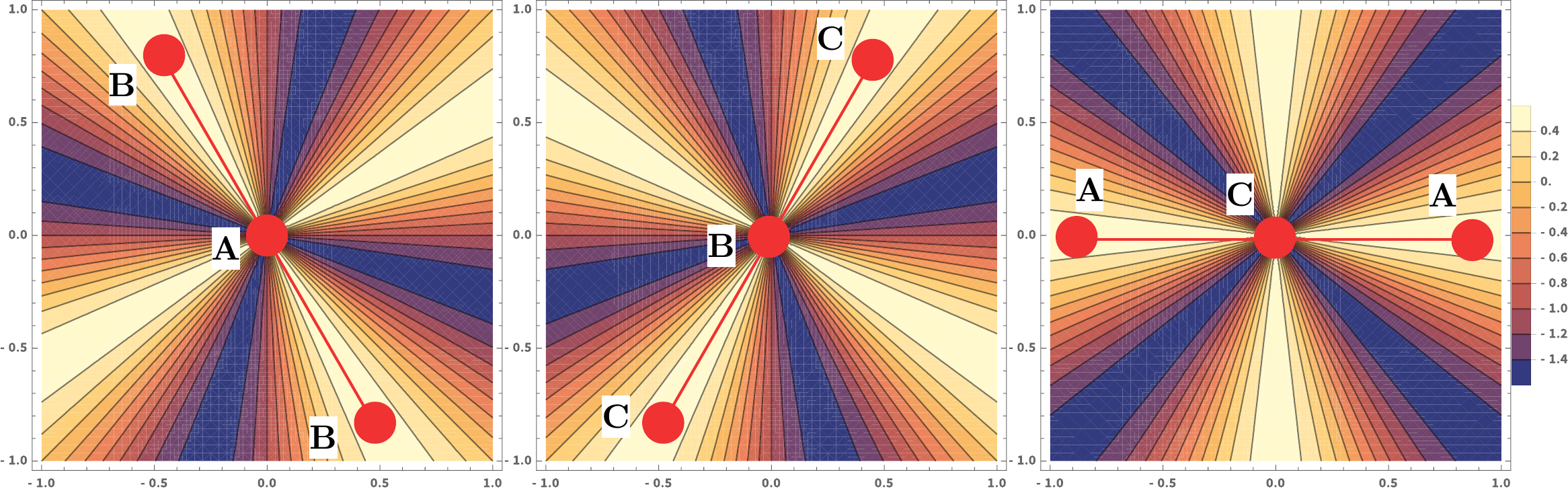}
    \caption{Distribution of $f_{mn}(\vb{r})$ with respect to $\vb{r}=(r_x,r_y,r_z=0)$, the red circles are the lattice sites, and the red lines are the directions of the NN hopping vectors. From left to right is $f_{A4,B4}(\vb{r})$, $f_{B4,C4}(\vb{r})$, and $f_{A4,C4}(\vb{r})$. In the kagome lattice defined by \cref{eq:direct_lat,eq:sub_lat}, the hopping vectors defined accordingly are then $\vb{r}_{A4,B4} = (a/4,-\sqrt{3}a/4,0)$, $\vb{r}_{B4,C4} = (a/4,\sqrt{3}a/4,0)$, and $\vb{r}_{A4,C4} = (a/2,0,0)$, which are shown by red lines in the corresponding panels, respectively.}
    \label{fig:fmn}
\end{figure}

However, the calculated EPC of the $B_{3u}$ phonon is still $\sqrt{3}$ times smaller than the EPC of the $B_{2u}$ phonon.
This is because function $f_{mn}(\vb{r})$ reaches its maximum with hopping vector $\vb{r}$ being the nearest neighboring (\cref{fig:fmn}), resulting in the first derivatives of $f_{mn}(\vb{r})$ being zero.
As a result, the only contribution left in \cref{eq:pdv} is the radial part, where the EPC of the $B_{3u}$ phonon is $\sqrt{3}$ times smaller than the EPC of the $B_{2u}$ phonon as discussed above.

\subsection{\label{app:ModelsSpringBall} Spring-ball model for kagome lattice}
To further investigate the phonon spectrum of the kagome lattice, we employ the spring-ball model. The kagome lattice formed by the springs and balls is described by the lattice vectors:
\begin{equation}\begin{aligned}
    (\vb{a},\vb{b},\vb{c}) \eq 
    \mqty( \sqrt{3}a/2 & 0 & 0 \\
    -a/2 & a & 0 \\
    0 & 0 & c)
\end{aligned}\label{app:eq:spring-ball-basis}
\end{equation}
where $a$ and $c$ denote the lattice constants.
The three kagome sites, labeled as $A$, $B$, and $C$, are positioned at the $3f$ Wyckoff positions (on the $z=0$ plane) as defined in \cref{eq:sub_lat}. 
In Cartesian coordinates 
\begin{equation}\begin{aligned}
    \vb{t}_A \eq (\sqrt{3}a/4, a/4, 0) \quad
    &\vb{t}_B \eq (\sqrt{3}a/4, -a/4, 0) \quad
    &\vb{t}_C \eq (0, a/2, 0) .
\end{aligned}\end{equation}
In this spring-ball model, every two nearest neighboring kagome sites are connected by a pre-stretched spring (characterized by $\eta$) with spring constant $\kappa$.
The potential energy of the spring connecting neighboring sites $i$ and $j$ is given by the expression
\begin{equation}\begin{aligned}\label{eq:Eij}
    E_{ij}(\vb{R}_j-\vb{R}_i) \eq \frac{\kappa}{2} \bqty{ \abs{\vb{t}_j + \vb{u}_j - \vb{t}_i - \vb{u}_i + \vb{R}_j - \vb{R}_i} - \eta \abs{\vb{t}_j - \vb{t}_i + \vb{R}_j - \vb{R}_i} }^2
\end{aligned}\end{equation}
where $\vb{u}_i = (u_{i,x}, u_{i,y}, u_{i,z})$ and $\vb{u}_j = (u_{j,x}, u_{j,y}, u_{j,z})$ represent the small displacements of the atoms $i$ and $j$ from their equilibrium positions in global Cartesian coordinate. Here, $\eta$ represents the pre-stretching parameter of the springs, which defines the equilibrium length of the spring as $\eta$ times the NN hopping vector length, denoted by $\abs{\vb{r}}$. In other words, the natural length of each spring in this model is scaled by the factor $\eta$ relative to the original NN bond length. 

\begin{figure}
    \centering
    \includegraphics[width=0.7\linewidth]{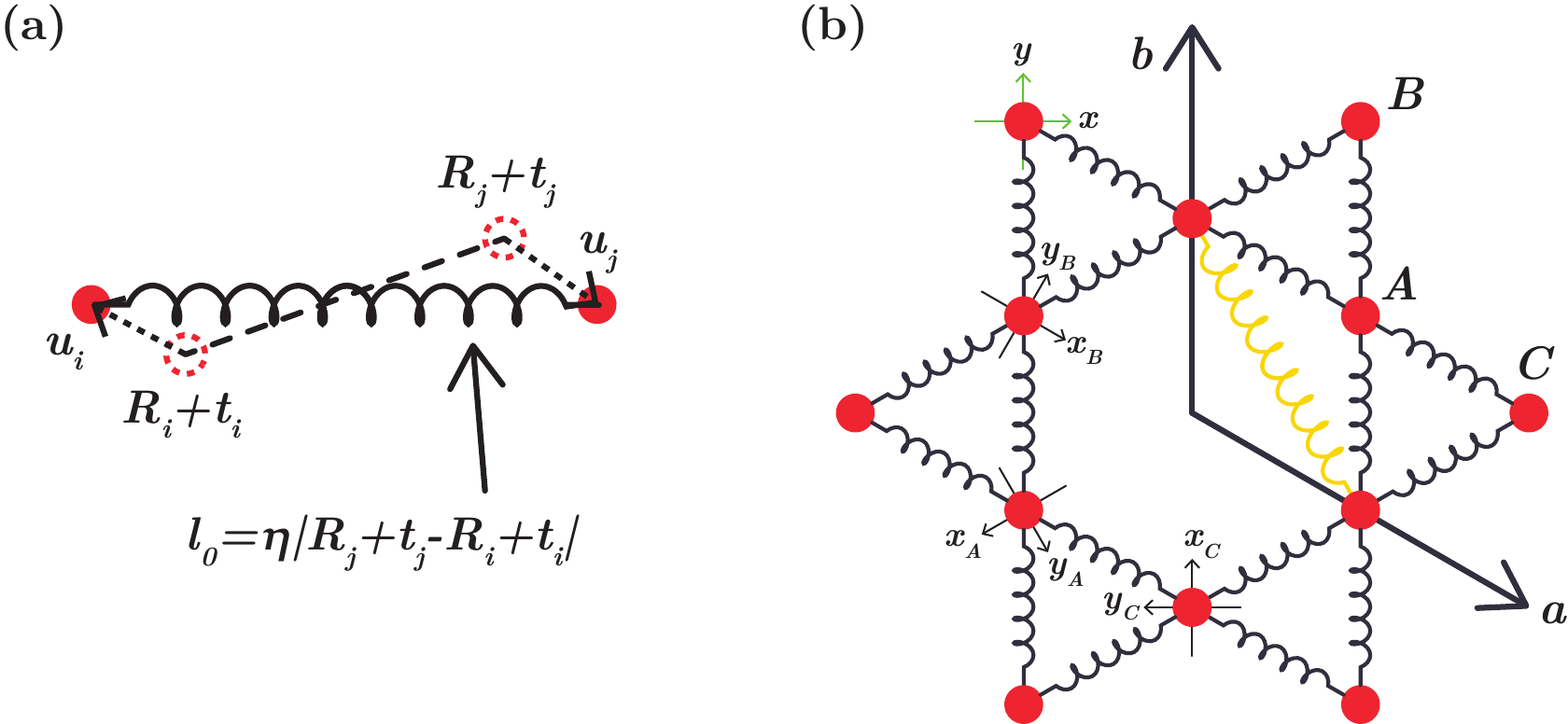}
    \caption{(a) Illustration of \cref{eq:Eij}, depicting the interaction or energy relationship between different sites. In (a), $l_0=\eta \abs{\vb{R}_j+\vb{t}_j-\vb{R}_i-\vb{t}_i}$ is the equilibrium length of the spring.
    (b) Spring-ball model for a 2D kagome system, with global Cartesian coordinate axes shown in green arrows and local axes for each site indicated by black arrows.}
    \label{fig:SpringBallIllustration}
\end{figure}

In practice, for the NN bonds, we have
\begin{equation}\begin{aligned}
    E_{AB}[(0,0,0)] \eq \frac{\kappa}{2} \bqty{ \pqty{(u_{B,x} - u_{A,x})^2 + (u_{B,y} - u_{A,y} -l\frac{a}{2})^2 + (u_{B,z}-u_{A,z})^2}^{1/2} - \eta \frac{a}{2}}^2
\end{aligned}\end{equation}
\begin{equation}\begin{aligned}
    E_{AC}[(0,0,0)] \eq \frac{\kappa}{2} \bqty{ \pqty{ (u_{C,x} - u_{A,x} - \frac{\sqrt{3}a}{4})^2 + (u_{C,y} - u_{A,y} + \frac{a}{4})^2 + (u_{C,z} - u_{A,z})}^{1/2} - \eta \frac{a}{2} }^2
\end{aligned}\end{equation}
\begin{equation}\begin{aligned}
    E_{BC}[(1,0,0)] \eq \frac{\kappa}{2} \bqty{ \pqty{ (u_{C,x} - u_{B,x} + \frac{\sqrt{3}a}{4})^2 + (u_{C,y} - u_{B,y} + \frac{a}{4})^2 + (u_{C,z} - u_{B,z})^2}^{1/2} - \eta \frac{a}{2} }^2.
\end{aligned}\end{equation}
The force constant matrix $F_{i\mu,j\nu}(\vb{R}_j - \vb{R}_i)$, which represents the interaction between sites $i$ and $j$, is calculated as the second derivative of the potential:
\begin{equation}\begin{aligned}
    F_{i\mu,j\nu}(\vb{R}_j-\vb{R}_i) \eq \eval{\pdv{E_{ij}(\vb{R}_j-\vb{R}_i)}{\vb{u}_{i\mu}}{\vb{u}_{j\nu}}}_{ \vb{u}_{i\mu} = 0, \vb{u}_{j\nu} = 0 }
\end{aligned}\end{equation}
therefore, for the NN bonds, 
\begin{equation}\begin{aligned}
    F_{AB}[(0,0,0)] \eq \sum_{\mu, \nu} \eval{\pdv{E_{AB}}{\vb{u}_{A\mu}}{\vb{u}_{B\nu}}}_{\vb{u}_{A\mu} = 0, \vb{u}_{B\nu} = 0}
    = \mqty(\dmat{ -2\kappa ( 1 - \eta),  - 2\kappa,  -2\kappa ( 1 - \eta) })
\end{aligned}\end{equation}
\begin{equation}\begin{aligned}
    F_{AC}[(0,0,0)] \eq \sum_{\mu, \nu} \eval{\pdv{E_{AC}}{\vb{u}_{A\mu}}{\vb{u}_{C\nu}}}_{\vb{u}_{A\mu} = 0, \vb{u}_{C\nu} = 0} =
    \begin{pmatrix}
        -\kappa ( 2- \eta/2) & \sqrt{3}\kappa\eta/2 & \\
        \sqrt{3}\kappa\eta/2 & -\kappa(2 - 3\eta/2) & \\
        & & -\kappa ( 2- \eta/2)
    \end{pmatrix}
\end{aligned}\end{equation}
\begin{equation}\begin{aligned}
    F_{BC}[(1,0,0)] \eq \sum_{\mu, \nu} \eval{\pdv{E_{BC}}{\vb{u}_{B\mu}}{\vb{u}_{C\nu}}}_{\vb{u}_{B\mu} = 0, \vb{u}_{C\nu} = 0} =
    \begin{pmatrix}
        -\kappa ( 2- \eta/2) & -\sqrt{3}\kappa\eta/2 & \\
        -\sqrt{3}\kappa\eta/2 & -\kappa(2 - 3\eta/2) & \\
        & & -\kappa ( 2- \eta/2)
    \end{pmatrix}.
\end{aligned}\end{equation}
From this, it is evident that the $z$-phonon is completely decoupled from the in-plane $x$- and $y$-phonons due to the mirror $z$ symmetry $\mathcal{M}_z$. As a result, we will limit our analysis to the system consisting only of the $x$- and $y$-phonons. 
The onsite energies in the global coordinate for the phonons are represented as
\begin{equation}\begin{aligned}
    F_{AA}[(0,0,0)] \eq \sum_{\mu, \nu} \eval{\pdv{E_{AA}}{\vb{u}_{A\mu}}{\vb{u}_{A\nu}}}_{\vb{u}_{A\mu} = 0, \vb{u}_{A\nu} = 0} =
    \begin{pmatrix}
        \kappa (8-5\eta) & -\sqrt{3} \kappa \eta \\
        -\sqrt{3} \kappa \eta & \kappa (8-3\eta)
    \end{pmatrix},
\end{aligned}\end{equation}
\begin{equation}\begin{aligned}
    F_{BB}[(0,0,0)] \eq \sum_{\mu, \nu} \eval{\pdv{E_{BB}}{\vb{u}_{B\mu}}{\vb{u}_{B\nu}}}_{\vb{u}_{B\mu} = 0, \vb{u}_{B\nu} = 0} =
    \begin{pmatrix}
        \kappa (8-5\eta) & \sqrt{3} \kappa \eta \\
        \sqrt{3} \kappa \eta & \kappa (8-3\eta)
    \end{pmatrix},
\end{aligned}\end{equation}
\begin{equation}\begin{aligned}
    F_{CC}[(0,0,0)] \eq \sum_{\mu, \nu} \eval{\pdv{E_{CC}}{\vb{u}_{C\mu}}{\vb{u}_{C\nu}}}_{\vb{u}_{C\mu} = 0, \vb{u}_{C\nu} = 0} =
    \mqty( \dmat{2\kappa(4-\eta), 2\kappa(4-3\eta)} ),
\end{aligned}\end{equation}

\begin{figure}[H]
    \centering
    \includegraphics[width=0.8\linewidth]{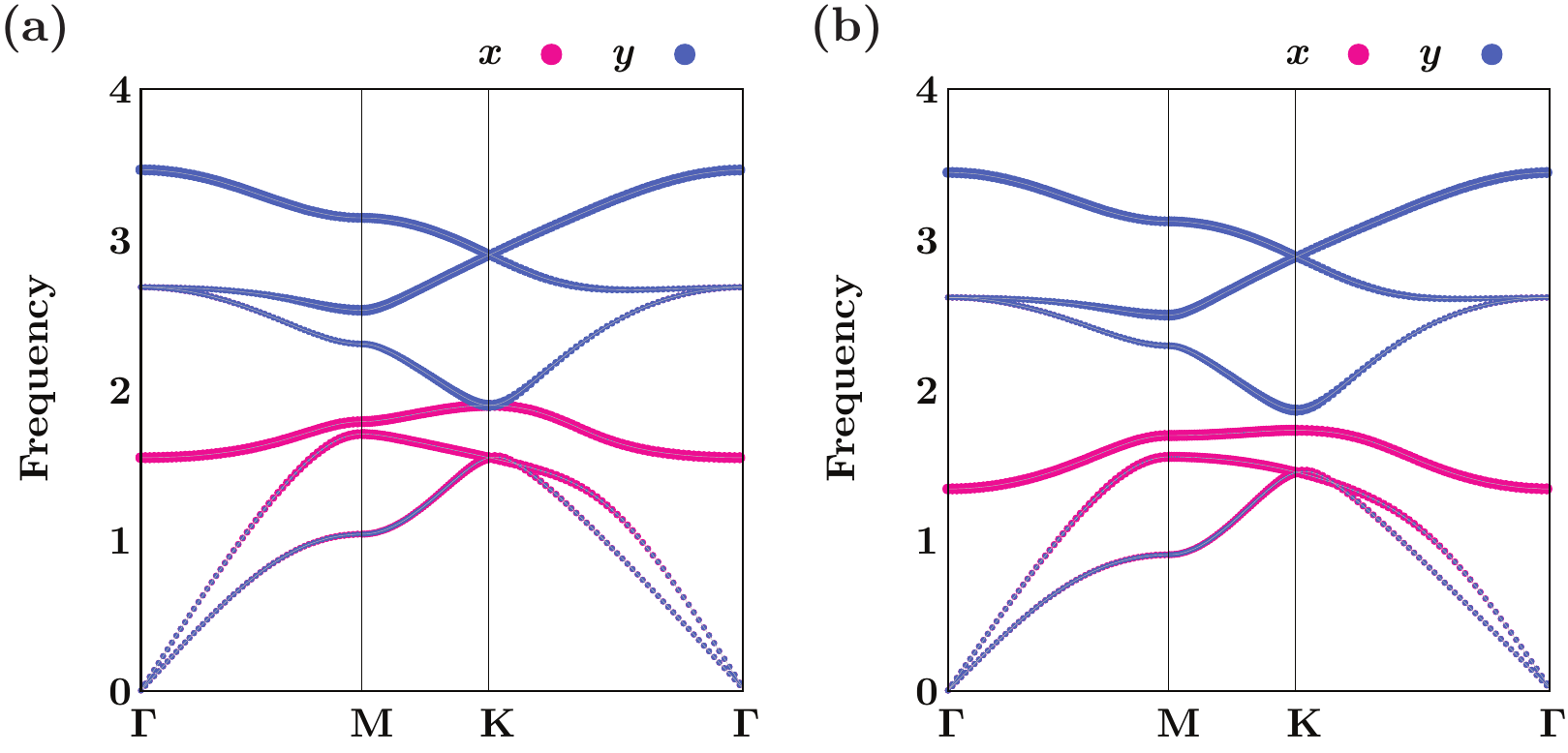}
    \caption{The phonon-mode-resolved spectrum of the kagome spring-ball model without (a) and with (b) considering NNN springs (the yellow spring in \cref{fig:SpringBallIllustration}{(b)}).}
    \label{fig:SpringBall}
\end{figure}

The local axes $(x_i,y_i)$ for each sublattice $\vb{t}_i$ defined in this kagome lattice are presented in \cref{fig:SpringBallIllustration}{(b)}. The rotation matrices from displacement vector $\vb{u}$ defined in global Cartesian coordinate to displacement vector $\vb{u}'$ defined in the local coordinate of each site are
\begin{equation}\begin{aligned}
    \mathcal{R}_{A} = \mathcal{R}^{p}(-\frac{5\pi}{6}) \eq \mqty( -\sqrt{3}/2 & 1/2 & 0 \\ -1/2 & -\sqrt{3}/2  & 0 \\ 0 & 0 & 1), \quad
    & \mathcal{R}_{B} = \mathcal{R}^{p}(\frac{\pi}{2}) \eq \mqty( \sqrt{3}/2 & 1/2 & 0 \\ -1/2 & \sqrt{3}/2  & 0 \\ 0 & 0 & 1), \quad
    & \mathcal{R}_{C} = \mathcal{R}^{p}(-\frac{\pi}{6}) \eq \mqty( 0 & -1 & 0 \\ 1 & 0 & 0 \\ 0 & 0 & 1) .
\end{aligned}\end{equation}
Since only one type of atom is considered in this model, the real-space dynamic matrix $D(\vb{R})$ is simply:
\begin{equation}\begin{aligned}
    D_{i\mu,j\nu}(\vb{R}) \eq \frac{F_{i\mu,j\nu}(\vb{R})}{\sqrt{M_iM_j}} = \frac{F_{i\mu,j\nu}(\vb{R})}{M}
\end{aligned}\end{equation}
where $M$ is the mass.
In the local coordinate system $(x_i,y_i)$ for site $\vb{t}_i$ (here, owing to the $\mathcal{M}_z$ symmetry, the $z$ axis can be neglected), as defined in \cref{fig:SpringBallIllustration}{(b)}, the dynamical matrix for the nearest-neighbor kagome spring-ball model in real space takes the following form
\begin{equation}\begin{aligned}
    D_{AB}[(0,0,0)] \eq \frac{1}{M}
    \begin{pmatrix}
        t_{xx} & t_{xy} \\
       -t_{xy} & t_{yy}
    \end{pmatrix} = D_{AC}[(0,0,0)] = D_{BC}[(0,0,0)]
\end{aligned}\end{equation}
where
\begin{equation}\begin{aligned}
    t_{xx} \eq \frac{\kappa}{2} (2-3\eta), \quad
    &t_{xy} \eq \frac{\sqrt{3}\kappa}{2} (2-\eta), \quad
    &t_{yy} \eq \frac{\kappa}{2} (2+\eta),
\end{aligned}\end{equation}
and the onsite energies are
\begin{equation}\begin{aligned}
    D_{AA}[(0,0,0)] \eq \frac{1}{M} \mqty( \dmat{\varepsilon_{x}, \varepsilon_{y}} ) = D_{BB}[(0,0,0)] = D_{CC}[(0,0,0)]
\end{aligned}\end{equation}
where
\begin{equation}\begin{aligned}
    \varepsilon_{x} \eq 2\kappa (4-3\eta), \quad
    &\varepsilon_{y} \eq 2\kappa (4-\eta).
\end{aligned}\end{equation}
This indicates that the $x$-phonon generally has lower onsite energy compared to the $y$-phonon, which arises from the fact that the horizontal movement involves lower energy and weaker coupling, leading to a smaller bandwidth for the $x$-phonon. 
Upon performing the Fourier transformation, the dynamical matrix in $\vb{q}$-space becomes:
\begin{equation}\begin{aligned}
    D_{i\mu,j\nu}(\vb{q}) \eq \sum_{\vb{R}} e^{i\vb{q}\cdot\vb{R}} D_{i\mu,j\nu}(\vb{R})
\end{aligned}\end{equation}
where $\vb{R} = \vb{R}_j - \vb{R}_i$ due to translational symmetry.
The Hamiltonian $H(\vb{q})$ of the kagome spring-ball model in local coordinate is then
\begin{equation}\begin{aligned}\label{eq:app-spring-ball}
    H(\vb{q}) \eq \frac{1}{M} \begin{pmatrix}
        \varepsilon_{x} & 0 & t_{AxBx}(\vb{q}) & t_{AxBy}(\vb{q}) & t_{AxCx}(\vb{q}) & t_{AxCy}(\vb{q}) \\
        & \varepsilon_{y} & -t_{AxBy}(\vb{q}) & t_{AyBy}(\vb{q}) & -t_{AxCy}(\vb{q}) & t_{AyCy}(\vb{q}) \\
        & & \varepsilon_{x} & 0 & t_{BxCx}(\vb{q}) & t_{BxCy}(\vb{q}) \\
        & & & \varepsilon_{y} & -t_{BxCy}(\vb{q}) & t_{ByCy}(\vb{q}) \\
        & & \dag & & \varepsilon_{x} & 0 \\
        & & & & & \varepsilon_{y}
    \end{pmatrix}
\end{aligned}\end{equation}
where 
\begin{equation}\begin{aligned}
    t_{AxBx}(\vb{q}) \eq t_{xx} (1 + e^{i q_b}), \quad
    &t_{AxBy}(\vb{q}) \eq t_{xy} (1 + e^{i q_b}), \quad
    &t_{AyBy}(\vb{q}) \eq t_{yy} (1 + e^{i q_b}), \\
    t_{AxCx}(\vb{q}) \eq t_{xx} (1 + e^{i q_a}), \quad
    &t_{AxCy}(\vb{q}) \eq t_{xy} (1 + e^{i q_a}), \quad
    &t_{AyCy}(\vb{q}) \eq t_{yy} (1 + e^{i q_a}), \\
    t_{BxCx}(\vb{q}) \eq t_{xx} (1 + e^{i (q_a+q_b)}), \quad
    &t_{BxCy}(\vb{q}) \eq t_{xy} (1 + e^{i (q_a+q_b)}), \quad
    &t_{ByCy}(\vb{q}) \eq t_{yy} (1 + e^{i (q_a+q_b)}).
\end{aligned}\end{equation}
Here, $\vb{q} = (q_a, q_b, q_c)$ is the phonon wavevector in direct coordinates in the BZ from \cref{app:eq:spring-ball-basis}. Without loss of generality, the eigenvalues of the spring-ball model \cref{eq:app-spring-ball} are computed using the parameters $\eta = 0.8$, $\kappa = 1$, and $M = 1$. 
As demonstrated by the mode-resolved phonon dispersion in \cref{fig:SpringBall}{(a)}, numerical analysis confirms that the local $x$ phonon exhibits a roughly two-times lower frequency compared to the local $y$ phonon. 
As discussed in \cref{app:ModelsGaussianEPCHoppingSK}, the real-space EPC of the local $x$ and $y$ phonons are approximated as equal in DFT. However, the contribution to $\lambda$ from the Ru local $x$ and $y$ phonons is approximately in a $4:1$ ratio (see \cref{fig:EPCUndoped}(a)). This aligns with the phonon frequency obtained from the spring-ball model, as $\lambda$ is proportional to $\frac{|g|^2}{\omega^2}$ (see \cref{app:eq:lambda_wqv_relation}), where $g$ represents the real-space EPC strength.

The observed degeneracy between the $x$ and $y$ phonon bands ($B_{3u}@3g$ and $B_{2u}@3g$ EBRs) at the $K$ point is accidental and can be lifted by including springs beyond the nearest neighbors. Specifically, when considering the spring connecting the next-nearest neighboring (NNN) atoms (represented by the yellow springs in \cref{fig:SpringBallIllustration}{(b)}), the phonon dispersion changes, as shown in \cref{fig:SpringBall}{(b)}.

\subsection{Tight-binding model and superfluid weight} \label{app:SFW}

In this section, we construct a simple kagome-lattice model with an attractive interaction to compute the superfluid weight. Our results indicate that the superfluid weight is predominantly determined by the conventional contribution, with the geometric contribution playing a negligible role.

\begin{figure}[b]
    \centering
    \includegraphics[width=\linewidth]{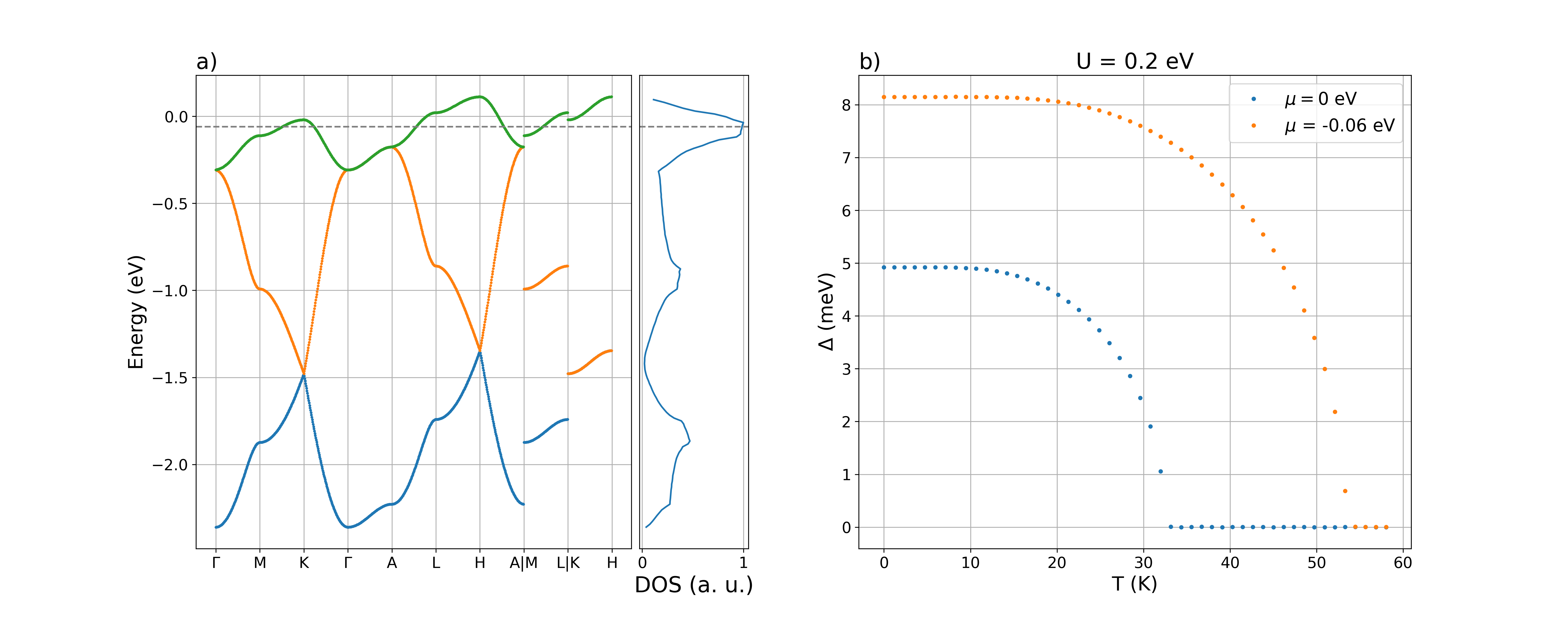}
    \caption{(a) The band structure and the density of states of the tight-binding model. The Fermi energy is set to zero. The dashed gray line indicates where the superconducting gap reaches a maximum when the chemical potential is tuned. (b) The superconducting gap $\Delta$ as a function of temperature with and without tuning the chemical potential. A slight change in the Fermi energy increases both the gap and the critical temperature by around 60\%.}
    \label{fig:tight_binding_band_structure_and_delta}
\end{figure}

The tight-binding model is constructed with an on-site energy $E = -0.9259$ eV, hopping amplitudes $t_{\text{NN}} = -0.39$ eV and $t_{\text{NNN}} = 0.048$ eV between nearest and next nearest neighbors in the kagome plane, respectively, and an out-of-plane hopping amplitude $t_z = -0.033$ eV between the kagome layers. The resulting band structure is shown in Fig. \ref{fig:tight_binding_band_structure_and_delta}(a). Using multiband Bardeen-Cooper-Schrieffer (BCS) mean-field theory with an attractive Hubbard-$U$ interaction~\cite{Peotta2015SuperfluidityA}, where $U = 0.2$ eV has been calibrated to yield a superconducting gap of $\Delta \approx 5$ meV at zero temperature, we find a mean-field $T_c$ of 33 K, see Fig. \ref{fig:tight_binding_band_structure_and_delta}(b). The gap $\Delta \approx 5$ meV is chosen to match the average gap on the $d_{x^2-y^2}$ FS, as shown in \cref{fig:Gap191}. 
By tuning the chemical potential so that the Fermi energy is lowered by 0.06 eV, roughly corresponding to a maximum of the density of states, the superconducting gap and the $T_c$ are increased by around 60\% to 8 meV and 54 K, respectively.
The overestimation of $T_c$ in the model calculation may arise from multiple factors. First, mean-field approximations are known to generally overestimate superconducting transition temperatures. Additionally, the model used is highly simplified, as \ch{LaRu3Si2} possesses multiple FSs, whereas the model calculation primarily focuses on the dominant contribution from $d_{x^2-y^2}$ orbitals. The superconducting gap on other FSs is significantly smaller.

\begin{figure}[b]
    \centering
    \includegraphics[width=0.5\linewidth]{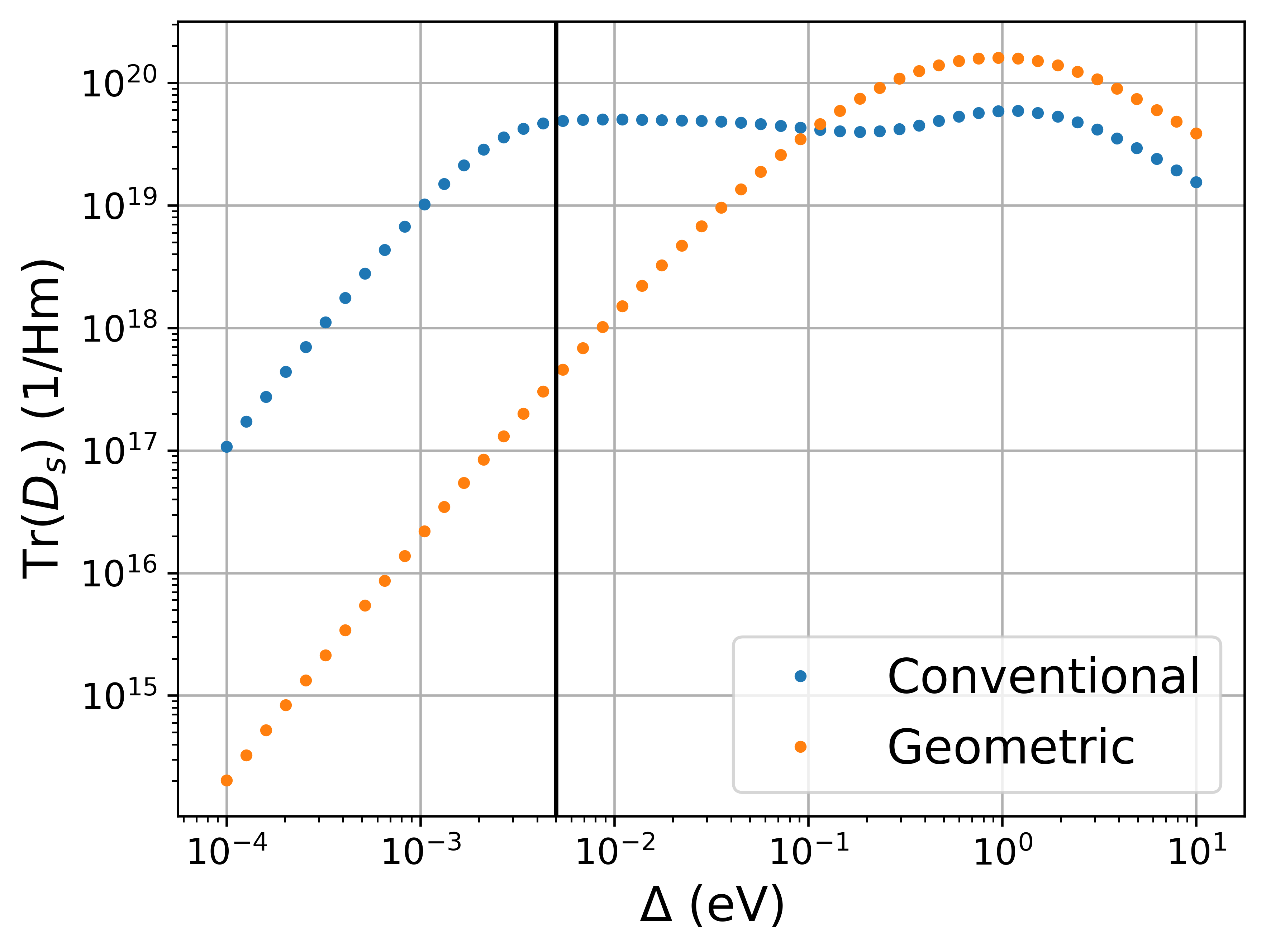}
    \caption{The traced conventional and geometric components of the superfluid weight for the tight-binding model in \cref{fig:tight_binding_band_structure_and_delta}{(a)} are shown, with the black line indicating $\Delta = 5 \,\mathrm{meV}$. The geometric contribution becomes comparable to the conventional part when $\Delta \sim 100 \,\mathrm{meV}$, at which point the superconducting gap is on the order of the flat-band bandwidth.}
    \label{fig:sfw}
\end{figure}

\begin{figure}[htb]
    \centering
    \includegraphics[width=\linewidth]{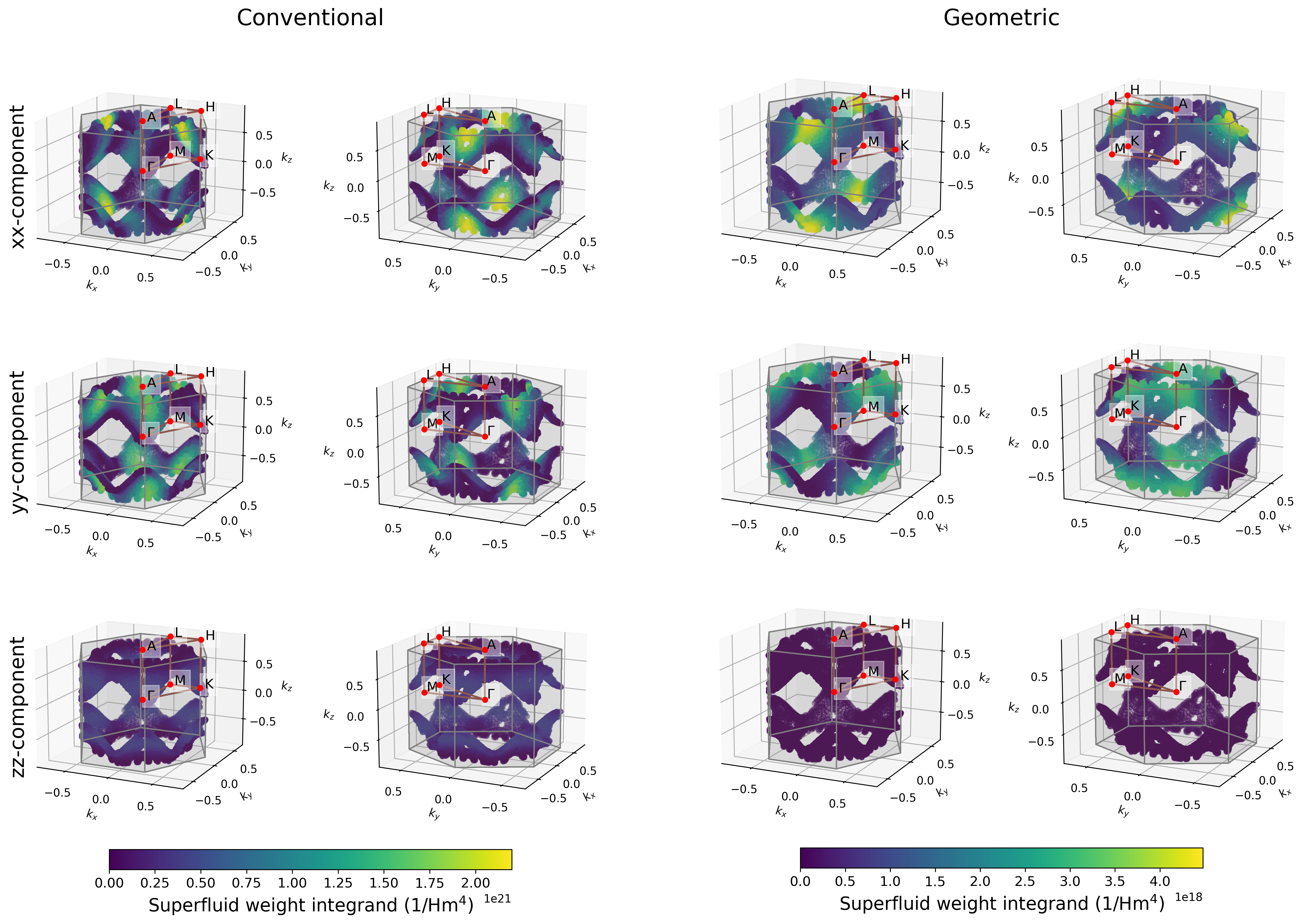}
    \caption{The diagonal components of the superfluid weight integrand around the Fermi surface. Each plot is shown from two different angles. Note the difference in scales between the conventional and geometric parts, with the conventional part being two orders of magnitude larger than the geometric part.}
    \label{fig:tight_binding_sfw}
\end{figure}

We calculate the superfluid weight tensor $D_s$ for this model, split into its conventional and geometric parts, as described in Refs.~\cite{Peotta2015SuperfluidityA, Huhtinen2022RevisitingA}, and find
\begin{align}
    D_s^\text{conv} &= \begin{pmatrix}
        1.96 \cdot 10^{19} & 7.42 \cdot 10^{17} & 0\\
        7.42 \cdot 10^{17} & 2.05 \cdot 10^{19} & 0 \\
        0 & 0 & 8.29 \cdot 10^{18}
    \end{pmatrix},\\
    D_s^\text{geom} &= \begin{pmatrix}
        1.95 \cdot 10^{17} & -2.03 \cdot 10^{15} & 0\\
        -2.03 \cdot 10^{15} & 1.93 \cdot 10^{17} & 0\\
        0 & 0 & 0
    \end{pmatrix},\\
    D_s = D_s^\text{conv} + D_s^\text{geom} &= \begin{pmatrix}
        1.98 \cdot 10^{19} & 7.40 \cdot 10^{17} & 0\\
        7.40 \cdot 10^{17} & 2.07 \cdot 10^{19} & 0\\
        0 & 0 & 8.29 \cdot 10^{18}
    \end{pmatrix}.
\end{align}
The components are indexed by the Cartesian coordinates $x, y, z$, and are expressed in SI-units 1/Hm. The superfluid weight is almost purely conventional, with a difference of two orders of magnitude between the conventional and geometric parts. As $D_s$ is defined as an integral over the Brillouin zone, in Fig. \ref{fig:tight_binding_sfw} we show how the integrand behaves around the Fermi surface. The $x$ and $y$ components especially show clear peaks in complementing directions around the $xy$ plane, while the contribution to the superfluid weight drops to zero in other directions. 
The large conventional part in the superfluid weight is likely due to the large bandwidth compared with the superconducting gap $\Delta$. 
As shown in \cref{fig:sfw}, when the superconducting gap $\Delta$ increases to the order of the quasi-flat-band bandwidth $\sim 100\,\mathrm{meV}$, the geometric contribution to the superfluid weight becomes comparable to the conventional part. From the figure one can also see that even if $\Delta \approx 5$ meV is an overestimate for this model, the geometric part would remain much smaller than the conventional part also in case of smaller $\Delta$.

By tuning the chemical potential to the maximal DOS, the total superfluid weight increases by around 60\% similarly to $\Delta$ and $T_c$. The relative increase for the geometric part is larger than for the conventional part such that it now accounts for around 2\% of the total superfluid weight (compared to 1\% above).

 \clearpage
\section{\label{app:Prediction} Predictions of new kagome flat-band superconductors in 1:3:2 family}

Compounds crystallizing in space group $P6/mmm$ (SG 191), such as \ch{LaIr3Ga2}~\cite{Gui2022LaIr3Ga2A} and \ch{YRu3Si2}~\cite{Gong2022SuperconductivityA}, belong to the 1:3:2 family, also referred to as the \ch{\textit{RT}3\textit{X}2} family, where \textit{T} represents a $4d$ or $5d$ transition metal. Several compounds in the 1:3:2 family have been experimentally confirmed to exhibit superconductivity with critical temperatures around $5 \,\mathrm{K}$, while some others are predicted by \textit{ab initio}~\cite{Barz1980TernaryA, Ku1980SuperconductingA, Chevalier1983SuperconductingA, Rauchschwalbe1984SuperconductivityA, Godart1987CoexistenceA, Escorne1994Type-IIA, Kishimoto2002MagneticA, Li2011AnomalousA, Mielke2021NodelessA, Gong2022SuperconductivityA, Gui2022LaIr3Ga2A, Chaudhary2023RoleA, Liu2024SuperconductivityA, Ma2024Dome-ShapedA, Ushioda2024Two-gapA, Li2012DistinctA, Li2016ChemicalA, Chakrabortty2023EffectA}. The shared structural and electronic features among \ch{\textit{RT}3\textit{X}2} compounds suggest a common underlying mechanism governing their superconducting properties.

{High-throughput \textit{ab initio} calculations were performed to study the kagome family \ch{\textit{RT}3\textit{X}2}, crystallizing in space group $P6/mmm$ (SG 191). The chemical space was explored by systematically substituting elements up to Bi in all possible combinations within the prototype structure. From the resulting compounds, we selected those with distances to the convex hull below 100 meV/atom (calculated within the PBE approximation~\cite{Perdew1996GeneralizedA} against the \textsc{Alexandria} database~\cite{Schmidt2021CrystalA, Schmidt2024ImprovingA}), metallic, and nonmagnetic. For the resulting 3063 candidate materials, we performed DFPT calculations to obtain the electron-phonon coupling constants. All calculations were performed using \textsc{Quantum Espresso} (version 7.1)~\cite{Giannozzi2009QUANTUMA,Giannozzi2017AdvancedA} with the Perdew-Burke-Ernzerhof generalized gradient approximation for solids (PBEsol)~\cite{Perdew2008RestoringA}, and we chose the PBEsol pseudopotentials from the \textsc{pseudodojo} project~\cite{Setten2018The-PseudoDojo:A}, specifically the stringent, scalar-relativistic norm-conserving set. Geometry optimizations were performed using uniform $\Gamma$-centered $\vb{k}$-point grids with a density of 1500 $\vb{k}$-points per reciprocal atom. Convergence thresholds for energies, forces, and stresses were set to $1\times10^{-8}$ a.u., $1\times10^{-6}$ a.u., and $5\times10^{-2}$ kBar, respectively. For electron-phonon calculations, we used the double-grid technique, using the previous k-grid as the coarse grid and a quadrupled $\vb{k}$-grid for the fine sampling. Phonon $\vb{q}$-points were sampled at half the initial $\vb{k}$-point grid. The Eliashberg function was obtained through double delta integration using Methfessel-Paxton smearing of $0.03$ Ry. Materials exhibiting imaginary frequencies at explicitly calculated $\vb{q}$-points were excluded from further analysis, while those developing instabilities during Fourier interpolation were retained but flagged as potentially unstable. In both cases, the materials might stabilize in a lower symmetry ground state, as is characteristic of \ch{LaRu3Si2} and other kagome phases~\cite{Feng2024CatalogueA}. We leave a more detailed study on the CDW phases and SC properties of these materials for future works. 
Superconducting transition temperatures were estimated using the McMillan formula [\cref{eq:McMillian-Tc}]~\cite{McMillan1968TransitionA,Dynes1972McMillantextquotesinglesA}, with $\mu^* = 0.10$.}

Among the resulting candidate materials, two groups can be identified based on their chemical compositions and electronic structures.
The first group closely resembles \ch{LaRu3Si2} and can be considered as its doped analogs. Examples include \ch{LaIr3Ga2}, \ch{CaRh3Ga2}, and others. A complete list of compounds exhibiting $T_c > 1$ K is provided in \cref{tab:PredictionTypeI}. The ones with $E_{\mathrm{hull}}<30$ meV/atom along with their electronic, phononic and EPC properties are presented in \cref{tab:StablePredictionTypeI} and \cref{fig:type-i-1}.
The presence of an FB near $E_F$ is determined by identifying a local DOS peak within the energy range of $E_F \pm 0.5\,\mathrm{eV}$.
The second group differs from the first in that its \textit{T} elements are post-transition metals, resulting in kagome bands derived from $p$ electrons rather than $d$ electrons. Consequently, their electronic structures exhibit distinct characteristics, as illustrated in \cref{fig:type-ii-1}. A complete list of materials in this category with $T_c > 1$ K is provided in \cref{tab:PredictionTypeII}. The ones with $E_{\mathrm{hull}}<30$ meV/atom are listed in \cref{tab:StablePredictionTypeII}.
Despite these differences, the EPC constants in both groups are predominantly contributed by low-frequency phonons, specifically the kagome-$x$ phonons. This suggests a universal property of kagome-type materials, where low-frequency phonon modes play a key role in electron-phonon coupling.
As discussed in \cref{app:ModelsSpringBall}, this characteristic can be readily understood through the analytical framework of the spring-ball model we have proposed. 

\subsection{\label{app:PredictionTypeI} Type I: analogs of \ch{LaRu3Si2}}

\begin{figure}
    \centering
    \caption{Electronic band, electronic DOS, phononic band, phononic DOS and Eliashberg spectral function of materials listed in \cref{tab:StablePredictionTypeI}.}
    \includegraphics[width=0.75\linewidth]{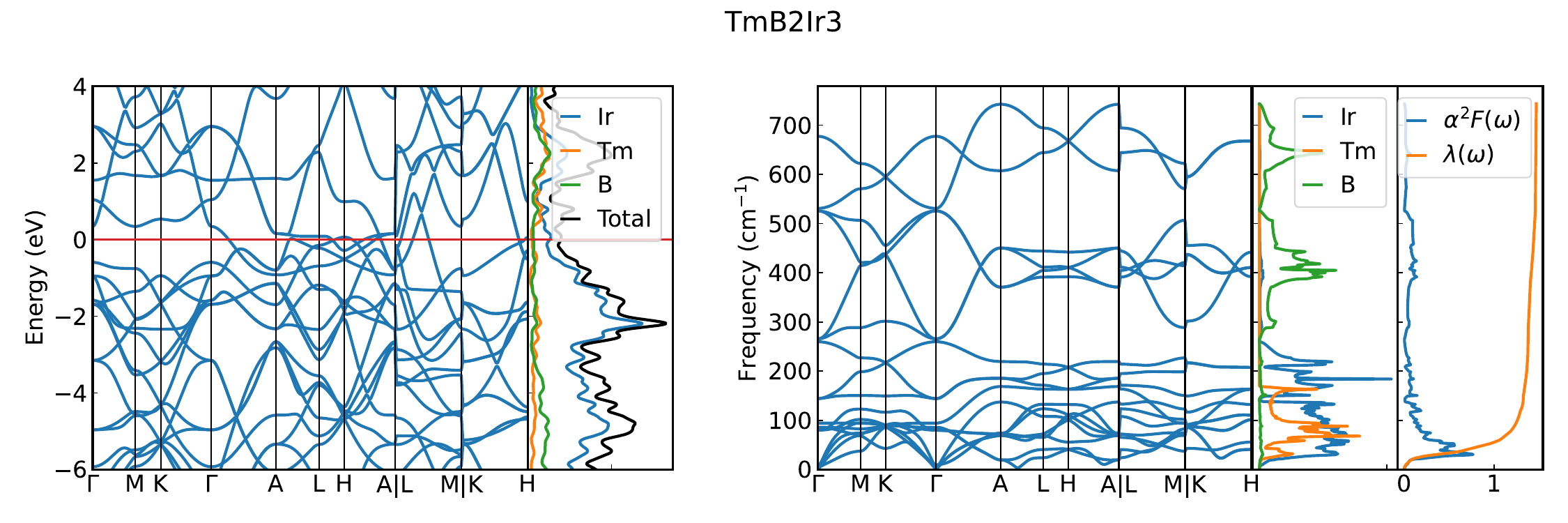}
    \includegraphics[width=0.75\linewidth]{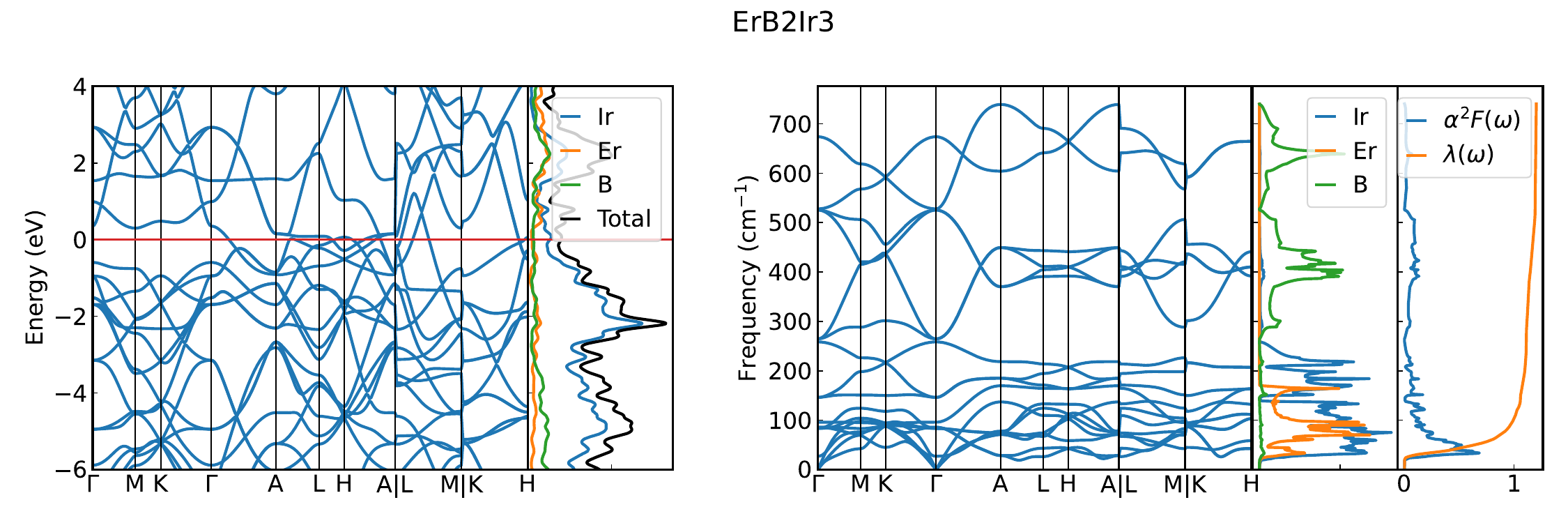}
    \includegraphics[width=0.75\linewidth]{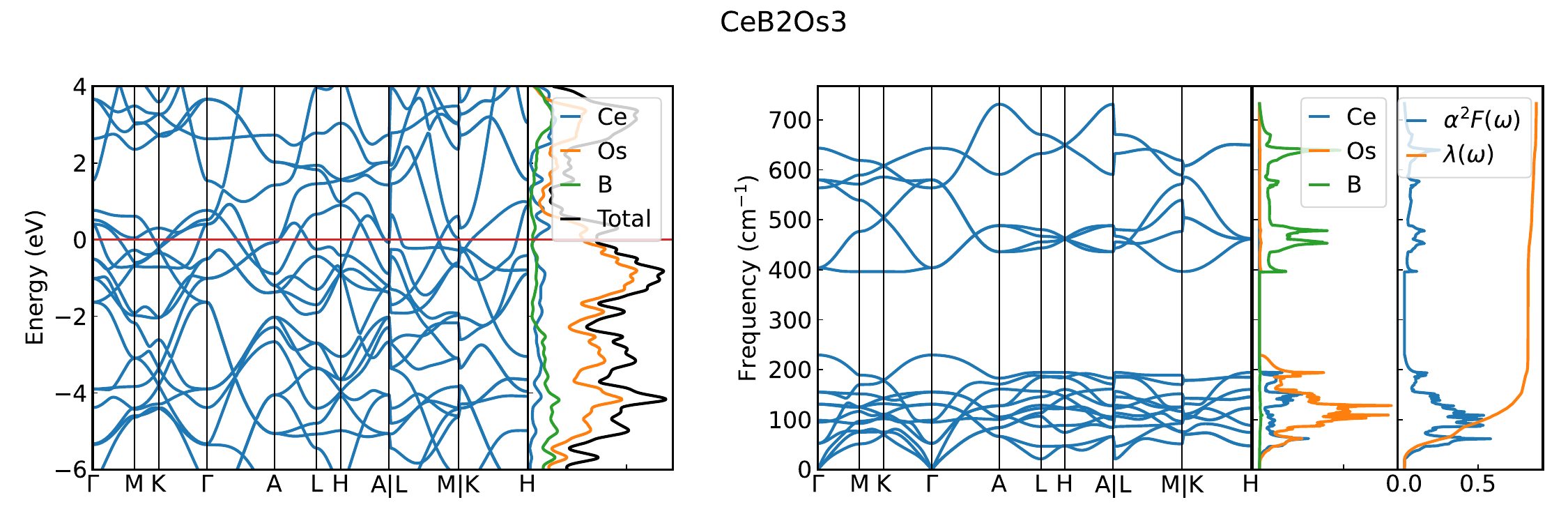}
    \includegraphics[width=0.75\linewidth]{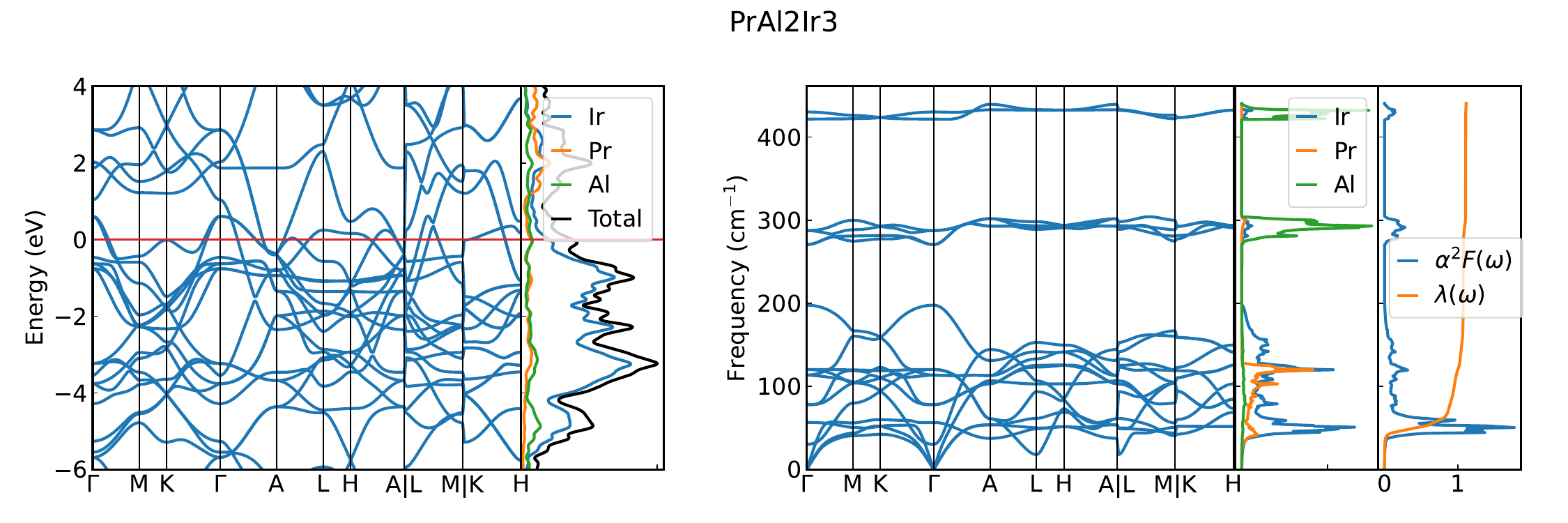}
    \includegraphics[width=0.75\linewidth]{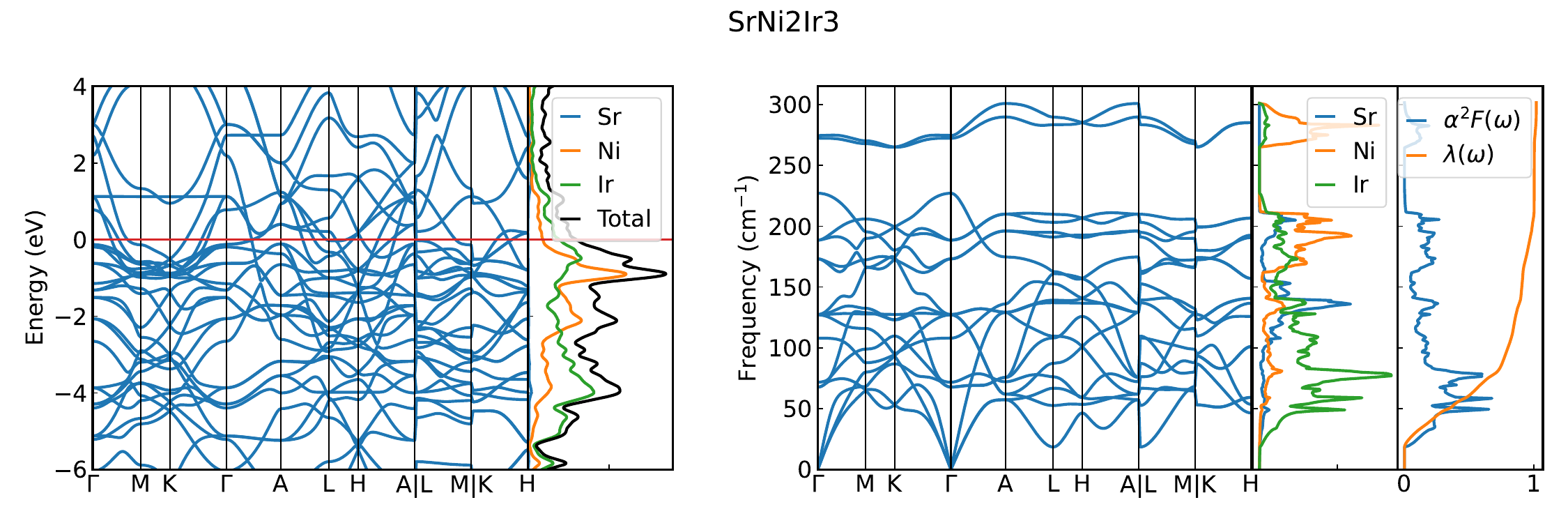}
    \label{fig:type-i-1}
\end{figure}

As shown by the phonon spectrum and EPC properties in \cref{fig:type-i-1}, nearly all materials in the first category of the \ch{\textit{RT}3\textit{X}2} family exhibit a common feature in their phonon spectra: the EPC is predominantly contributed by a low-frequency mode, identified as the local $x$-phonon mode of the kagome lattice.
In addition to this characteristic of EPC, these compounds universally feature a quasi-FB formed by kagome $d_{x^2 - y^2}$ electrons near $E_F$. This quasi-FB contributes to the Fermi surface and leads to a pronounced peak in the DOS near $E_F$, similar to what is observed in \ch{LaRu3Si2}.

Notably, the broad range of available transition metal substitutions provides significant opportunities for tuning the electronic properties of these compounds, effectively acting as a doping mechanism. In most cases, such substitutions lead to electron doping, which shifts the FB associated with the \textit{T} $d_{x^2 - y^2}$ orbitals away from $E_F$. While electron doping may enhance structural stability, it often reduces the DOS at $E_F$, potentially leading to a decrease in $T_c$, as discussed in \cref{app:ResultsDFTElectronPhononLaRu3Si2Doped} and observed in Refs.~\cite{Li2012DistinctA, Li2016ChemicalA, Chakrabortty2023EffectA}.

\subsection{\label{app:PredictionTypeII} Type II: kagome superconductors with $p$ kagome bands}

\begin{figure}
    \centering
    \caption{Electronic band, electronic DOS, phononic band, phononic DOS and Eliashberg spectral function of materials listed in \cref{tab:StablePredictionTypeII}.}
    \includegraphics[width=0.75\linewidth]{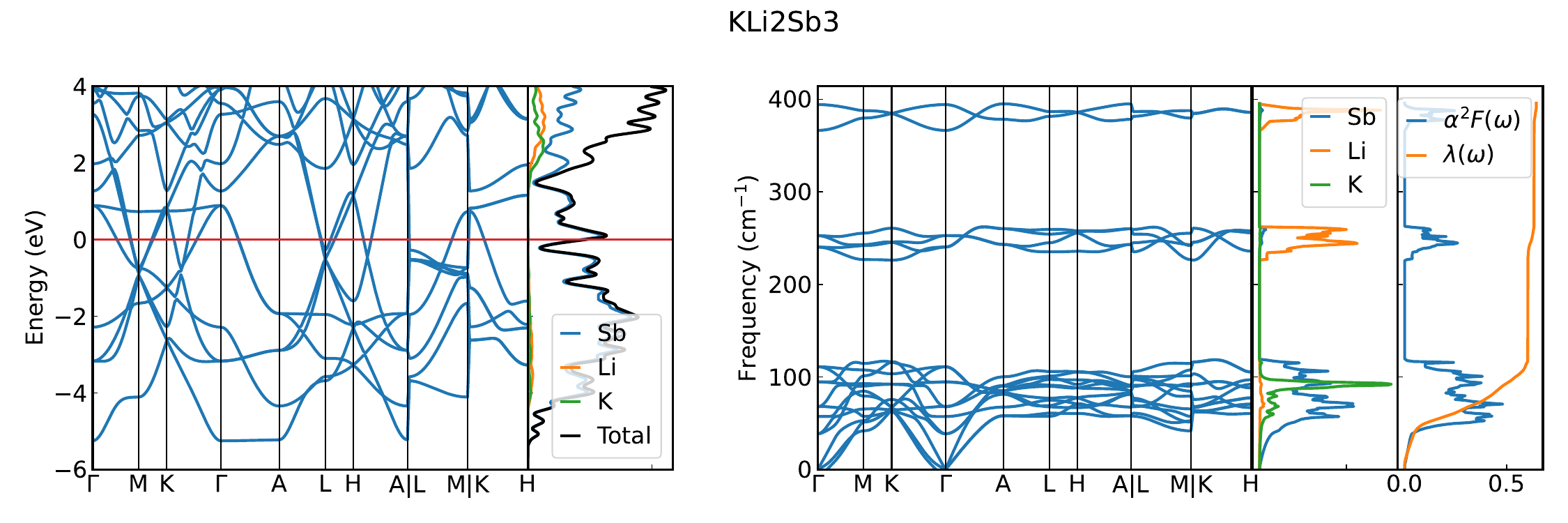}
    \includegraphics[width=0.75\linewidth]{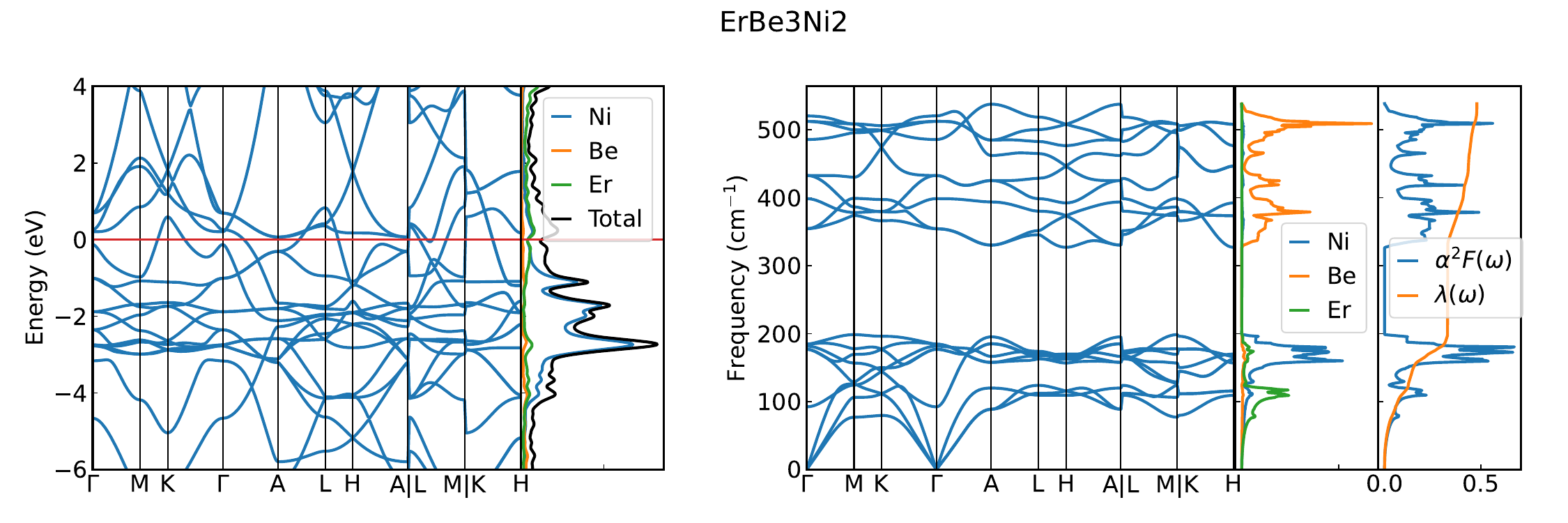}
    \includegraphics[width=0.75\linewidth]{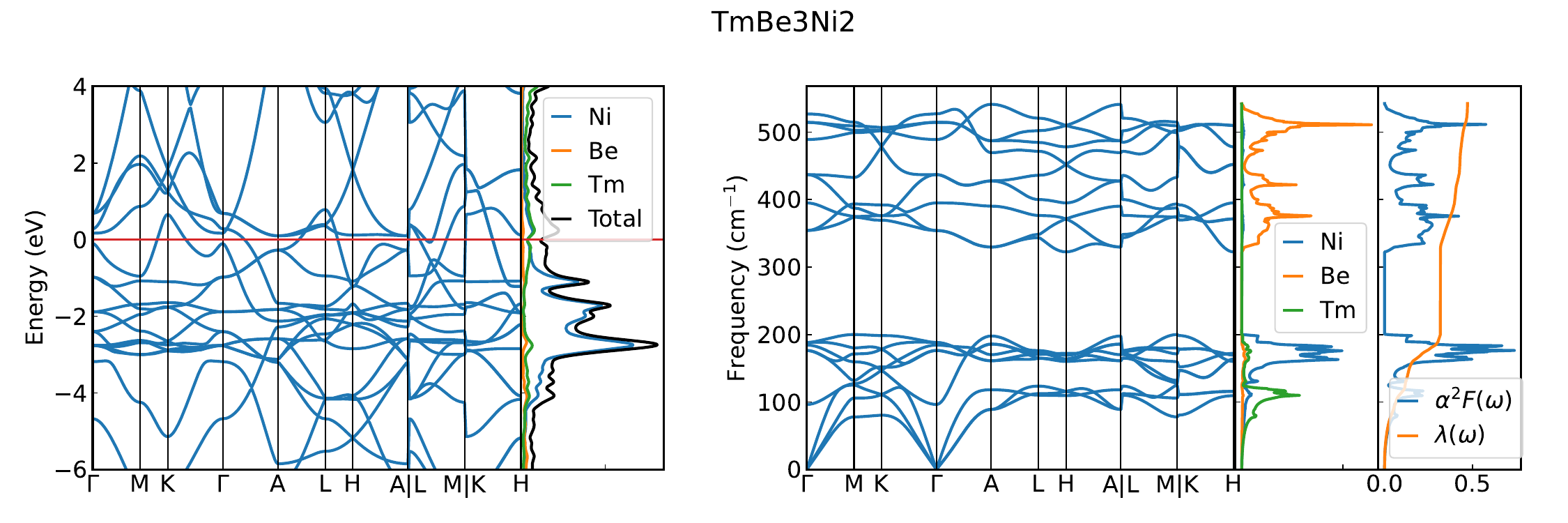}
    \includegraphics[width=0.75\linewidth]{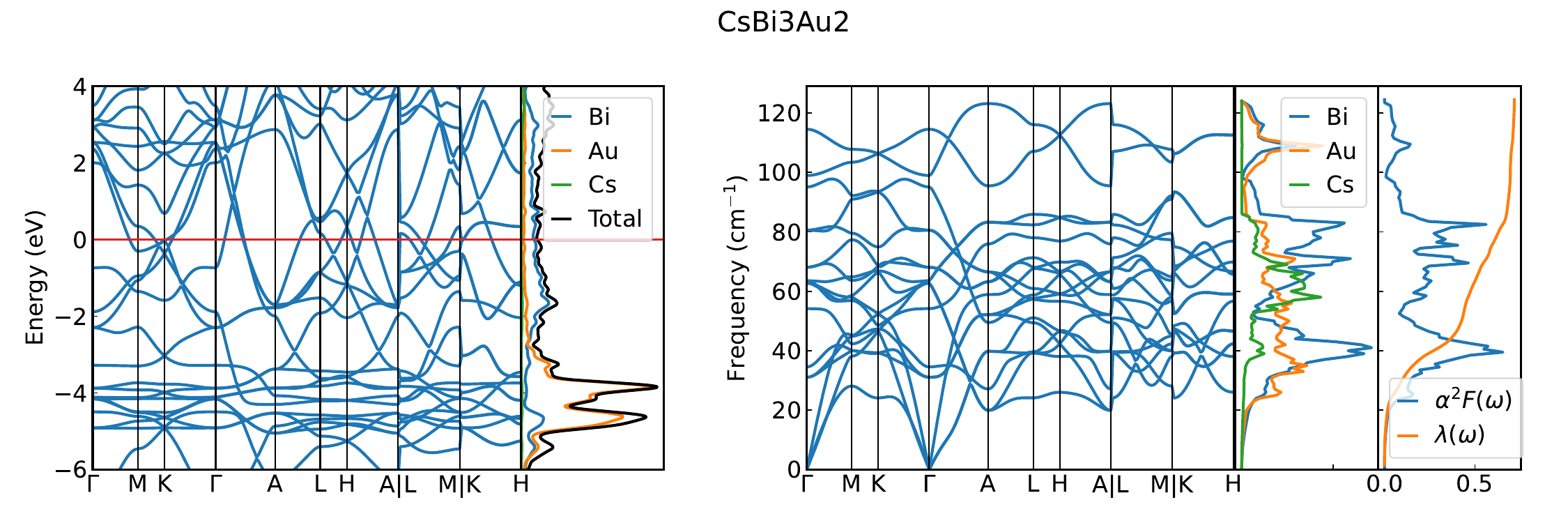}
    \includegraphics[width=0.75\linewidth]{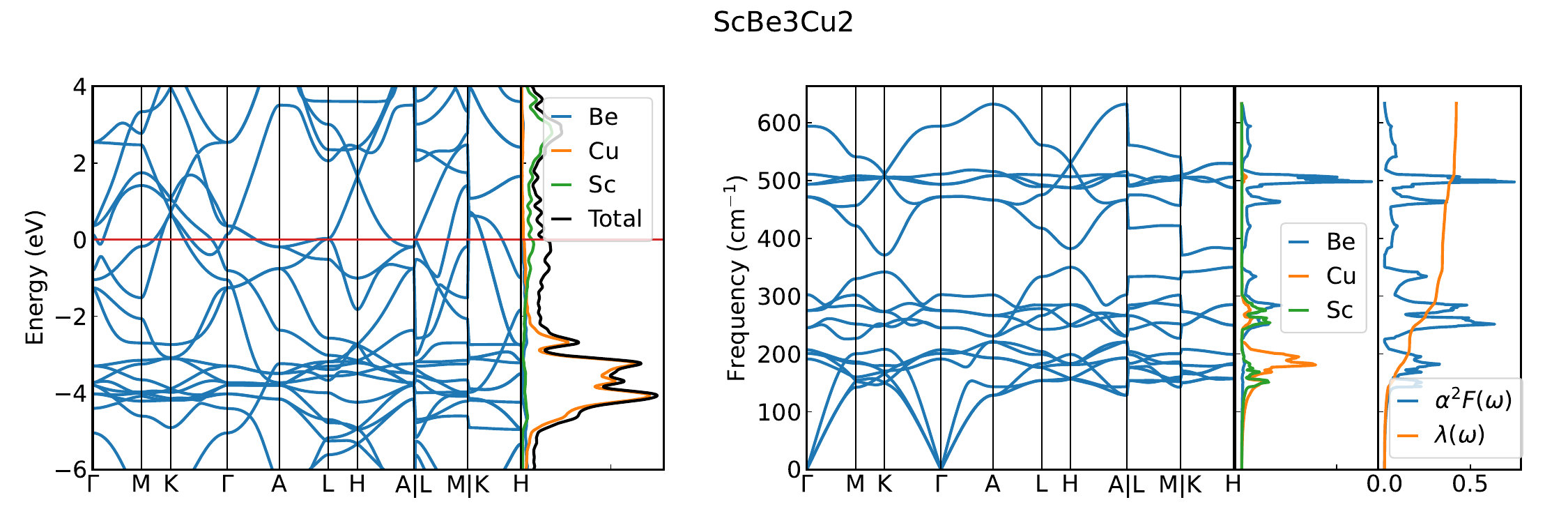}
\label{fig:type-ii-1}
\end{figure}

The second category, however, is fundamentally different from the first. As introduced earlier, the kagome bands in this group are formed by the $p$ electrons rather than the $d$ electrons, which may result in weaker electron-electron correlations due to the more itinerant nature of $p$ orbitals.
As shown by the electronic properties in \cref{fig:type-ii-1}, most compounds in this category host a quasi-FB at the Fermi level originating from kagome $p_z$ orbitals. These flat bands are typically only partially flat on the $k_z = \pi$ plane. Nevertheless, they still contribute to a DOS peak near $E_F$. Notably, the $T_c$ in this category are on average higher than those in the first group, with typical values around 10 K and a maximum reaching 15 K.
However, compounds in this category generally exhibit lower thermodynamic stability compared to those in the first group, with their distances to the convex hull typically around 50 meV/atom. Despite differences in stability and electronic structure, the EPC properties in this category share key similarities with those of the first group. In particular, the EPC constant $\lambda$ remains predominantly determined by low-frequency phonons, as shown in the EPC properties in \cref{fig:type-ii-1}.

\begin{center}
\begin{longtable}{|c|c|c|c|c|c|c|c|}
\caption{Predicted compounds with $E_{\mathrm{hull}} < 30$ meV/atom in Type I.}
\label{tab:StablePredictionTypeI} \\
\hline \multicolumn{1}{|c|}{Formula} & \multicolumn{1}{|c|}{Material ID} & \multicolumn{1}{c|}{$E_{\mathrm{hull}}$ (meV/atom)} & \multicolumn{1}{c|}{$\omega_{\mathrm{log}}$ (K)} & \multicolumn{1}{c|}{$D(E_F)$} & \multicolumn{1}{c|}{$\lambda$} & \multicolumn{1}{c|}{$T_c$ (K)} & \multicolumn{1}{c|}{FB @ $E_F$} \\ \hline
\endfirsthead
\multicolumn{8}{c}{{\bfseries \tablename\ \thetable{} -- continued from previous page}} \\
\hline \multicolumn{1}{|c|}{Formula} & \multicolumn{1}{|c|}{Material ID} & \multicolumn{1}{c|}{$E_{\mathrm{hull}}$ (meV/atom)} & \multicolumn{1}{c|}{$\omega_{\mathrm{log}}$ (K)} & \multicolumn{1}{c|}{$D(E_F)$} & \multicolumn{1}{c|}{$\lambda$} & \multicolumn{1}{c|}{$T_c$ (K)} & \multicolumn{1}{c|}{FB @ $E_F$} \\ \hline
\endhead
\hline \multicolumn{8}{|r|}{{Continued on next page}} \\ \hline
\endfoot
\hline \hline
\endlastfoot
\ch{\green{Tm}\blue{B}2\red{Ir}3} & agm002137253 & $   0$ & $  73$ & $ 13.3$ & $  1.45$ & $  9.2$ & No \\
\hline
\ch{\green{Er}\blue{B}2\red{Ir}3} & agm003202452 & $   0$ & $  85$ & $ 13.4$ & $  1.20$ & $  8.5$ & No \\
\hline
\ch{\green{Ce}\blue{B}2\red{Os}3} & agm002199385 & $  27$ & $ 131$ & $ 27.1$ & $  0.89$ & $  8.1$ & Yes \\
\hline
\ch{\green{Pr}\blue{Al}2\red{Ir}3} & agm002195389 & $  25$ & $  89$ & $ 22.4$ & $  1.11$ & $  7.9$ & Yes \\
\hline
\ch{\green{Sr}\blue{Ni}2\red{Ir}3} & agm002226323 & $   0$ & $  94$ & $ 43.7$ & $  1.02$ & $  7.3$ & Yes \\
\hline
\ch{\green{Lu}\blue{B}2\red{Rh}3} & agm003195416 & $  21$ & $ 127$ & $ 14.6$ & $  0.85$ & $  7.2$ & No \\
\hline
\ch{\green{La}\blue{Ga}2\red{Ir}3} & agm002274867 & $   0$ & $  80$ & $ 22.1$ & $  1.10$ & $  6.9$ & Yes \\
\hline
\ch{\green{Tb}\blue{B}2\red{Ru}3} & agm003202167 & $   0$ & $ 137$ & $ 26.7$ & $  0.81$ & $  6.9$ & Yes \\
\hline
\ch{\green{Y}\blue{B}2\red{Ru}3} & agm003202168 & $   0$ & $ 165$ & $ 26.6$ & $  0.74$ & $  6.8$ & Yes \\
\hline
\ch{\green{Sm}\blue{B}2\red{Ir}3} & agm002137232 & $   3$ & $ 112$ & $ 14.5$ & $  0.88$ & $  6.8$ & No \\
\hline
\ch{\green{Pr}\blue{Ga}2\red{Ir}3} & agm002274922 & $   0$ & $  90$ & $ 21.9$ & $  1.00$ & $  6.7$ & Yes \\
\hline
\ch{\green{Pm}\blue{B}2\red{Ru}3} & agm071840373 & $  12$ & $ 162$ & $ 28.2$ & $  0.74$ & $  6.7$ & Yes \\
\hline
\ch{\green{Sm}\blue{B}2\red{Ru}3} & agm003195414 & $   7$ & $ 157$ & $ 27.7$ & $  0.74$ & $  6.5$ & Yes \\
\hline
\ch{\green{Nd}\blue{B}2\red{Ru}3} & agm003195435 & $  21$ & $ 166$ & $ 28.8$ & $  0.71$ & $  6.3$ & Yes \\
\hline
\ch{\green{Tm}\blue{B}2\red{Rh}3} & agm002137195 & $  13$ & $ 144$ & $ 14.8$ & $  0.75$ & $  6.2$ & No \\
\hline
\ch{\green{La}\blue{B}2\red{Ru}3} & agm003195432 & $  23$ & $ 166$ & $ 28.8$ & $  0.71$ & $  6.2$ & Yes \\
\hline
\ch{\green{Ce}\blue{B}2\red{Ru}3} & agm003202683 & $   0$ & $ 178$ & $ 29.8$ & $  0.67$ & $  5.9$ & Yes \\
\hline
\ch{\green{Ca}\blue{Ga}2\red{Rh}3} & agm002260176 & $   0$ & $ 117$ & $ 23.9$ & $  0.80$ & $  5.8$ & Yes \\
\hline
\ch{\green{La}\blue{B}2\red{Rh}3} & agm003202760 & $  17$ & $ 180$ & $ 18.1$ & $  0.66$ & $  5.7$ & No \\
\hline
\ch{\green{Er}\blue{B}2\red{Rh}3} & agm003195426 & $   8$ & $ 152$ & $ 14.9$ & $  0.70$ & $  5.5$ & No \\
\hline
\ch{\green{Ho}\blue{B}2\red{Rh}3} & agm003195417 & $   5$ & $ 158$ & $ 15.1$ & $  0.67$ & $  5.1$ & No \\
\hline
\ch{\green{La}\blue{B}2\red{Ir}3} & agm003195415 & $  23$ & $ 130$ & $ 14.9$ & $  0.72$ & $  5.1$ & No \\
\hline
\ch{\green{Dy}\blue{B}2\red{Rh}3} & agm003195430 & $   2$ & $ 162$ & $ 15.4$ & $  0.66$ & $  5.0$ & No \\
\hline
\ch{\green{Tb}\blue{B}2\red{Rh}3} & agm002137193 & $   1$ & $ 167$ & $ 15.7$ & $  0.64$ & $  4.7$ & No \\
\hline
\ch{\green{Y}\blue{B}2\red{Rh}3} & agm003202164 & $   0$ & $ 175$ & $ 15.5$ & $  0.62$ & $  4.6$ & No \\
\hline
\ch{\green{Sm}\blue{Au}2\red{Zn}3} & agm002324461 & $   0$ & $  90$ & $ 17.4$ & $  0.80$ & $  4.4$ & No \\
\hline
\ch{\green{Tb}\blue{Au}2\red{Zn}3} & agm002324476 & $   0$ & $  90$ & $ 16.4$ & $  0.77$ & $  4.0$ & No \\
\hline
\ch{\green{La}\blue{Au}2\red{Zn}3} & agm002246263 & $   0$ & $ 103$ & $ 20.4$ & $  0.71$ & $  3.9$ & No \\
\hline
\ch{\green{K}\blue{Ge}2\red{Pt}3} & agm038799382 & $  25$ & $  93$ & $ 17.0$ & $  0.74$ & $  3.8$ & No \\
\hline
\ch{\green{Er}\blue{Ni}2\red{Ir}3} & agm002225978 & $  25$ & $ 127$ & $ 34.1$ & $  0.62$ & $  3.2$ & No \\
\hline
\ch{\green{Ho}\blue{Ni}2\red{Ir}3} & agm002226083 & $  18$ & $ 131$ & $ 34.1$ & $  0.61$ & $  3.2$ & No \\
\hline
\ch{\green{Sr}\blue{Ga}2\red{Rh}3} & agm002275043 & $   0$ & $ 132$ & $ 22.5$ & $  0.60$ & $  3.1$ & Yes \\
\hline
\ch{\green{Dy}\blue{Ni}2\red{Ir}3} & agm002225968 & $  12$ & $ 136$ & $ 34.2$ & $  0.58$ & $  2.8$ & No \\
\hline
\ch{\green{Tb}\blue{Ni}2\red{Ir}3} & agm002226335 & $   5$ & $ 140$ & $ 34.4$ & $  0.56$ & $  2.7$ & No \\
\hline
\ch{\green{Y}\blue{Ni}2\red{Ir}3} & agm002226370 & $  11$ & $ 143$ & $ 34.5$ & $  0.55$ & $  2.6$ & No \\
\hline
\ch{\green{Pm}\blue{Au}2\red{Zn}3} & agm031465753 & $   0$ & $ 110$ & $ 17.7$ & $  0.60$ & $  2.6$ & No \\
\hline
\ch{\green{Pr}\blue{Au}2\red{Zn}3} & agm002324445 & $   0$ & $ 114$ & $ 18.3$ & $  0.59$ & $  2.5$ & No \\
\hline
\ch{\green{K}\blue{In}2\red{Au}3} & agm040038476 & $  27$ & $  54$ & $ 12.5$ & $  0.76$ & $  2.3$ & No \\
\hline
\ch{\green{Nd}\blue{Au}2\red{Zn}3} & agm002324425 & $   0$ & $ 114$ & $ 18.0$ & $  0.57$ & $  2.3$ & No \\
\hline
\ch{\green{Ba}\blue{Li}2\red{Cd}3} & agm071678116 & $  16$ & $ 103$ & $ 25.9$ & $  0.57$ & $  2.1$ & Yes \\
\hline
\ch{\green{Sm}\blue{Ni}2\red{Ir}3} & agm002226250 & $   0$ & $ 149$ & $ 34.8$ & $  0.52$ & $  2.1$ & No \\
\hline
\ch{\green{Pm}\blue{Ni}2\red{Ir}3} & agm005853112 & $   0$ & $ 151$ & $ 34.9$ & $  0.50$ & $  2.0$ & No \\
\hline
\ch{\green{La}\blue{Ni}2\red{Ir}3} & agm002167274 & $   0$ & $ 152$ & $ 36.4$ & $  0.50$ & $  1.8$ & No \\
\hline
\ch{\green{Dy}\blue{Ni}2\red{Rh}3} & agm002167305 & $  23$ & $ 143$ & $ 40.7$ & $  0.50$ & $  1.7$ & No \\
\hline
\ch{\green{Tb}\blue{Ag}2\red{Zn}3} & agm002360031 & $  14$ & $ 108$ & $ 17.4$ & $  0.53$ & $  1.7$ & No \\
\hline
\ch{\green{Pr}\blue{Ni}2\red{Ir}3} & agm002226175 & $   0$ & $ 157$ & $ 35.4$ & $  0.48$ & $  1.7$ & No \\
\hline
\ch{\green{Sr}\blue{Li}2\red{Cd}3} & agm071725069 & $  23$ & $ 102$ & $ 23.2$ & $  0.54$ & $  1.7$ & No \\
\hline
\ch{\green{Sr}\blue{Au}2\red{Zn}3} & agm002324467 & $  18$ & $ 121$ & $ 17.0$ & $  0.51$ & $  1.6$ & No \\
\hline
\ch{\green{Tb}\blue{Ni}2\red{Rh}3} & agm002167312 & $  17$ & $ 150$ & $ 40.9$ & $  0.48$ & $  1.6$ & No \\
\hline
\ch{\green{Na}\blue{B}2\red{Ir}3} & agm005806208 & $   2$ & $ 185$ & $ 15.3$ & $  0.46$ & $  1.6$ & No \\
\hline
\ch{\green{La}\blue{Ag}2\red{Zn}3} & agm002324396 & $   0$ & $ 118$ & $ 19.5$ & $  0.51$ & $  1.6$ & No \\
\hline
\ch{\green{Pm}\blue{Pt}2\red{Rh}3} & agm031020160 & $  28$ & $ 159$ & $ 35.3$ & $  0.47$ & $  1.5$ & No \\
\hline
\ch{\green{Ce}\blue{Ni}2\red{Ir}3} & agm002227628 & $   0$ & $ 160$ & $ 35.9$ & $  0.46$ & $  1.4$ & No \\
\hline
\ch{\green{Ce}\blue{Be}2\red{Co}3} & agm038805054 & $  28$ & $ 201$ & $ 45.5$ & $  0.44$ & $  1.4$ & Yes \\
\hline
\ch{\green{Na}\blue{Au}2\red{Zn}3} & agm039946149 & $  14$ & $ 114$ & $ 14.1$ & $  0.50$ & $  1.4$ & No \\
\hline
\ch{\green{Pr}\blue{Pt}2\red{Rh}3} & agm002235386 & $  28$ & $ 165$ & $ 35.6$ & $  0.45$ & $  1.3$ & No \\
\hline
\ch{\green{La}\blue{Pt}2\red{Rh}3} & agm002180579 & $  22$ & $ 160$ & $ 36.6$ & $  0.45$ & $  1.3$ & No \\
\hline
\ch{\green{Ba}\blue{Pd}2\red{Zn}3} & agm002360061 & $  28$ & $ 117$ & $ 19.8$ & $  0.49$ & $  1.3$ & No \\
\hline
\ch{\green{Ba}\blue{Ga}2\red{Rh}3} & agm002337544 & $   0$ & $ 135$ & $ 20.8$ & $  0.47$ & $  1.3$ & Yes \\
\hline
\ch{\green{Sm}\blue{Ag}2\red{Zn}3} & agm002360030 & $   2$ & $ 120$ & $ 17.7$ & $  0.48$ & $  1.2$ & No \\
\hline
\ch{\green{Sm}\blue{Ni}2\red{Rh}3} & agm002167311 & $   1$ & $ 167$ & $ 41.4$ & $  0.44$ & $  1.2$ & No \\
\hline
\ch{\green{La}\blue{Pd}2\red{Zn}3} & agm002246275 & $   0$ & $ 145$ & $ 23.6$ & $  0.45$ & $  1.1$ & No \\
\hline
\ch{\green{La}\blue{Pd}2\red{Pt}3} & agm002234138 & $   0$ & $ 103$ & $ 27.7$ & $  0.47$ & $  1.0$ & Yes \\
\hline
\ch{\green{Pr}\blue{Ag}2\red{Zn}3} & agm002360029 & $   0$ & $ 126$ & $ 17.8$ & $  0.45$ & $  1.0$ & No \\
\hline
\ch{\green{Sr}\blue{Ag}2\red{Zn}3} & agm002360072 & $   0$ & $ 124$ & $ 18.1$ & $  0.45$ & $  1.0$ & No \\
\hline
\end{longtable}
\end{center}

\begin{figure}
    \centering
    \includegraphics[width=0.75\linewidth]{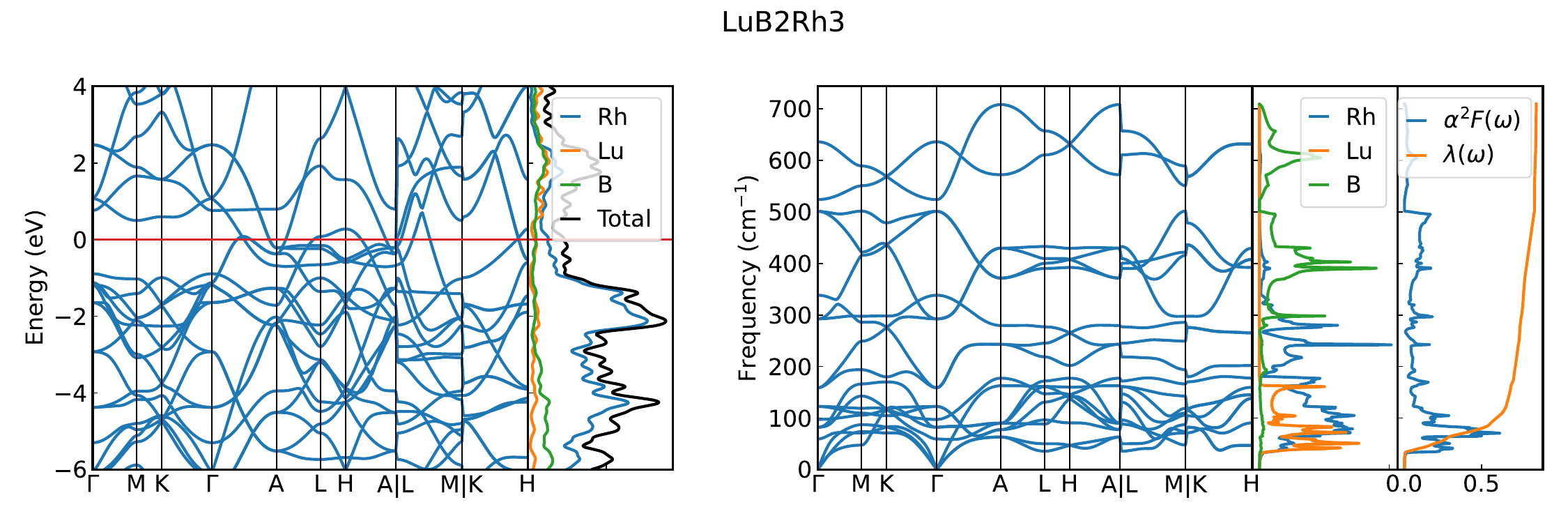}
    \includegraphics[width=0.75\linewidth]{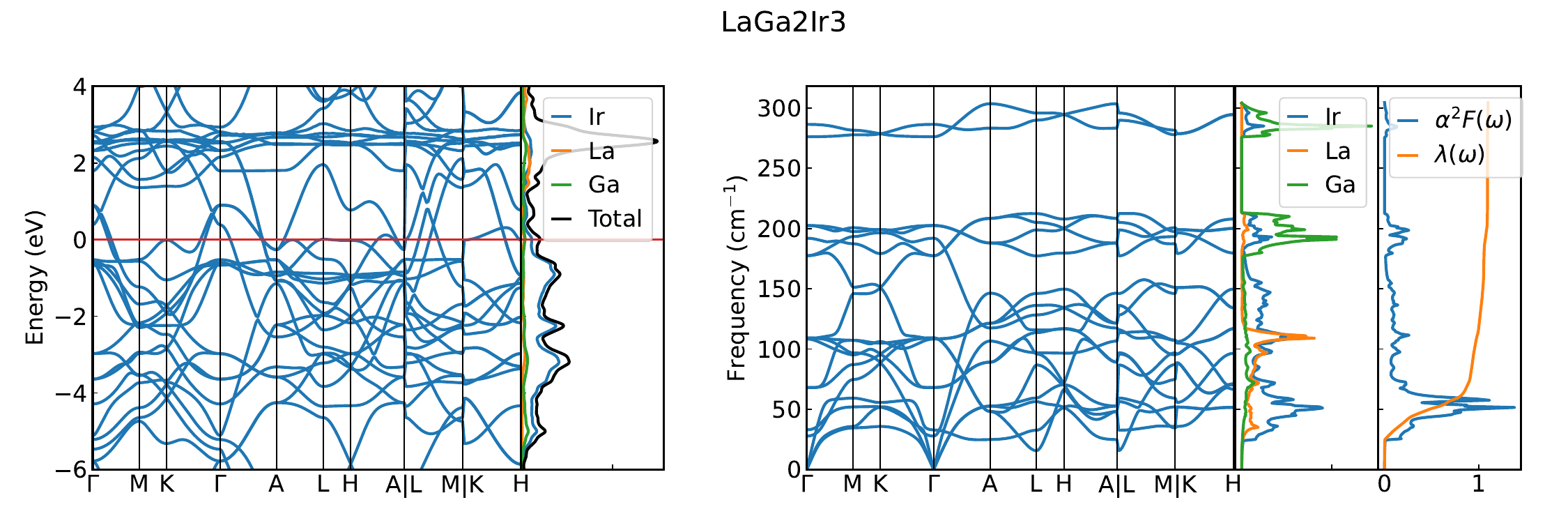}
    \includegraphics[width=0.75\linewidth]{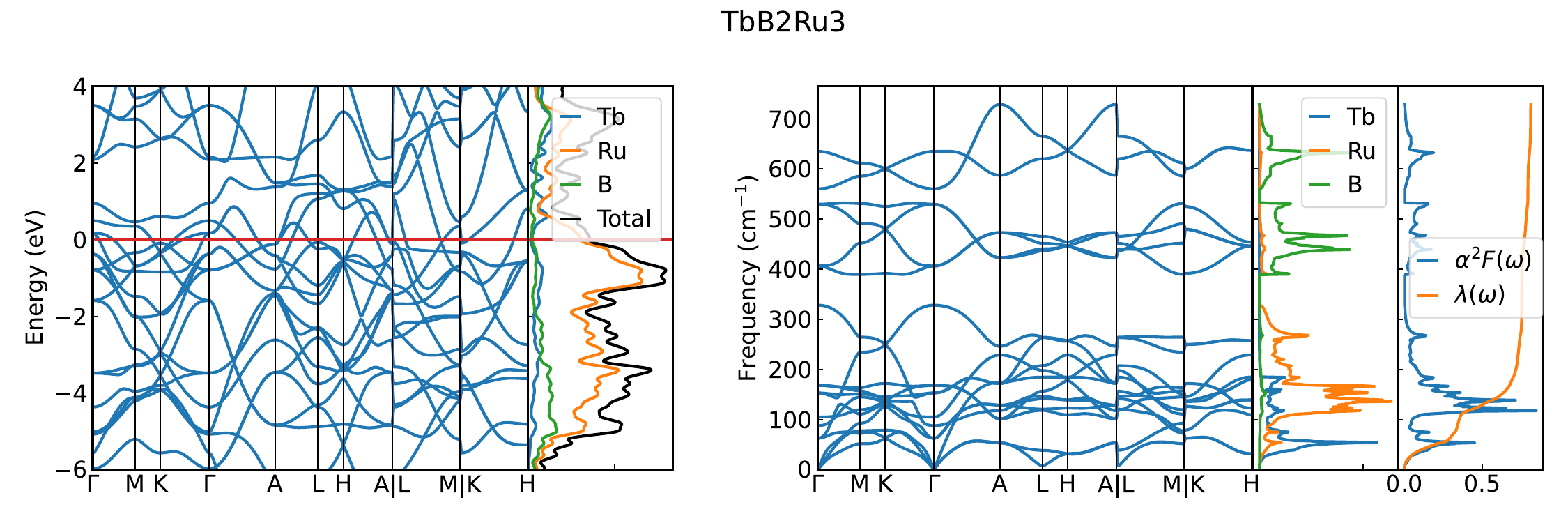}
    \includegraphics[width=0.75\linewidth]{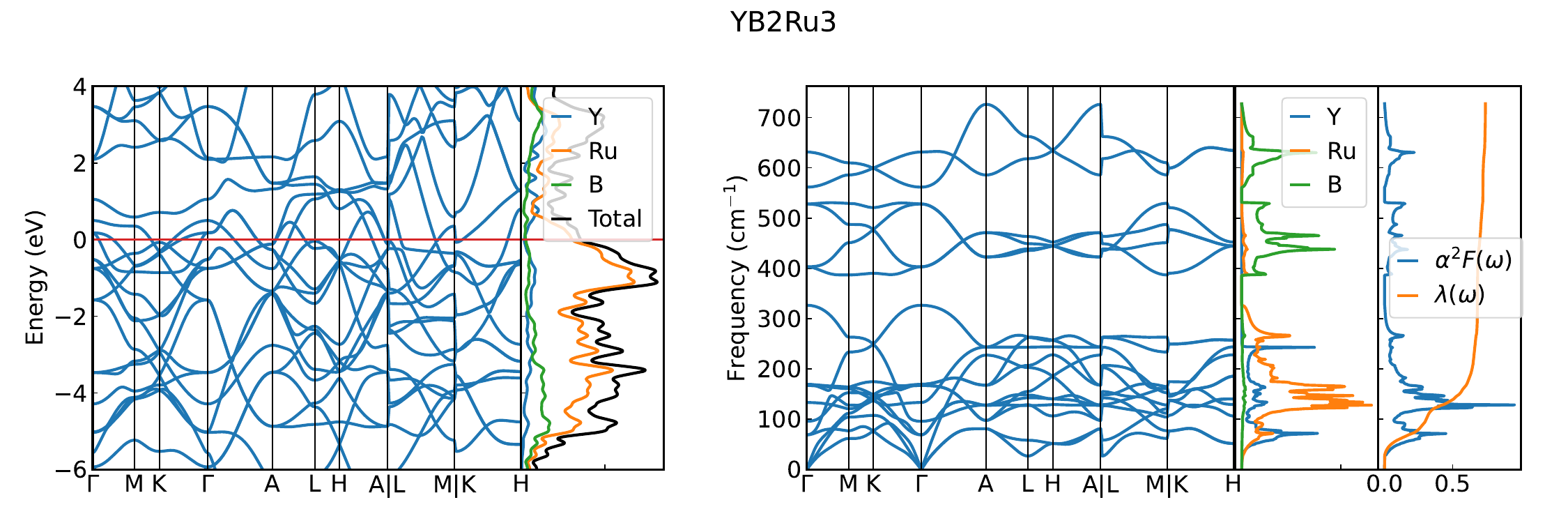}
    \includegraphics[width=0.75\linewidth]{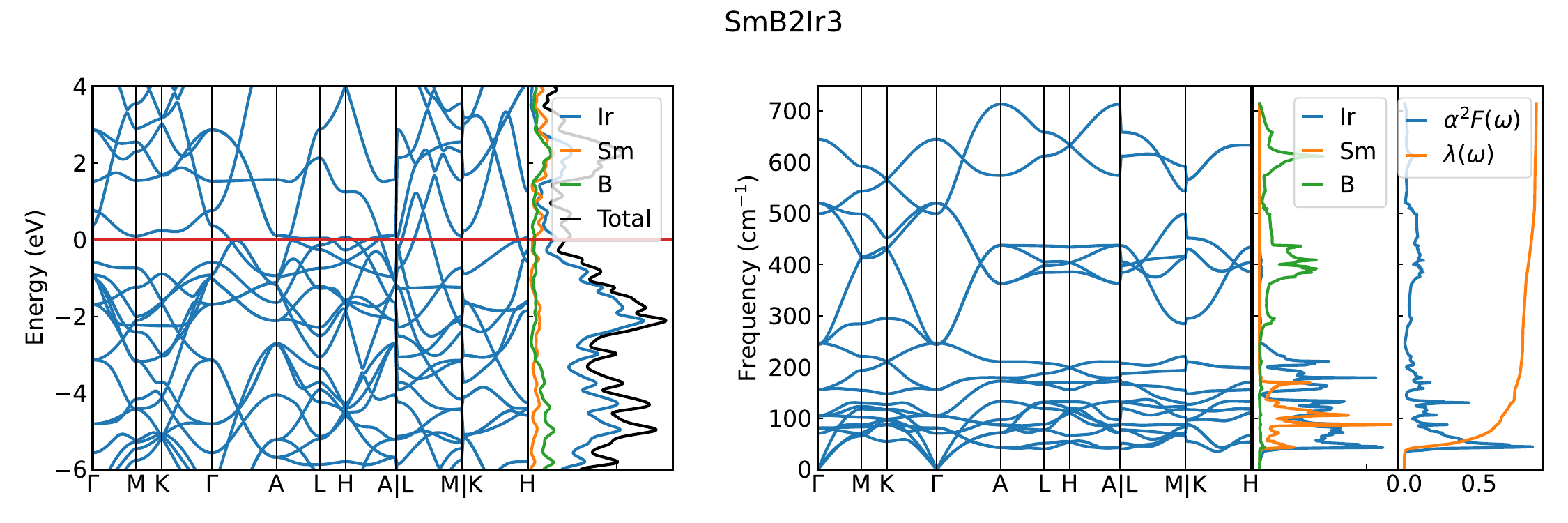}
\label{fig:type-i-2}
\end{figure}

\begin{figure}
    \centering
    \includegraphics[width=0.75\linewidth]{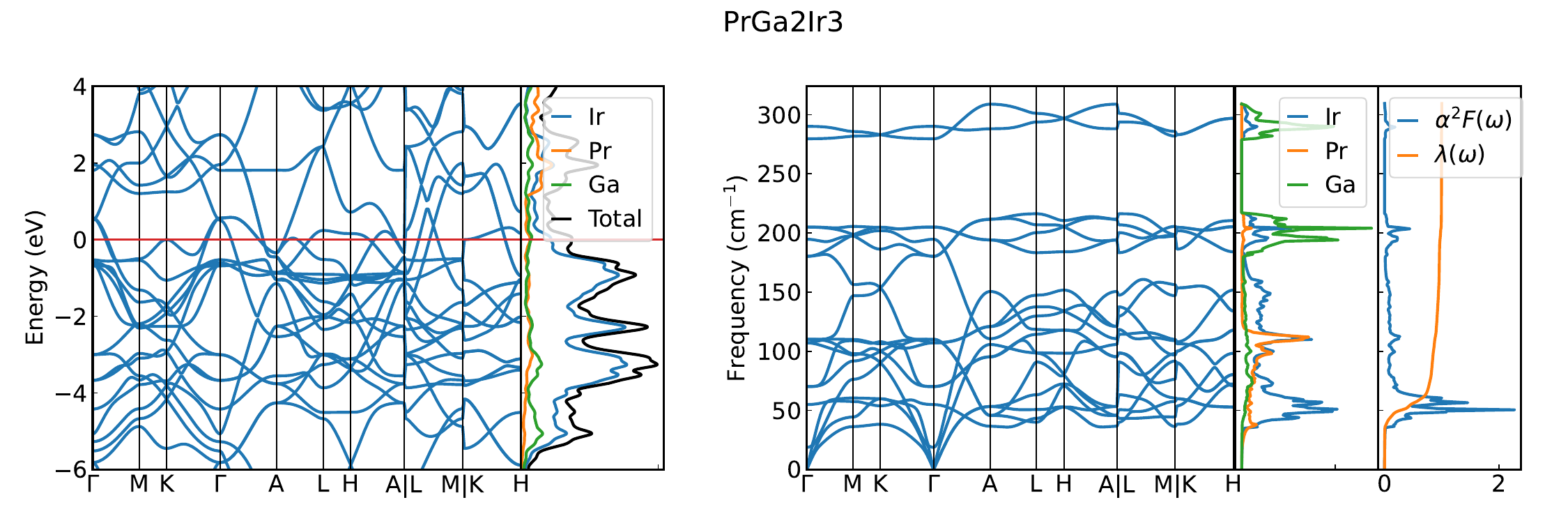}
    \includegraphics[width=0.75\linewidth]{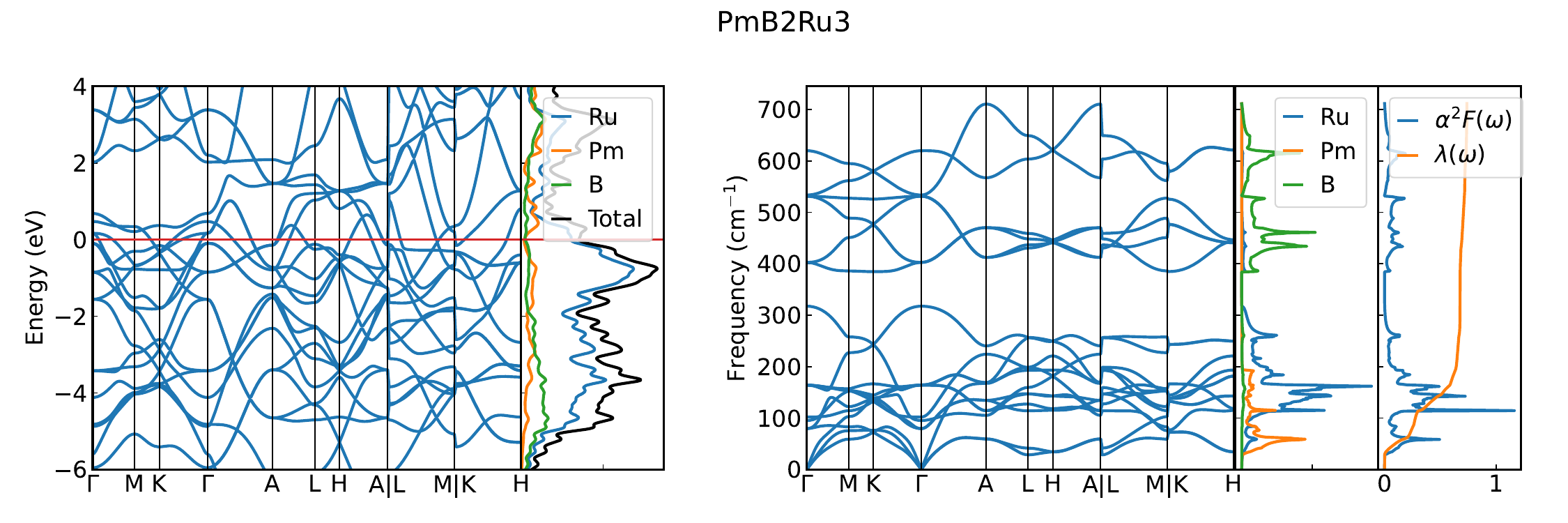}
    \includegraphics[width=0.75\linewidth]{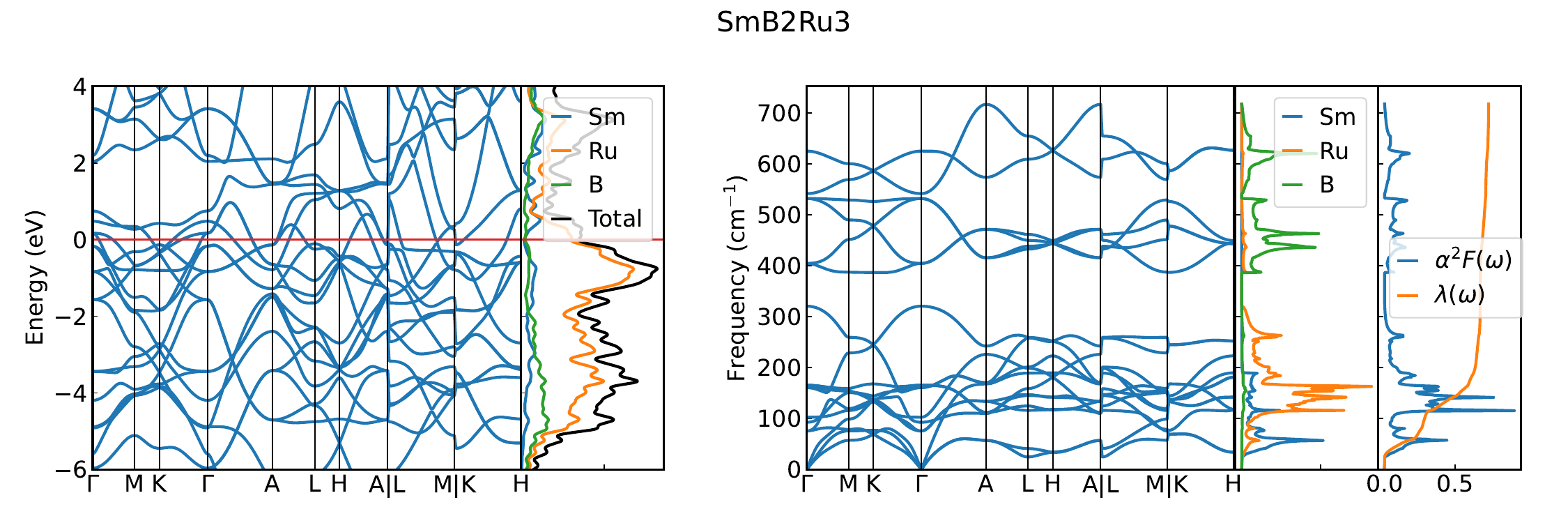}
    \includegraphics[width=0.75\linewidth]{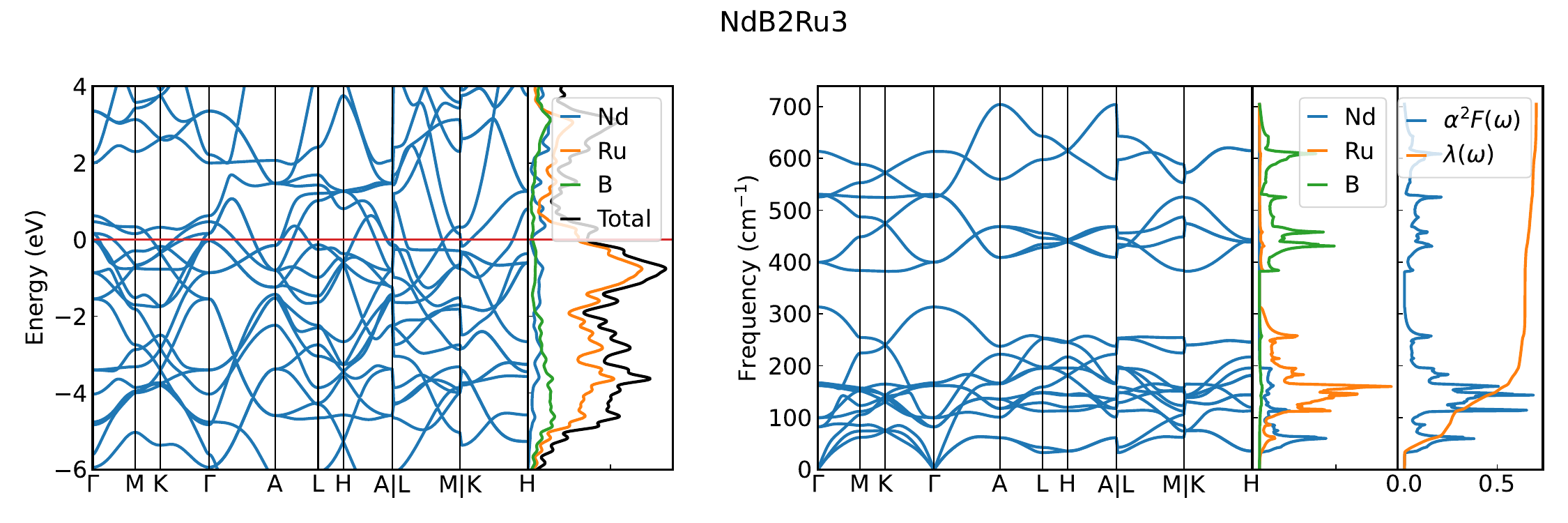}
    \includegraphics[width=0.75\linewidth]{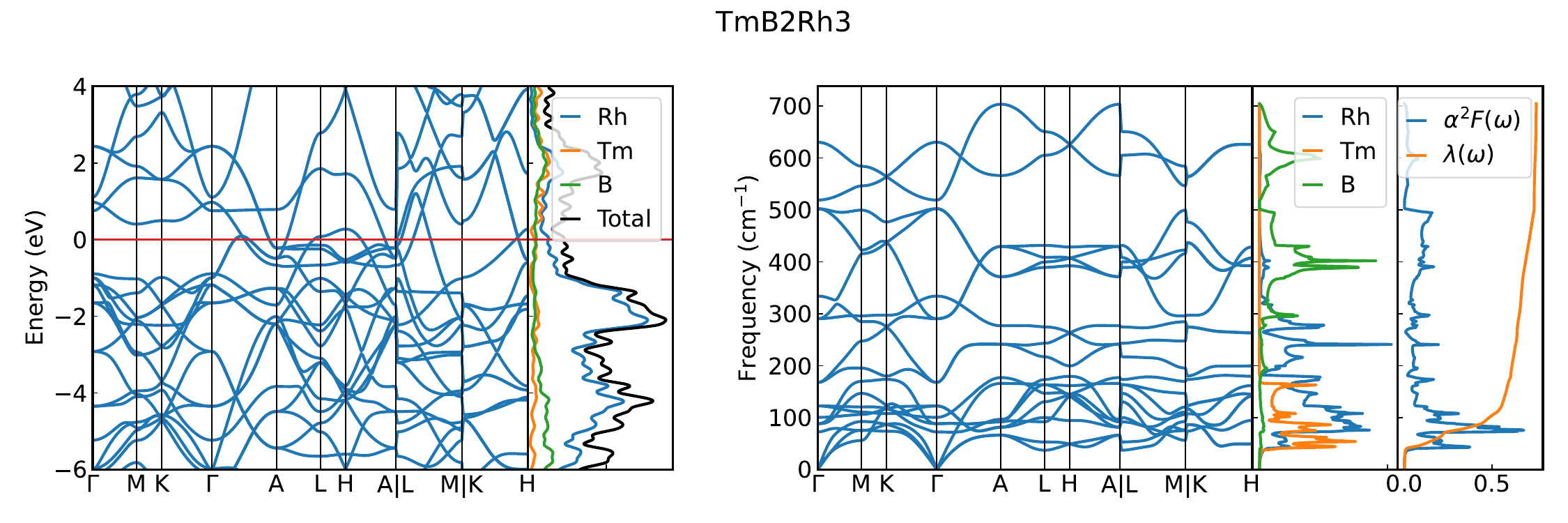}
\label{fig:type-i-3}
\end{figure}

\begin{figure}
    \centering
    \includegraphics[width=0.75\linewidth]{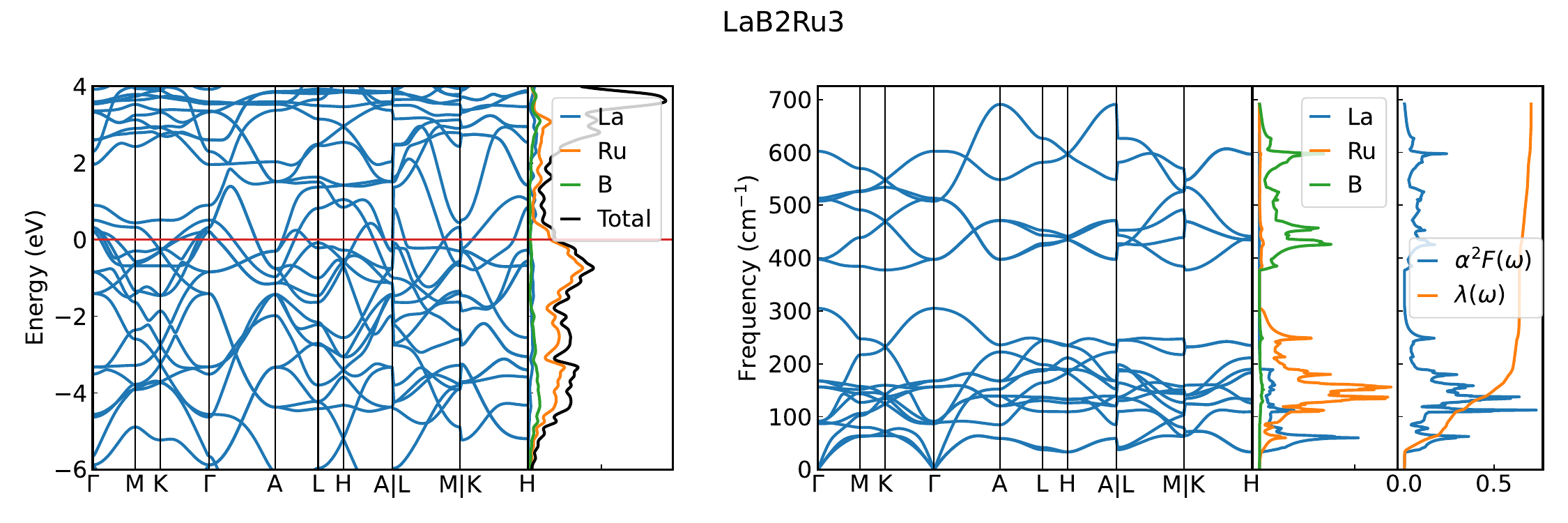}
    \includegraphics[width=0.75\linewidth]{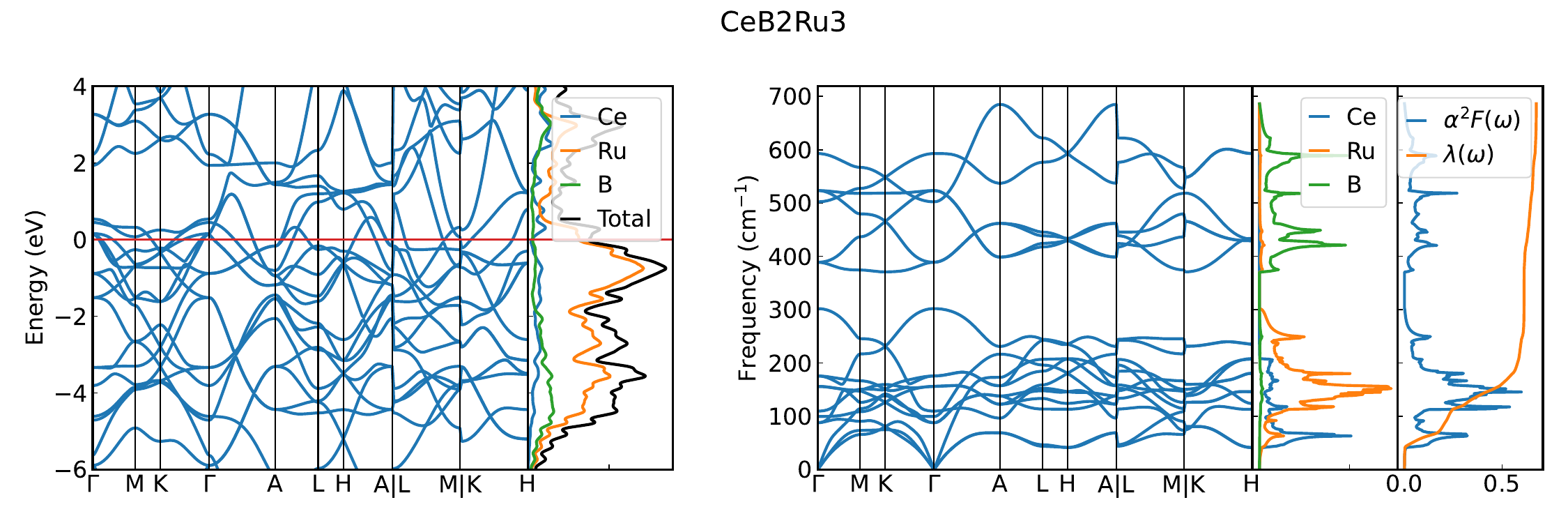}
    \includegraphics[width=0.75\linewidth]{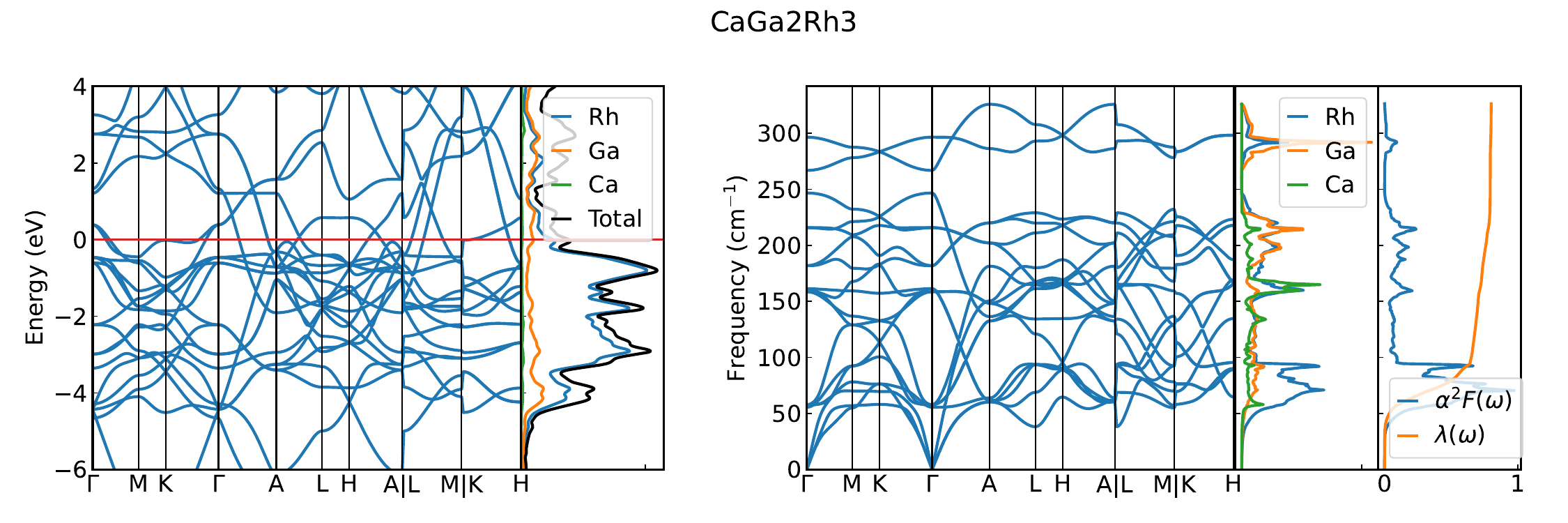}
    \includegraphics[width=0.75\linewidth]{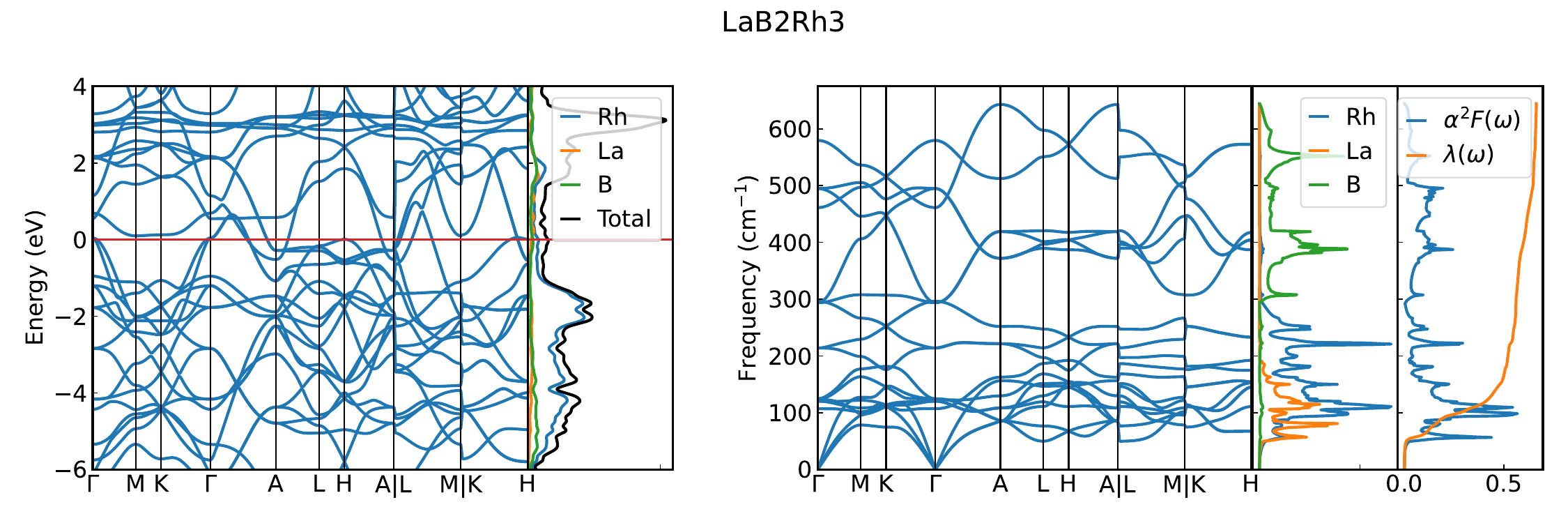}
    \includegraphics[width=0.75\linewidth]{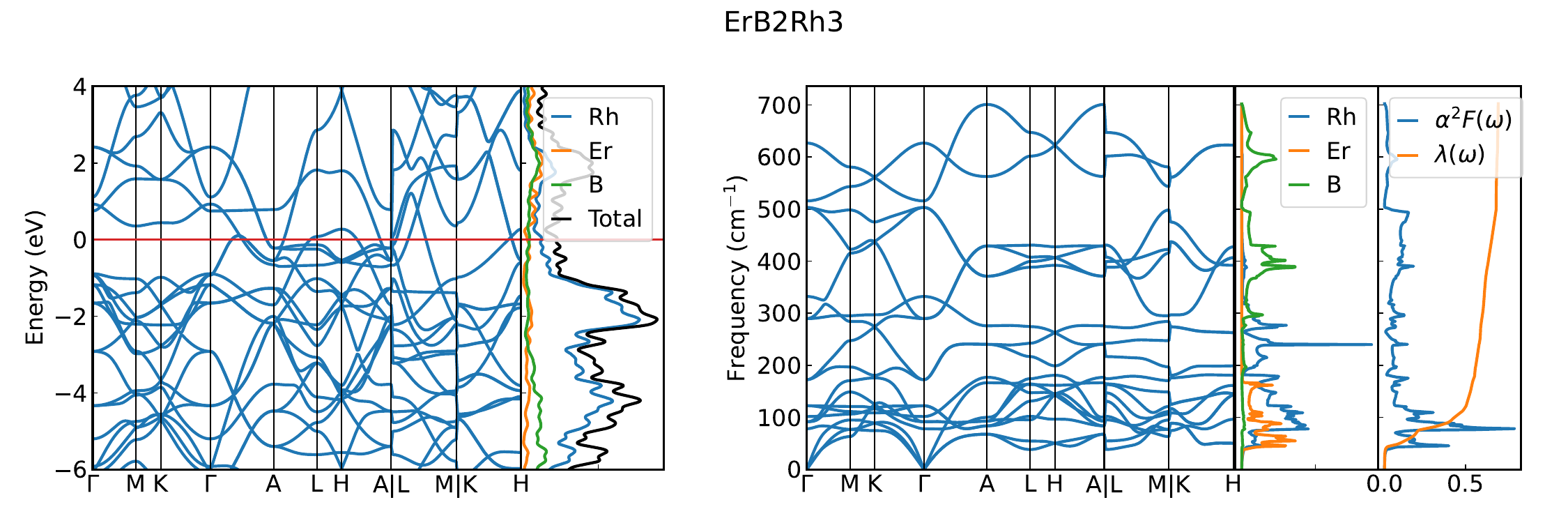}
\label{fig:type-i-4}
\end{figure}

\begin{figure}
    \centering
    \includegraphics[width=0.75\linewidth]{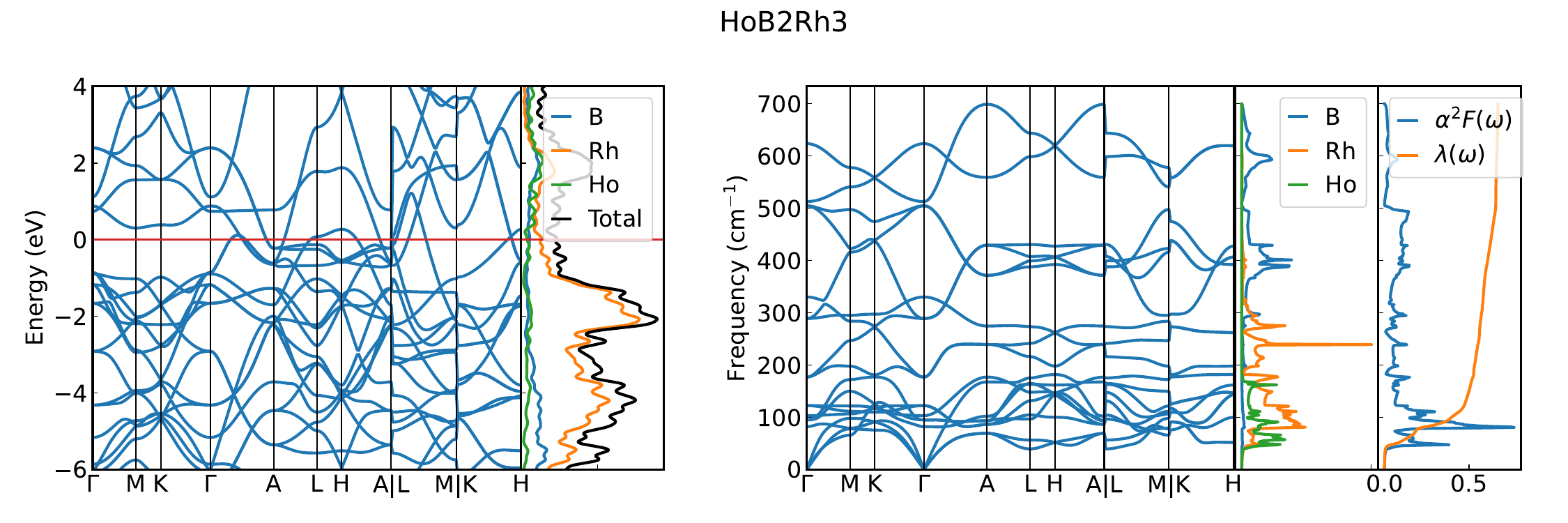}
    \includegraphics[width=0.75\linewidth]{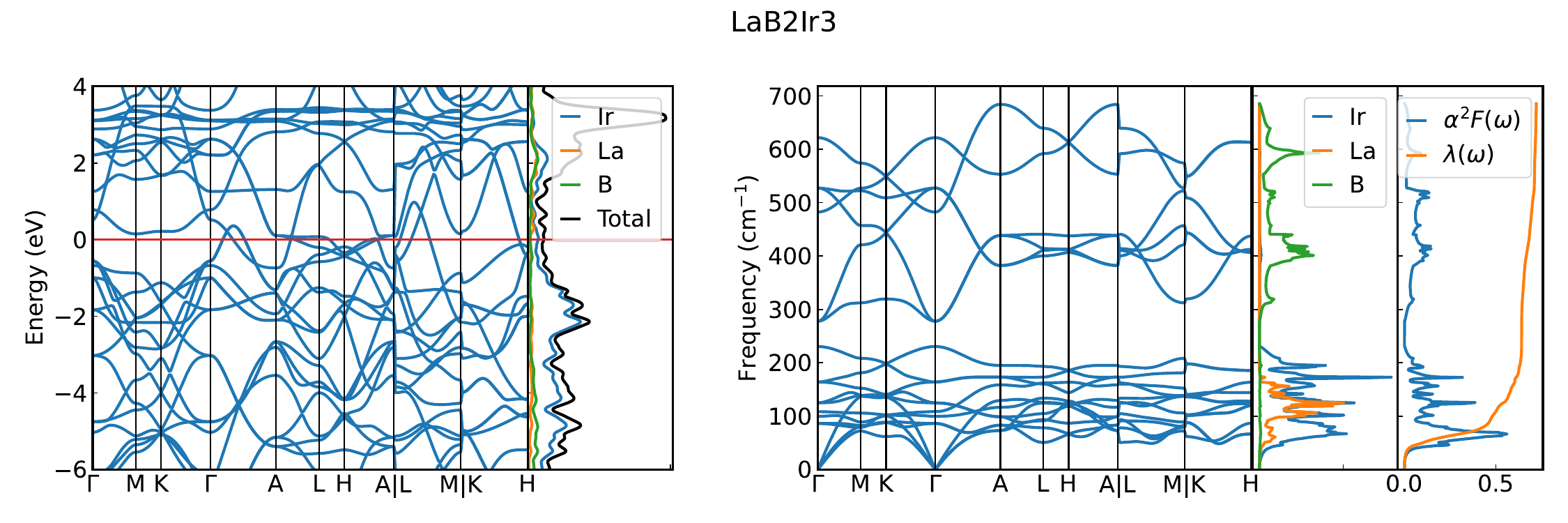}
    \includegraphics[width=0.75\linewidth]{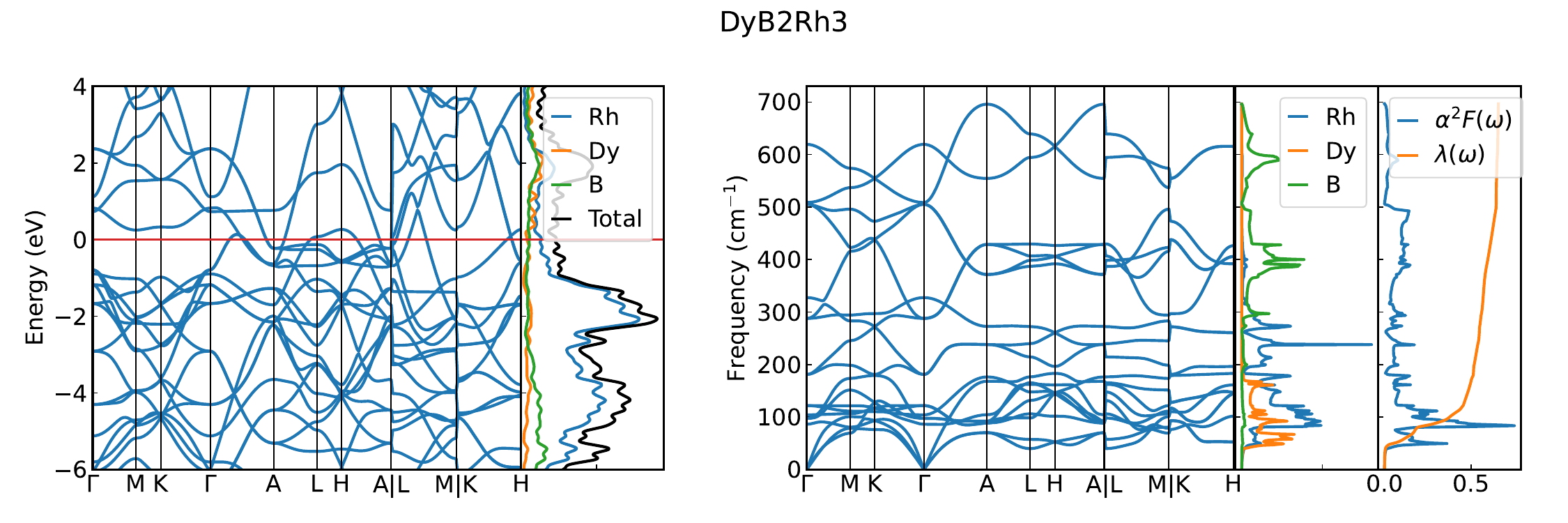}
    \includegraphics[width=0.75\linewidth]{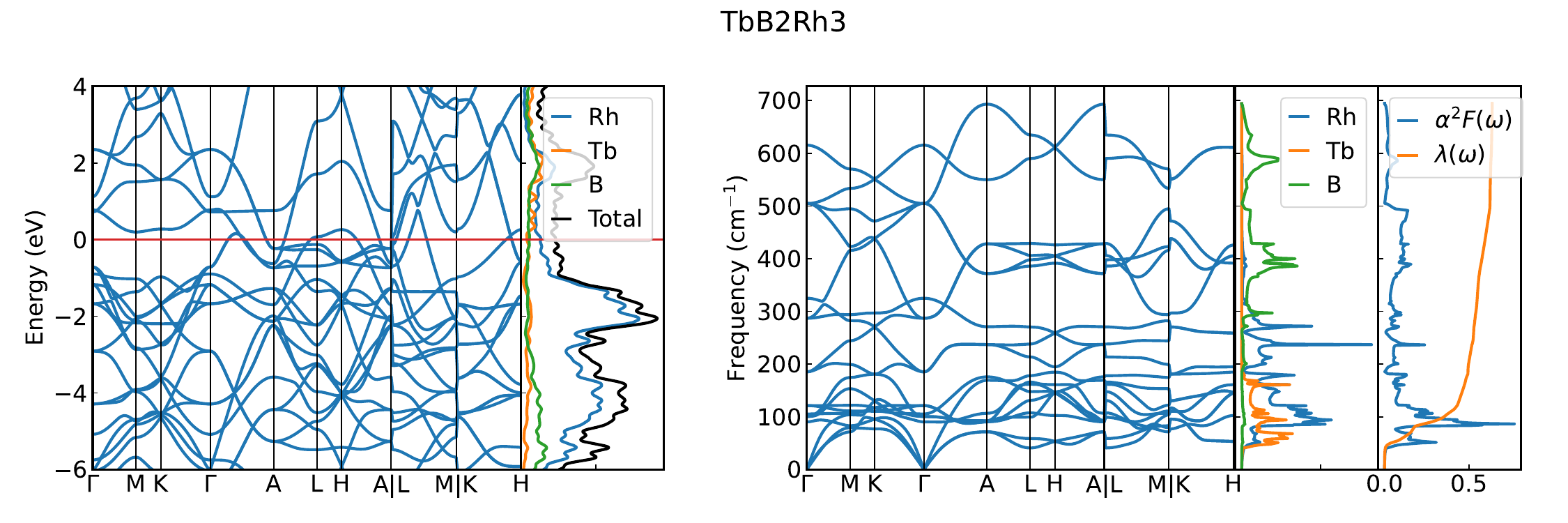}
    \includegraphics[width=0.75\linewidth]{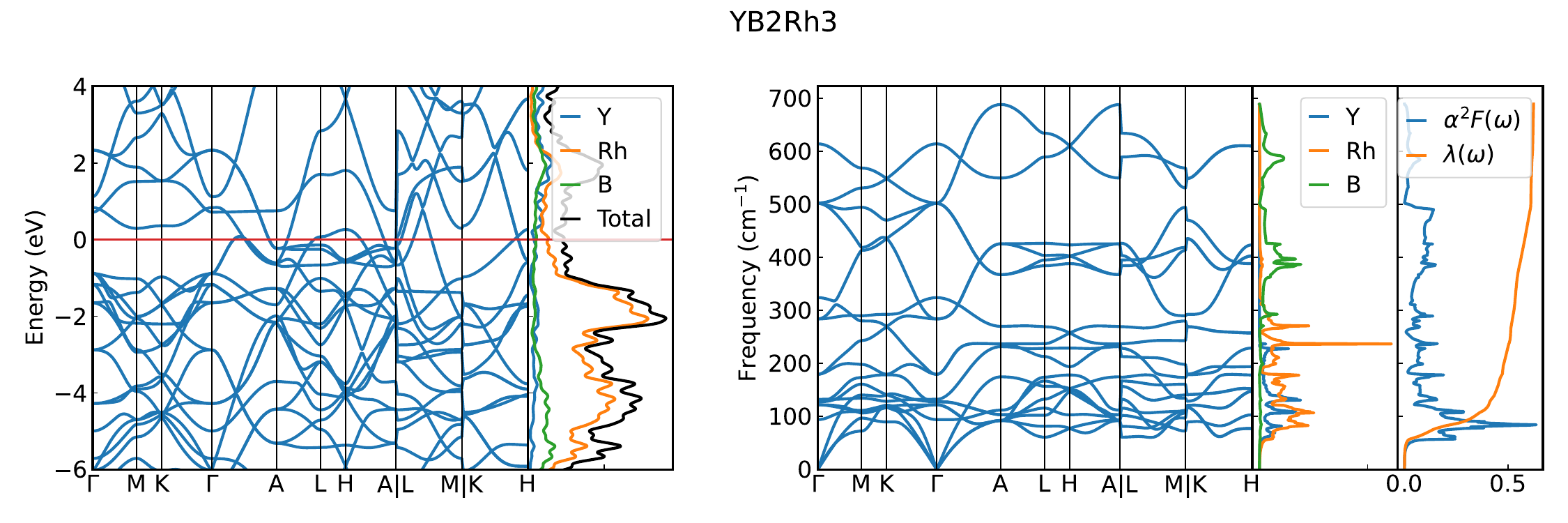}
\label{fig:type-i-5}
\end{figure}

\begin{figure}
    \centering
    \includegraphics[width=0.75\linewidth]{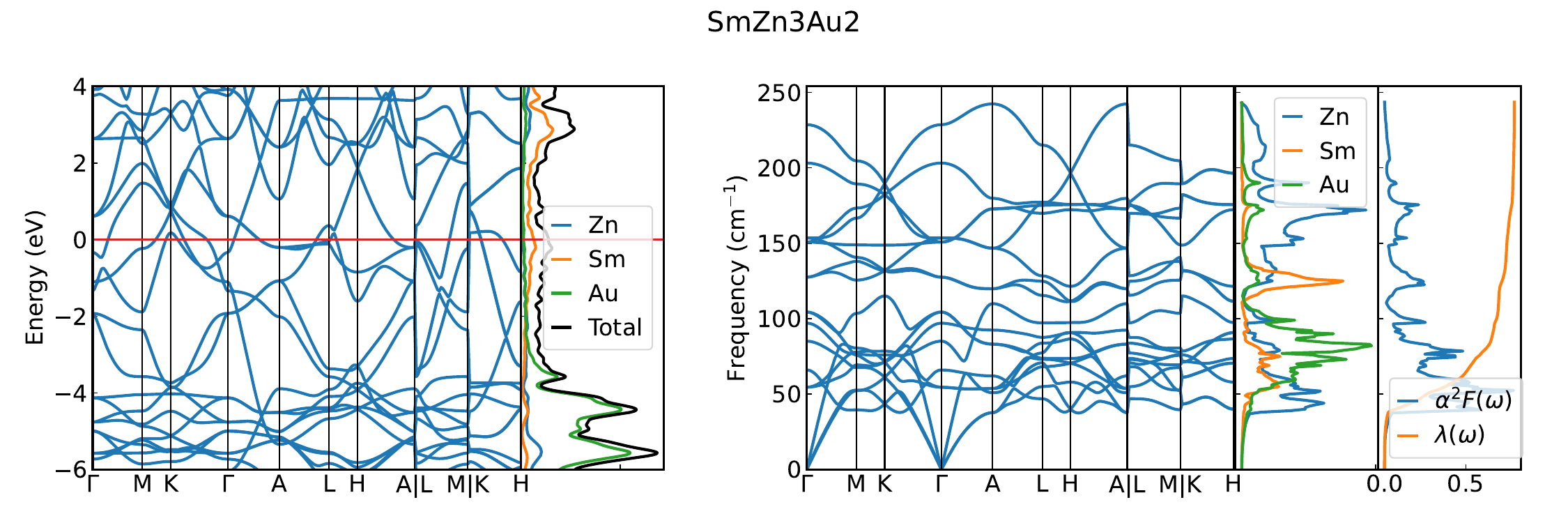}
    \includegraphics[width=0.75\linewidth]{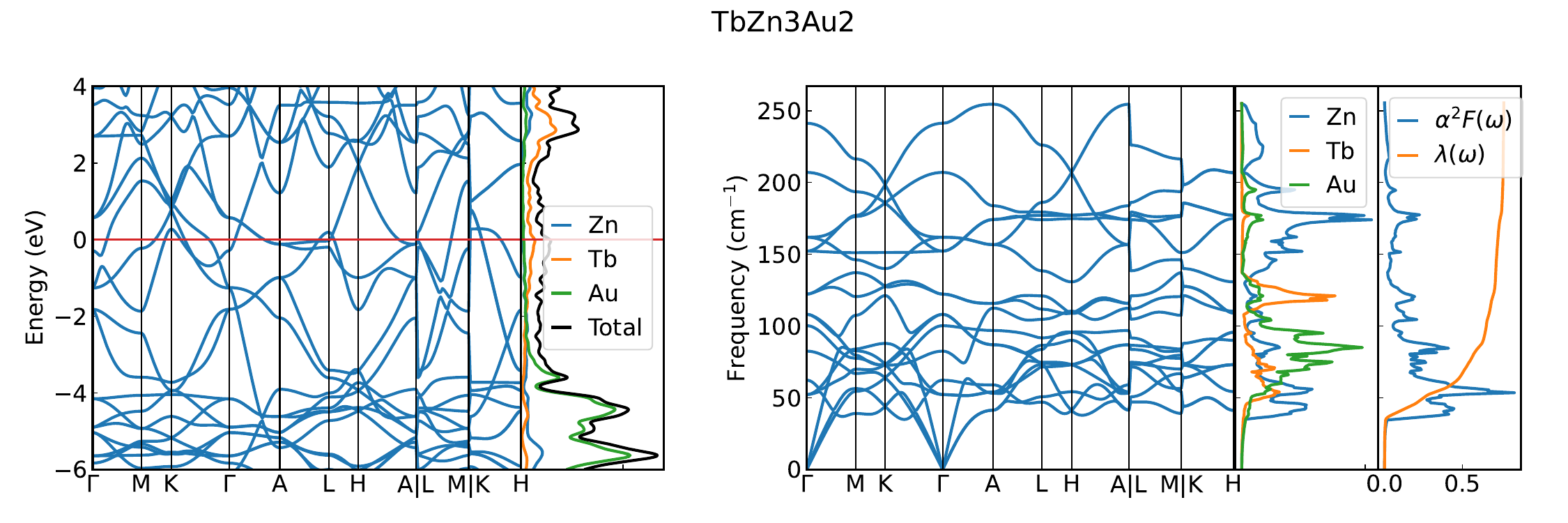}
    \includegraphics[width=0.75\linewidth]{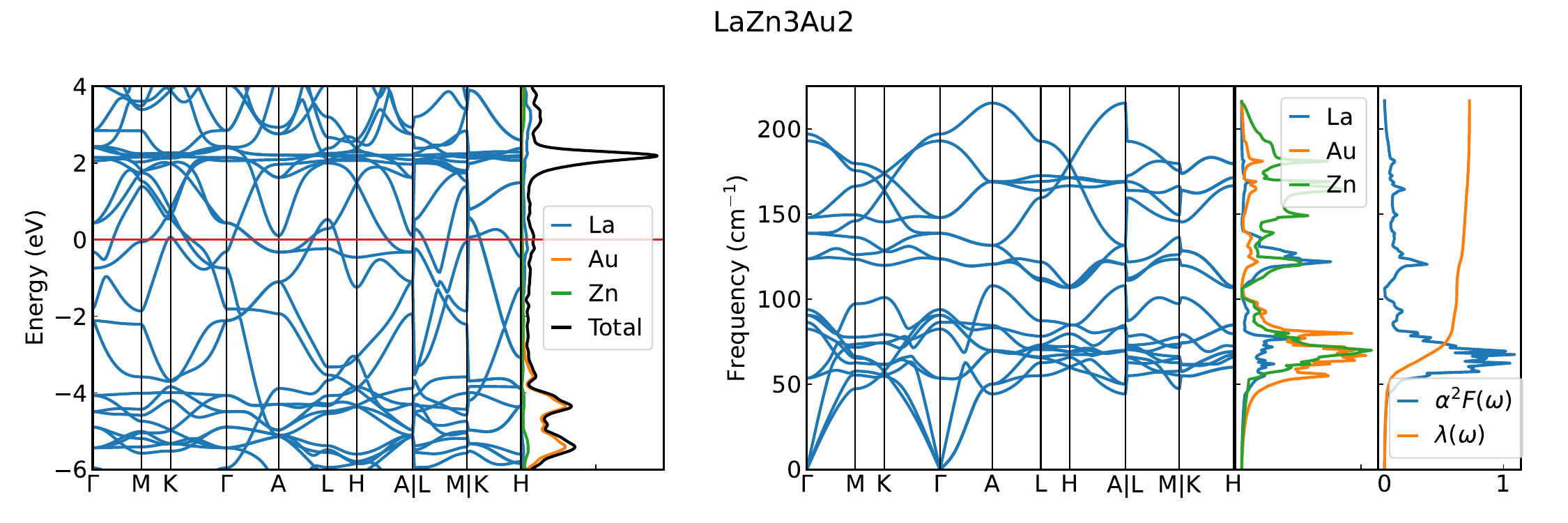}
    \includegraphics[width=0.75\linewidth]{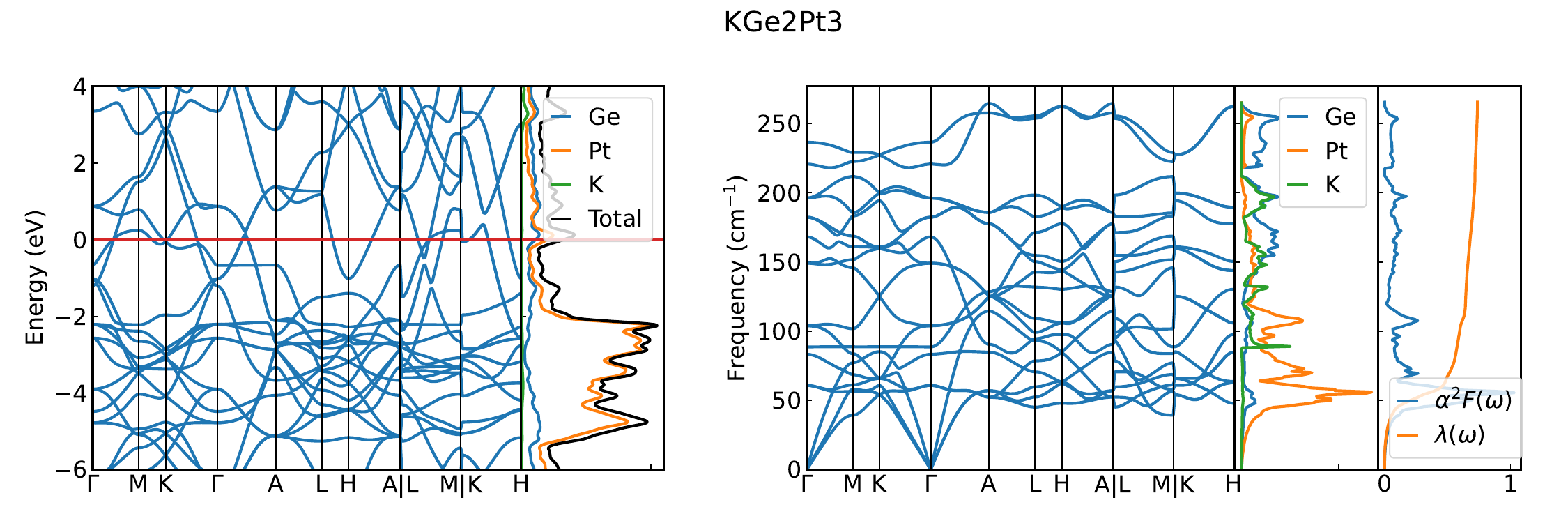}
    \includegraphics[width=0.75\linewidth]{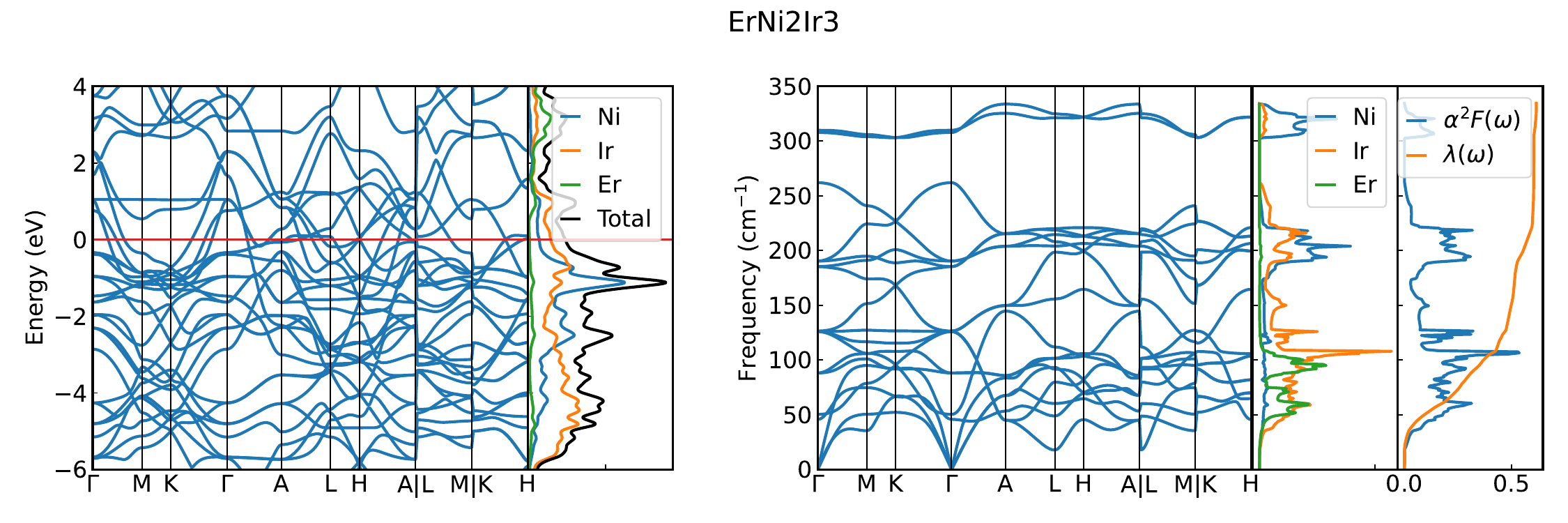}
\label{fig:type-i-6}
\end{figure}

\begin{figure}
    \centering
    \includegraphics[width=0.75\linewidth]{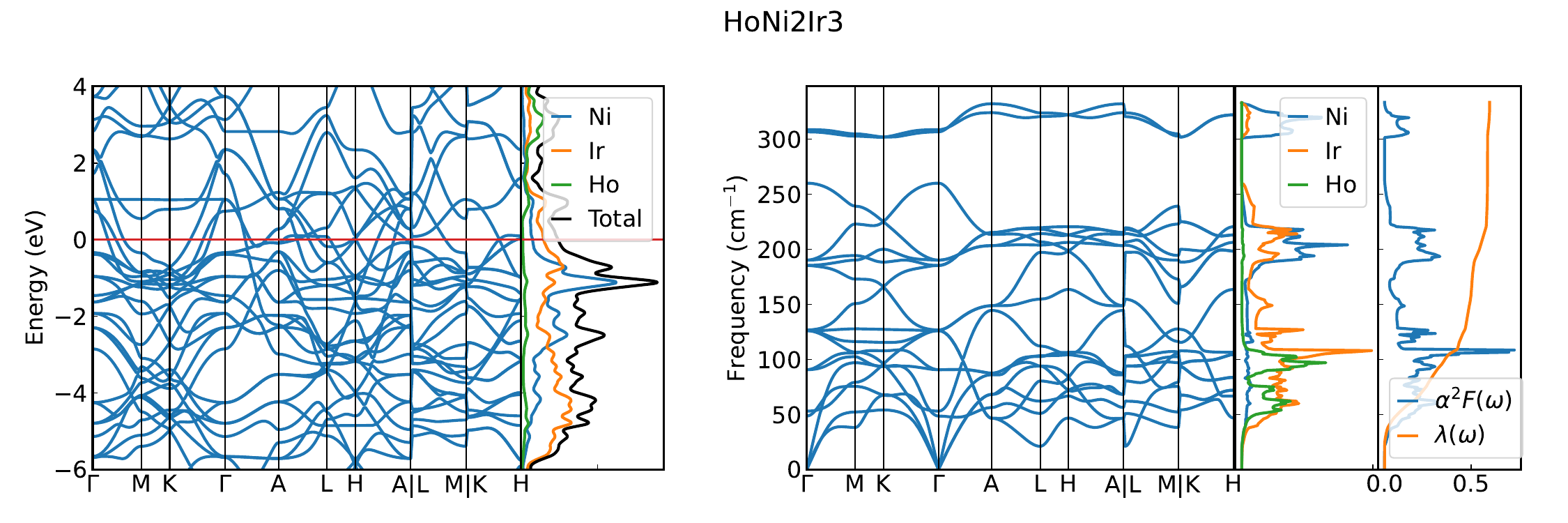}
    \includegraphics[width=0.75\linewidth]{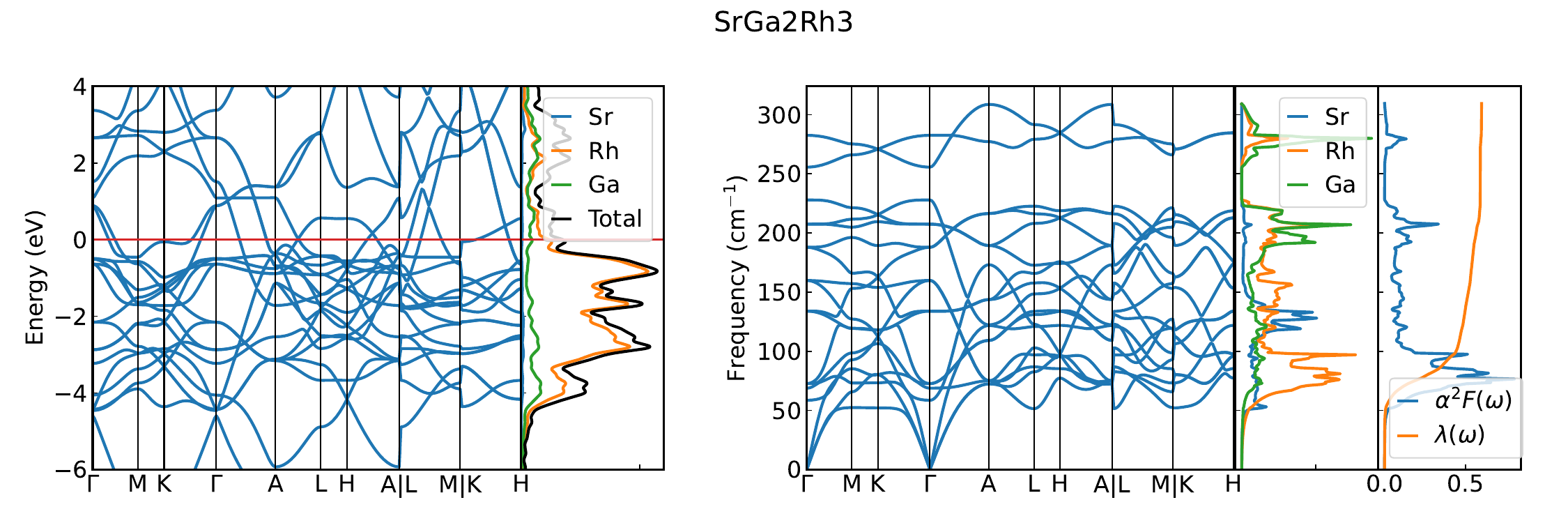}
    \includegraphics[width=0.75\linewidth]{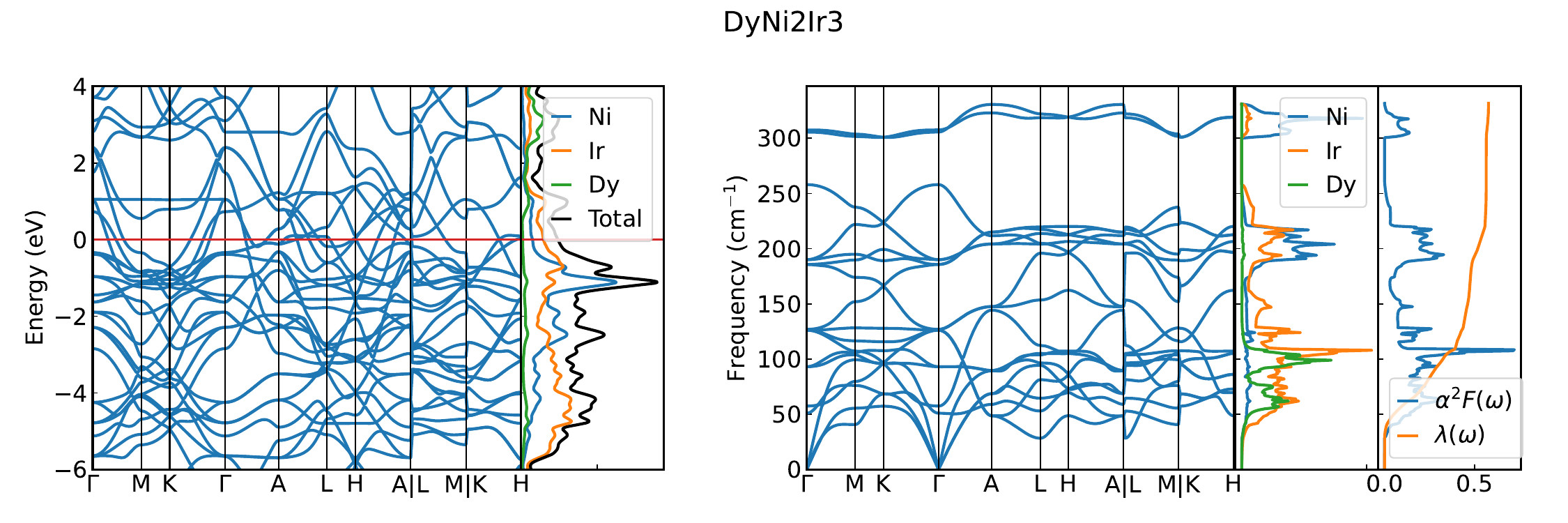}
    \includegraphics[width=0.75\linewidth]{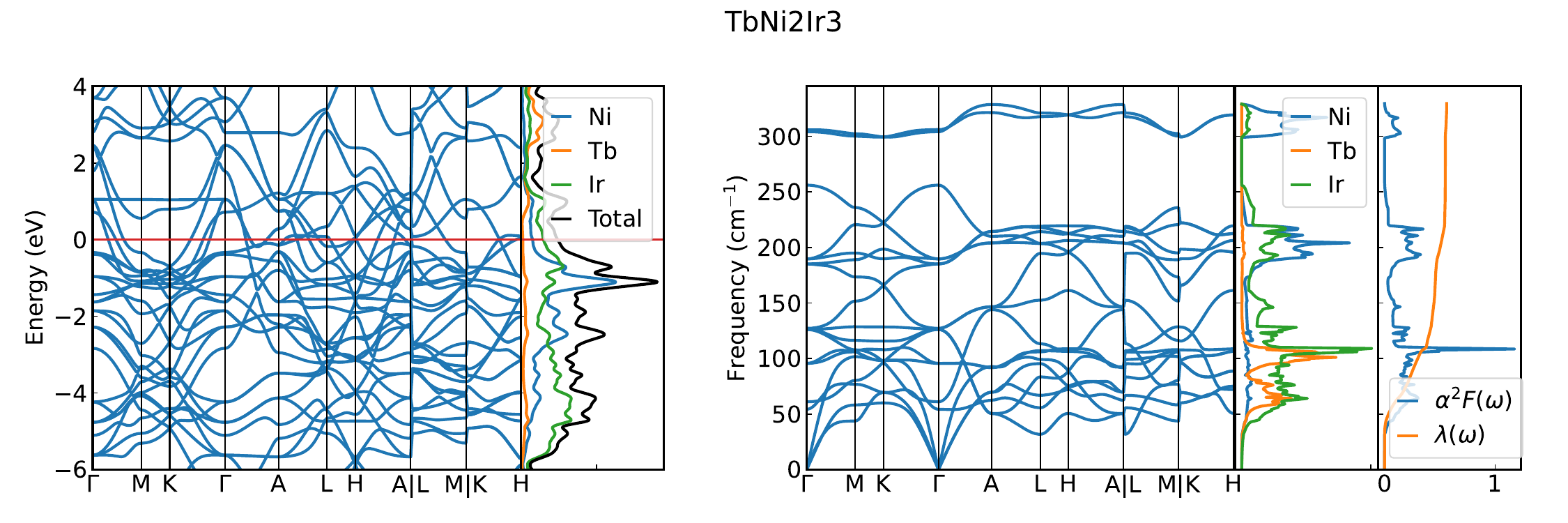}
    \includegraphics[width=0.75\linewidth]{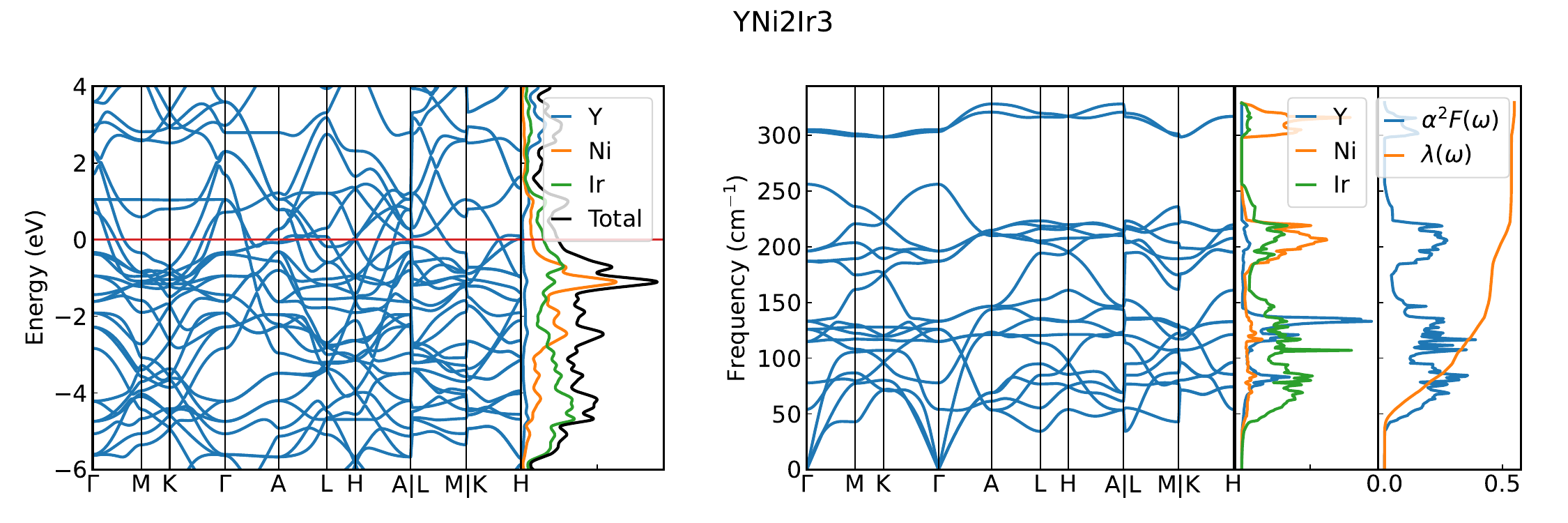}
\label{fig:type-i-7}
\end{figure}

\begin{figure}
    \centering
    \includegraphics[width=0.75\linewidth]{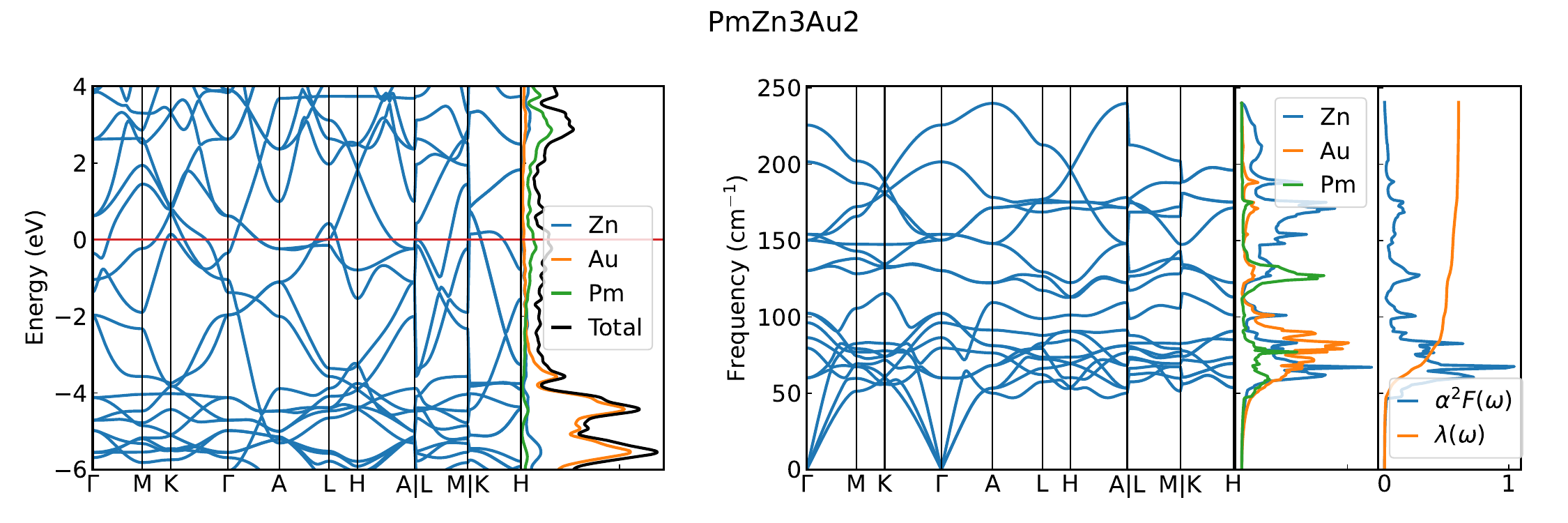}
    \includegraphics[width=0.75\linewidth]{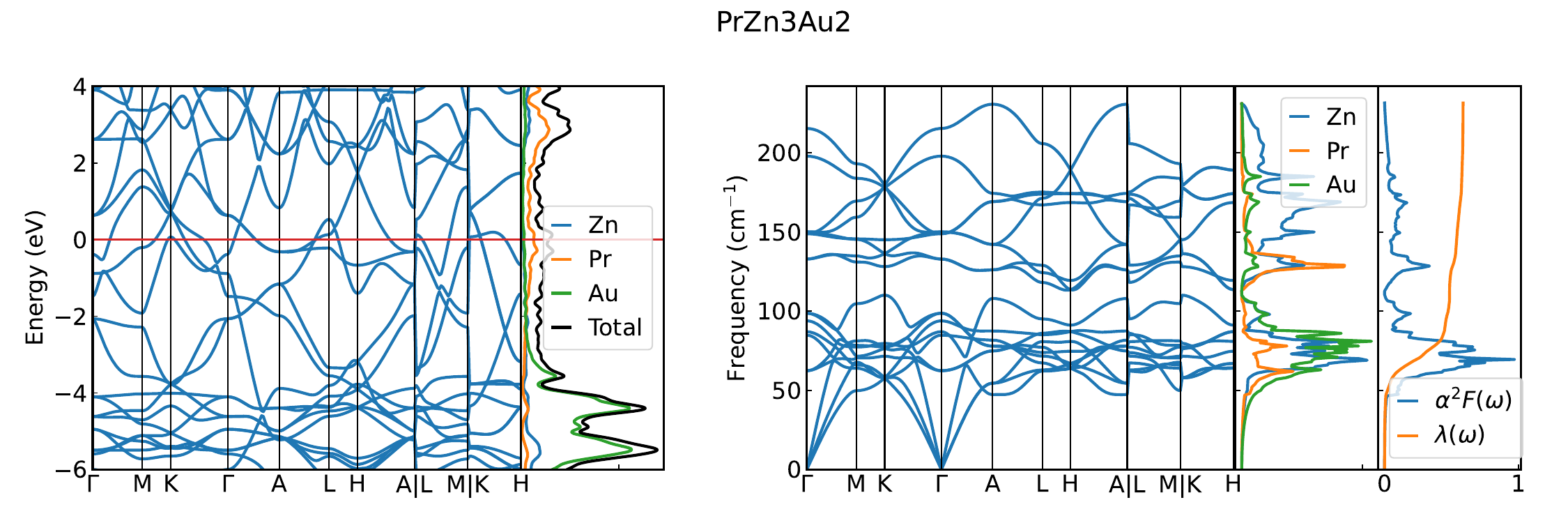}
    \includegraphics[width=0.75\linewidth]{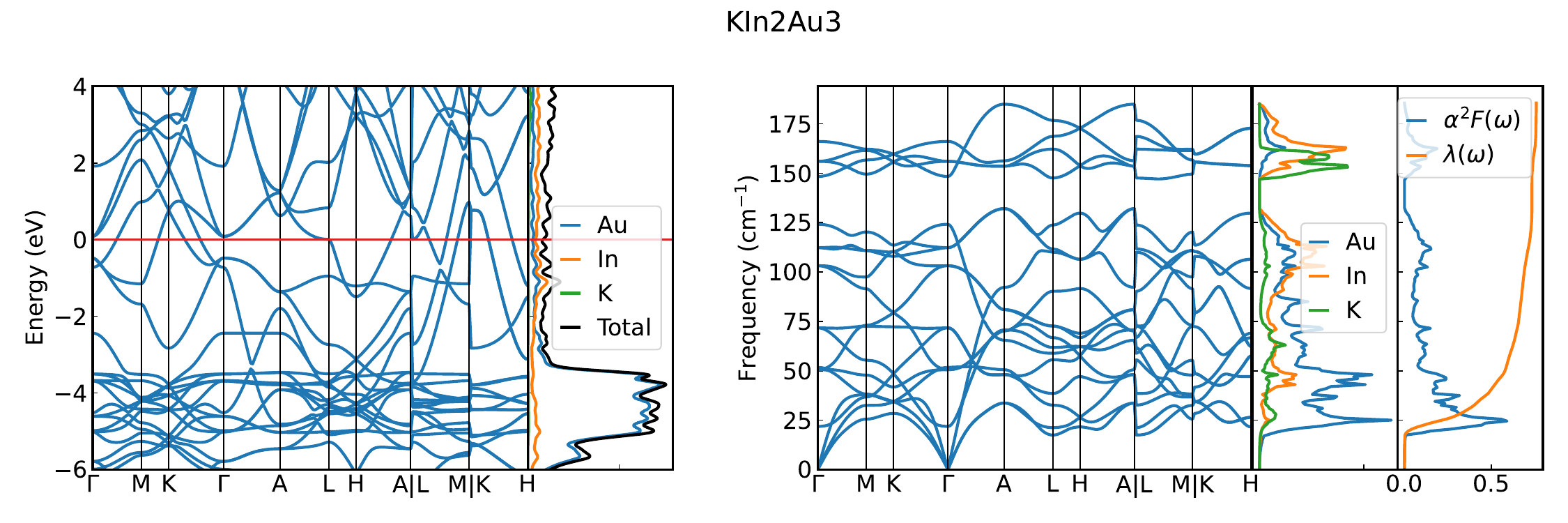}
    \includegraphics[width=0.75\linewidth]{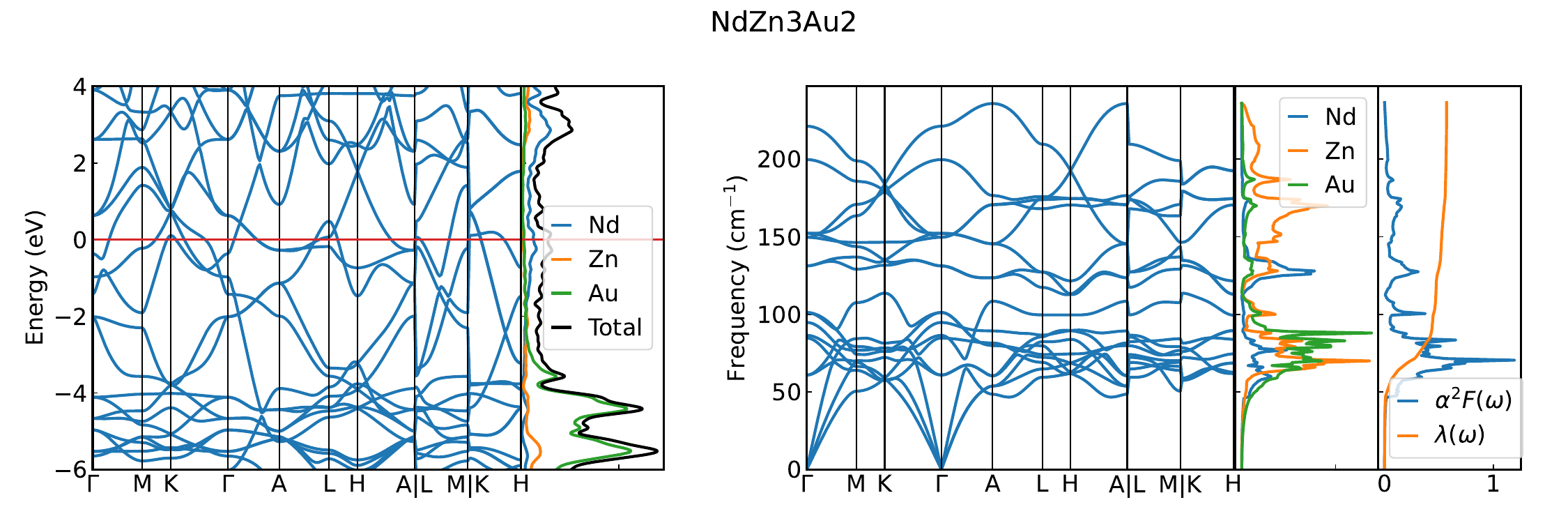}
    \includegraphics[width=0.75\linewidth]{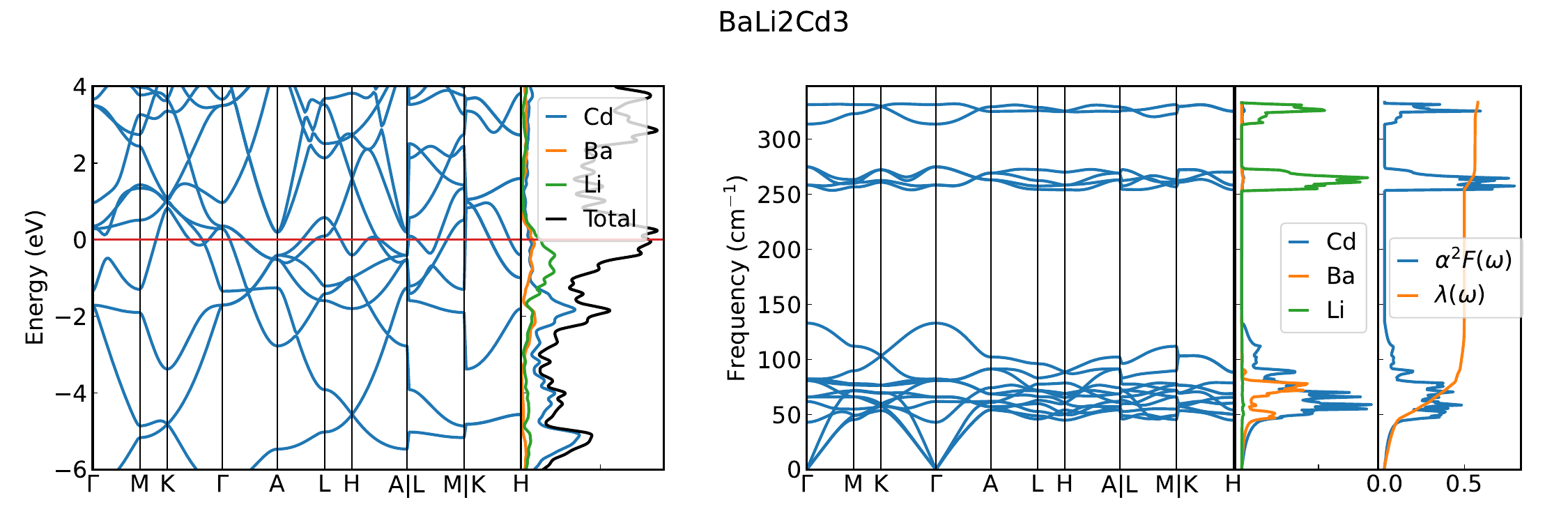}
\label{fig:type-i-8}
\end{figure}

\begin{figure}
    \centering
    \includegraphics[width=0.75\linewidth]{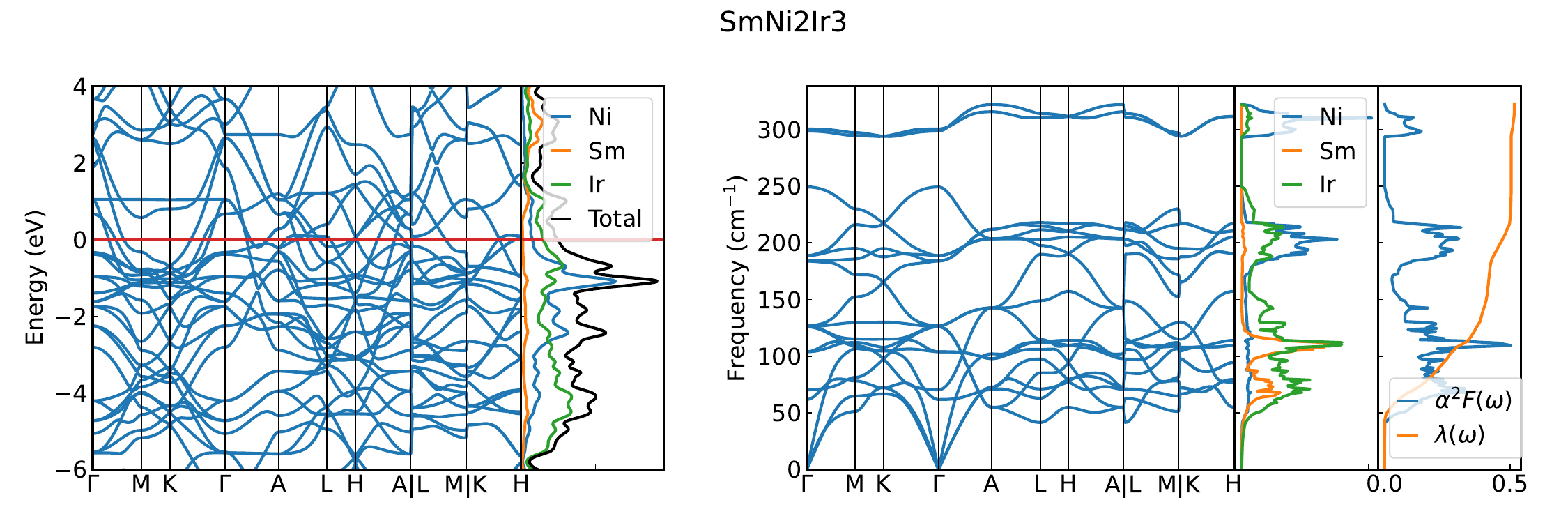}
    \includegraphics[width=0.75\linewidth]{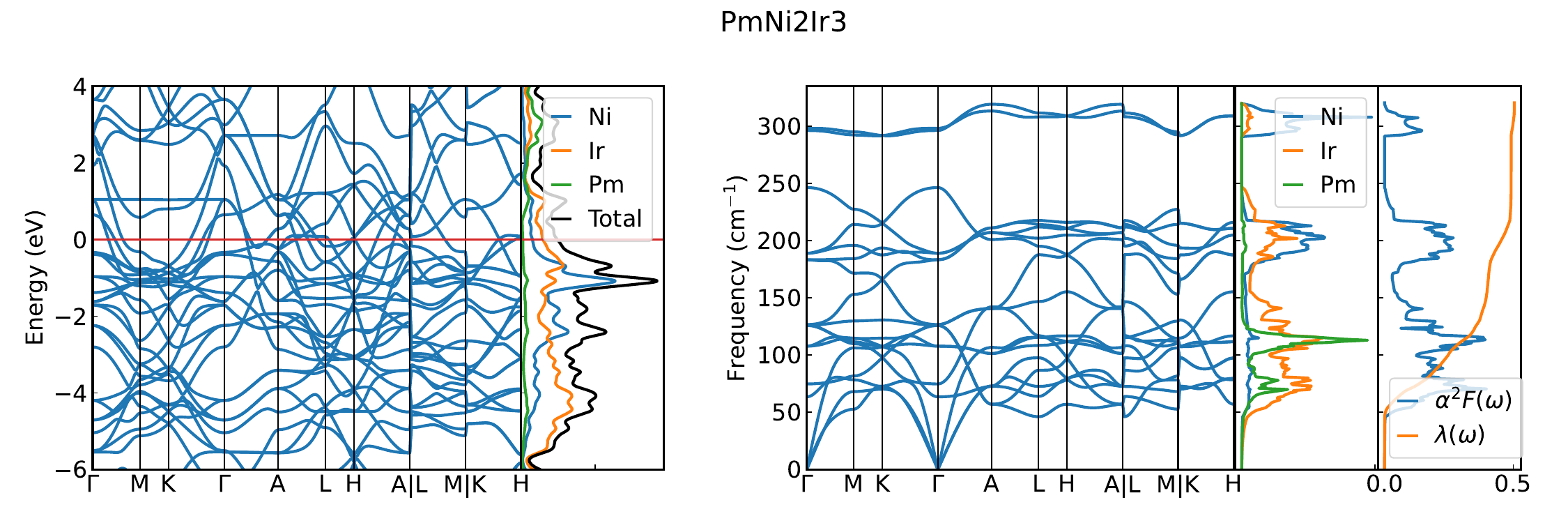}
    \includegraphics[width=0.75\linewidth]{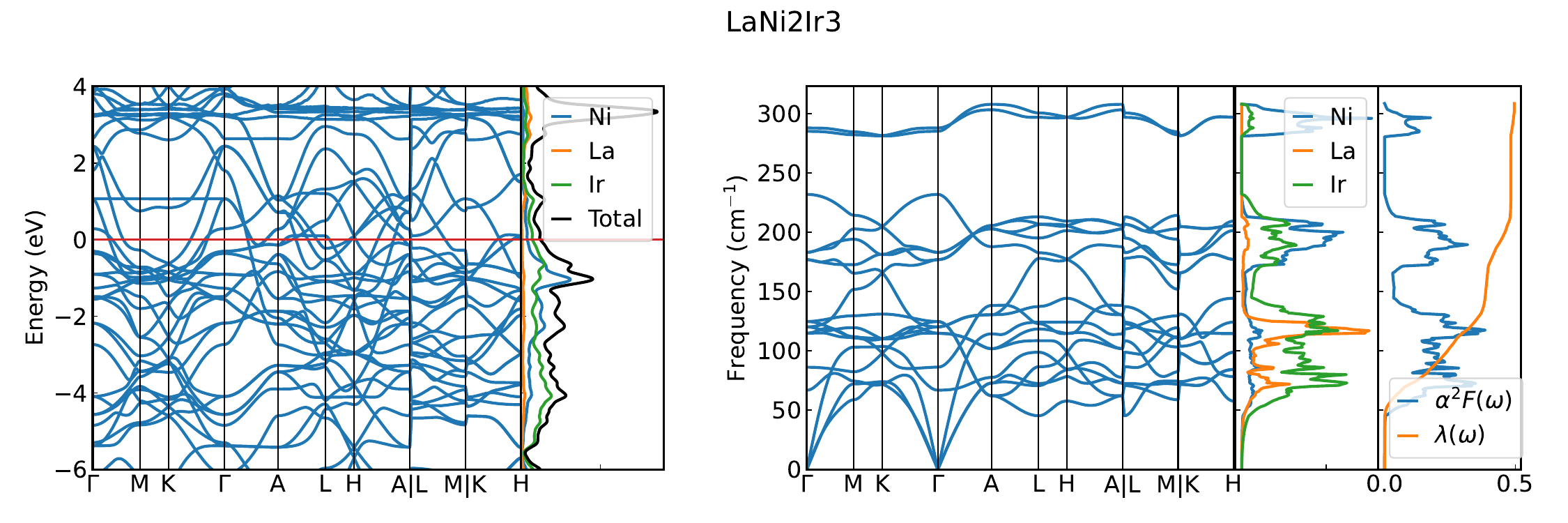}
    \includegraphics[width=0.75\linewidth]{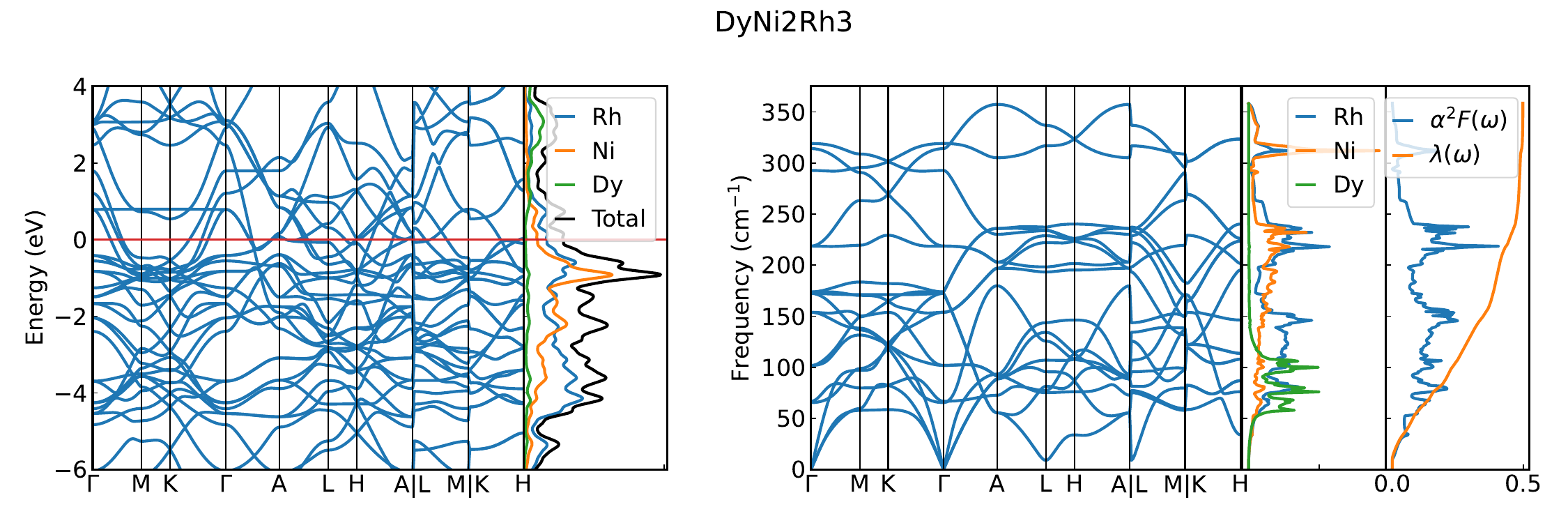}
    \includegraphics[width=0.75\linewidth]{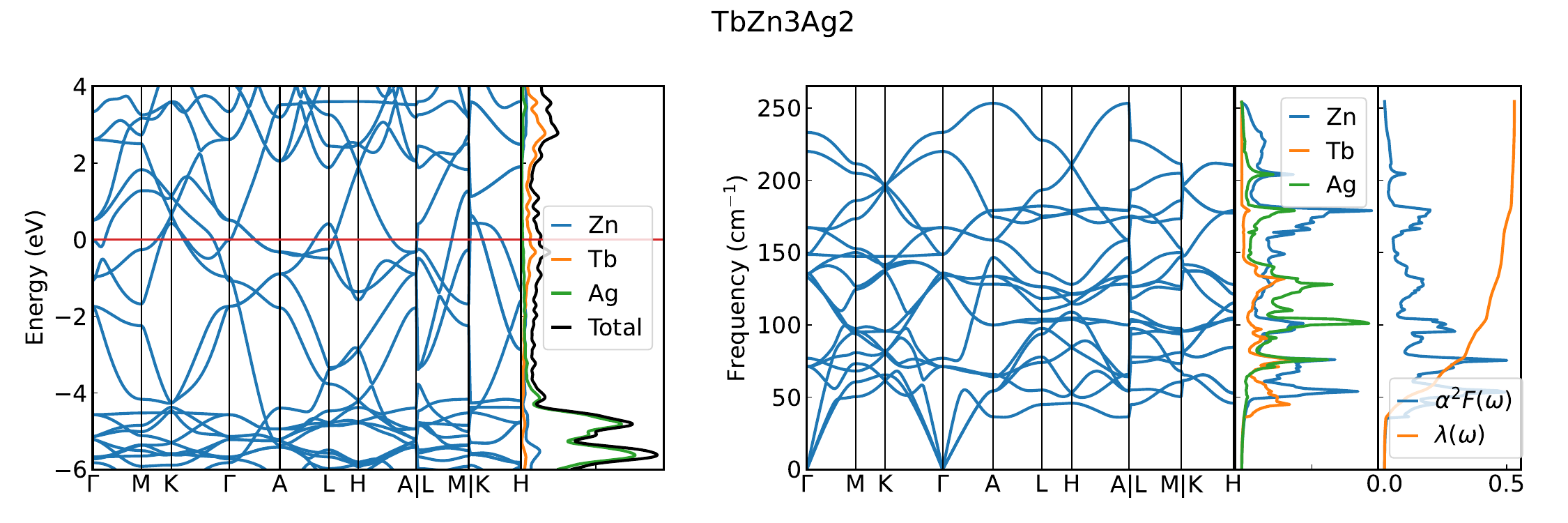}
\label{fig:type-i-9}
\end{figure}

\begin{figure}
    \centering
    \includegraphics[width=0.75\linewidth]{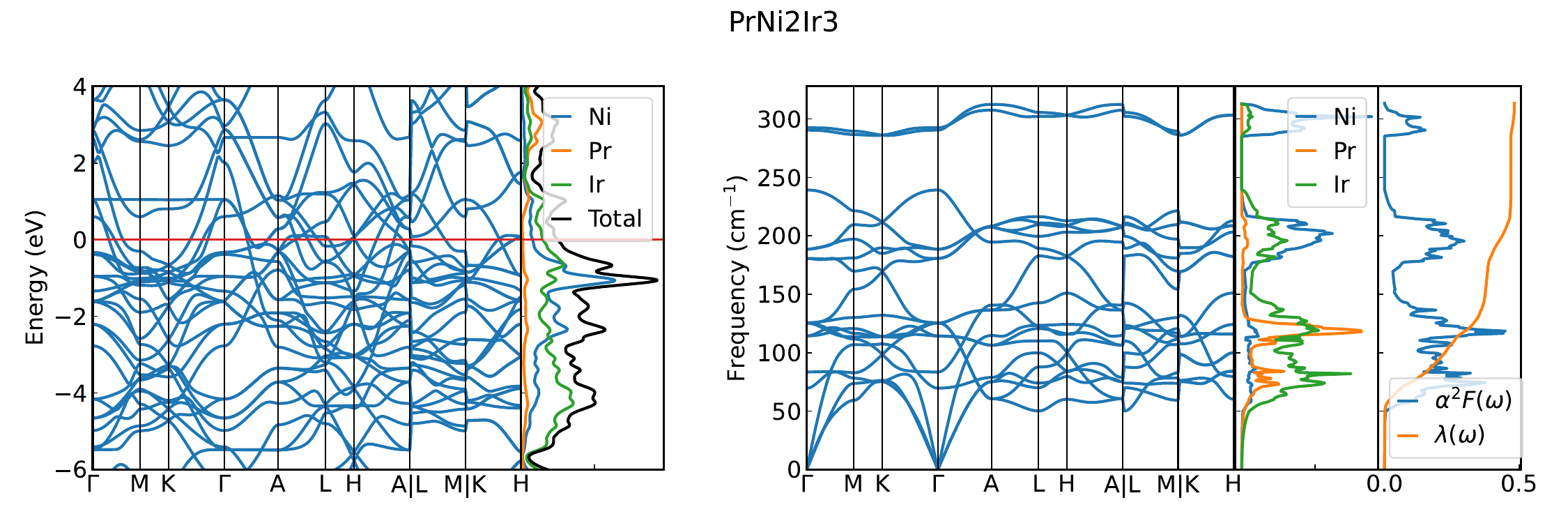}
    \includegraphics[width=0.75\linewidth]{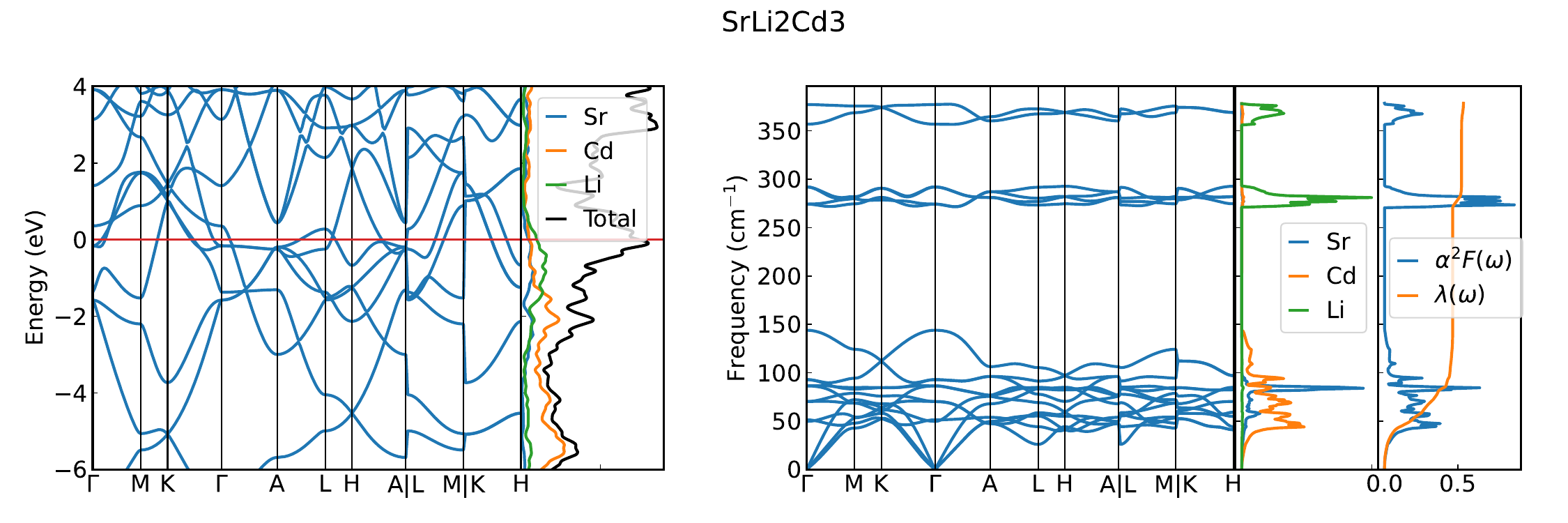}
    \includegraphics[width=0.75\linewidth]{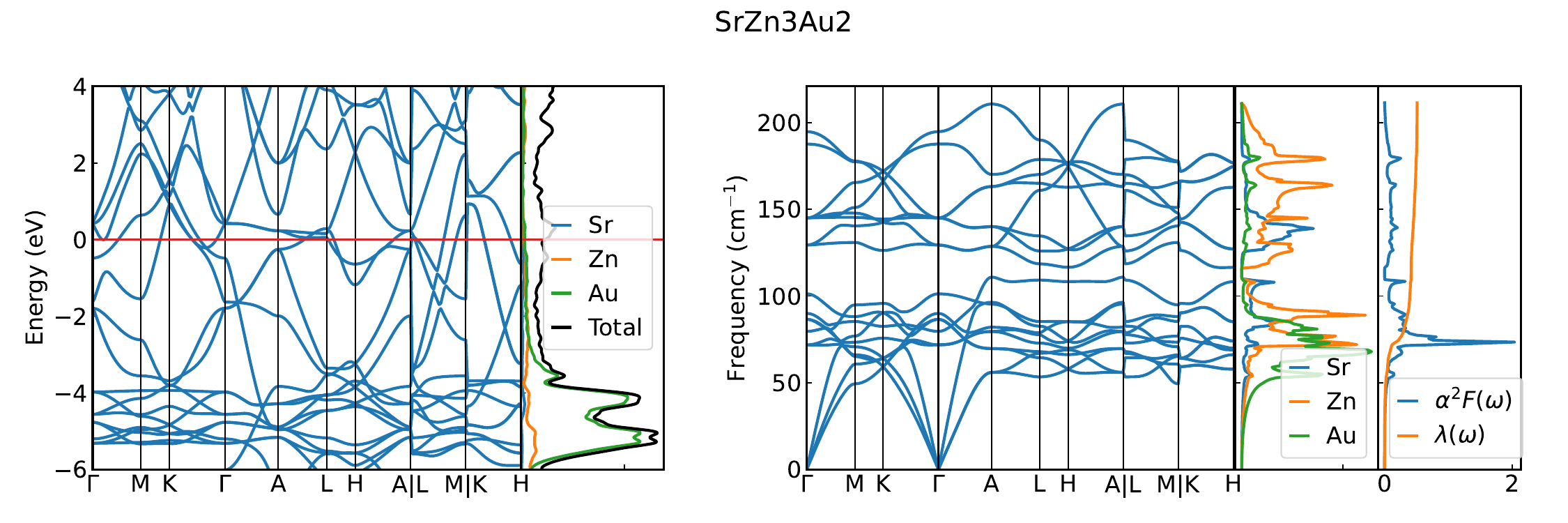}
    \includegraphics[width=0.75\linewidth]{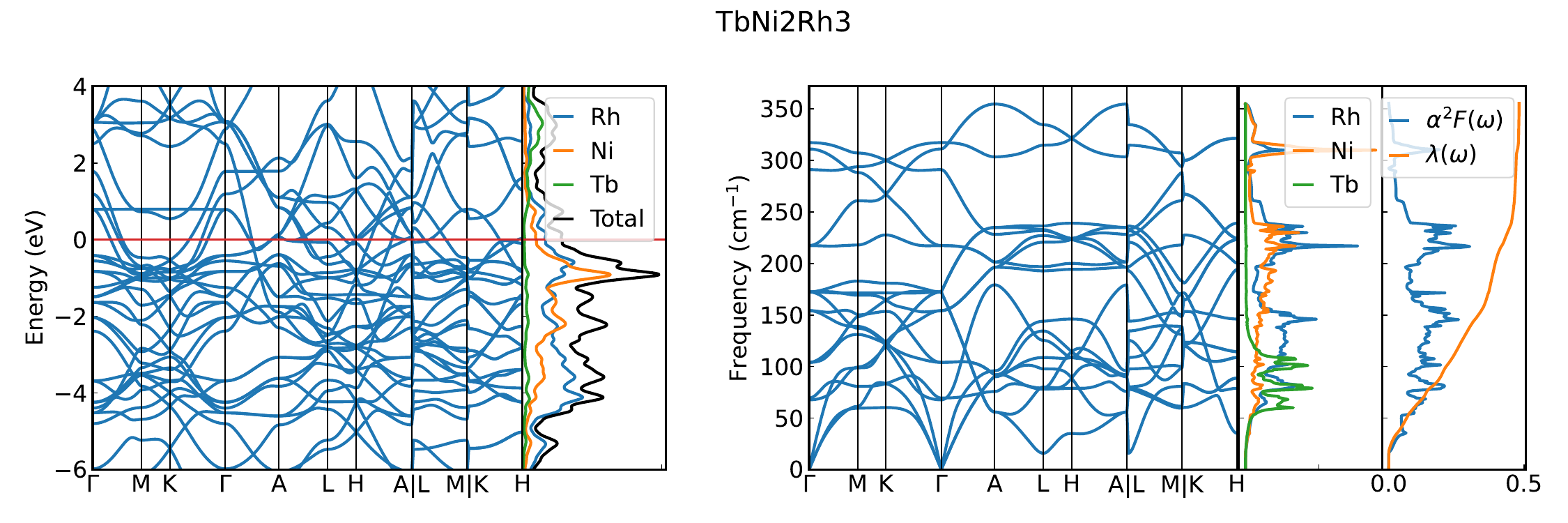}
    \includegraphics[width=0.75\linewidth]{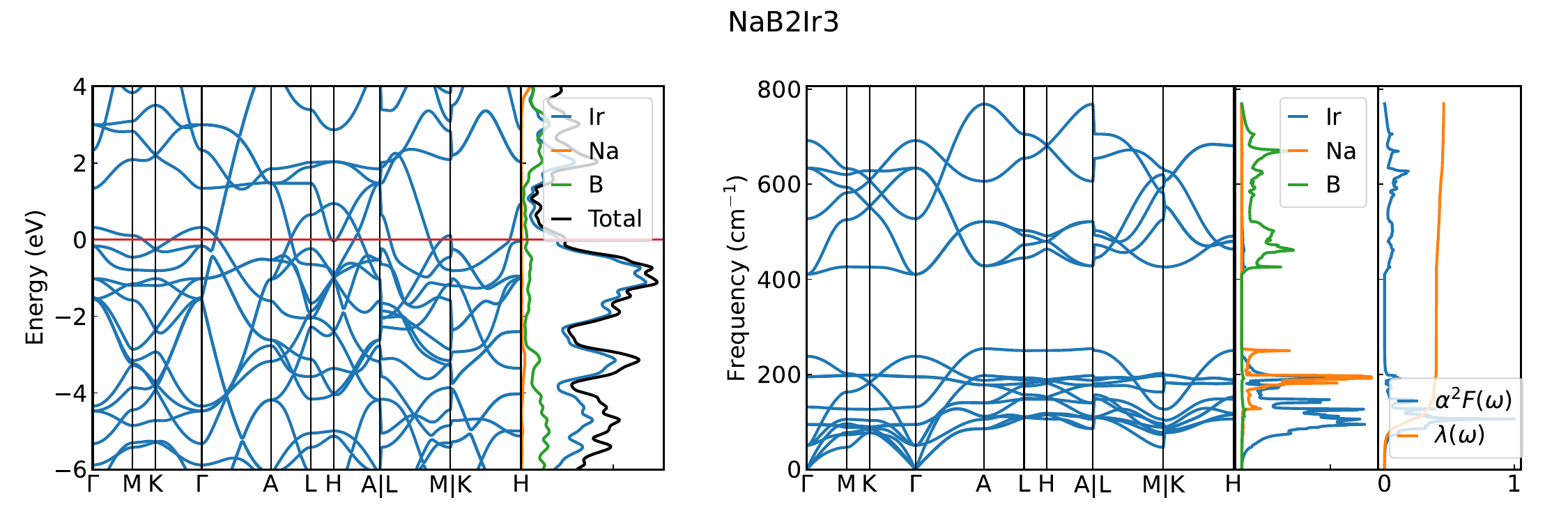}
\label{fig:type-i-10}
\end{figure}

\begin{figure}
    \centering
    \includegraphics[width=0.75\linewidth]{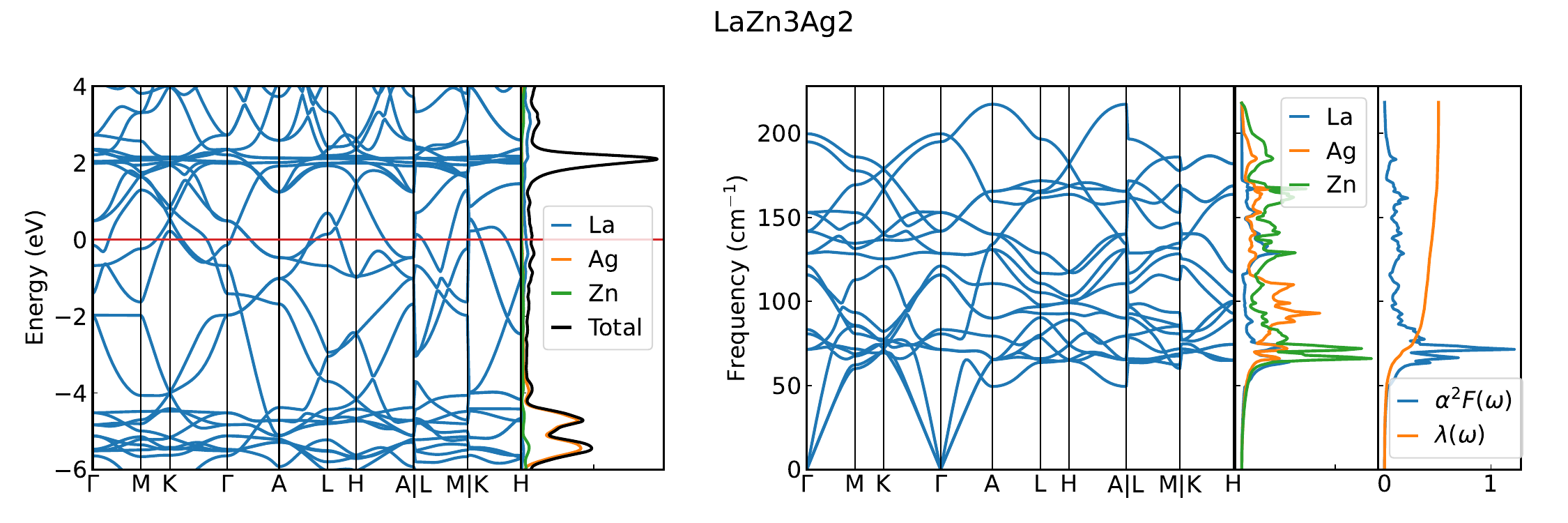}
    \includegraphics[width=0.75\linewidth]{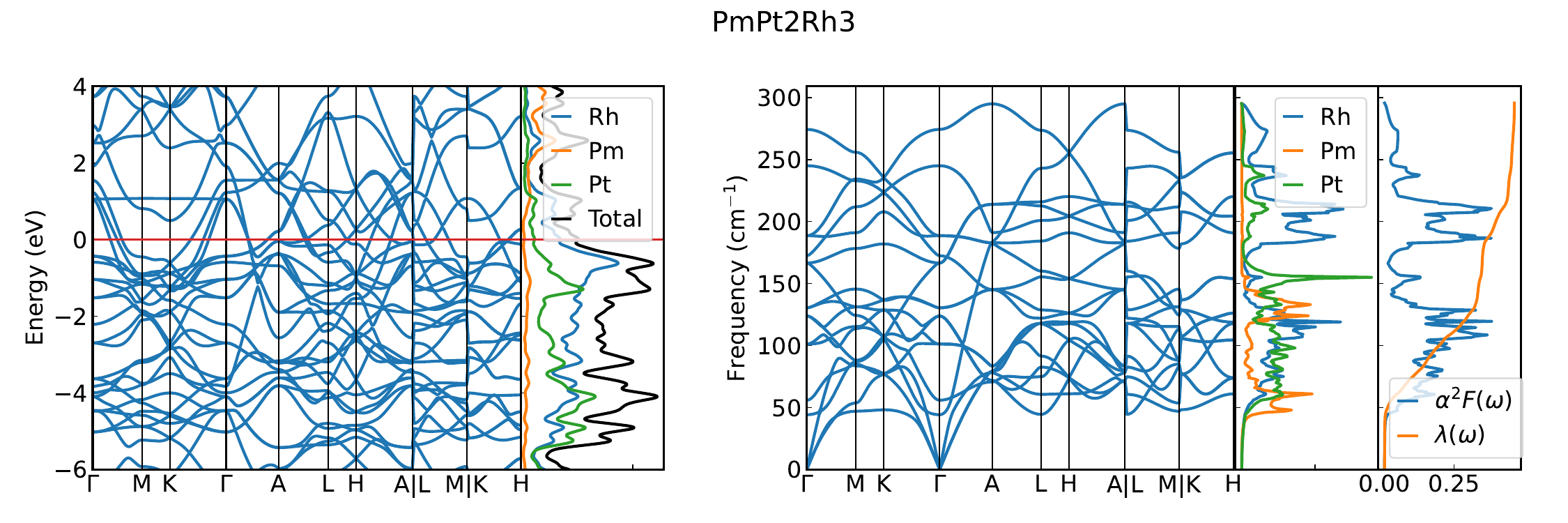}
    \includegraphics[width=0.75\linewidth]{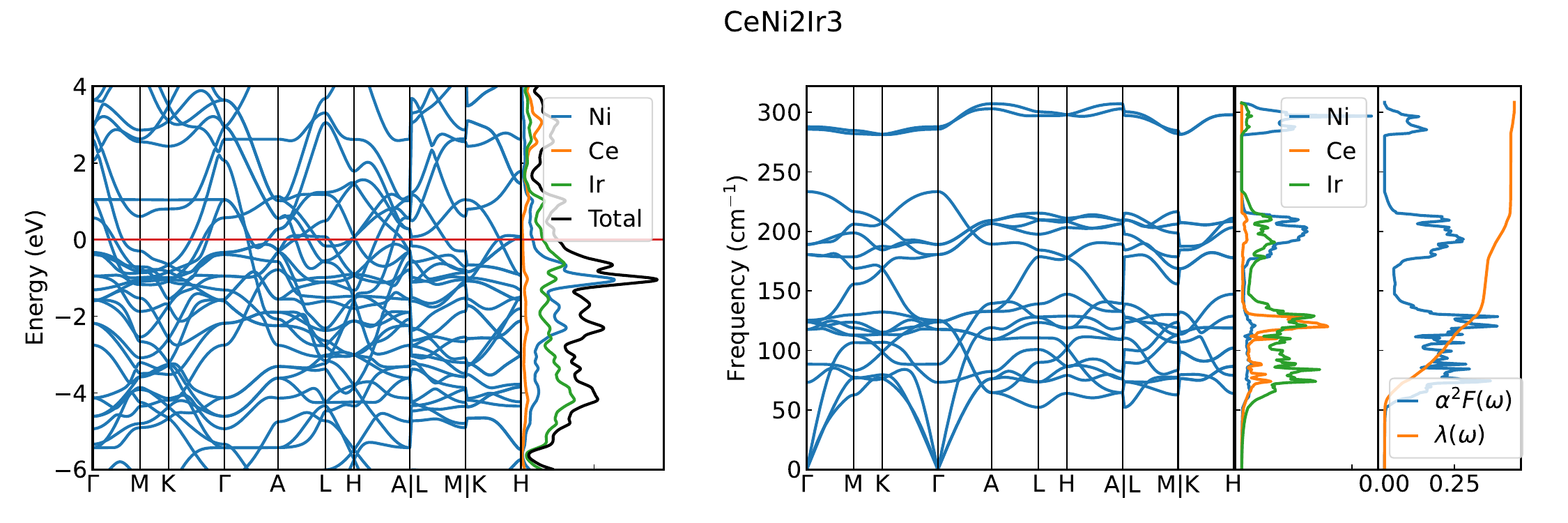}
    \includegraphics[width=0.75\linewidth]{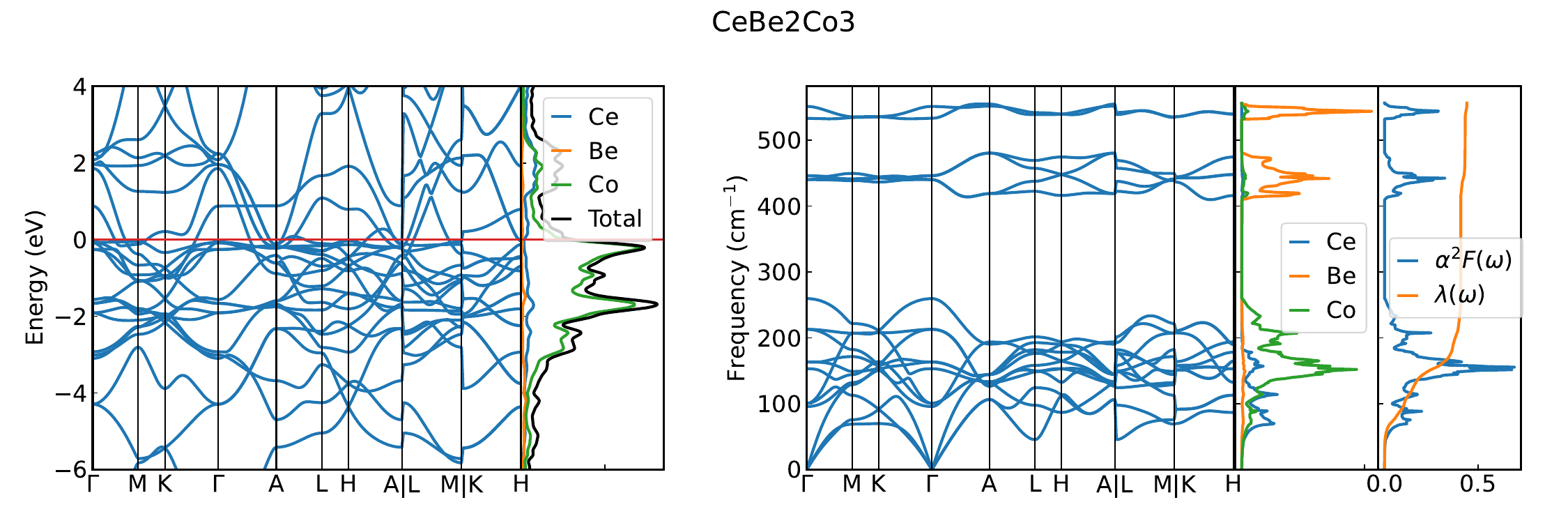}
    \includegraphics[width=0.75\linewidth]{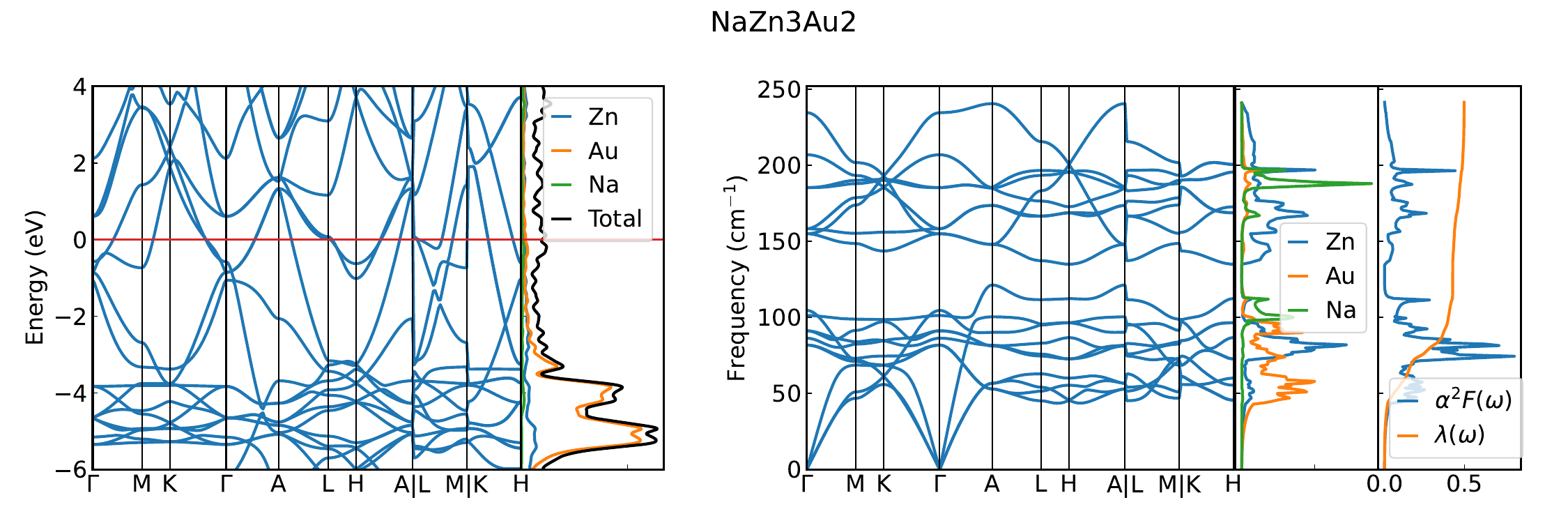}
\label{fig:type-i-11}
\end{figure}

\begin{figure}
    \centering
    \includegraphics[width=0.75\linewidth]{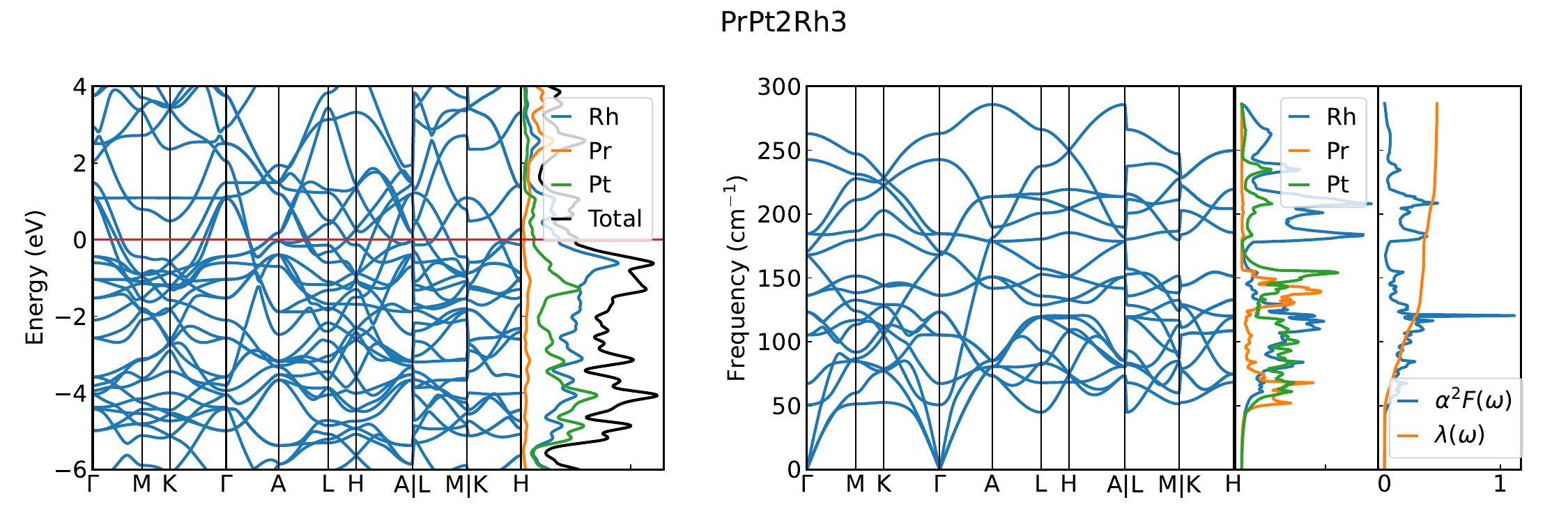}
    \includegraphics[width=0.75\linewidth]{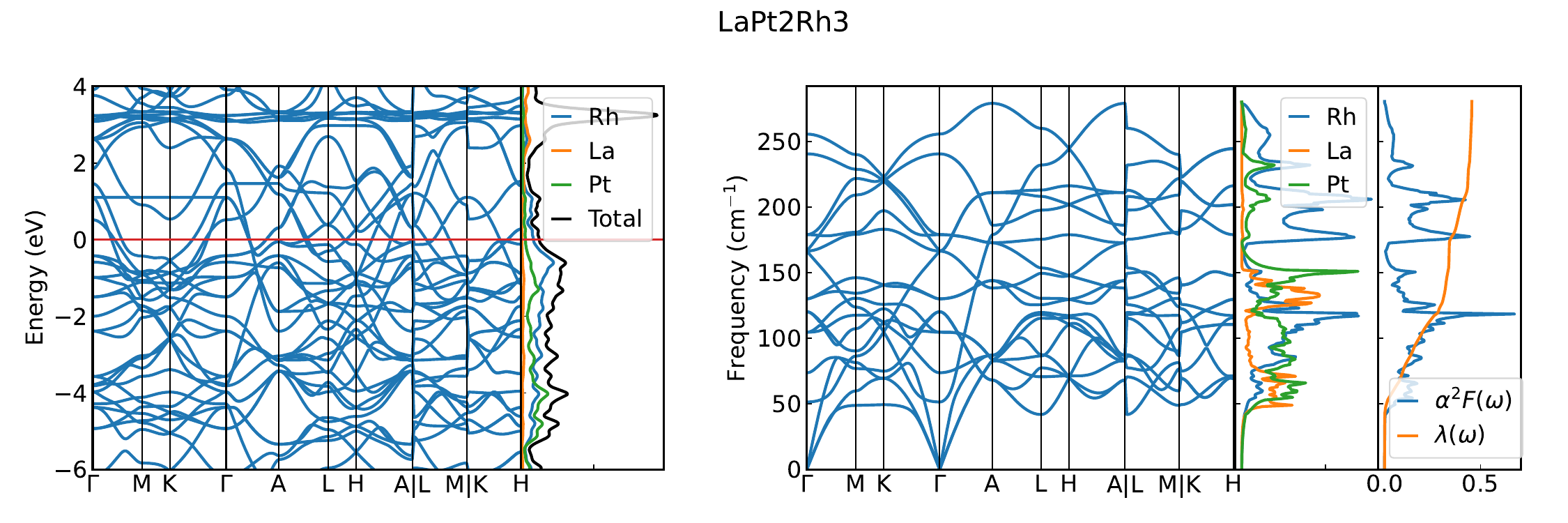}
    \includegraphics[width=0.75\linewidth]{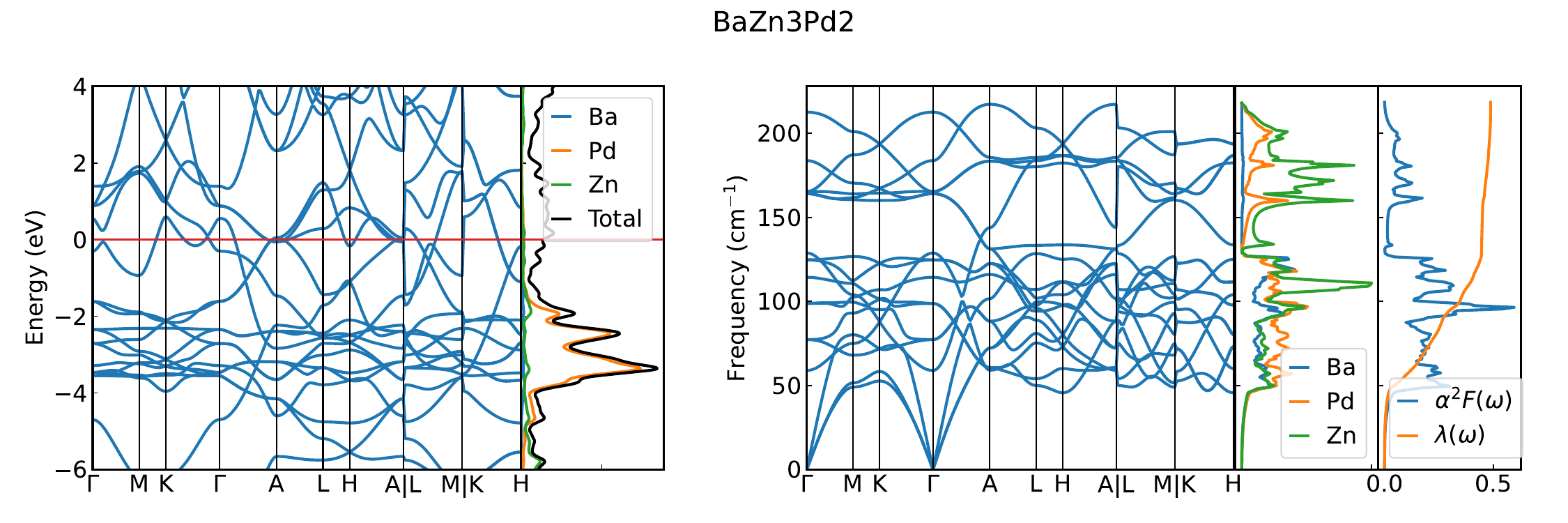}
    \includegraphics[width=0.75\linewidth]{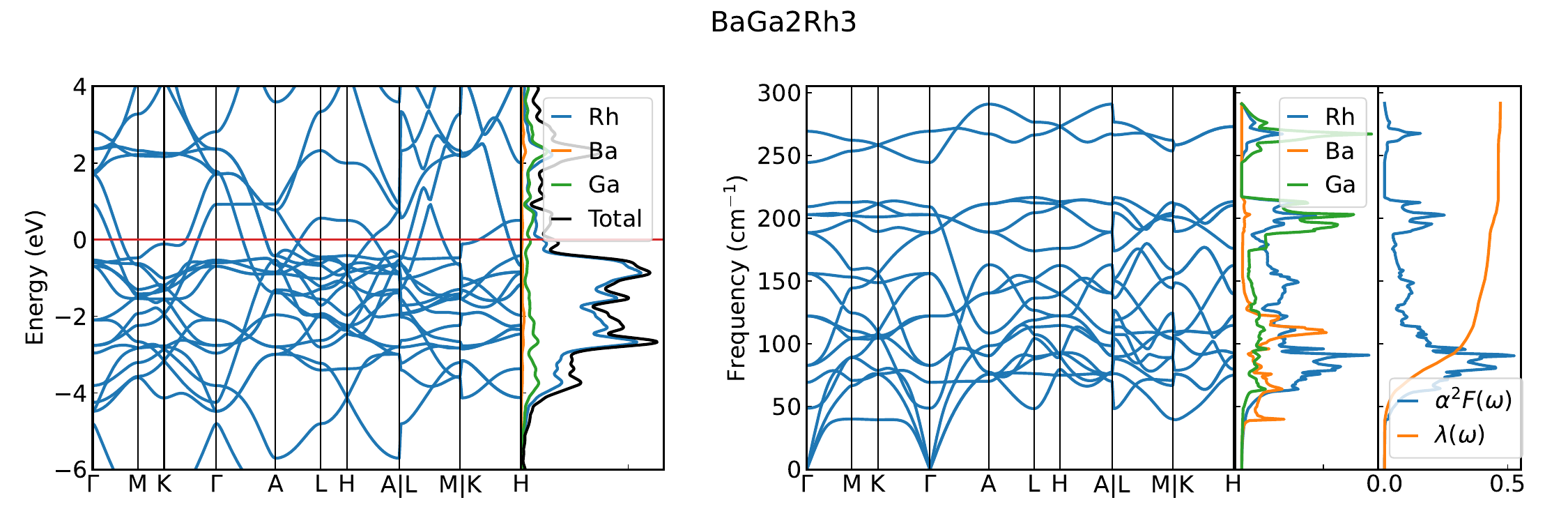}
    \includegraphics[width=0.75\linewidth]{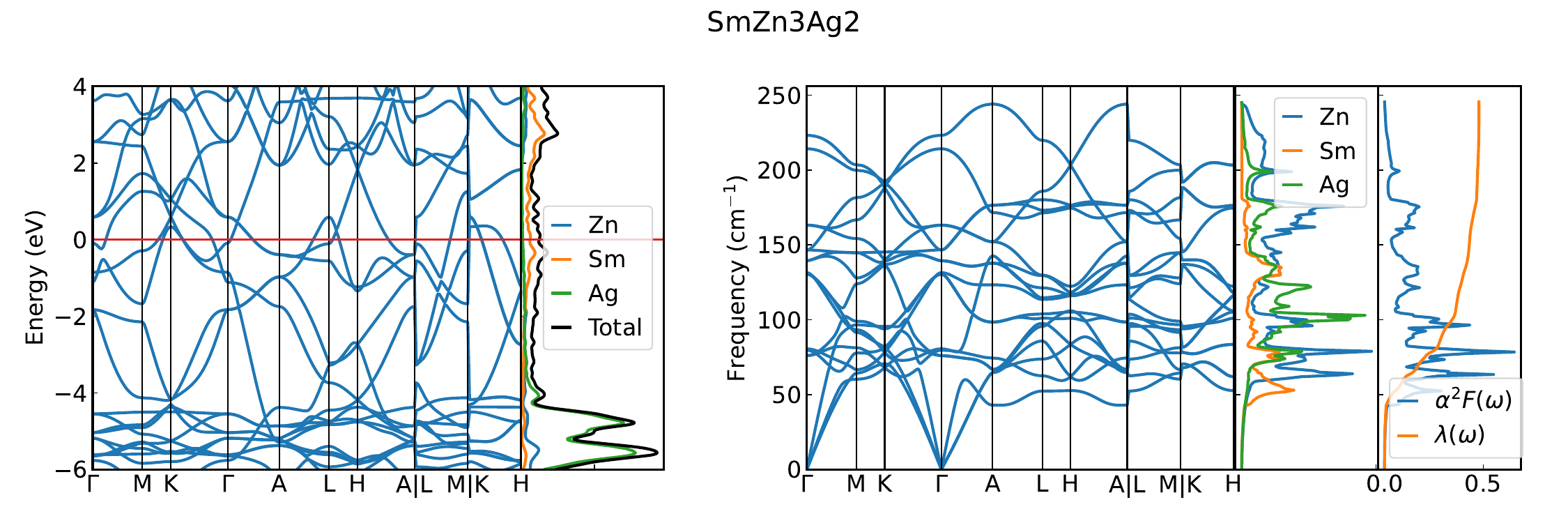}
\label{fig:type-i-12}
\end{figure}

\begin{figure}
    \centering
    \includegraphics[width=0.75\linewidth]{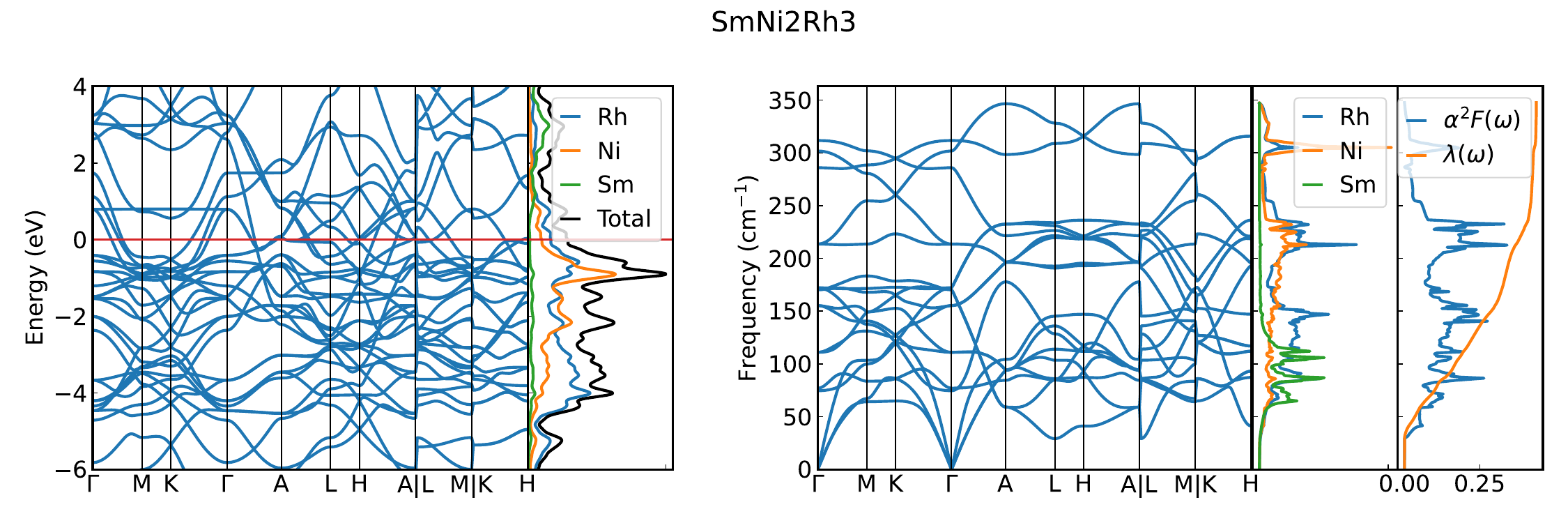}
    \includegraphics[width=0.75\linewidth]{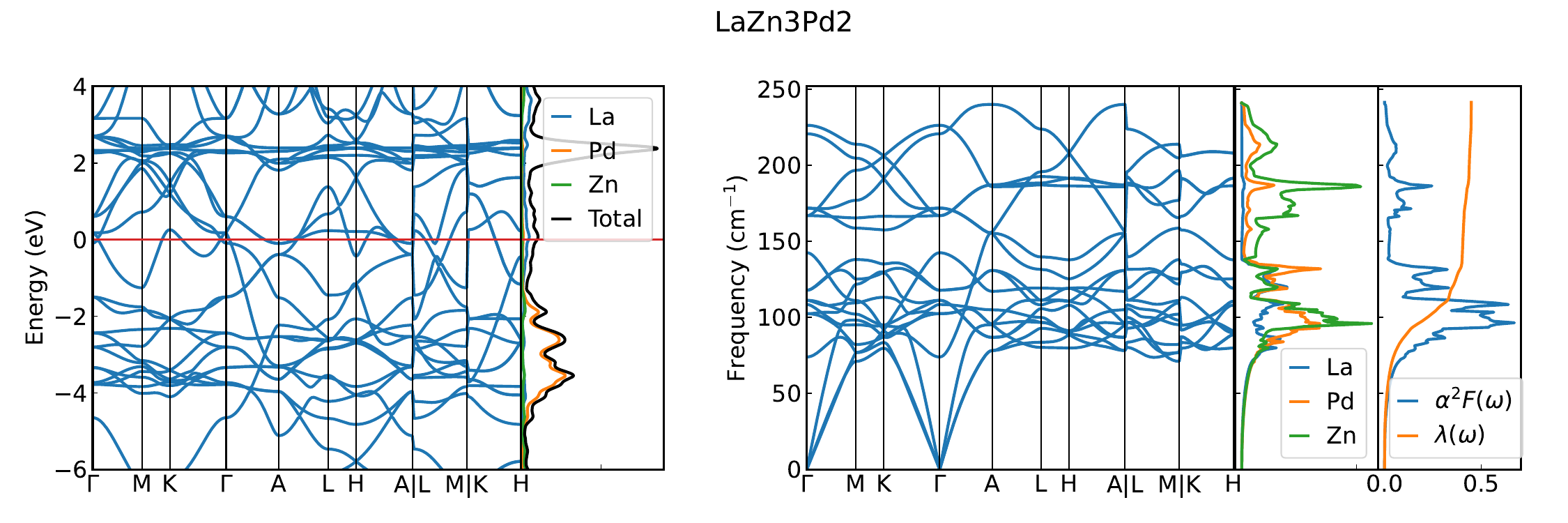}
    \includegraphics[width=0.75\linewidth]{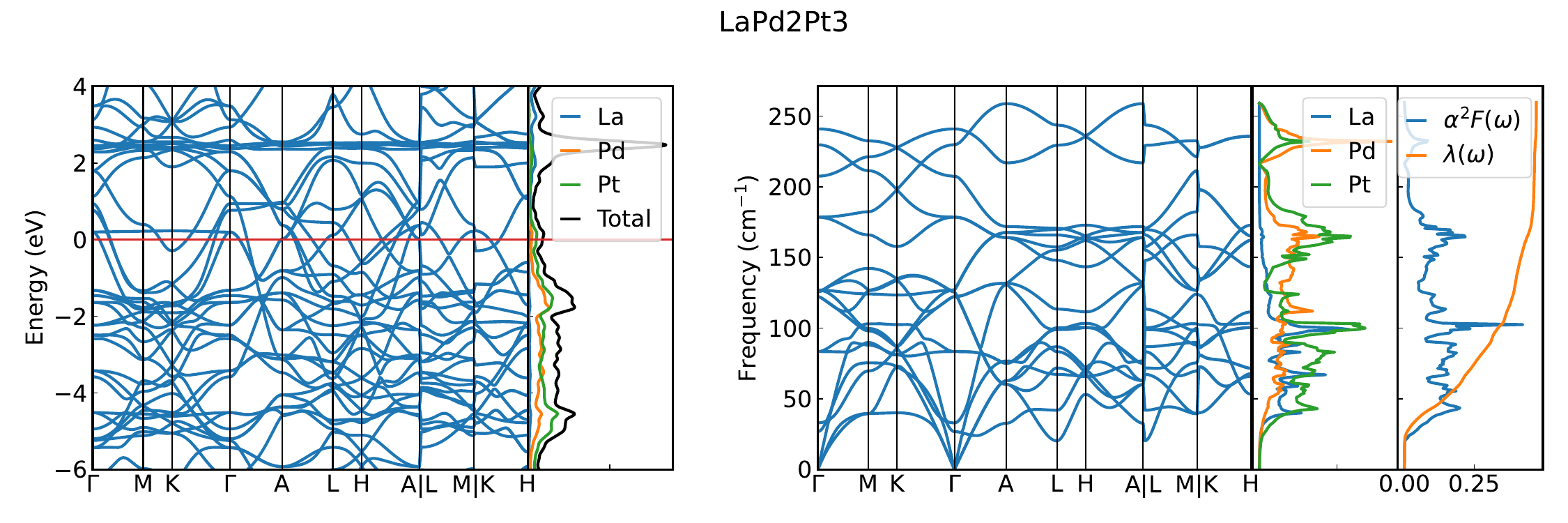}
    \includegraphics[width=0.75\linewidth]{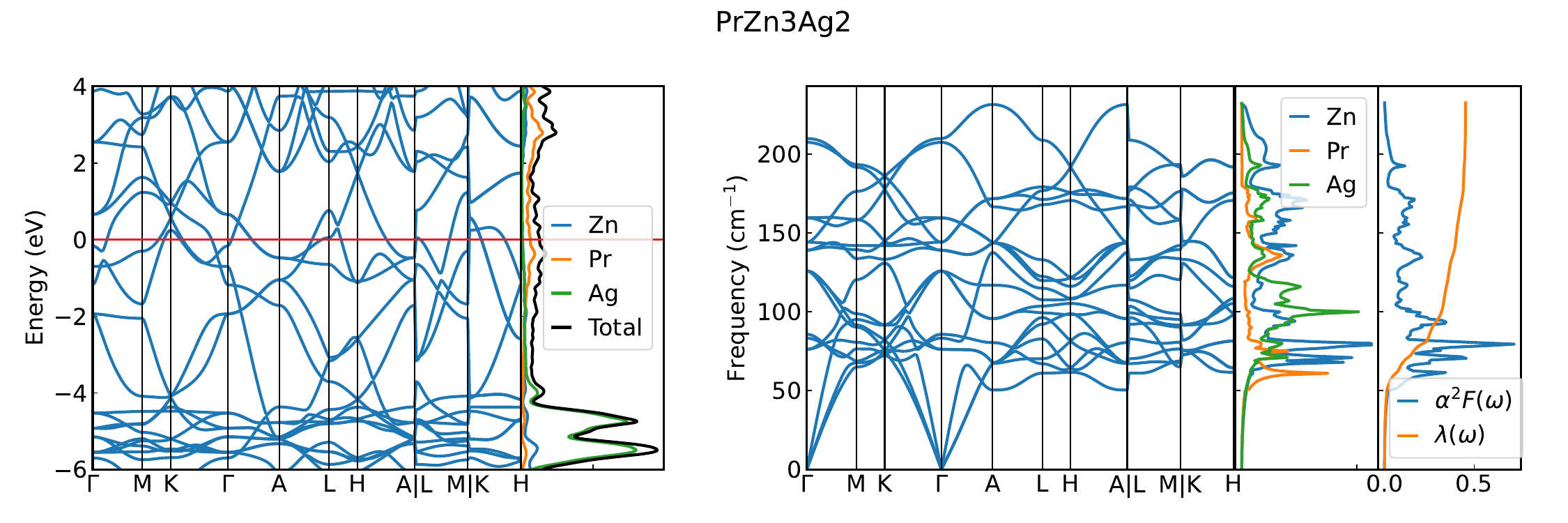}
    \includegraphics[width=0.75\linewidth]{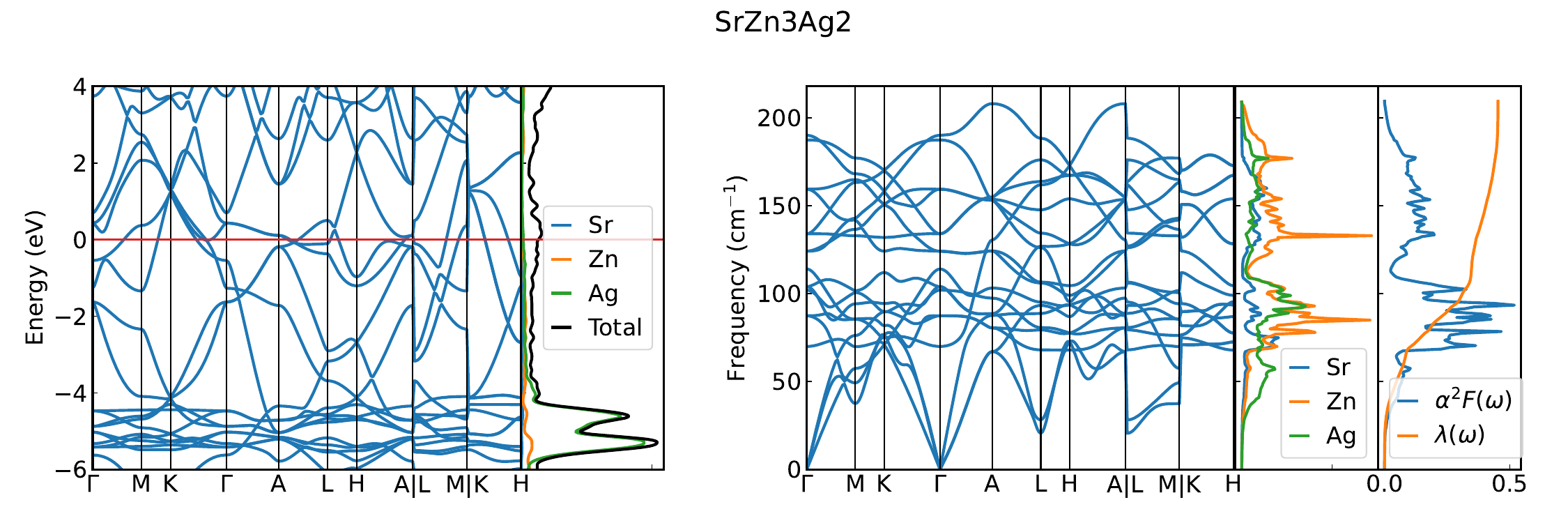}
\label{fig:type-i-13}
\end{figure}
 \begin{center}
\begin{longtable}{|c|c|c|c|c|c|c|c|}
\caption{Predicted compounds with $E_{\mathrm{hull}} < 30$ meV/atom in Type II.}
\label{tab:StablePredictionTypeII} \\
\hline \multicolumn{1}{|c|}{Formula} & \multicolumn{1}{|c|}{Material ID} & \multicolumn{1}{c|}{$E_{\mathrm{hull}}$ (meV/atom)} & \multicolumn{1}{c|}{$\omega_{\mathrm{log}}$ (K)} & \multicolumn{1}{c|}{$D(E_F)$} & \multicolumn{1}{c|}{$\lambda$} & \multicolumn{1}{c|}{$T_c$ (K)}  & \multicolumn{1}{c|}{FB @ $E_F$} \\ \hline
\endfirsthead
\multicolumn{8}{c}{{\bfseries \tablename\ \thetable{} -- continued from previous page}} \\
\hline \multicolumn{1}{|c|}{Formula} & \multicolumn{1}{|c|}{Material ID} & \multicolumn{1}{c|}{$E_{\mathrm{hull}}$ (meV/atom)} & \multicolumn{1}{c|}{$\omega_{\mathrm{log}}$ (K)} & \multicolumn{1}{c|}{$D(E_F)$} & \multicolumn{1}{c|}{$\lambda$} & \multicolumn{1}{c|}{$T_c$ (K)}  & \multicolumn{1}{c|}{FB @ $E_F$} \\ \hline
\endhead
\hline \multicolumn{8}{|r|}{{Continued on next page}} \\ \hline
\endfoot
\hline \hline
\endlastfoot
\ch{\green{K}\blue{Li}2\red{Sb}3} & agm071907720 & $  27$ & $ 102$ & $ 12.9$ & $  0.64$ & $  3.0$ & No \\
\hline
\ch{\green{Er}\blue{Ni}2\red{Be}3} & agm071708349 & $  27$ & $ 268$ & $ 21.8$ & $  0.48$ & $  2.8$ & Yes \\
\hline
\ch{\green{Tm}\blue{Ni}2\red{Be}3} & agm072045138 & $  22$ & $ 274$ & $ 21.8$ & $  0.47$ & $  2.7$ & Yes \\
\hline
\ch{\green{Cs}\blue{Au}2\red{Bi}3} & agm071689831 & $  27$ & $  68$ & $ 16.9$ & $  0.72$ & $  2.6$ & No \\
\hline
\ch{\green{Sc}\blue{Cu}2\red{Be}3} & agm071792821 & $   3$ & $ 365$ & $ 17.9$ & $  0.42$ & $  2.0$ & No \\
\hline
\ch{\green{Ti}\blue{Ni}2\red{Be}3} & agm071725295 & $  21$ & $ 335$ & $ 22.8$ & $  0.42$ & $  1.9$ & Yes \\
\hline
\ch{\green{Zr}\blue{Ni}2\red{Be}3} & agm072043899 & $   0$ & $ 334$ & $ 22.6$ & $  0.42$ & $  1.8$ & Yes \\
\hline
\ch{\green{Hf}\blue{Ni}2\red{Be}3} & agm071827366 & $   0$ & $ 321$ & $ 21.6$ & $  0.42$ & $  1.8$ & Yes \\
\hline
\ch{\green{Ba}\blue{Li}2\red{In}3} & agm038771285 & $  21$ & $ 111$ & $ 17.1$ & $  0.51$ & $  1.5$ & No \\
\hline
\ch{\green{La}\blue{Pd}2\red{Ga}3} & agm003202957 & $  22$ & $ 149$ & $ 17.0$ & $  0.47$ & $  1.4$ & No \\
\hline
\ch{\green{Ca}\blue{Pd}2\red{Al}3} & agm002133833 & $  20$ & $ 198$ & $ 19.6$ & $  0.44$ & $  1.4$ & No \\
\hline
\ch{\green{Sc}\blue{Ni}2\red{Be}3} & agm072095816 & $   0$ & $ 346$ & $ 23.3$ & $  0.39$ & $  1.4$ & Yes \\
\hline
\ch{\green{Rb}\blue{Mg}2\red{In}3} & agm035460572 & $  28$ & $  97$ & $ 16.2$ & $  0.50$ & $  1.3$ & No \\
\hline
\ch{\green{Ba}\blue{Li}2\red{Pb}3} & agm005704465 & $   0$ & $  81$ & $ 15.6$ & $  0.52$ & $  1.2$ & No \\
\hline
\ch{\green{Pr}\blue{Ni}2\red{Al}3} & agm003195351 & $  27$ & $ 195$ & $ 23.3$ & $  0.42$ & $  1.1$ & Yes \\
\hline
\ch{\green{Zr}\blue{Cu}2\red{Be}3} & agm071593344 & $  16$ & $ 319$ & $ 14.4$ & $  0.38$ & $  1.1$ & No \\
\hline
\ch{\green{Tm}\blue{Pd}2\red{Al}3} & agm002133979 & $   0$ & $ 174$ & $ 16.0$ & $  0.43$ & $  1.1$ & Yes \\
\hline
\ch{\green{Nd}\blue{Ni}2\red{Al}3} & agm003195352 & $  18$ & $ 195$ & $ 22.9$ & $  0.42$ & $  1.1$ & Yes \\
\hline
\ch{\green{Ba}\blue{Li}2\red{Al}3} & agm071675708 & $   8$ & $ 221$ & $ 16.8$ & $  0.41$ & $  1.1$ & No \\
\hline
\ch{\green{Hf}\blue{Cu}2\red{Be}3} & agm071848908 & $  28$ & $ 269$ & $ 13.5$ & $  0.39$ & $  1.0$ & No \\
\hline
\end{longtable}
\end{center}

\begin{figure}
    \centering
    \includegraphics[width=0.75\linewidth]{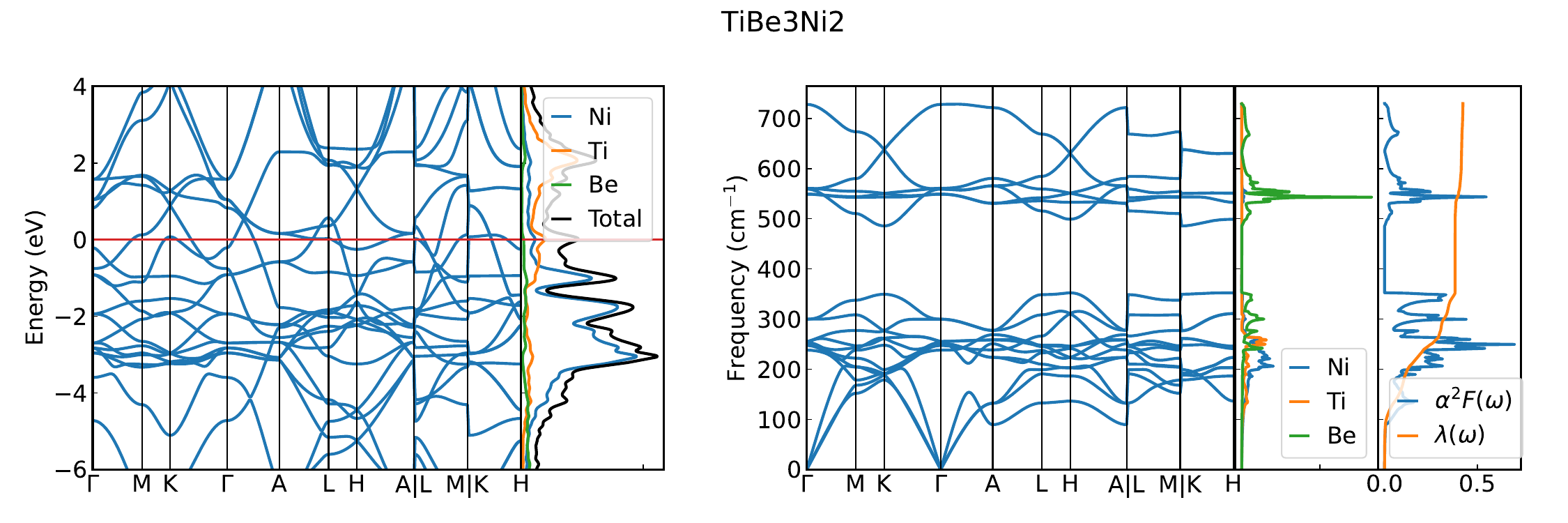}
    \includegraphics[width=0.75\linewidth]{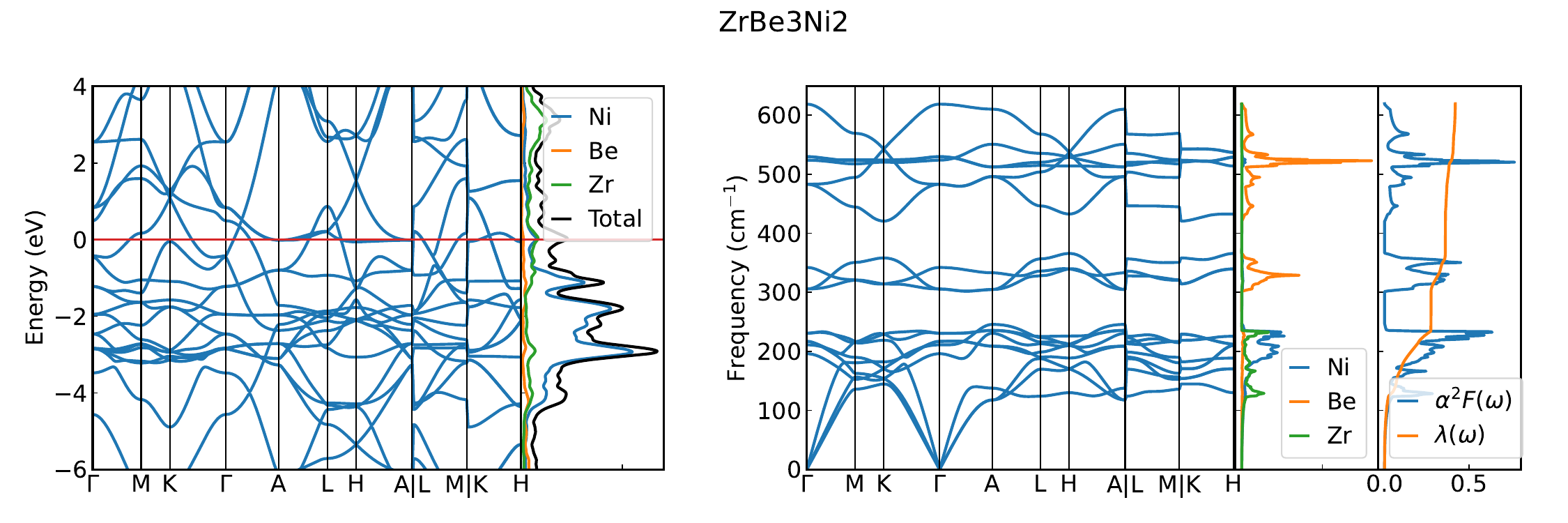}
    \includegraphics[width=0.75\linewidth]{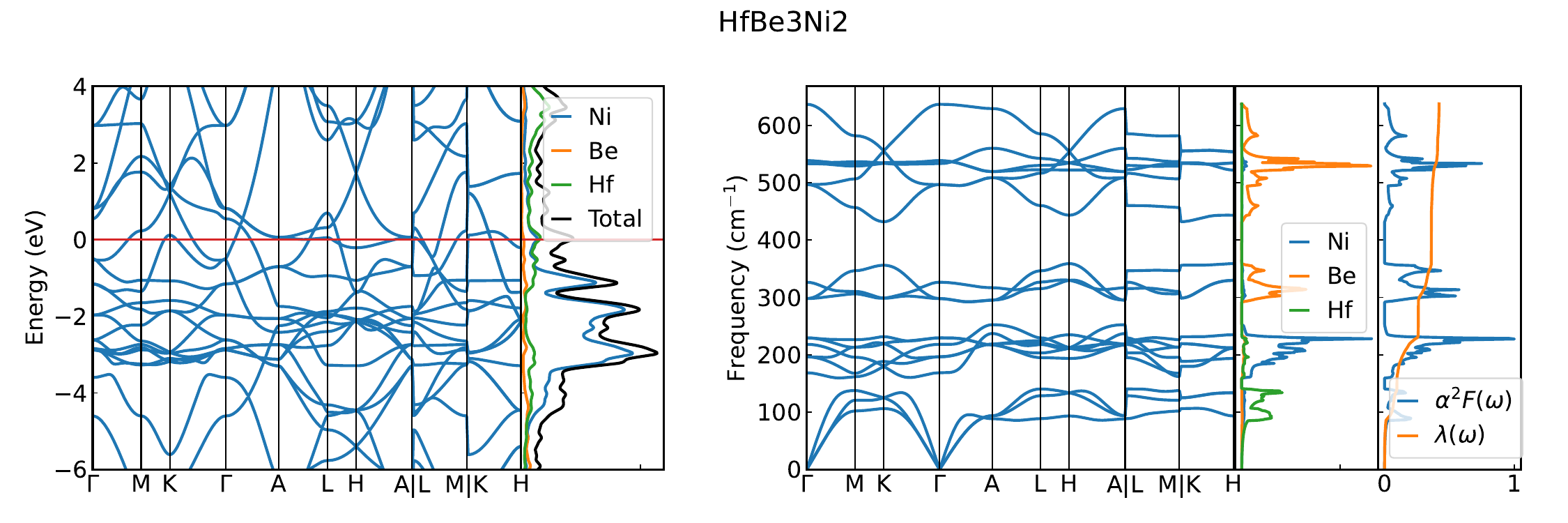}
    \includegraphics[width=0.75\linewidth]{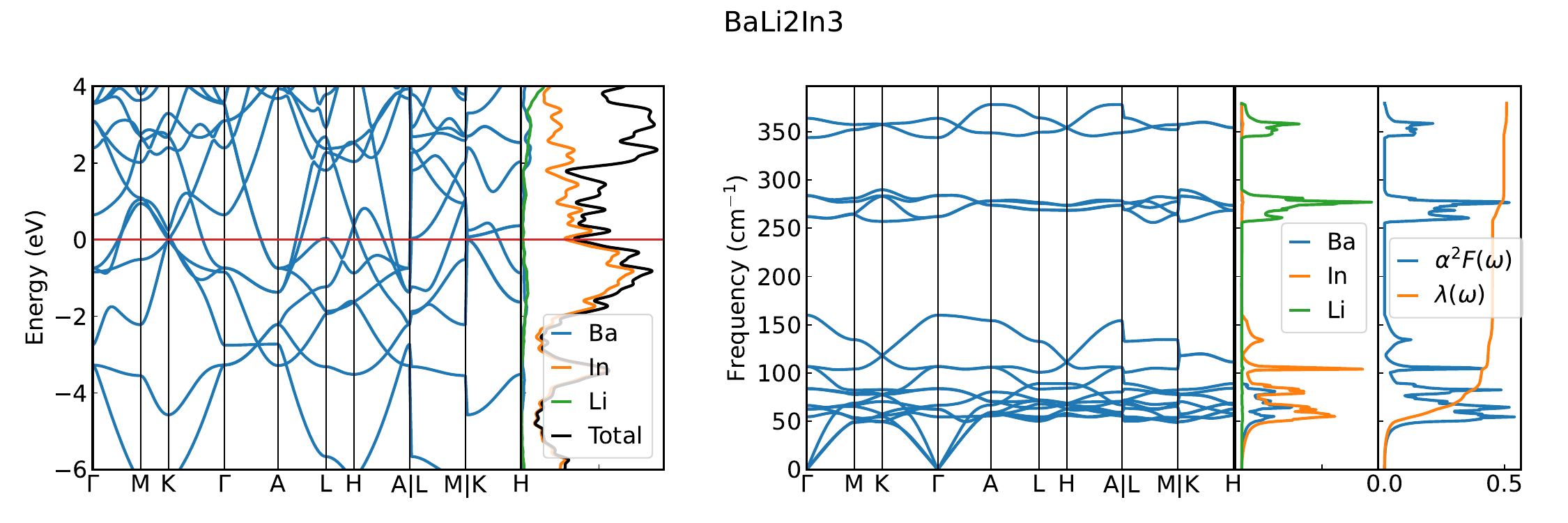}
    \includegraphics[width=0.75\linewidth]{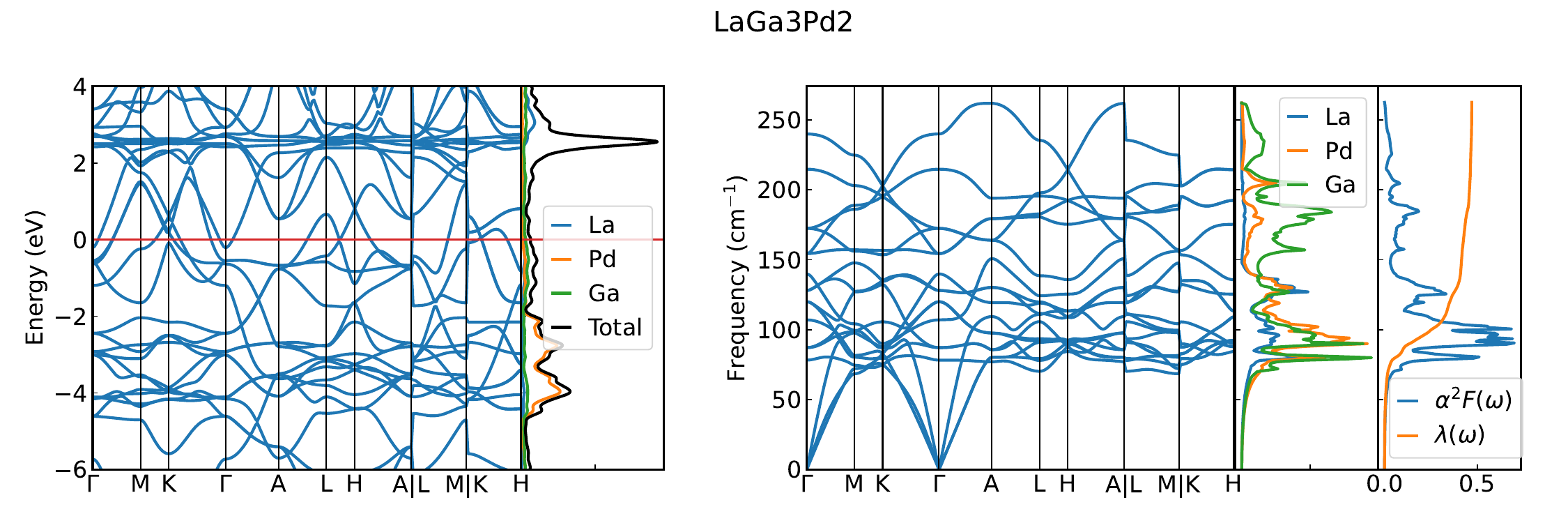}
\label{fig:type-ii-2}
\end{figure}

\begin{figure}
    \centering
    \includegraphics[width=0.75\linewidth]{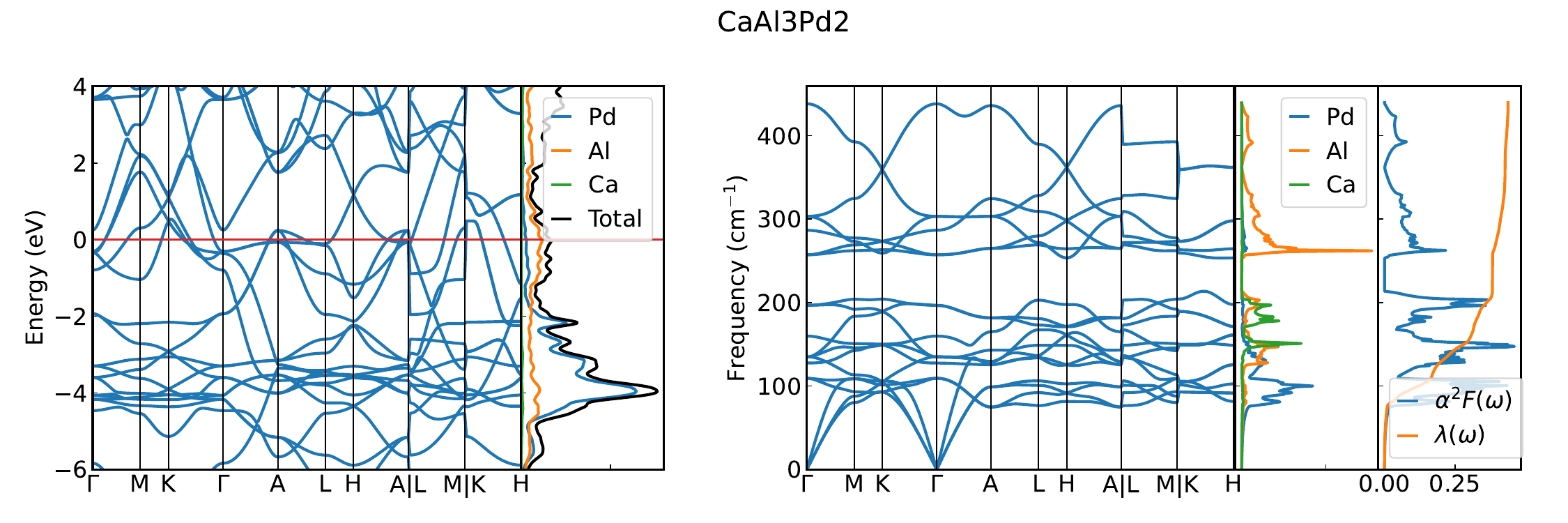}
    \includegraphics[width=0.75\linewidth]{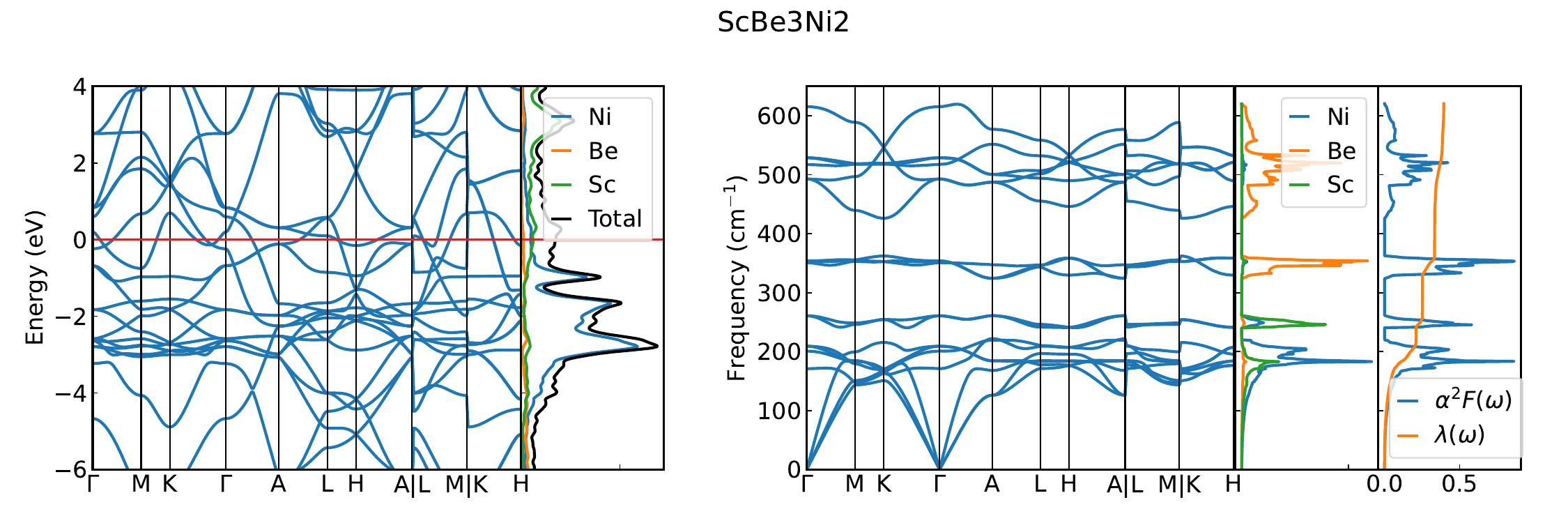}
    \includegraphics[width=0.75\linewidth]{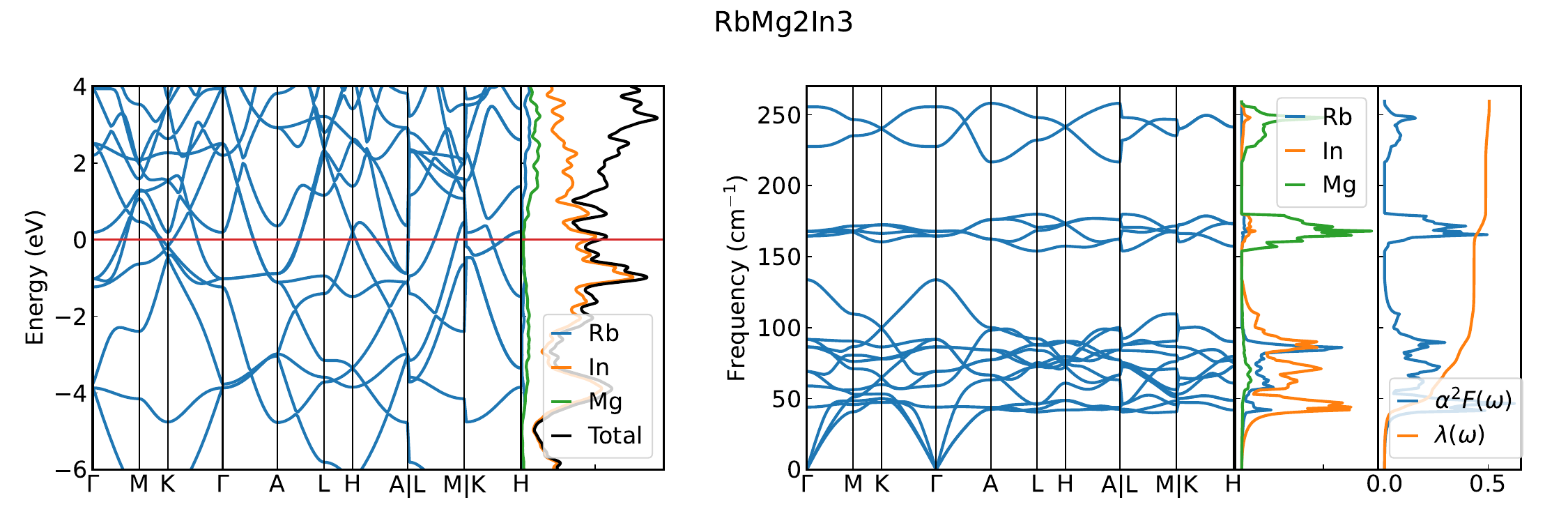}
    \includegraphics[width=0.75\linewidth]{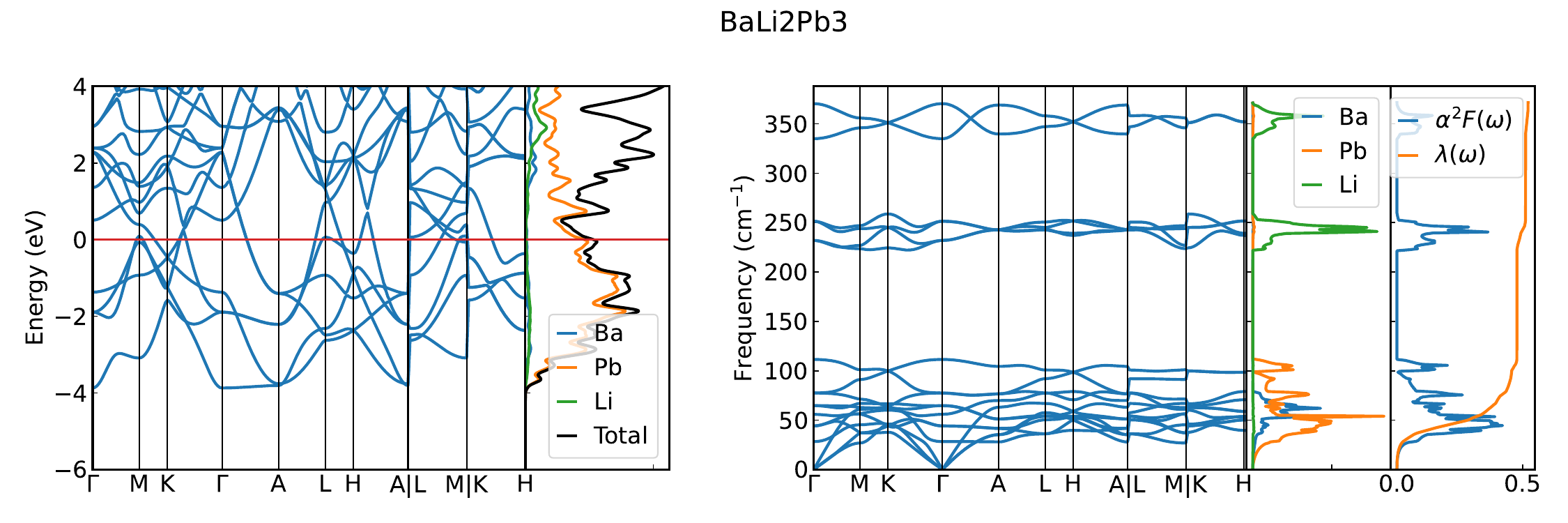}
    \includegraphics[width=0.75\linewidth]{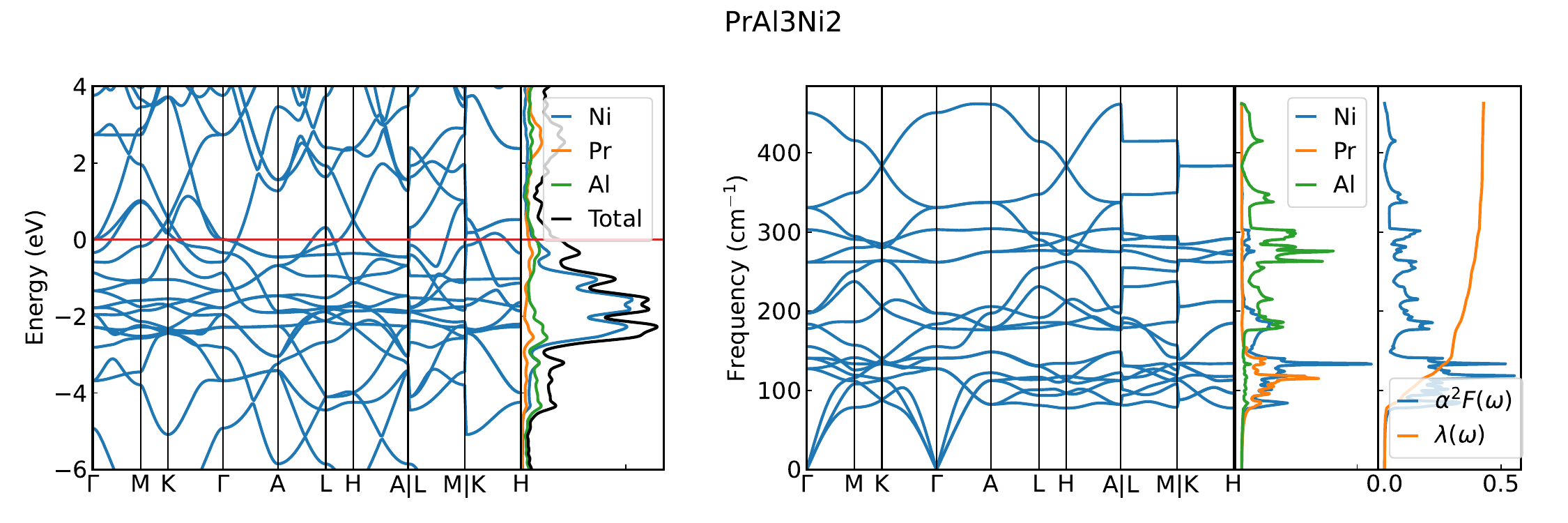}
\label{fig:type-ii-3}
\end{figure}

\begin{figure}
    \centering
    \includegraphics[width=0.75\linewidth]{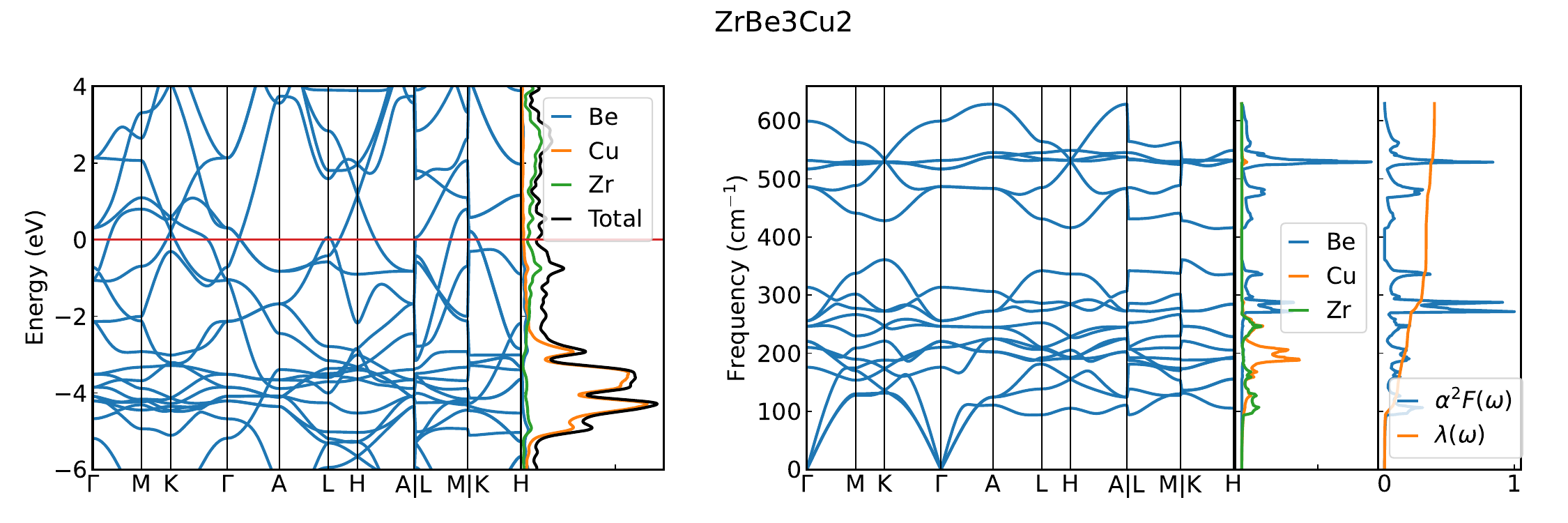}
    \includegraphics[width=0.75\linewidth]{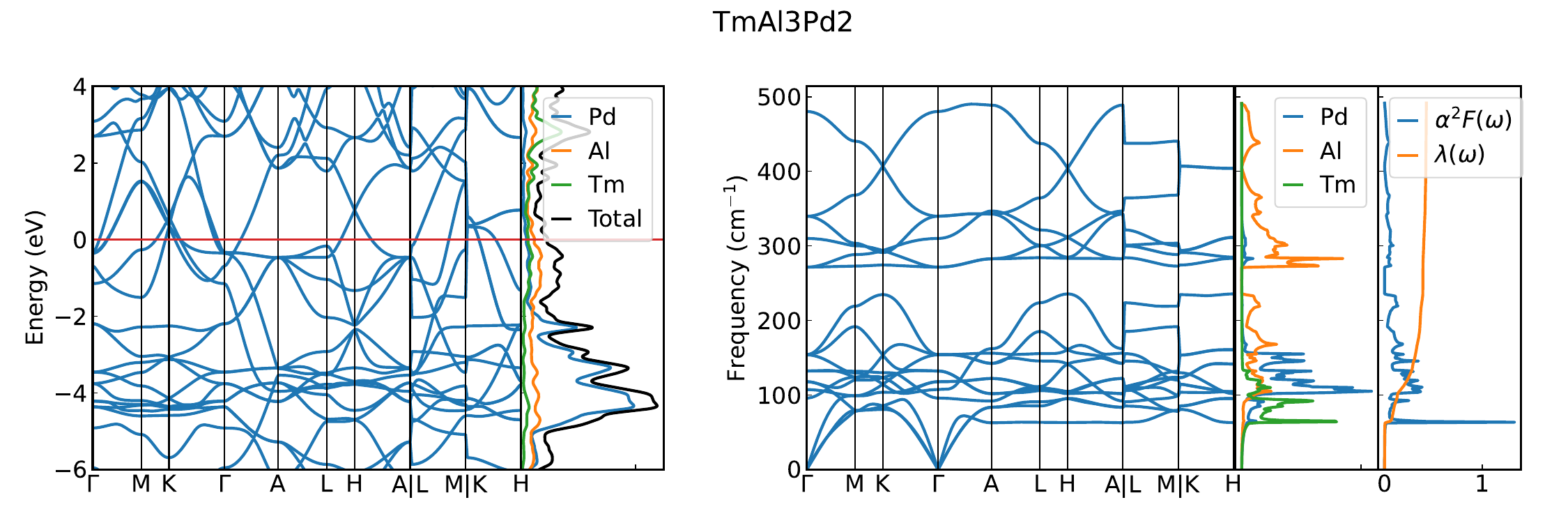}
    \includegraphics[width=0.75\linewidth]{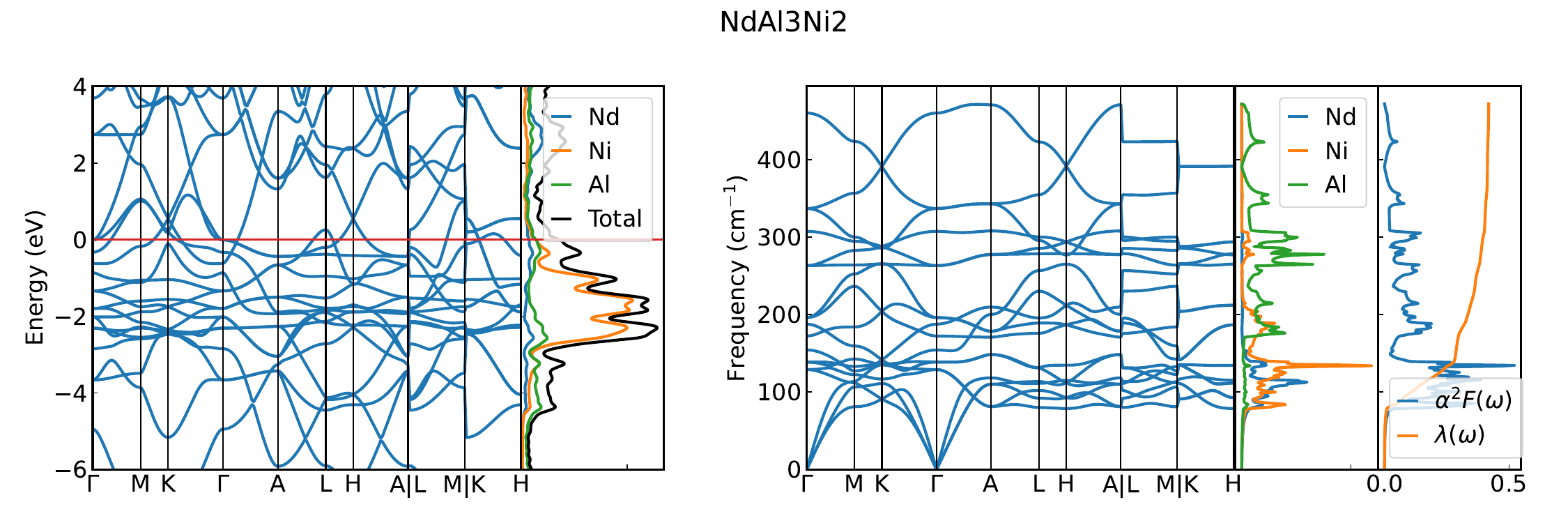}
    \includegraphics[width=0.75\linewidth]{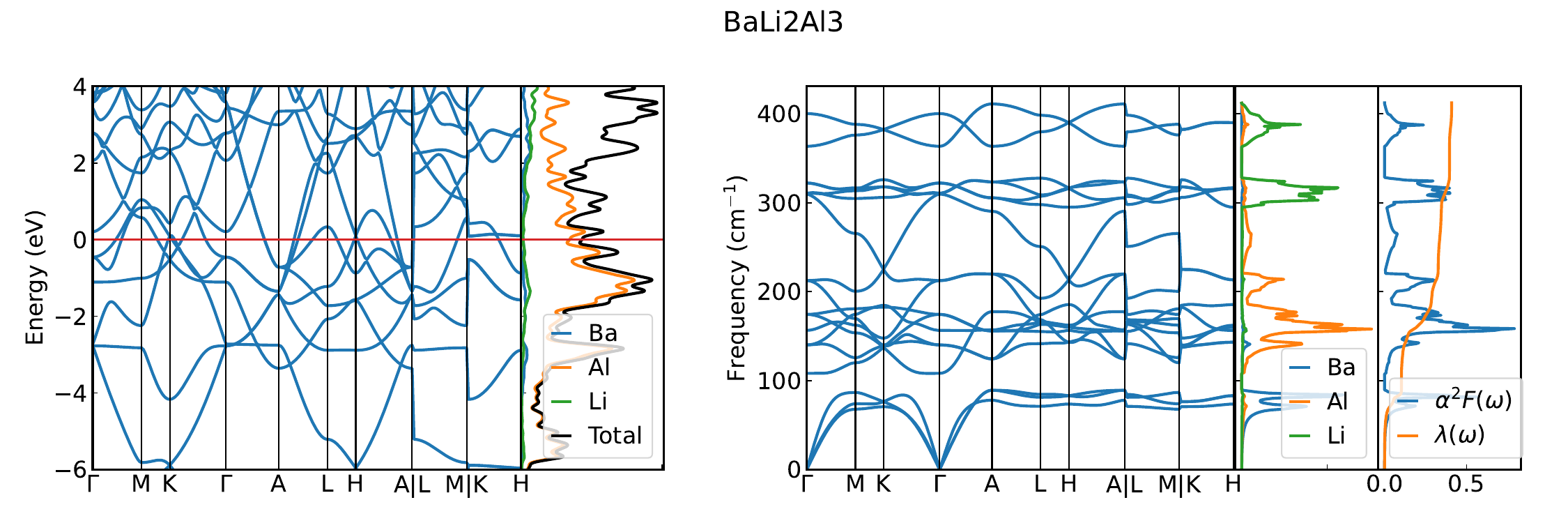}
    \includegraphics[width=0.75\linewidth]{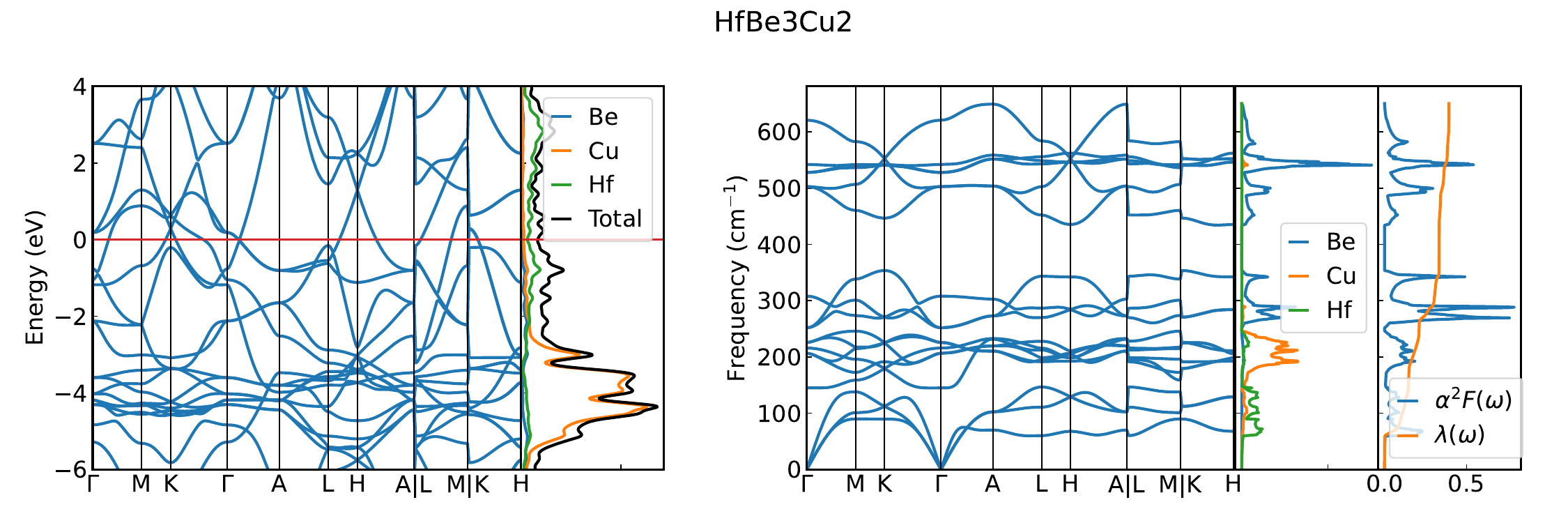}
\label{fig:type-ii-4}
\end{figure} 
\clearpage
\begin{center}
\begin{longtable}{|c|c|c|c|c|c|c|}
\caption{Predicted compounds with $E_{\mathrm{hull}} < 100$ meV/atom in Type I.}
\label{tab:PredictionTypeI} \\
\hline \multicolumn{1}{|c|}{Formula} & \multicolumn{1}{|c|}{Material ID} & \multicolumn{1}{c|}{$E_{\mathrm{hull}}$ (meV/atom)} & \multicolumn{1}{c|}{$\omega_{\mathrm{log}}$ (K)} & \multicolumn{1}{c|}{$D(E_F)$} & \multicolumn{1}{c|}{$\lambda$} & \multicolumn{1}{c|}{$T_c$ (K)} \\ \hline
\endfirsthead
\multicolumn{7}{c}{{\bfseries \tablename\ \thetable{} -- continued from previous page}} \\
\hline \multicolumn{1}{|c|}{Formula} & \multicolumn{1}{|c|}{Material ID} & \multicolumn{1}{c|}{$E_{\mathrm{hull}}$ (meV/atom)} & \multicolumn{1}{c|}{$\omega_{\mathrm{log}}$ (K)} & \multicolumn{1}{c|}{$D(E_F)$} & \multicolumn{1}{c|}{$\lambda$} & \multicolumn{1}{c|}{$T_c$ (K)} \\ \hline
\endhead
\hline \multicolumn{7}{|r|}{{Continued on next page}} \\ \hline
\endfoot
\hline \hline
\endlastfoot
\ch{\green{Tm}\blue{B}2\red{Ir}3} & agm002137253 & $   0$ & $  73$ & $ 13.3$ & $  1.45$ & $  9.2$ \\
\hline
\ch{\green{Ca}\blue{Ni}2\red{Ir}3} & agm002203799 & $  59$ & $  90$ & $ 41.8$ & $  1.18$ & $  8.7$ \\
\hline
\ch{\green{Er}\blue{B}2\red{Ir}3} & agm003202452 & $   0$ & $  85$ & $ 13.4$ & $  1.20$ & $  8.5$ \\
\hline
\ch{\green{Ce}\blue{B}2\red{Os}3} & agm002199385 & $  27$ & $ 131$ & $ 27.1$ & $  0.89$ & $  8.1$ \\
\hline
\ch{\green{Nd}\blue{Al}2\red{Ir}3} & agm002195286 & $  37$ & $  88$ & $ 22.5$ & $  1.13$ & $  7.9$ \\
\hline
\ch{\green{Pr}\blue{Al}2\red{Ir}3} & agm002195389 & $  25$ & $  89$ & $ 22.4$ & $  1.11$ & $  7.9$ \\
\hline
\ch{\green{Sr}\blue{Ni}2\red{Ir}3} & agm002226323 & $   0$ & $  94$ & $ 43.7$ & $  1.02$ & $  7.3$ \\
\hline
\ch{\green{Lu}\blue{B}2\red{Rh}3} & agm003195416 & $  21$ & $ 127$ & $ 14.6$ & $  0.85$ & $  7.2$ \\
\hline
\ch{\green{Sr}\blue{Cu}2\red{Ir}3} & agm002266899 & $  54$ & $  66$ & $ 39.0$ & $  1.29$ & $  7.2$ \\
\hline
\ch{\green{Ce}\blue{Al}2\red{Ir}3} & agm002133322 & $  87$ & $  94$ & $ 22.2$ & $  1.01$ & $  7.1$ \\
\hline
\ch{\green{La}\blue{Ga}2\red{Ir}3} & agm002274867 & $   0$ & $  80$ & $ 22.1$ & $  1.10$ & $  6.9$ \\
\hline
\ch{\green{Tb}\blue{B}2\red{Ru}3} & agm003202167 & $   0$ & $ 137$ & $ 26.7$ & $  0.81$ & $  6.9$ \\
\hline
\ch{\green{Y}\blue{B}2\red{Ru}3} & agm003202168 & $   0$ & $ 165$ & $ 26.6$ & $  0.74$ & $  6.8$ \\
\hline
\ch{\green{Sm}\blue{B}2\red{Ir}3} & agm002137232 & $   3$ & $ 112$ & $ 14.5$ & $  0.88$ & $  6.8$ \\
\hline
\ch{\green{Pr}\blue{Ga}2\red{Ir}3} & agm002274922 & $   0$ & $  90$ & $ 21.9$ & $  1.00$ & $  6.7$ \\
\hline
\ch{\green{Pm}\blue{B}2\red{Ru}3} & agm071840373 & $  12$ & $ 162$ & $ 28.2$ & $  0.74$ & $  6.7$ \\
\hline
\ch{\green{Pm}\blue{Pd}2\red{Ir}3} & agm032116159 & $  97$ & $  90$ & $ 32.8$ & $  0.99$ & $  6.6$ \\
\hline
\ch{\green{Sm}\blue{B}2\red{Ru}3} & agm003195414 & $   7$ & $ 157$ & $ 27.7$ & $  0.74$ & $  6.5$ \\
\hline
\ch{\green{Nd}\blue{B}2\red{Ru}3} & agm003195435 & $  21$ & $ 166$ & $ 28.8$ & $  0.71$ & $  6.3$ \\
\hline
\ch{\green{Tm}\blue{B}2\red{Rh}3} & agm002137195 & $  13$ & $ 144$ & $ 14.8$ & $  0.75$ & $  6.2$ \\
\hline
\ch{\green{La}\blue{B}2\red{Ru}3} & agm003195432 & $  23$ & $ 166$ & $ 28.8$ & $  0.71$ & $  6.2$ \\
\hline
\ch{\green{Pr}\blue{B}2\red{Ru}3} & agm003195427 & $  30$ & $ 171$ & $ 29.3$ & $  0.70$ & $  6.2$ \\
\hline
\ch{\green{Ce}\blue{B}2\red{Ru}3} & agm003202683 & $   0$ & $ 178$ & $ 29.8$ & $  0.67$ & $  5.9$ \\
\hline
\ch{\green{Nd}\blue{Pd}2\red{Ir}3} & agm002307003 & $  86$ & $  99$ & $ 32.9$ & $  0.88$ & $  5.9$ \\
\hline
\ch{\green{Ca}\blue{Ga}2\red{Rh}3} & agm002260176 & $   0$ & $ 117$ & $ 23.9$ & $  0.80$ & $  5.8$ \\
\hline
\ch{\green{La}\blue{B}2\red{Rh}3} & agm003202760 & $  17$ & $ 180$ & $ 18.1$ & $  0.66$ & $  5.7$ \\
\hline
\ch{\green{Rb}\blue{Ge}2\red{Pt}3} & agm034485715 & $  61$ & $  81$ & $ 18.0$ & $  0.96$ & $  5.6$ \\
\hline
\ch{\green{Er}\blue{B}2\red{Rh}3} & agm003195426 & $   8$ & $ 152$ & $ 14.9$ & $  0.70$ & $  5.5$ \\
\hline
\ch{\green{Pr}\blue{Be}2\red{Os}3} & agm033617485 & $  93$ & $ 146$ & $ 32.6$ & $  0.71$ & $  5.5$ \\
\hline
\ch{\green{Nd}\blue{Be}2\red{Os}3} & agm041486382 & $  96$ & $ 149$ & $ 32.3$ & $  0.70$ & $  5.4$ \\
\hline
\ch{\green{Pr}\blue{Pd}2\red{Ir}3} & agm002307010 & $  73$ & $ 106$ & $ 33.0$ & $  0.80$ & $  5.2$ \\
\hline
\ch{\green{Ho}\blue{B}2\red{Rh}3} & agm003195417 & $   5$ & $ 158$ & $ 15.1$ & $  0.67$ & $  5.1$ \\
\hline
\ch{\green{La}\blue{B}2\red{Ir}3} & agm003195415 & $  23$ & $ 130$ & $ 14.9$ & $  0.72$ & $  5.1$ \\
\hline
\ch{\green{La}\blue{Pd}2\red{Ir}3} & agm002234136 & $  71$ & $ 106$ & $ 33.6$ & $  0.79$ & $  5.1$ \\
\hline
\ch{\green{Dy}\blue{B}2\red{Rh}3} & agm003195430 & $   2$ & $ 162$ & $ 15.4$ & $  0.66$ & $  5.0$ \\
\hline
\ch{\green{Na}\blue{Ge}2\red{Pt}3} & agm031905555 & $  31$ & $  70$ & $ 15.1$ & $  0.96$ & $  4.9$ \\
\hline
\ch{\green{Tb}\blue{B}2\red{Rh}3} & agm002137193 & $   1$ & $ 167$ & $ 15.7$ & $  0.64$ & $  4.7$ \\
\hline
\ch{\green{Tb}\blue{Rh}2\red{Cu}3} & agm002267053 & $  83$ & $ 103$ & $ 34.8$ & $  0.77$ & $  4.6$ \\
\hline
\ch{\green{Y}\blue{B}2\red{Rh}3} & agm003202164 & $   0$ & $ 175$ & $ 15.5$ & $  0.62$ & $  4.6$ \\
\hline
\ch{\green{Sm}\blue{Au}2\red{Zn}3} & agm002324461 & $   0$ & $  90$ & $ 17.4$ & $  0.80$ & $  4.4$ \\
\hline
\ch{\green{Dy}\blue{Rh}2\red{Cu}3} & agm002267047 & $  84$ & $ 105$ & $ 34.6$ & $  0.74$ & $  4.4$ \\
\hline
\ch{\green{Cs}\blue{Mg}2\red{Hg}3} & agm072076899 & $  69$ & $  44$ & $ 21.5$ & $  1.20$ & $  4.3$ \\
\hline
\ch{\green{Pr}\blue{Rh}2\red{Cu}3} & agm002267051 & $  76$ & $ 109$ & $ 35.3$ & $  0.72$ & $  4.3$ \\
\hline
\ch{\green{Sr}\blue{Al}2\red{Rh}3} & agm002195511 & $  67$ & $ 123$ & $ 21.4$ & $  0.69$ & $  4.3$ \\
\hline
\ch{\green{Ce}\blue{Ru}2\red{Cu}3} & agm031763759 & $  63$ & $ 130$ & $ 36.2$ & $  0.67$ & $  4.2$ \\
\hline
\ch{\green{Ho}\blue{Rh}2\red{Cu}3} & agm002267049 & $  85$ & $ 106$ & $ 34.5$ & $  0.73$ & $  4.2$ \\
\hline
\ch{\green{Ba}\blue{Ga}2\red{Au}3} & agm002337523 & $  62$ & $  70$ & $ 17.7$ & $  0.88$ & $  4.2$ \\
\hline
\ch{\green{K}\blue{Ga}2\red{Hg}3} & agm071870928 & $  87$ & $  61$ & $ 15.4$ & $  0.95$ & $  4.1$ \\
\hline
\ch{\green{Pr}\blue{Ru}2\red{Zn}3} & agm002324454 & $  55$ & $ 127$ & $ 33.6$ & $  0.67$ & $  4.1$ \\
\hline
\ch{\green{Nd}\blue{Ru}2\red{Zn}3} & agm002324453 & $  46$ & $ 127$ & $ 33.3$ & $  0.67$ & $  4.0$ \\
\hline
\ch{\green{Ba}\blue{Ni}2\red{Rh}3} & agm002296558 & $  62$ & $ 141$ & $ 55.3$ & $  0.64$ & $  4.0$ \\
\hline
\ch{\green{Tb}\blue{Au}2\red{Zn}3} & agm002324476 & $   0$ & $  90$ & $ 16.4$ & $  0.77$ & $  4.0$ \\
\hline
\ch{\green{Tm}\blue{Rh}2\red{Cu}3} & agm002267055 & $  85$ & $ 106$ & $ 34.2$ & $  0.71$ & $  4.0$ \\
\hline
\ch{\green{Sm}\blue{Pt}2\red{Ir}3} & agm002319875 & $  70$ & $ 116$ & $ 31.1$ & $  0.69$ & $  4.0$ \\
\hline
\ch{\green{La}\blue{Au}2\red{Zn}3} & agm002246263 & $   0$ & $ 103$ & $ 20.4$ & $  0.71$ & $  3.9$ \\
\hline
\ch{\green{Pm}\blue{Ru}2\red{Zn}3} & agm071694789 & $  39$ & $ 126$ & $ 33.0$ & $  0.66$ & $  3.9$ \\
\hline
\ch{\green{Sm}\blue{Ru}2\red{Zn}3} & agm002324455 & $  37$ & $ 125$ & $ 32.8$ & $  0.66$ & $  3.9$ \\
\hline
\ch{\green{K}\blue{Ge}2\red{Pt}3} & agm038799382 & $  25$ & $  93$ & $ 17.0$ & $  0.74$ & $  3.8$ \\
\hline
\ch{\green{Ce}\blue{Ni}2\red{Os}3} & agm002296416 & $  87$ & $ 153$ & $ 37.8$ & $  0.61$ & $  3.8$ \\
\hline
\ch{\green{Ba}\blue{Li}2\red{Zn}3} & agm072080557 & $  59$ & $ 126$ & $ 23.4$ & $  0.65$ & $  3.7$ \\
\hline
\ch{\green{Tm}\blue{Ni}2\red{Ir}3} & agm002227683 & $  34$ & $ 118$ & $ 34.0$ & $  0.66$ & $  3.7$ \\
\hline
\ch{\green{Pm}\blue{Pt}2\red{Ir}3} & agm034807046 & $  59$ & $ 120$ & $ 31.1$ & $  0.65$ & $  3.6$ \\
\hline
\ch{\green{Ba}\blue{Hg}2\red{Au}3} & agm035054481 & $  69$ & $  48$ & $ 17.1$ & $  1.00$ & $  3.6$ \\
\hline
\ch{\green{Y}\blue{Ru}2\red{Zn}3} & agm002324493 & $  36$ & $ 130$ & $ 32.6$ & $  0.63$ & $  3.6$ \\
\hline
\ch{\green{Ca}\blue{Cu}2\red{Rh}3} & agm002260829 & $  96$ & $ 134$ & $ 44.3$ & $  0.62$ & $  3.6$ \\
\hline
\ch{\green{Ho}\blue{Li}2\red{Zn}3} & agm071727858 & $  96$ & $ 117$ & $ 21.3$ & $  0.65$ & $  3.6$ \\
\hline
\ch{\green{Y}\blue{Li}2\red{Zn}3} & agm071921673 & $  86$ & $ 136$ & $ 21.8$ & $  0.62$ & $  3.5$ \\
\hline
\ch{\green{Tb}\blue{Ru}2\red{Zn}3} & agm002324456 & $  36$ & $ 125$ & $ 32.2$ & $  0.64$ & $  3.5$ \\
\hline
\ch{\green{Pr}\blue{Ni}2\red{Os}3} & agm002226176 & $  96$ & $ 159$ & $ 37.1$ & $  0.58$ & $  3.4$ \\
\hline
\ch{\green{Dy}\blue{Ru}2\red{Zn}3} & agm002360069 & $  39$ & $ 125$ & $ 31.9$ & $  0.63$ & $  3.4$ \\
\hline
\ch{\green{Dy}\blue{Li}2\red{Zn}3} & agm071883793 & $  89$ & $ 122$ & $ 21.2$ & $  0.63$ & $  3.3$ \\
\hline
\ch{\green{La}\blue{Al}2\red{Ag}3} & agm002194988 & $  72$ & $  96$ & $ 23.8$ & $  0.68$ & $  3.3$ \\
\hline
\ch{\green{Nd}\blue{Pt}2\red{Ir}3} & agm002294665 & $  60$ & $ 124$ & $ 31.2$ & $  0.62$ & $  3.3$ \\
\hline
\ch{\green{Er}\blue{Ni}2\red{Ir}3} & agm002225978 & $  25$ & $ 127$ & $ 34.1$ & $  0.62$ & $  3.2$ \\
\hline
\ch{\green{Ba}\blue{Al}2\red{Ag}3} & agm002194986 & $  79$ & $  91$ & $ 21.2$ & $  0.69$ & $  3.2$ \\
\hline
\ch{\green{Ho}\blue{Ru}2\red{Zn}3} & agm002360071 & $  44$ & $ 125$ & $ 31.6$ & $  0.62$ & $  3.2$ \\
\hline
\ch{\green{Nd}\blue{Ni}2\red{Os}3} & agm002226149 & $  90$ & $ 162$ & $ 36.6$ & $  0.57$ & $  3.2$ \\
\hline
\ch{\green{Ho}\blue{Ni}2\red{Ir}3} & agm002226083 & $  18$ & $ 131$ & $ 34.1$ & $  0.61$ & $  3.2$ \\
\hline
\ch{\green{Sm}\blue{Ni}2\red{Os}3} & agm002226251 & $  99$ & $ 165$ & $ 35.9$ & $  0.56$ & $  3.1$ \\
\hline
\ch{\green{Sr}\blue{Ga}2\red{Rh}3} & agm002275043 & $   0$ & $ 132$ & $ 22.5$ & $  0.60$ & $  3.1$ \\
\hline
\ch{\green{Rb}\blue{In}2\red{Hg}3} & agm072085691 & $  58$ & $  54$ & $ 17.2$ & $  0.86$ & $  3.1$ \\
\hline
\ch{\green{Er}\blue{Ru}2\red{Zn}3} & agm002360070 & $  48$ & $ 125$ & $ 31.3$ & $  0.61$ & $  3.0$ \\
\hline
\ch{\green{Pr}\blue{Pt}2\red{Ir}3} & agm002308019 & $  56$ & $ 129$ & $ 31.3$ & $  0.60$ & $  3.0$ \\
\hline
\ch{\green{Tb}\blue{Li}2\red{Zn}3} & agm071791439 & $  82$ & $ 130$ & $ 21.1$ & $  0.60$ & $  3.0$ \\
\hline
\ch{\green{Tm}\blue{Ru}2\red{Zn}3} & agm033727218 & $  52$ & $ 125$ & $ 31.0$ & $  0.60$ & $  2.9$ \\
\hline
\ch{\green{La}\blue{Pt}2\red{Ir}3} & agm002218184 & $  38$ & $ 128$ & $ 32.4$ & $  0.59$ & $  2.9$ \\
\hline
\ch{\green{Dy}\blue{Ni}2\red{Ir}3} & agm002225968 & $  12$ & $ 136$ & $ 34.2$ & $  0.58$ & $  2.8$ \\
\hline
\ch{\green{Ba}\blue{Tl}2\red{Ag}3} & agm043468796 & $  86$ & $  60$ & $ 20.1$ & $  0.79$ & $  2.8$ \\
\hline
\ch{\green{Rb}\blue{Al}2\red{Ag}3} & agm071885861 & $  99$ & $  83$ & $ 14.7$ & $  0.68$ & $  2.8$ \\
\hline
\ch{\green{K}\blue{Ga}2\red{Au}3} & agm031668111 & $  54$ & $  77$ & $ 12.5$ & $  0.70$ & $  2.8$ \\
\hline
\ch{\green{Sm}\blue{Pd}2\red{Rh}3} & agm002235382 & $  85$ & $ 145$ & $ 37.7$ & $  0.56$ & $  2.7$ \\
\hline
\ch{\green{Pb}\blue{Ga}2\red{Au}3} & agm039509009 & $  57$ & $  51$ & $ 10.4$ & $  0.82$ & $  2.7$ \\
\hline
\ch{\green{Tb}\blue{Ni}2\red{Ir}3} & agm002226335 & $   5$ & $ 140$ & $ 34.4$ & $  0.56$ & $  2.7$ \\
\hline
\ch{\green{Pb}\blue{Al}2\red{Au}3} & agm034909433 & $  73$ & $  67$ & $ 11.3$ & $  0.72$ & $  2.6$ \\
\hline
\ch{\green{Sm}\blue{V}2\red{Co}3} & agm071991445 & $  77$ & $ 234$ & $ 40.2$ & $  0.49$ & $  2.6$ \\
\hline
\ch{\green{La}\blue{In}2\red{Ag}3} & agm002132859 & $  98$ & $  79$ & $ 25.3$ & $  0.68$ & $  2.6$ \\
\hline
\ch{\green{Ce}\blue{Pt}2\red{Ir}3} & agm002261886 & $  46$ & $ 131$ & $ 31.6$ & $  0.57$ & $  2.6$ \\
\hline
\ch{\green{Ba}\blue{Cu}2\red{Rh}3} & agm002371184 & $  51$ & $ 132$ & $ 46.4$ & $  0.57$ & $  2.6$ \\
\hline
\ch{\green{Y}\blue{Ni}2\red{Ir}3} & agm002226370 & $  11$ & $ 143$ & $ 34.5$ & $  0.55$ & $  2.6$ \\
\hline
\ch{\green{Sm}\blue{Si}2\red{Ni}3} & agm002316118 & $  63$ & $ 172$ & $ 30.6$ & $  0.52$ & $  2.6$ \\
\hline
\ch{\green{Pm}\blue{Au}2\red{Zn}3} & agm031465753 & $   0$ & $ 110$ & $ 17.7$ & $  0.60$ & $  2.6$ \\
\hline
\ch{\green{Pm}\blue{Ga}2\red{Cu}3} & agm033452294 & $  66$ & $ 123$ & $ 21.3$ & $  0.58$ & $  2.5$ \\
\hline
\ch{\green{Pm}\blue{Pd}2\red{Rh}3} & agm041108863 & $  73$ & $ 149$ & $ 37.7$ & $  0.54$ & $  2.5$ \\
\hline
\ch{\green{Pr}\blue{Au}2\red{Zn}3} & agm002324445 & $   0$ & $ 114$ & $ 18.3$ & $  0.59$ & $  2.5$ \\
\hline
\ch{\green{K}\blue{Hg}2\red{Au}3} & agm035388847 & $  73$ & $  60$ & $ 15.8$ & $  0.74$ & $  2.5$ \\
\hline
\ch{\green{Na}\blue{Zn}2\red{Au}3} & agm043268940 & $  43$ & $  74$ & $ 15.1$ & $  0.68$ & $  2.5$ \\
\hline
\ch{\green{Nd}\blue{Li}2\red{Cd}3} & agm071982672 & $  60$ & $  96$ & $ 24.5$ & $  0.62$ & $  2.5$ \\
\hline
\ch{\green{Dy}\blue{Pt}2\red{Rh}3} & agm044065542 & $  77$ & $ 143$ & $ 35.0$ & $  0.54$ & $  2.5$ \\
\hline
\ch{\green{Ba}\blue{Ag}2\red{Zn}3} & agm002384834 & $  34$ & $ 112$ & $ 19.5$ & $  0.59$ & $  2.4$ \\
\hline
\ch{\green{K}\blue{In}2\red{Au}3} & agm040038476 & $  27$ & $  54$ & $ 12.5$ & $  0.76$ & $  2.3$ \\
\hline
\ch{\green{Nd}\blue{V}2\red{Os}3} & agm071889712 & $  85$ & $ 190$ & $ 26.3$ & $  0.50$ & $  2.3$ \\
\hline
\ch{\green{Nd}\blue{Au}2\red{Zn}3} & agm002324425 & $   0$ & $ 114$ & $ 18.0$ & $  0.57$ & $  2.3$ \\
\hline
\ch{\green{Sr}\blue{Li}2\red{Zn}3} & agm071926568 & $  35$ & $ 152$ & $ 21.9$ & $  0.52$ & $  2.3$ \\
\hline
\ch{\green{Ba}\blue{Al}2\red{Zn}3} & agm002327363 & $  75$ & $ 142$ & $ 15.4$ & $  0.53$ & $  2.3$ \\
\hline
\ch{\green{La}\blue{Ga}2\red{Cu}3} & agm002267000 & $  84$ & $ 131$ & $ 22.7$ & $  0.55$ & $  2.3$ \\
\hline
\ch{\green{Sm}\blue{Li}2\red{Zn}3} & agm071929341 & $  59$ & $ 141$ & $ 20.7$ & $  0.53$ & $  2.3$ \\
\hline
\ch{\green{Pr}\blue{V}2\red{Os}3} & agm072092776 & $  76$ & $ 191$ & $ 26.3$ & $  0.49$ & $  2.2$ \\
\hline
\ch{\green{Tb}\blue{Pt}2\red{Rh}3} & agm039629032 & $  64$ & $ 147$ & $ 35.0$ & $  0.52$ & $  2.2$ \\
\hline
\ch{\green{Ba}\blue{In}2\red{Ag}3} & agm002247479 & $  47$ & $  74$ & $ 21.6$ & $  0.64$ & $  2.1$ \\
\hline
\ch{\green{Ba}\blue{Ga}2\red{Ag}3} & agm002274762 & $  47$ & $ 106$ & $ 21.6$ & $  0.57$ & $  2.1$ \\
\hline
\ch{\green{Pr}\blue{Mg}2\red{Zn}3} & agm039915784 & $  73$ & $  98$ & $ 18.3$ & $  0.58$ & $  2.1$ \\
\hline
\ch{\green{Ba}\blue{Li}2\red{Cd}3} & agm071678116 & $  16$ & $ 103$ & $ 25.9$ & $  0.57$ & $  2.1$ \\
\hline
\ch{\green{Sm}\blue{Cu}2\red{Ir}3} & agm002266833 & $  80$ & $ 124$ & $ 32.9$ & $  0.54$ & $  2.1$ \\
\hline
\ch{\green{Sm}\blue{Ni}2\red{Ir}3} & agm002226250 & $   0$ & $ 149$ & $ 34.8$ & $  0.52$ & $  2.1$ \\
\hline
\ch{\green{Tb}\blue{V}2\red{Co}3} & agm071683364 & $  61$ & $ 239$ & $ 39.4$ & $  0.46$ & $  2.1$ \\
\hline
\ch{\green{Ca}\blue{Li}2\red{Zn}3} & agm071598618 & $  37$ & $ 158$ & $ 21.2$ & $  0.51$ & $  2.1$ \\
\hline
\ch{\green{Nd}\blue{Li}2\red{Zn}3} & agm072072237 & $  46$ & $ 147$ & $ 20.3$ & $  0.51$ & $  2.0$ \\
\hline
\ch{\green{Nd}\blue{Pd}2\red{Rh}3} & agm002235380 & $  65$ & $ 156$ & $ 37.8$ & $  0.51$ & $  2.0$ \\
\hline
\ch{\green{Dy}\blue{Be}2\red{Zn}3} & agm030890617 & $  97$ & $ 165$ & $ 12.6$ & $  0.50$ & $  2.0$ \\
\hline
\ch{\green{K}\blue{Al}2\red{Zn}3} & agm071765022 & $  82$ & $ 147$ & $ 12.6$ & $  0.51$ & $  2.0$ \\
\hline
\ch{\green{Pr}\blue{Li}2\red{Zn}3} & agm071732143 & $  39$ & $ 148$ & $ 20.2$ & $  0.51$ & $  2.0$ \\
\hline
\ch{\green{La}\blue{Li}2\red{Zn}3} & agm071648186 & $  42$ & $ 149$ & $ 22.1$ & $  0.51$ & $  2.0$ \\
\hline
\ch{\green{Pm}\blue{Ni}2\red{Ir}3} & agm005853112 & $   0$ & $ 151$ & $ 34.9$ & $  0.50$ & $  2.0$ \\
\hline
\ch{\green{Ho}\blue{V}2\red{Co}3} & agm071696855 & $  69$ & $ 241$ & $ 38.9$ & $  0.45$ & $  1.9$ \\
\hline
\ch{\green{Pr}\blue{Pd}2\red{Rh}3} & agm002235381 & $  53$ & $ 159$ & $ 37.9$ & $  0.50$ & $  1.9$ \\
\hline
\ch{\green{Ca}\blue{Ga}2\red{Zn}3} & agm002331397 & $  89$ & $ 140$ & $ 12.7$ & $  0.51$ & $  1.9$ \\
\hline
\ch{\green{Tm}\blue{V}2\red{Co}3} & agm071696573 & $  81$ & $ 240$ & $ 38.5$ & $  0.45$ & $  1.9$ \\
\hline
\ch{\green{Ba}\blue{Ga}2\red{Cu}3} & agm002371224 & $  89$ & $ 130$ & $ 20.8$ & $  0.52$ & $  1.9$ \\
\hline
\ch{\green{Er}\blue{V}2\red{Co}3} & agm071982099 & $  73$ & $ 241$ & $ 38.7$ & $  0.45$ & $  1.9$ \\
\hline
\ch{\green{La}\blue{Pd}2\red{Rh}3} & agm002180591 & $  55$ & $ 155$ & $ 38.6$ & $  0.50$ & $  1.9$ \\
\hline
\ch{\green{Ca}\blue{Li}2\red{Cd}3} & agm072084042 & $  58$ & $  96$ & $ 22.1$ & $  0.57$ & $  1.9$ \\
\hline
\ch{\green{La}\blue{Ni}2\red{Ir}3} & agm002167274 & $   0$ & $ 152$ & $ 36.4$ & $  0.50$ & $  1.8$ \\
\hline
\ch{\green{Ba}\blue{Ga}2\red{Cd}3} & agm071730727 & $  89$ & $ 110$ & $ 17.3$ & $  0.54$ & $  1.8$ \\
\hline
\ch{\green{Sr}\blue{Ge}2\red{Ni}3} & agm002295246 & $  71$ & $ 149$ & $ 26.3$ & $  0.49$ & $  1.8$ \\
\hline
\ch{\green{Rb}\blue{Tl}2\red{Pt}3} & agm034136002 & $  94$ & $  61$ & $ 16.3$ & $  0.64$ & $  1.8$ \\
\hline
\ch{\green{Nd}\blue{Mg}2\red{Zn}3} & agm033357793 & $  79$ & $  98$ & $ 18.0$ & $  0.55$ & $  1.7$ \\
\hline
\ch{\green{Nd}\blue{Cu}2\red{Ir}3} & agm002268171 & $  61$ & $ 133$ & $ 32.7$ & $  0.51$ & $  1.7$ \\
\hline
\ch{\green{La}\blue{Li}2\red{Cd}3} & agm071989287 & $  47$ & $ 104$ & $ 26.3$ & $  0.54$ & $  1.7$ \\
\hline
\ch{\green{Dy}\blue{Ni}2\red{Rh}3} & agm002167305 & $  23$ & $ 143$ & $ 40.7$ & $  0.50$ & $  1.7$ \\
\hline
\ch{\green{Na}\blue{Al}2\red{Zn}3} & agm071985931 & $  50$ & $ 138$ & $ 11.0$ & $  0.50$ & $  1.7$ \\
\hline
\ch{\green{La}\blue{Hg}2\red{Ag}3} & agm071674034 & $  84$ & $  59$ & $ 22.9$ & $  0.64$ & $  1.7$ \\
\hline
\ch{\green{Y}\blue{V}2\red{Co}3} & agm072080688 & $  60$ & $ 252$ & $ 39.2$ & $  0.44$ & $  1.7$ \\
\hline
\ch{\green{Tb}\blue{Ag}2\red{Zn}3} & agm002360031 & $  14$ & $ 108$ & $ 17.4$ & $  0.53$ & $  1.7$ \\
\hline
\ch{\green{Pr}\blue{Ni}2\red{Ir}3} & agm002226175 & $   0$ & $ 157$ & $ 35.4$ & $  0.48$ & $  1.7$ \\
\hline
\ch{\green{Sr}\blue{Li}2\red{Cd}3} & agm071725069 & $  23$ & $ 102$ & $ 23.2$ & $  0.54$ & $  1.7$ \\
\hline
\ch{\green{Ba}\blue{Ag}2\red{Hg}3} & agm034076976 & $  70$ & $  62$ & $ 19.2$ & $  0.62$ & $  1.6$ \\
\hline
\ch{\green{Sr}\blue{Au}2\red{Zn}3} & agm002324467 & $  18$ & $ 121$ & $ 17.0$ & $  0.51$ & $  1.6$ \\
\hline
\ch{\green{Tb}\blue{Ni}2\red{Rh}3} & agm002167312 & $  17$ & $ 150$ & $ 40.9$ & $  0.48$ & $  1.6$ \\
\hline
\ch{\green{Sm}\blue{Pt}2\red{Rh}3} & agm002235389 & $  39$ & $ 157$ & $ 35.2$ & $  0.48$ & $  1.6$ \\
\hline
\ch{\green{Na}\blue{B}2\red{Ir}3} & agm005806208 & $   2$ & $ 185$ & $ 15.3$ & $  0.46$ & $  1.6$ \\
\hline
\ch{\green{La}\blue{Ag}2\red{Zn}3} & agm002324396 & $   0$ & $ 118$ & $ 19.5$ & $  0.51$ & $  1.6$ \\
\hline
\ch{\green{Pr}\blue{Cu}2\red{Ir}3} & agm002268180 & $  49$ & $ 135$ & $ 32.7$ & $  0.49$ & $  1.5$ \\
\hline
\ch{\green{K}\blue{Ag}2\red{Zn}3} & agm039598974 & $  40$ & $ 108$ & $ 14.8$ & $  0.52$ & $  1.5$ \\
\hline
\ch{\green{Rb}\blue{Zn}2\red{Au}3} & agm041423042 & $  61$ & $  76$ & $ 13.4$ & $  0.57$ & $  1.5$ \\
\hline
\ch{\green{Pm}\blue{Pt}2\red{Rh}3} & agm031020160 & $  28$ & $ 159$ & $ 35.3$ & $  0.47$ & $  1.5$ \\
\hline
\ch{\green{Ce}\blue{Ni}2\red{Ir}3} & agm002227628 & $   0$ & $ 160$ & $ 35.9$ & $  0.46$ & $  1.4$ \\
\hline
\ch{\green{Ce}\blue{Be}2\red{Co}3} & agm038805054 & $  28$ & $ 201$ & $ 45.5$ & $  0.44$ & $  1.4$ \\
\hline
\ch{\green{Nd}\blue{Pt}2\red{Rh}3} & agm002235378 & $  32$ & $ 161$ & $ 35.4$ & $  0.46$ & $  1.4$ \\
\hline
\ch{\green{Ba}\blue{Cu}2\red{Zn}3} & agm002334916 & $  55$ & $ 135$ & $ 20.3$ & $  0.48$ & $  1.4$ \\
\hline
\ch{\green{Na}\blue{Au}2\red{Zn}3} & agm039946149 & $  14$ & $ 114$ & $ 14.1$ & $  0.50$ & $  1.4$ \\
\hline
\ch{\green{Cs}\blue{Au}2\red{Hg}3} & agm071934027 & $  33$ & $  54$ & $ 12.2$ & $  0.61$ & $  1.4$ \\
\hline
\ch{\green{Tb}\blue{Si}2\red{Ni}3} & agm002316120 & $  55$ & $ 195$ & $ 28.5$ & $  0.44$ & $  1.3$ \\
\hline
\ch{\green{Pr}\blue{Pt}2\red{Rh}3} & agm002235386 & $  28$ & $ 165$ & $ 35.6$ & $  0.45$ & $  1.3$ \\
\hline
\ch{\green{Na}\blue{Ge}2\red{Cu}3} & agm032410195 & $  76$ & $ 161$ & $ 13.3$ & $  0.45$ & $  1.3$ \\
\hline
\ch{\green{Sr}\blue{Al}2\red{Ag}3} & agm002133724 & $  42$ & $ 126$ & $ 18.8$ & $  0.48$ & $  1.3$ \\
\hline
\ch{\green{La}\blue{Pt}2\red{Rh}3} & agm002180579 & $  22$ & $ 160$ & $ 36.6$ & $  0.45$ & $  1.3$ \\
\hline
\ch{\green{Ba}\blue{Pd}2\red{Zn}3} & agm002360061 & $  28$ & $ 117$ & $ 19.8$ & $  0.49$ & $  1.3$ \\
\hline
\ch{\green{La}\blue{Mg}2\red{Zn}3} & agm002286631 & $  57$ & $ 111$ & $ 18.8$ & $  0.49$ & $  1.3$ \\
\hline
\ch{\green{K}\blue{Zn}2\red{Cd}3} & agm032429899 & $  68$ & $ 102$ & $ 16.4$ & $  0.50$ & $  1.3$ \\
\hline
\ch{\green{Sr}\blue{Al}2\red{Zn}3} & agm002248477 & $  78$ & $ 160$ & $ 13.4$ & $  0.45$ & $  1.3$ \\
\hline
\ch{\green{Sr}\blue{Ga}2\red{Zn}3} & agm002360053 & $  95$ & $ 149$ & $ 13.3$ & $  0.46$ & $  1.3$ \\
\hline
\ch{\green{Ba}\blue{Ga}2\red{Rh}3} & agm002337544 & $   0$ & $ 135$ & $ 20.8$ & $  0.47$ & $  1.3$ \\
\hline
\ch{\green{Sr}\blue{Ga}2\red{Cu}3} & agm002267004 & $  49$ & $ 144$ & $ 19.6$ & $  0.46$ & $  1.3$ \\
\hline
\ch{\green{Sm}\blue{Ag}2\red{Zn}3} & agm002360030 & $   2$ & $ 120$ & $ 17.7$ & $  0.48$ & $  1.2$ \\
\hline
\ch{\green{Ba}\blue{Li}2\red{Ag}3} & agm071772019 & $  42$ & $  69$ & $ 20.3$ & $  0.55$ & $  1.2$ \\
\hline
\ch{\green{K}\blue{Zn}2\red{Au}3} & agm032323358 & $  42$ & $  86$ & $ 14.5$ & $  0.51$ & $  1.2$ \\
\hline
\ch{\green{Sr}\blue{Ga}2\red{Ag}3} & agm002213111 & $  32$ & $ 115$ & $ 18.6$ & $  0.48$ & $  1.2$ \\
\hline
\ch{\green{K}\blue{Au}2\red{Hg}3} & agm072065774 & $  57$ & $  59$ & $ 11.9$ & $  0.57$ & $  1.2$ \\
\hline
\ch{\green{Sm}\blue{Ni}2\red{Rh}3} & agm002167311 & $   1$ & $ 167$ & $ 41.4$ & $  0.44$ & $  1.2$ \\
\hline
\ch{\green{Dy}\blue{Si}2\red{Ni}3} & agm002316112 & $  53$ & $ 199$ & $ 27.9$ & $  0.42$ & $  1.1$ \\
\hline
\ch{\green{Rb}\blue{Au}2\red{Hg}3} & agm071959257 & $  39$ & $  58$ & $ 11.7$ & $  0.57$ & $  1.1$ \\
\hline
\ch{\green{Ce}\blue{Pt}2\red{Rh}3} & agm033056941 & $  55$ & $ 166$ & $ 35.9$ & $  0.44$ & $  1.1$ \\
\hline
\ch{\green{La}\blue{Cu}2\red{Ir}3} & agm002207746 & $  35$ & $ 138$ & $ 31.6$ & $  0.45$ & $  1.1$ \\
\hline
\ch{\green{Ba}\blue{Sb}2\red{Rh}3} & agm071960227 & $  46$ & $ 126$ & $ 19.6$ & $  0.46$ & $  1.1$ \\
\hline
\ch{\green{La}\blue{Pd}2\red{Zn}3} & agm002246275 & $   0$ & $ 145$ & $ 23.6$ & $  0.45$ & $  1.1$ \\
\hline
\ch{\green{Rb}\blue{Cd}2\red{Au}3} & agm039823197 & $  45$ & $  66$ & $ 14.5$ & $  0.54$ & $  1.1$ \\
\hline
\ch{\green{Sr}\blue{Zn}2\red{Au}3} & agm002324322 & $  50$ & $  83$ & $ 16.5$ & $  0.51$ & $  1.1$ \\
\hline
\ch{\green{Tm}\blue{V}2\red{Ru}3} & agm071671242 & $  96$ & $ 223$ & $ 23.4$ & $  0.41$ & $  1.1$ \\
\hline
\ch{\green{La}\blue{V}2\red{Ru}3} & agm071966397 & $  41$ & $ 206$ & $ 24.1$ & $  0.41$ & $  1.0$ \\
\hline
\ch{\green{Pm}\blue{Zn}2\red{Ag}3} & agm072053067 & $  47$ & $ 101$ & $ 21.9$ & $  0.48$ & $  1.0$ \\
\hline
\ch{\green{La}\blue{Pd}2\red{Pt}3} & agm002234138 & $   0$ & $ 103$ & $ 27.7$ & $  0.47$ & $  1.0$ \\
\hline
\ch{\green{Pr}\blue{Ag}2\red{Zn}3} & agm002360029 & $   0$ & $ 126$ & $ 17.8$ & $  0.45$ & $  1.0$ \\
\hline
\ch{\green{Sr}\blue{Ag}2\red{Zn}3} & agm002360072 & $   0$ & $ 124$ & $ 18.1$ & $  0.45$ & $  1.0$ \\
\hline
\end{longtable}
\end{center} \clearpage
\begin{center}
\begin{longtable}{|c|c|c|c|c|c|c|}
\caption{Predicted compounds with $E_{\mathrm{hull}} < 100$ meV/atom in Type II.}
\label{tab:PredictionTypeII} \\
\hline \multicolumn{1}{|c|}{Formula} & \multicolumn{1}{|c|}{Material ID} & \multicolumn{1}{c|}{$E_{\mathrm{hull}}$ (meV/atom)} & \multicolumn{1}{c|}{$\omega_{\mathrm{log}}$ (K)} & \multicolumn{1}{c|}{$D(E_F)$} & \multicolumn{1}{c|}{$\lambda$} & \multicolumn{1}{c|}{$T_c$ (K)} \\ \hline
\endfirsthead
\multicolumn{7}{c}{{\bfseries \tablename\ \thetable{} -- continued from previous page}} \\
\hline \multicolumn{1}{|c|}{Formula} & \multicolumn{1}{|c|}{Material ID} & \multicolumn{1}{c|}{$E_{\mathrm{hull}}$ (meV/atom)} & \multicolumn{1}{c|}{$\omega_{\mathrm{log}}$ (K)} & \multicolumn{1}{c|}{$D(E_F)$} & \multicolumn{1}{c|}{$\lambda$} & \multicolumn{1}{c|}{$T_c$ (K)} \\ \hline
\endhead
\hline \multicolumn{7}{|r|}{{Continued on next page}} \\ \hline
\endfoot
\hline \hline
\endlastfoot
\ch{\green{Zr}\blue{Pd}2\red{Be}3} & agm071721414 & $  60$ & $ 189$ & $ 21.8$ & $  1.04$ & $ 15.0$ \\
\hline
\ch{\green{Hf}\blue{Pd}2\red{Be}3} & agm071647583 & $  71$ & $ 202$ & $ 20.3$ & $  0.89$ & $ 12.3$ \\
\hline
\ch{\green{Hf}\blue{Pt}2\red{Be}3} & agm071755690 & $  49$ & $ 202$ & $ 15.9$ & $  0.88$ & $ 12.0$ \\
\hline
\ch{\green{Ti}\blue{Pt}2\red{Be}3} & agm071816678 & $  73$ & $ 193$ & $ 17.9$ & $  0.89$ & $ 11.8$ \\
\hline
\ch{\green{Zr}\blue{Pt}2\red{Be}3} & agm072017833 & $  34$ & $ 215$ & $ 16.9$ & $  0.81$ & $ 10.9$ \\
\hline
\ch{\green{La}\blue{Si}2\red{Al}3} & agm002134055 & $  86$ & $ 160$ & $ 19.7$ & $  0.87$ & $  9.4$ \\
\hline
\ch{\green{La}\blue{Rh}2\red{Ga}3} & agm002151213 & $  65$ & $ 120$ & $ 27.1$ & $  0.93$ & $  7.9$ \\
\hline
\ch{\green{La}\blue{Pt}2\red{Ga}3} & agm002151180 & $  87$ & $ 100$ & $ 18.8$ & $  1.04$ & $  7.9$ \\
\hline
\ch{\green{Ba}\blue{Pd}2\red{Al}3} & agm002248752 & $  96$ & $ 118$ & $ 24.8$ & $  0.93$ & $  7.8$ \\
\hline
\ch{\green{Tb}\blue{Pd}2\red{Be}3} & agm071773951 & $  85$ & $ 216$ & $ 20.6$ & $  0.69$ & $  7.6$ \\
\hline
\ch{\green{Tm}\blue{Pd}2\red{Be}3} & agm071781809 & $  78$ & $ 229$ & $ 20.4$ & $  0.68$ & $  7.6$ \\
\hline
\ch{\green{Ho}\blue{Pd}2\red{Be}3} & agm071784762 & $  80$ & $ 223$ & $ 20.5$ & $  0.68$ & $  7.5$ \\
\hline
\ch{\green{Er}\blue{Pd}2\red{Be}3} & agm071892668 & $  79$ & $ 228$ & $ 20.4$ & $  0.67$ & $  7.4$ \\
\hline
\ch{\green{La}\blue{Rh}2\red{Al}3} & agm002134004 & $  58$ & $ 145$ & $ 27.7$ & $  0.81$ & $  7.3$ \\
\hline
\ch{\green{Cs}\blue{Ga}2\red{Bi}3} & agm071652597 & $  78$ & $  60$ & $ 19.3$ & $  1.39$ & $  7.1$ \\
\hline
\ch{\green{Ce}\blue{Rh}2\red{Ga}3} & agm002151212 & $  54$ & $ 125$ & $ 25.3$ & $  0.86$ & $  7.1$ \\
\hline
\ch{\green{Sc}\blue{Pd}2\red{Be}3} & agm072096804 & $  61$ & $ 257$ & $ 22.0$ & $  0.63$ & $  7.0$ \\
\hline
\ch{\green{Ba}\blue{Pd}2\red{Ga}3} & agm002213204 & $  79$ & $ 114$ & $ 22.7$ & $  0.88$ & $  6.8$ \\
\hline
\ch{\green{Pr}\blue{Rh}2\red{Al}3} & agm002195978 & $  43$ & $ 148$ & $ 25.8$ & $  0.77$ & $  6.7$ \\
\hline
\ch{\green{Pr}\blue{Rh}2\red{Ga}3} & agm002213223 & $  63$ & $ 131$ & $ 25.0$ & $  0.78$ & $  6.1$ \\
\hline
\ch{\green{La}\blue{Pt}2\red{Al}3} & agm002133904 & $  68$ & $ 137$ & $ 21.0$ & $  0.75$ & $  5.9$ \\
\hline
\ch{\green{Nd}\blue{Rh}2\red{Al}3} & agm002248776 & $  45$ & $ 159$ & $ 25.4$ & $  0.68$ & $  5.4$ \\
\hline
\ch{\green{Ca}\blue{Ir}2\red{Al}3} & agm005704535 & $  99$ & $  89$ & $ 22.8$ & $  0.86$ & $  5.1$ \\
\hline
\ch{\green{Pr}\blue{Pt}2\red{Al}3} & agm002195965 & $  56$ & $ 149$ & $ 18.9$ & $  0.67$ & $  4.7$ \\
\hline
\ch{\green{Tm}\blue{Ru}2\red{Al}3} & agm033899183 & $  97$ & $ 158$ & $ 36.0$ & $  0.65$ & $  4.7$ \\
\hline
\ch{\green{La}\blue{Ir}2\red{Al}3} & agm002133903 & $  88$ & $ 141$ & $ 23.3$ & $  0.68$ & $  4.7$ \\
\hline
\ch{\green{Lu}\blue{Pt}2\red{Al}3} & agm002195858 & $  78$ & $ 139$ & $ 15.3$ & $  0.68$ & $  4.7$ \\
\hline
\ch{\green{Tm}\blue{Pt}2\red{Al}3} & agm002196091 & $  70$ & $ 142$ & $ 15.6$ & $  0.67$ & $  4.6$ \\
\hline
\ch{\green{Nd}\blue{Pt}2\red{Al}3} & agm002248749 & $  54$ & $ 150$ & $ 18.4$ & $  0.65$ & $  4.5$ \\
\hline
\ch{\green{Er}\blue{Pt}2\red{Al}3} & agm002195729 & $  65$ & $ 143$ & $ 15.8$ & $  0.66$ & $  4.5$ \\
\hline
\ch{\green{Dy}\blue{Pt}2\red{Al}3} & agm002195723 & $  61$ & $ 146$ & $ 16.2$ & $  0.65$ & $  4.4$ \\
\hline
\ch{\green{Tb}\blue{Pt}2\red{Al}3} & agm002196075 & $  59$ & $ 148$ & $ 16.5$ & $  0.65$ & $  4.4$ \\
\hline
\ch{\green{Pm}\blue{Rh}2\red{Al}3} & agm032802255 & $  46$ & $ 167$ & $ 25.1$ & $  0.62$ & $  4.4$ \\
\hline
\ch{\green{Ho}\blue{Pt}2\red{Al}3} & agm002195840 & $  64$ & $ 145$ & $ 16.0$ & $  0.65$ & $  4.4$ \\
\hline
\ch{\green{Sm}\blue{Pt}2\red{Al}3} & agm002248953 & $  55$ & $ 151$ & $ 17.5$ & $  0.64$ & $  4.3$ \\
\hline
\ch{\green{Pm}\blue{Pt}2\red{Al}3} & agm038766191 & $  37$ & $ 151$ & $ 17.9$ & $  0.64$ & $  4.3$ \\
\hline
\ch{\green{Pm}\blue{Cu}2\red{Be}3} & agm071902574 & $  86$ & $ 303$ & $ 19.7$ & $  0.52$ & $  4.3$ \\
\hline
\ch{\green{Tm}\blue{Cu}2\red{Be}3} & agm071590064 & $  47$ & $ 303$ & $ 18.7$ & $  0.52$ & $  4.3$ \\
\hline
\ch{\green{Ho}\blue{Cu}2\red{Be}3} & agm071706171 & $  53$ & $ 308$ & $ 19.2$ & $  0.51$ & $  4.3$ \\
\hline
\ch{\green{Dy}\blue{Cu}2\red{Be}3} & agm071966243 & $  56$ & $ 305$ & $ 19.4$ & $  0.51$ & $  4.3$ \\
\hline
\ch{\green{Cs}\blue{Ga}2\red{In}3} & agm071615201 & $  75$ & $  96$ & $ 18.6$ & $  0.76$ & $  4.2$ \\
\hline
\ch{\green{Er}\blue{Cu}2\red{Be}3} & agm072008289 & $  50$ & $ 305$ & $ 19.0$ & $  0.51$ & $  4.2$ \\
\hline
\ch{\green{Sm}\blue{Cu}2\red{Be}3} & agm071702232 & $  79$ & $ 304$ & $ 19.7$ & $  0.51$ & $  4.1$ \\
\hline
\ch{\green{K}\blue{Li}2\red{Tl}3} & agm071807539 & $  74$ & $  68$ & $ 23.0$ & $  0.88$ & $  4.1$ \\
\hline
\ch{\green{Y}\blue{Pt}2\red{Al}3} & agm002196108 & $  57$ & $ 156$ & $ 16.3$ & $  0.62$ & $  4.1$ \\
\hline
\ch{\green{Ti}\blue{Cu}2\red{Be}3} & agm071800885 & $  64$ & $ 163$ & $ 16.1$ & $  0.61$ & $  4.1$ \\
\hline
\ch{\green{Sr}\blue{Rh}2\red{Ga}3} & agm002275198 & $  85$ & $ 147$ & $ 25.0$ & $  0.63$ & $  4.0$ \\
\hline
\ch{\green{Sm}\blue{Rh}2\red{Al}3} & agm002248777 & $  48$ & $ 173$ & $ 24.8$ & $  0.59$ & $  3.9$ \\
\hline
\ch{\green{Rb}\blue{Tl}2\red{Pb}3} & agm071712400 & $  45$ & $  55$ & $ 21.9$ & $  0.98$ & $  3.9$ \\
\hline
\ch{\green{La}\blue{Ga}2\red{Al}3} & agm002327397 & $  92$ & $ 133$ & $ 18.6$ & $  0.64$ & $  3.8$ \\
\hline
\ch{\green{K}\blue{Tl}2\red{Pb}3} & agm071986780 & $  61$ & $  54$ & $ 21.1$ & $  0.97$ & $  3.8$ \\
\hline
\ch{\green{Cs}\blue{Hg}2\red{Tl}3} & agm033730936 & $  56$ & $  49$ & $ 21.8$ & $  1.03$ & $  3.8$ \\
\hline
\ch{\green{Y}\blue{Cu}2\red{Be}3} & agm071894160 & $  56$ & $ 331$ & $ 20.0$ & $  0.49$ & $  3.8$ \\
\hline
\ch{\green{Sr}\blue{Pd}2\red{Ga}3} & agm002151226 & $  52$ & $ 135$ & $ 20.3$ & $  0.64$ & $  3.7$ \\
\hline
\ch{\green{Ba}\blue{Li}2\red{Ga}3} & agm071816344 & $  31$ & $ 147$ & $ 16.4$ & $  0.61$ & $  3.7$ \\
\hline
\ch{\green{Ba}\blue{Hg}2\red{Al}3} & agm031653555 & $  89$ & $ 133$ & $ 16.8$ & $  0.63$ & $  3.6$ \\
\hline
\ch{\green{Sr}\blue{Li}2\red{Tl}3} & agm072059304 & $  45$ & $  57$ & $ 18.7$ & $  0.90$ & $  3.5$ \\
\hline
\ch{\green{Rb}\blue{Hg}2\red{Pb}3} & agm071980026 & $  56$ & $  54$ & $ 18.8$ & $  0.92$ & $  3.5$ \\
\hline
\ch{\green{Rb}\blue{Li}2\red{Sb}3} & agm071828172 & $  30$ & $  93$ & $ 13.4$ & $  0.71$ & $  3.5$ \\
\hline
\ch{\green{K}\blue{Hg}2\red{In}3} & agm003420560 & $  47$ & $  67$ & $ 17.3$ & $  0.82$ & $  3.4$ \\
\hline
\ch{\green{Pr}\blue{Ir}2\red{Al}3} & agm002195964 & $  73$ & $ 158$ & $ 21.7$ & $  0.58$ & $  3.4$ \\
\hline
\ch{\green{Tb}\blue{Ni}2\red{Be}3} & agm071909953 & $  43$ & $ 257$ & $ 22.2$ & $  0.51$ & $  3.4$ \\
\hline
\ch{\green{La}\blue{Co}2\red{Ga}3} & agm002143780 & $  69$ & $ 153$ & $ 39.6$ & $  0.59$ & $  3.4$ \\
\hline
\ch{\green{Cs}\blue{Au}2\red{In}3} & agm071742875 & $  67$ & $  76$ & $ 21.1$ & $  0.76$ & $  3.4$ \\
\hline
\ch{\green{La}\blue{Co}2\red{Al}3} & agm002133847 & $  97$ & $ 171$ & $ 39.9$ & $  0.56$ & $  3.3$ \\
\hline
\ch{\green{Na}\blue{Li}2\red{Sb}3} & agm071795043 & $  74$ & $  99$ & $ 12.3$ & $  0.68$ & $  3.3$ \\
\hline
\ch{\green{Cs}\blue{Hg}2\red{In}3} & agm003420557 & $  45$ & $  71$ & $ 18.6$ & $  0.77$ & $  3.3$ \\
\hline
\ch{\green{Dy}\blue{Ni}2\red{Be}3} & agm071796331 & $  36$ & $ 261$ & $ 22.0$ & $  0.50$ & $  3.3$ \\
\hline
\ch{\green{Sm}\blue{Ru}2\red{Al}3} & agm031813136 & $  65$ & $ 170$ & $ 34.1$ & $  0.56$ & $  3.2$ \\
\hline
\ch{\green{La}\blue{Ru}2\red{Ga}3} & agm032677294 & $  67$ & $ 142$ & $ 32.6$ & $  0.59$ & $  3.2$ \\
\hline
\ch{\green{Rb}\blue{Au}2\red{In}3} & agm071921243 & $  67$ & $  75$ & $ 19.3$ & $  0.75$ & $  3.2$ \\
\hline
\ch{\green{Rb}\blue{Ga}2\red{Pb}3} & agm071627791 & $  93$ & $  74$ & $ 18.6$ & $  0.74$ & $  3.1$ \\
\hline
\ch{\green{Sn}\blue{Ta}2\red{Ga}3} & agm071785002 & $  88$ & $ 138$ & $ 18.2$ & $  0.59$ & $  3.1$ \\
\hline
\ch{\green{Ho}\blue{Ni}2\red{Be}3} & agm072053668 & $  31$ & $ 263$ & $ 21.9$ & $  0.49$ & $  3.1$ \\
\hline
\ch{\green{Nd}\blue{Ir}2\red{Al}3} & agm002251375 & $  69$ & $ 160$ & $ 21.6$ & $  0.56$ & $  3.0$ \\
\hline
\ch{\green{K}\blue{Li}2\red{Sb}3} & agm071907720 & $  27$ & $ 102$ & $ 12.9$ & $  0.64$ & $  3.0$ \\
\hline
\ch{\green{Y}\blue{Ni}2\red{Be}3} & agm071893162 & $  34$ & $ 280$ & $ 22.5$ & $  0.48$ & $  2.9$ \\
\hline
\ch{\green{Ba}\blue{Au}2\red{Sn}3} & agm003411324 & $  97$ & $  73$ & $ 16.7$ & $  0.73$ & $  2.9$ \\
\hline
\ch{\green{Tm}\blue{Pt}2\red{Be}3} & agm071624759 & $  78$ & $ 221$ & $ 15.8$ & $  0.51$ & $  2.9$ \\
\hline
\ch{\green{La}\blue{Ru}2\red{Al}3} & agm041897259 & $  91$ & $ 169$ & $ 32.8$ & $  0.54$ & $  2.9$ \\
\hline
\ch{\green{Sm}\blue{Ru}2\red{Ga}3} & agm002275193 & $  65$ & $ 148$ & $ 33.5$ & $  0.57$ & $  2.9$ \\
\hline
\ch{\green{Sr}\blue{Pd}2\red{Al}3} & agm002134079 & $  32$ & $ 174$ & $ 22.0$ & $  0.54$ & $  2.9$ \\
\hline
\ch{\green{Tb}\blue{Pt}2\red{Be}3} & agm071683421 & $  87$ & $ 213$ & $ 15.4$ & $  0.51$ & $  2.9$ \\
\hline
\ch{\green{Er}\blue{Ni}2\red{Be}3} & agm071708349 & $  27$ & $ 268$ & $ 21.8$ & $  0.48$ & $  2.8$ \\
\hline
\ch{\green{Pr}\blue{Ru}2\red{Al}3} & agm031668506 & $  64$ & $ 173$ & $ 33.0$ & $  0.54$ & $  2.8$ \\
\hline
\ch{\green{Ba}\blue{Li}2\red{Tl}3} & agm072080483 & $  30$ & $  75$ & $ 19.7$ & $  0.71$ & $  2.8$ \\
\hline
\ch{\green{Cs}\blue{Ag}2\red{In}3} & agm072020533 & $  85$ & $  78$ & $ 20.4$ & $  0.70$ & $  2.8$ \\
\hline
\ch{\green{Dy}\blue{Pt}2\red{Be}3} & agm071800260 & $  84$ & $ 217$ & $ 15.4$ & $  0.50$ & $  2.8$ \\
\hline
\ch{\green{Er}\blue{Pt}2\red{Be}3} & agm071999384 & $  80$ & $ 220$ & $ 15.6$ & $  0.50$ & $  2.7$ \\
\hline
\ch{\green{Sr}\blue{Ni}2\red{Ga}3} & agm002294810 & $  85$ & $ 153$ & $ 25.9$ & $  0.55$ & $  2.7$ \\
\hline
\ch{\green{Pm}\blue{Ir}2\red{Al}3} & agm033291134 & $  55$ & $ 161$ & $ 21.5$ & $  0.54$ & $  2.7$ \\
\hline
\ch{\green{Tm}\blue{Ni}2\red{Be}3} & agm072045138 & $  22$ & $ 274$ & $ 21.8$ & $  0.47$ & $  2.7$ \\
\hline
\ch{\green{Cs}\blue{Au}2\red{Bi}3} & agm071689831 & $  27$ & $  68$ & $ 16.9$ & $  0.72$ & $  2.6$ \\
\hline
\ch{\green{Pm}\blue{Zn}2\red{Al}3} & agm003420545 & $  88$ & $ 137$ & $ 17.0$ & $  0.56$ & $  2.5$ \\
\hline
\ch{\green{Sm}\blue{Ir}2\red{Al}3} & agm043463842 & $  67$ & $ 161$ & $ 21.5$ & $  0.53$ & $  2.5$ \\
\hline
\ch{\green{Al}\blue{Ni}2\red{Be}3} & agm071967410 & $  79$ & $ 242$ & $ 15.2$ & $  0.48$ & $  2.5$ \\
\hline
\ch{\green{Na}\blue{Pd}2\red{Ga}3} & agm071700982 & $  65$ & $ 135$ & $ 15.8$ & $  0.56$ & $  2.5$ \\
\hline
\ch{\green{Tb}\blue{Rh}2\red{Al}3} & agm002195979 & $  53$ & $ 185$ & $ 23.9$ & $  0.51$ & $  2.4$ \\
\hline
\ch{\green{Pr}\blue{Ru}2\red{Ga}3} & agm002275191 & $  59$ & $ 153$ & $ 32.5$ & $  0.53$ & $  2.4$ \\
\hline
\ch{\green{Nd}\blue{Ru}2\red{Ga}3} & agm002275190 & $  53$ & $ 153$ & $ 32.8$ & $  0.53$ & $  2.4$ \\
\hline
\ch{\green{Ba}\blue{Cd}2\red{Al}3} & agm043790175 & $  94$ & $ 163$ & $ 16.8$ & $  0.52$ & $  2.4$ \\
\hline
\ch{\green{Sr}\blue{Li}2\red{Ga}3} & agm071848698 & $  41$ & $ 163$ & $ 15.0$ & $  0.52$ & $  2.3$ \\
\hline
\ch{\green{K}\blue{Li}2\red{Pb}3} & agm071988679 & $  81$ & $  80$ & $ 17.0$ & $  0.64$ & $  2.3$ \\
\hline
\ch{\green{Ca}\blue{Zn}2\red{Al}3} & agm003420537 & $  59$ & $ 177$ & $ 16.6$ & $  0.50$ & $  2.3$ \\
\hline
\ch{\green{Na}\blue{Ag}2\red{Al}3} & agm005806252 & $  46$ & $ 166$ & $ 14.2$ & $  0.51$ & $  2.2$ \\
\hline
\ch{\green{Ca}\blue{Rh}2\red{Ga}3} & agm002260183 & $  44$ & $ 163$ & $ 24.1$ & $  0.51$ & $  2.2$ \\
\hline
\ch{\green{Ca}\blue{Rh}2\red{Al}3} & agm042201279 & $  71$ & $ 180$ & $ 23.3$ & $  0.50$ & $  2.2$ \\
\hline
\ch{\green{Pr}\blue{Co}2\red{Al}3} & agm002195637 & $  82$ & $ 187$ & $ 38.3$ & $  0.49$ & $  2.2$ \\
\hline
\ch{\green{Nd}\blue{Zn}2\red{Al}3} & agm003420540 & $  84$ & $ 150$ & $ 16.9$ & $  0.52$ & $  2.2$ \\
\hline
\ch{\green{Dy}\blue{Rh}2\red{Al}3} & agm002195973 & $  55$ & $ 186$ & $ 23.6$ & $  0.49$ & $  2.2$ \\
\hline
\ch{\green{Dy}\blue{Ir}2\red{Al}3} & agm002195722 & $  78$ & $ 158$ & $ 21.6$ & $  0.51$ & $  2.1$ \\
\hline
\ch{\green{Tb}\blue{Ir}2\red{Al}3} & agm002196074 & $  76$ & $ 159$ & $ 21.6$ & $  0.51$ & $  2.1$ \\
\hline
\ch{\green{Ho}\blue{Ir}2\red{Al}3} & agm002195839 & $  81$ & $ 158$ & $ 21.6$ & $  0.51$ & $  2.1$ \\
\hline
\ch{\green{Sr}\blue{Li}2\red{Pb}3} & agm005853412 & $  34$ & $  65$ & $ 14.8$ & $  0.66$ & $  2.1$ \\
\hline
\ch{\green{K}\blue{Li}2\red{In}3} & agm071928819 & $  91$ & $ 115$ & $ 19.9$ & $  0.55$ & $  2.0$ \\
\hline
\ch{\green{Ho}\blue{Rh}2\red{Al}3} & agm002195976 & $  57$ & $ 188$ & $ 23.3$ & $  0.48$ & $  2.0$ \\
\hline
\ch{\green{Ca}\blue{Ru}2\red{Ga}3} & agm041449476 & $  32$ & $ 161$ & $ 32.3$ & $  0.50$ & $  2.0$ \\
\hline
\ch{\green{Sc}\blue{Cu}2\red{Be}3} & agm071792821 & $   3$ & $ 365$ & $ 17.9$ & $  0.42$ & $  2.0$ \\
\hline
\ch{\green{Lu}\blue{Au}2\red{Al}3} & agm002195856 & $  94$ & $ 102$ & $ 11.5$ & $  0.56$ & $  2.0$ \\
\hline
\ch{\green{Ti}\blue{Ni}2\red{Be}3} & agm071725295 & $  21$ & $ 335$ & $ 22.8$ & $  0.42$ & $  1.9$ \\
\hline
\ch{\green{Er}\blue{Ir}2\red{Al}3} & agm002195728 & $  84$ & $ 157$ & $ 21.6$ & $  0.50$ & $  1.9$ \\
\hline
\ch{\green{Lu}\blue{Ir}2\red{Al}3} & agm002195857 & $  84$ & $ 152$ & $ 21.4$ & $  0.50$ & $  1.9$ \\
\hline
\ch{\green{La}\blue{Ni}2\red{Al}3} & agm002133938 & $  36$ & $ 181$ & $ 24.8$ & $  0.48$ & $  1.9$ \\
\hline
\ch{\green{Pr}\blue{Zn}2\red{Al}3} & agm003420542 & $  80$ & $ 160$ & $ 16.9$ & $  0.49$ & $  1.9$ \\
\hline
\ch{\green{Y}\blue{Ir}2\red{Al}3} & agm002196107 & $  71$ & $ 165$ & $ 21.6$ & $  0.49$ & $  1.9$ \\
\hline
\ch{\green{Tm}\blue{Ir}2\red{Al}3} & agm002196090 & $  88$ & $ 156$ & $ 21.6$ & $  0.50$ & $  1.9$ \\
\hline
\ch{\green{Zr}\blue{Ni}2\red{Be}3} & agm072043899 & $   0$ & $ 334$ & $ 22.6$ & $  0.42$ & $  1.8$ \\
\hline
\ch{\green{Pr}\blue{Co}2\red{Ga}3} & agm002205430 & $  58$ & $ 173$ & $ 37.8$ & $  0.48$ & $  1.8$ \\
\hline
\ch{\green{Hf}\blue{Ni}2\red{Be}3} & agm071827366 & $   0$ & $ 321$ & $ 21.6$ & $  0.42$ & $  1.8$ \\
\hline
\ch{\green{Ba}\blue{Mg}2\red{Al}3} & agm043575535 & $  48$ & $ 195$ & $ 17.8$ & $  0.47$ & $  1.8$ \\
\hline
\ch{\green{Y}\blue{Rh}2\red{Al}3} & agm002196109 & $  47$ & $ 200$ & $ 23.8$ & $  0.46$ & $  1.7$ \\
\hline
\ch{\green{K}\blue{Zn}2\red{Al}3} & agm034193883 & $  59$ & $ 191$ & $ 13.2$ & $  0.46$ & $  1.7$ \\
\hline
\ch{\green{Ca}\blue{Au}2\red{Al}3} & agm032393163 & $  45$ & $ 159$ & $ 12.4$ & $  0.48$ & $  1.6$ \\
\hline
\ch{\green{Er}\blue{Rh}2\red{Al}3} & agm002195974 & $  59$ & $ 189$ & $ 23.0$ & $  0.46$ & $  1.6$ \\
\hline
\ch{\green{Tm}\blue{Rh}2\red{Al}3} & agm002195981 & $  62$ & $ 188$ & $ 22.7$ & $  0.46$ & $  1.6$ \\
\hline
\ch{\green{Tm}\blue{Tc}2\red{Al}3} & agm071773881 & $  83$ & $ 167$ & $ 29.2$ & $  0.47$ & $  1.6$ \\
\hline
\ch{\green{Er}\blue{Tc}2\red{Al}3} & agm072023859 & $  81$ & $ 168$ & $ 29.3$ & $  0.46$ & $  1.5$ \\
\hline
\ch{\green{Zr}\blue{Ni}2\red{Al}3} & agm003199116 & $  62$ & $ 162$ & $ 16.8$ & $  0.47$ & $  1.5$ \\
\hline
\ch{\green{In}\blue{Ta}2\red{Ga}3} & agm071596010 & $  64$ & $ 166$ & $ 18.1$ & $  0.46$ & $  1.5$ \\
\hline
\ch{\green{Ho}\blue{Tc}2\red{Al}3} & agm071792469 & $  80$ & $ 169$ & $ 29.4$ & $  0.46$ & $  1.5$ \\
\hline
\ch{\green{Sr}\blue{Mg}2\red{Al}3} & agm034675852 & $  54$ & $ 200$ & $ 17.2$ & $  0.45$ & $  1.5$ \\
\hline
\ch{\green{Ba}\blue{Li}2\red{In}3} & agm038771285 & $  21$ & $ 111$ & $ 17.1$ & $  0.51$ & $  1.5$ \\
\hline
\ch{\green{Rb}\blue{Li}2\red{Sn}3} & agm072039152 & $  97$ & $ 123$ & $ 16.1$ & $  0.50$ & $  1.5$ \\
\hline
\ch{\green{Tb}\blue{Co}2\red{Al}3} & agm002195638 & $  58$ & $ 191$ & $ 37.6$ & $  0.45$ & $  1.5$ \\
\hline
\ch{\green{Hf}\blue{Ni}2\red{Al}3} & agm033574488 & $  69$ & $ 145$ & $ 15.5$ & $  0.48$ & $  1.5$ \\
\hline
\ch{\green{La}\blue{Pd}2\red{Ga}3} & agm003202957 & $  22$ & $ 149$ & $ 17.0$ & $  0.47$ & $  1.4$ \\
\hline
\ch{\green{Sr}\blue{Zn}2\red{Al}3} & agm002248994 & $  40$ & $ 196$ & $ 16.3$ & $  0.44$ & $  1.4$ \\
\hline
\ch{\green{Tm}\blue{Au}2\red{Al}3} & agm002196089 & $  75$ & $ 115$ & $ 11.3$ & $  0.50$ & $  1.4$ \\
\hline
\ch{\green{Cs}\blue{Mg}2\red{Tl}3} & agm039487723 & $  38$ & $  73$ & $ 19.0$ & $  0.56$ & $  1.4$ \\
\hline
\ch{\green{Cs}\blue{Mg}2\red{In}3} & agm034268822 & $  41$ & $  99$ & $ 17.2$ & $  0.51$ & $  1.4$ \\
\hline
\ch{\green{Ca}\blue{Pd}2\red{Al}3} & agm002133833 & $  20$ & $ 198$ & $ 19.6$ & $  0.44$ & $  1.4$ \\
\hline
\ch{\green{Sc}\blue{Ni}2\red{Be}3} & agm072095816 & $   0$ & $ 346$ & $ 23.3$ & $  0.39$ & $  1.4$ \\
\hline
\ch{\green{Y}\blue{Tc}2\red{Al}3} & agm071966873 & $  78$ & $ 184$ & $ 29.9$ & $  0.44$ & $  1.4$ \\
\hline
\ch{\green{Pm}\blue{Co}2\red{Ga}3} & agm034920347 & $  46$ & $ 177$ & $ 37.8$ & $  0.44$ & $  1.3$ \\
\hline
\ch{\green{Dy}\blue{Co}2\red{Al}3} & agm002195633 & $  61$ & $ 193$ & $ 37.3$ & $  0.44$ & $  1.3$ \\
\hline
\ch{\green{La}\blue{Ag}2\red{Ga}3} & agm041761825 & $  59$ & $ 114$ & $ 13.9$ & $  0.49$ & $  1.3$ \\
\hline
\ch{\green{Sm}\blue{Tc}2\red{Al}3} & agm071655756 & $  93$ & $ 176$ & $ 29.7$ & $  0.44$ & $  1.3$ \\
\hline
\ch{\green{Pm}\blue{Tc}2\red{Al}3} & agm071704127 & $  95$ & $ 179$ & $ 29.7$ & $  0.44$ & $  1.3$ \\
\hline
\ch{\green{Rb}\blue{Mg}2\red{In}3} & agm035460572 & $  28$ & $  97$ & $ 16.2$ & $  0.50$ & $  1.3$ \\
\hline
\ch{\green{K}\blue{Zn}2\red{In}3} & agm039657883 & $  92$ & $ 106$ & $ 14.9$ & $  0.49$ & $  1.2$ \\
\hline
\ch{\green{Lu}\blue{Rh}2\red{Al}3} & agm002195977 & $  58$ & $ 189$ & $ 22.1$ & $  0.43$ & $  1.2$ \\
\hline
\ch{\green{In}\blue{Nb}2\red{Ga}3} & agm071672771 & $  88$ & $ 206$ & $ 19.1$ & $  0.43$ & $  1.2$ \\
\hline
\ch{\green{Nb}\blue{Ni}2\red{Be}3} & agm071623885 & $  44$ & $ 317$ & $ 18.0$ & $  0.39$ & $  1.2$ \\
\hline
\ch{\green{Ba}\blue{Cu}2\red{Ga}3} & agm002266652 & $  60$ & $ 151$ & $ 15.1$ & $  0.45$ & $  1.2$ \\
\hline
\ch{\green{Ho}\blue{Co}2\red{Al}3} & agm002195635 & $  64$ & $ 192$ & $ 37.0$ & $  0.43$ & $  1.2$ \\
\hline
\ch{\green{Sr}\blue{Li}2\red{In}3} & agm034085442 & $  41$ & $ 110$ & $ 15.8$ & $  0.48$ & $  1.2$ \\
\hline
\ch{\green{Ba}\blue{Li}2\red{Pb}3} & agm005704465 & $   0$ & $  81$ & $ 15.6$ & $  0.52$ & $  1.2$ \\
\hline
\ch{\green{Ba}\blue{Au}2\red{In}3} & agm033628198 & $  59$ & $  86$ & $ 15.6$ & $  0.51$ & $  1.2$ \\
\hline
\ch{\green{Er}\blue{Au}2\red{Al}3} & agm002195727 & $  67$ & $ 121$ & $ 11.3$ & $  0.47$ & $  1.2$ \\
\hline
\ch{\green{Pr}\blue{Ni}2\red{Al}3} & agm003195351 & $  27$ & $ 195$ & $ 23.3$ & $  0.42$ & $  1.1$ \\
\hline
\ch{\green{Er}\blue{Co}2\red{Al}3} & agm002195634 & $  67$ & $ 192$ & $ 36.6$ & $  0.42$ & $  1.1$ \\
\hline
\ch{\green{La}\blue{Zn}2\red{Al}3} & agm002196113 & $  49$ & $ 175$ & $ 17.7$ & $  0.43$ & $  1.1$ \\
\hline
\ch{\green{Zr}\blue{Cu}2\red{Be}3} & agm071593344 & $  16$ & $ 319$ & $ 14.4$ & $  0.38$ & $  1.1$ \\
\hline
\ch{\green{Tm}\blue{Pd}2\red{Al}3} & agm002133979 & $   0$ & $ 174$ & $ 16.0$ & $  0.43$ & $  1.1$ \\
\hline
\ch{\green{Pm}\blue{Li}2\red{Ga}3} & agm071788452 & $  97$ & $ 148$ & $ 13.8$ & $  0.44$ & $  1.1$ \\
\hline
\ch{\green{Pr}\blue{Ag}2\red{Ga}3} & agm034145266 & $  74$ & $ 119$ & $ 12.8$ & $  0.46$ & $  1.1$ \\
\hline
\ch{\green{Nd}\blue{Ni}2\red{Al}3} & agm003195352 & $  18$ & $ 195$ & $ 22.9$ & $  0.42$ & $  1.1$ \\
\hline
\ch{\green{Ba}\blue{Li}2\red{Al}3} & agm071675708 & $   8$ & $ 221$ & $ 16.8$ & $  0.41$ & $  1.1$ \\
\hline
\ch{\green{Hf}\blue{Cu}2\red{Be}3} & agm071848908 & $  28$ & $ 269$ & $ 13.5$ & $  0.39$ & $  1.0$ \\
\hline
\ch{\green{Tm}\blue{Co}2\red{Al}3} & agm002195640 & $  70$ & $ 191$ & $ 36.2$ & $  0.42$ & $  1.0$ \\
\hline
\ch{\green{Rb}\blue{Li}2\red{Bi}3} & agm071755728 & $  47$ & $  94$ & $ 13.7$ & $  0.48$ & $  1.0$ \\
\hline
\ch{\green{Y}\blue{Co}2\red{Al}3} & agm002195641 & $  61$ & $ 209$ & $ 37.5$ & $  0.41$ & $  1.0$ \\
\hline
\end{longtable}
\end{center}   

\end{document}